\documentclass[12pt]{iopart}

\usepackage{iopams}  
\usepackage{epsfig}
\usepackage{epsf}
\usepackage{graphicx}
\usepackage{rotating}
\usepackage{epic}
\usepackage{curves}
\usepackage{rotate}

\begin{document}

\newcommand{\BBbar}{\mbox{$B^0$-$\bar{B}^0$ }}
\newcommand{\PhiOne}{\mbox{$ \sin 2 \phi_1\; $}}
\newcommand{\PhiOneValue}{\mbox{$+0.45^{+0.43}_{-0.44}$(stat.)$^{+0.07}_{-0.09}$(syst.)}}

\begin{minipage}[t]{6in}
\begin{flushright}
CERN-TH/2001-034\\
hep-ph/0102159\\
\vspace*{1.0cm}
\end{flushright}
\end{minipage}

\title[CP Violation, UK Phenomenology Workshop 2000]{\begin{center}
{Present and Future CP Measurements}
\footnote{To Appear in Journal of Physics G; Contribution 
of Working Group 4 to the UK Phenomenology Workshop on Heavy Flavour and
CP Violation, Durham, 17 - 22 September 2000} 
\\
\vspace{1cm}
{\small UK Phenomenology Workshop}\\
{\small September 2000 Durham}
\vspace{1cm}
\end{center}}

\author{Tobias Hurth {\it CERN}, Choong Sun Kim {\it University
of Yonsei}, Claire Shepherd-Themistocleous {\it University of Cambridge},\\ 
Fergus Wilson {\it University of Bristol} (Convenors),
Farrukh Azfar {\it University of Oxford},   
Roger Barlow {\it University of Manchester},
Martin Beneke {\it University of Aachen}, Noel Cottingham 
{\it University of Bristol},  Glen Cowan {\it Royal Holloway, University
of London},\\  Amol Dighe
{\it  CERN},  Paolo Gambino 
{\it CERN},  Val Gibson 
{\it University of Cambridge}, 
              Yoshihito Iwasaki {\it KEK},
Shaaban Khalil 
              {\it                           University of Sussex},
Victoria Martin 
              {\it                           University of Edinburgh},\\
Matthew Martin 
{\it                                         University of Oxford},
Fabrizio Salvatore \\
{\it                         Royal Holloway, University of London},
James Weatherall 
{\it                                         University of Manchester}, 
Daniel Wyler 
{\it                                         University of Zurich}} 

{\large 
\begin{abstract}
We review theoretical and experimental results on CP violation
summarizing the discussions in the working group on CP violation 
at the UK phenomenology workshop 2000 in Durham.
\end{abstract}
}
\maketitle

\section{Introduction}

In the standard $SU(2)\times U(1)$ gauge theory of Glashow, Salam and
Weinberg predictions involving fermion masses and hadronic flavour
changing weak transitions require a {\it prior} knowledge of the mass
generation mechanism. The simplest
method of giving mass to the fermions in the theory makes use of
Yukawa interactions involving the doublet Higgs field. An as yet 
unconfirmed mechanism. These
interactions give rise to the Cabibbo-Kobayashi-Maskawa (CKM) matrix:
Quarks of different flavour are mixed in the charged weak currents by
means of an unitary matrix $V$.  However, both the electromagnetic
current and the weak neutral current remain flavour diagonal.  Second
order weak processes such as mixing and CP--violation are even less
secure theoretically, since they can be affected by both beyond the
Standard Model virtual contributions, as well as new physics direct
contributions.  Our present understanding of CP--violation is based on
the three-family Kobayashi-Maskawa model of quarks, some of whose
charged-current couplings have phases. Over the past decade, new data
have enabled considerable refinement of our knowledge of the
parameters of this matrix $V$. Recent data based on the analysis of
leptons with high center-of-mass momentum in B meson decays,
indicate that the $b \to u$ transition matrix element is nonzero.  The
complex phase of this matrix element is very important for the
successful description of CP--violation within the framework of the
CKM matrix.  Results of experiments searching for the difference
between CP violating decays of kaons to pairs of neutral and charged
pions have been presented by FNAL and CERN, which support our
understanding of CP violation through the CKM matrix.  The top quark
enters into several constraints on CKM parameters through loop
diagrams, so that such an analysis necessarily implies a favored range
of top quark masses.  Over the past decade or so, many methods have
been proposed for obtaining the three interior angles of the unitarity
triangle of the matrix $V$, $\alpha$, $\beta$ and $\gamma$. Presently
these CP phases are being measured in a variety of experiments at
B-factories, KEK-B, SLAC-B, HERA-B and will be measured at LHC-B and
B-TeV.  As always, the hope is that these measurements will reveal the
presence of physics beyond the Standard Model.

The small visible branching ratio of B decays to CP eigenstates, $\cal
O$(10$^{-5}$), requires a large number of B mesons to be produced in
order to study CP violation in enough detail. The two complementary
methods are $e^+e^-$ colliders tuned to the $\Upsilon$(4S) resonance
or high energy hadron machines where the $b\overline{b}$
cross--section is large.

Current $e^+e^-$ colliders have achieved peak luminosities of $\sim
3\times10^{33}$cm$^{-2}$s$^{-1}$ producing $b\overline{b}$ pairs at
the rate of $\sim 1$\ Hz, although raw data rates are considerably
higher. The B mesons are produced in a coherent state and it is
necessary to measure the time separation of both decay vertices to
measure CP asymmetry in $B^0\overline{B^0}$ mixing. To this end both,
both PEP-II and KEK use asymmetric beam energies to boost the distance
between the decay vertices. Yet with typical separations of only
$\sim250\ \mu$m, the detectors used to resolve the vertices must still
be as close as possible to the beam line to achieve suitable vertex
accuracies of $\sim 50$-$100\ \mu$m. Symmetric energy machines can use
the decays of charged B mesons for CP investigations. In all cases, the
known initial beam energies, even in the presence of initial state
radiation, is an important constraint, improving reconstructed B mass
resolutions by an order of magnitude. The B$_S$ system can be
investigated by moving to the $\Upsilon(5S)$ resonance, although with
a large reduction in cross--section.

The hadron colliders have the advantage of much higher production
rates, ${\cal O}$($10^4$) Hz at the Tevatron and 10 times greater
again at the LHC as well as producing both B and B$_S$ mesons without
then need for altering beam energies. As the centre of mass energy
increases, the ratio of the $b\overline{b}$ to inelastic cross-section
increases. The challenge is however to cope with the very high rates
of background events and the large numbers of tracks in all events.
These very high interaction rates require the use of sophisticated
triggers operating at high rates. The beam crossing rate at the LHC 
for example is 40MHz while B decays that are interesting occur at 
a rate of a few Hz. 

The accurate reconstruction of decay vertices and the ability to
cleanly identify hadrons are major design requirements of both current
and future detectors. Good vertex resolution requires the use of high
precision vertex detectors very close to the beam pipe while the high
particle fluxes place stringent requirements on the radiation hardness
of the devices used. The large number of B hadrons produced means that,
particularly at hadron machines, it becomes possible to make use of
B decays with small cross sections where good differentiation between
charged pions and kaons is required. High precision measurements in
these final states will be the forte of the future experiments.  This
has prompted the development of high precision Cherenkov counters. 
The use of final states containing neutral particles requires 
the use of finely segmented calorimeters and will be  particularly
difficult at hadron colliders given the high backgrounds that will 
be encountered.

A complete review on the present status of CP measurements and 
their future prospects is beyond the scope of the present report.
In this report we summarize the discussions in the working group 
on CP violation 
at the UK phenomenology workshop 2000 in Durham.
In the following we give a short outline of the various topics
which were addressed. In Section 2 we describe 
the measurements of $sin 2 \beta$\,  by the BaBar and the Belle 
experiment and discuss  some prospects for the 
future. We also summarize 
CP violation measurements from $B$ decays observed at CDF 
during the last Tevatron run (Run-I). Prospects for measuring 
CP violation in the 
period between March 2001 and March 
2003 (Run-II) at the upgraded CDF-II detector are also summarized.
It is expected that the current generation of experiments 
will make the first observation of CP violation in the B  system.
We also  concentrate on the anticipated performance of the 
next generation experiments,
namely LHCb at CERN and BTeV at Fermilab,
which will measure CP violating observables with 
extremely high precision, thereby thoroughly testing the Standard Model 
description of CP violation  and searching for new physics beyond.
In section 3 we focus on the influence of new physics on
CKM phenomenology and on CP-violating observables. 
We also review various  methods for determining the weak phase
angle $\gamma$.
Section 4 is devoted to the CP violation parameter  
$\epsilon'/\epsilon$ within and beyond the SM.
In Section 5  discussions on specific  $B_s-$decays and on 
the width difference in the $B_d$-$\bar{B}_d$ system are included.
In Section 6 we give an overview of the B-Physics trigger strategy 
for Run 2 at the Tevatron, with an emphasis on CP-Violation.
Also the LHCb trigger strategy is reviewed. In addition a method
to separate B events from continuum background in BaBar is presented.
In Section 7 the phenomenological impact of the  QCD-improved
factorization
approach is discussed while Section 8 deals with statistical issues
relevant to heavy flavour physics including confidence level and
the new technique due to Feldman an Cousins.


\section{What ought we to be measuring? }

\newcommand\stwob{\ensuremath{\sin 2\beta}}
\newcommand\ups{$\Upsilon$(4S)}
\def\babar{\mbox{\sl B\hspace{-0.4em} {\small\sl A}\hspace{-0.4em} 
\sl B\hspace{-0.4em} {\small\sl A\hspace{-0.1em}R}}}
\def\BzBzb {\ensuremath{B^0 \Bbar^0}}
\def\Bbar  {\ensuremath{{\kern 0.2em\overline{\kern -0.2em B}}}}
\def\Bzb   {\ensuremath{\Bbar^0}}
\def\sb{\ensuremath{\sin\! 2 \beta   }}
\def\CP                 {\ensuremath{C\!P}}
\def\KS    {\ensuremath{K^0_{\scriptscriptstyle S}}}
\def\Bz    {\ensuremath{B^0}}
\def\jpsi  {\ensuremath{{J\mskip -3mu/\mskip -2mu\psi\mskip 2mu}}}
\def\psip {\ensuremath{\psi^{\prime}}}

\subsection{\bf The \babar\,   measurement of \sb \, 
and its future prospects~}

{\it James Weatherall, Univ. Manchester (representing the \babar\ Collaboration)}

\subsubsection{\bf Introduction}
The \babar\ experiment consists of an asymmetric electron-positron collider
operating at the \ups\ resonance.  More details on the detector can be found
in~\cite{Verderi:2000pv}.  The aim is to overconstrain the
unitarity triangle by measuring its sides and angles.  The analysis reviewed
here measures \sb\ by studying time-dependent \CP\ violating asymmetries in
$\Bz \rightarrow \jpsi \KS$ and $\Bz \rightarrow \psip \KS$ decays.

\subsubsection{\bf Overview of the \sb\ analysis}
There are five main parts to measuring the \CP\ violating asymmetry:

\begin{itemize}
\item Selection of signal \CP\ events
\item Measurement of the distance ${\rm \Delta} z$ between the two \Bz\ decay 
vertices along the \ups\ boost axis
\item Determination of the flavour of the tag-side B
\item Measurement of dilution factors for the different tagging categories
\item Extraction of \sb\ via an unbinned maximum likelihood fit
\end{itemize}

\begin{figure}[htbp]
\begin{center}
\epsfig{file=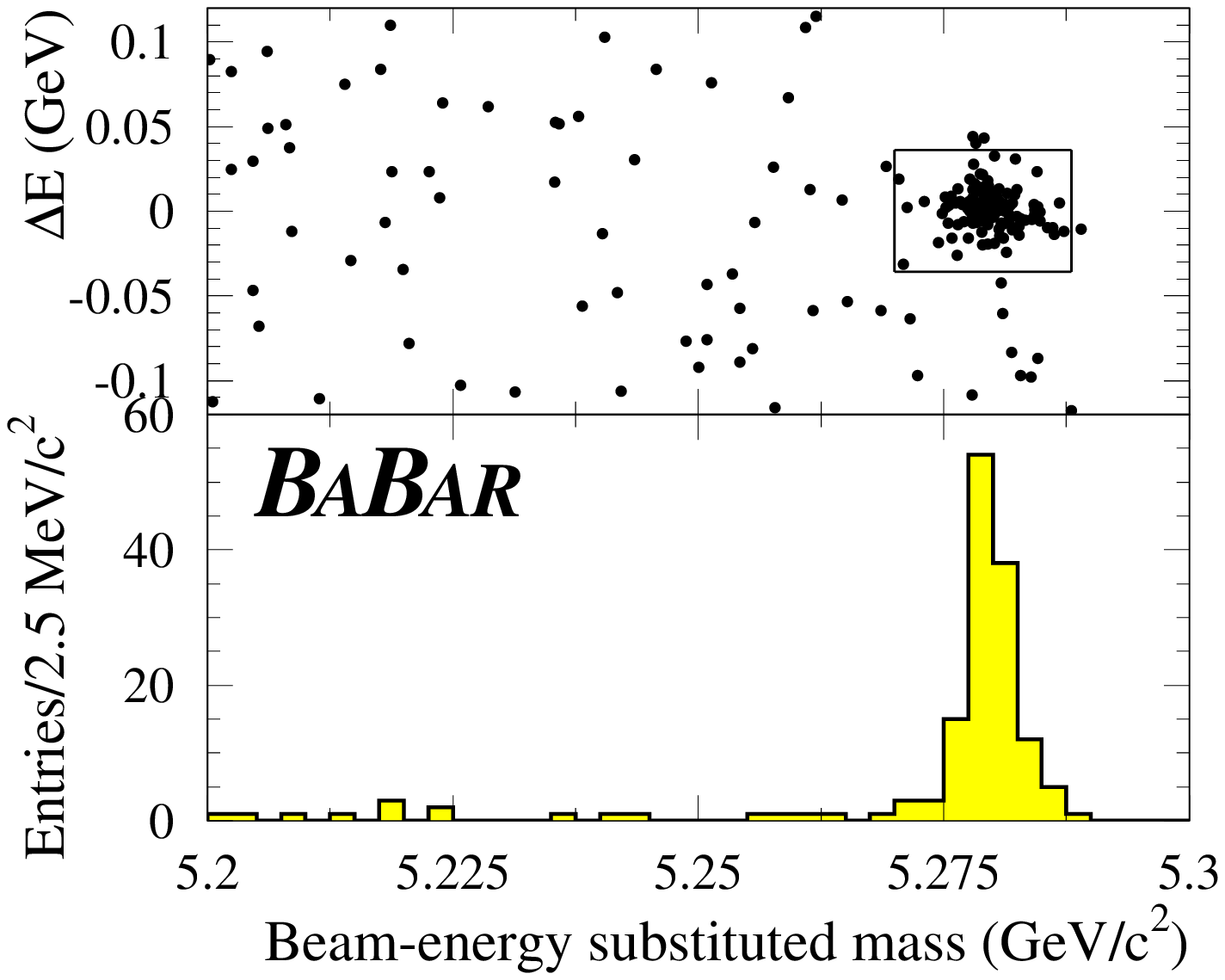,width=5cm}
\epsfig{file=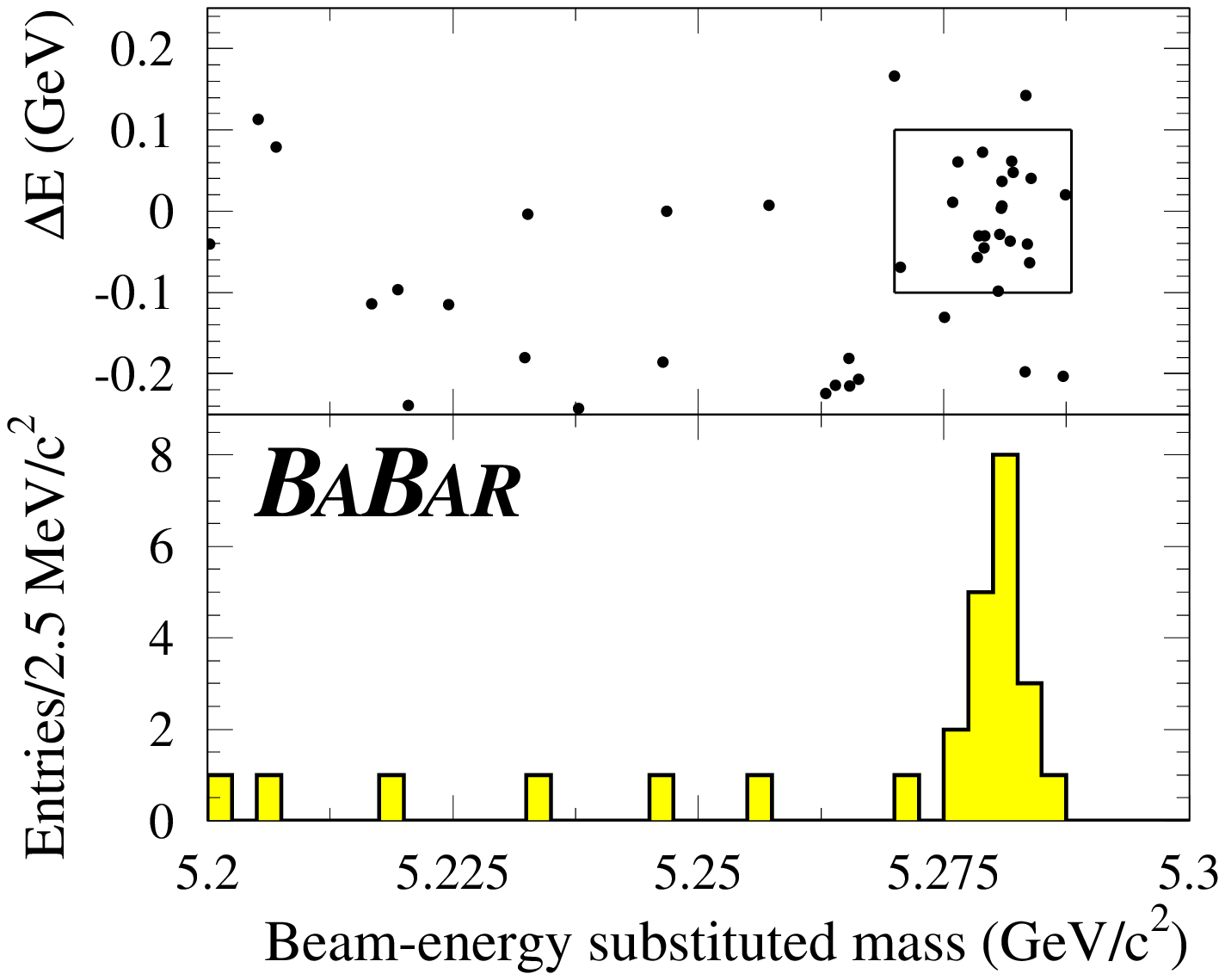,width=5cm}
\epsfig{file=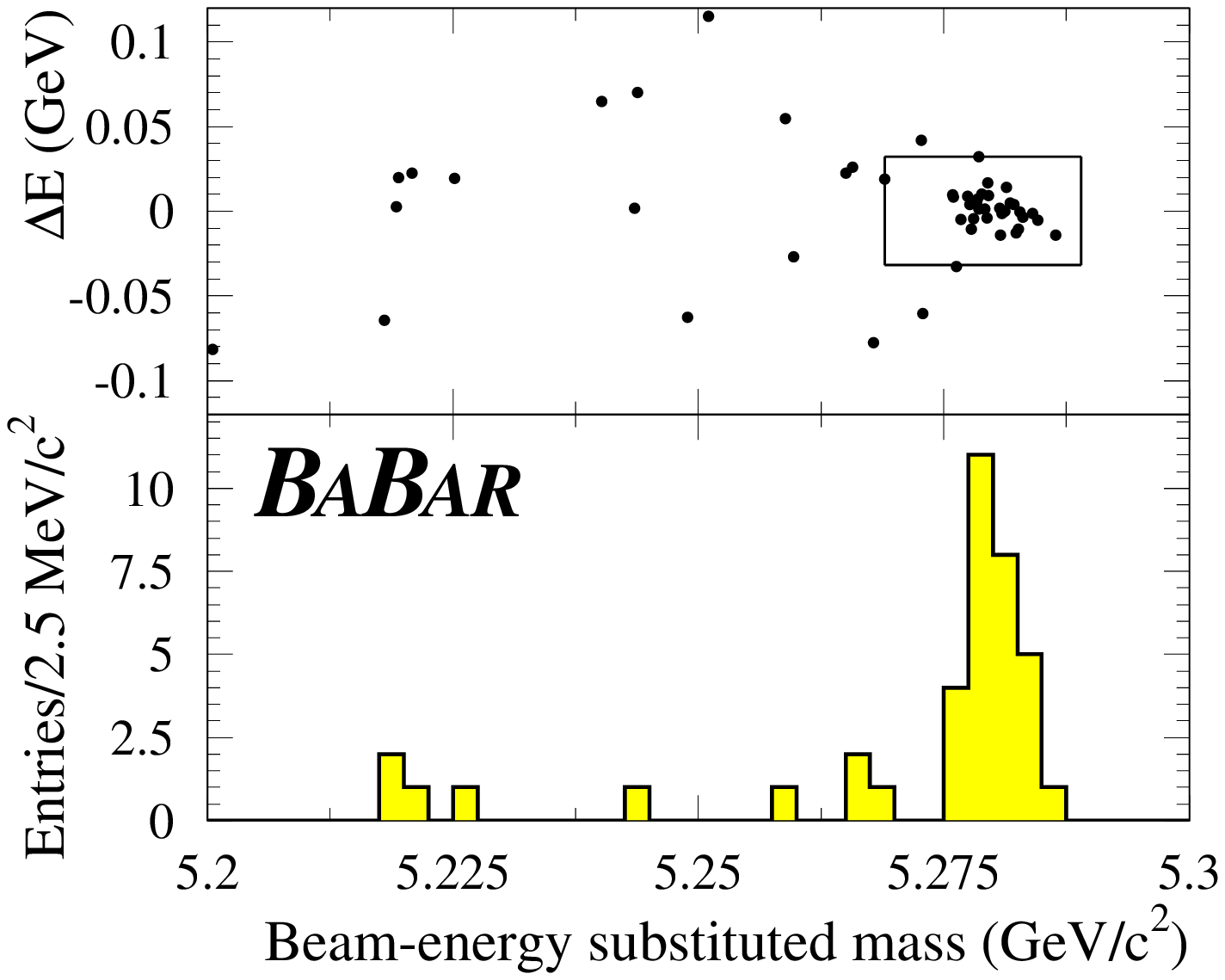,width=5cm}
\end{center}
\caption{\CP\ signal event distributions for $J/\psi K_S(\pi^+\pi^-)$ 
(left), $J/\psi K_S(\pi^0\pi^0)$ (middle) and $\psip\ K_S (\pi^+\pi^-)$ 
(right).}
\label{fig:sigsam}
\end{figure}

\paragraph{\bf Event Selection}
The sample used for the analysis is 9.8 fb$^{-1}$ of data recorded between 
January and July 2000 of
which 0.8 fb$^{-1}$ was recorded 40 MeV below the \ups\ resonance.  Particle
identification uses mainly the CsI calorimeter for electrons, the Instrumented
Flux Return for muons and the DIRC for kaons.  Extra information is provided 
by dE/dx measured in the tracking system.  
The selection for the \CP\ events proceeds as follows.
Pairs of electrons or muons coming
from a common vertex are combined to form \jpsi\ and \psip\ candidates.  
The \psip\ is also reconstructed from its decay into \jpsi\ $\pi^+\pi^-$.
The $K_S$ candidates are made from either a pair of charged tracks or a pair
of $\pi^0$ candidates.  In addition there are various event shape and 
topological cuts designed to reduce continuum and $B\overline{B}$ background.
Full details of the selection can be found in~\cite{Hitlin:2000tm}.
The final event sample is shown in figure~\ref{fig:sigsam}.

There are two other $B$ decay samples.  One consists of fully reconstructed
semileptonic ($B^0 \to D^{*-}l^+\nu_l$) and hadronic 
($B^0 \to D^{(*)-}\pi^+,D^{(*)-}\rho^+,D^{(*)-}a_1^+$) decays as well as a
control sample of $B^+ \to \overline{D}^{(*)0}\pi^+$ events.  The selection
of this sample is described in~\cite{Aubert:2000sz} and~\cite{Aubert:2000vt}.
The other is a charmonium control sample containing fully reconstructed 
neutral or charged $B$ candidates in two-body decay modes with a \jpsi\ in
the final state (e.g. $B^+ \to \jpsi\ K^+, B^0 \to \jpsi\ K^{*0}
(K^{*0} \to K^+\pi^-)$).

\paragraph{\bf Measuring ${\rm \Delta} z$}
The time-dependent decay rate for the $B_{\CP}$ is given by
\begin{equation}
\label{eq:TimeDep}
f_\pm(\, \Gamma, \, {\rm \Delta} m_d, \, {\cal {D}} \sin{ 2 \beta }, \, 
t \, )  = {\frac{1}{4}}\, \Gamma \, {\rm e}^{ - \Gamma \left| t \right| }
\, \left[  \, 1 \, \pm \, {\cal {D}} \sin{ 2 \beta } \times \sin{ {\rm \Delta}
m_d \, t } \,  \right]
\end{equation}
where the + or - sign indicates whether the $B_{tag}$ was tagged as a $B^0$
or $\Bzb$ respectively.  The dilution factor $\cal{D}$ is given by
${\cal D}=1-2w$, where $w$ is the mistag fraction (the probability that the 
$B_{tag}$ is identified incorrectly).  To account for finite detector 
resolution, the time distribution must be convoluted with a resolution 
function:

\begin{eqnarray}
{\cal {R}}( {\rm \Delta}z ; \hat {a} \,  ) &=& \sum^{2}_{i=1} \, \frac{f_i}
{\sigma_i\sqrt{2\pi}} \, {\rm exp} \left(  - ( {\rm \Delta}z-\delta_i)^2/2
{\sigma_i}^2   \right) \ \ \ , 
\end{eqnarray}

\noindent
which is just the sum of two Gaussians where the $f_i$, $\delta_i$ and 
$\sigma_i$ are the normalizations, biases and widths of the distributions.
In practice two scale factors ${\cal S}_1$ and ${\cal S}_2$ are introduced 
such that $\sigma_i={\cal S}_i\times\sigma_{\Delta t}$ where 
$\sigma_{\Delta t}$ is an event-by-event calculated error on $\Delta t$.
They take account of underestimating the uncertainty on $\Delta t$
due to effects such as hard scattering
and possible underestimation of the amount of material traversed by the 
particles.
The resolution function parameters are obtained from a maximum likelihood 
fit to the hadronic $B^0$ sample and are shown in table~\ref{tab:res}.  The 
$f_w$ parameter represents the width of a third Gaussian component, included
to accommodate a small ($\sim$1\%) 
fraction of events which have very large values 
of $\Delta z$, mostly caused by vertex reconstruction problems.  This 
Gaussian is unbiased with a fixed width of 8 ps.
Further details can be found in~\cite{Aubert:2000sz}.

\begin{table}
\vspace{0.3cm}
\begin{center}
\begin{tabular}{|cc|cc|} \hline
\multicolumn{2}{|c|}{Parameter} & \multicolumn{2}{c|}{Value} \\ \hline \hline
$\delta_1$   & (ps) & $-0.20\pm0.06$ & from fit \\ 
${\cal S}_1$ &      & $1.33\pm0.14$  & from fit \\ 
$f_w$        & (\%) & $1.6\pm0.6$ & from fit \\ 
$f_1$        & (\%) & $75$ & fixed \\ 
$\delta_2$   & (ps) & $0$ & fixed \\ 
${\cal S}_2$ &      & $2.1$ & fixed \\ 
\hline
\end{tabular}
\end{center}
\caption{Resolution function parameters.  Those, labeled 'from fit' are 
measured from data and those marked 'fixed' are determined from Monte Carlo.} 
\label{tab:res}
\end{table}

\paragraph{\bf B flavour tagging}
Each event with a \CP\ candidate is assigned a $B^0$ or $\Bzb$ tag 
if it satisfies the criteria for one of the several tagging categories.  The
figure of merit for each tagging category is the effective tagging efficiency
$Q_i = \epsilon_i(1-2w_i)^2$ where $\epsilon_i$ is the fraction of events 
assigned to category $i$ and $w_i$ is the probability of mis-tagging an event
in this category.  The statistical error on \sb\ is proportional to 
$1/\sqrt{Q}$ where $Q=\sum_iQ_i$.  
There are five tagging categories: {\tt Electron, Muon,
Kaon, NT1 and NT2}.  

The first three require the presence of a fast lepton 
and/or one or more charged kaons in the event and depend on the correlation 
between the charge of a primary lepton or kaon and the flavour of the $b$ 
quark.  If an event is not assigned to either the {\tt Electron} or {\tt Muon}
categories, it is assigned to the {\tt Kaon} category if the sum of the 
charges of all the identified kaons in the event is different from zero.  If
both lepton and kaon tags are available but inconsistent the event is rejected
from both categories.  

NT1 and NT2 are categories from a neural network algorithm, 
this approach being motivated by 
the potential flavour-tagging power carried by non-identified leptons and 
kaons, correlations between leptons and kaons and more generally the momentum
spectrum of charged particles from $B$ meson decays.  The output of the neural
network tagger $x_{NT}$ can be mapped onto the interval [-1,1] with 
$x_{NT}<0$ representing a $B^0$ tag and $x_{NT}>0$ a $\Bzb$ tag.  Events with
$|x_{NT}|>0.5$ are classified in the {\tt NT1} category and events with 
$0.2<|x_{NT}|<0.5$ in the {\tt NT2} category.  Events with $|x_{NT}|<0.2$ are
excluded from the final analysis sample.

\paragraph{\bf Measurement of tagging performance}
The effective tagging efficiencies and mistag fractions
for all the categories are measured from data using a maximum likelihood fit
to the time distributions of the $B^0$ hadronic event sample.  The procedure
uses events which have one $B$ fully reconstructed in a flavour eigenstate 
mode.  The tagging algorithms are then applied to the rest of the event, which
represents the potential $B_{tag}$.  Events are classified as {\em mixed} or
{\em unmixed} depending on whether the $B_{tag}$ is tagged with the same or
opposite flavour as the $B_{CP}$.  One can express the time-integrated 
fraction of mixed events $\chi$ as a function of the \BzBzb\ mixing 
probability, $\chi = \chi_d+(1-2\chi_d)w$
where $\chi_d=\frac{1}{2}x^2_d/(1+x^2_d)$, with $x_d=\Delta m_d/\Gamma$.  Thus
an experimental value of the mistag fraction $w$ can be deduced from the data.

\begin{figure}[htbp]
\begin{center}
\epsfig{file=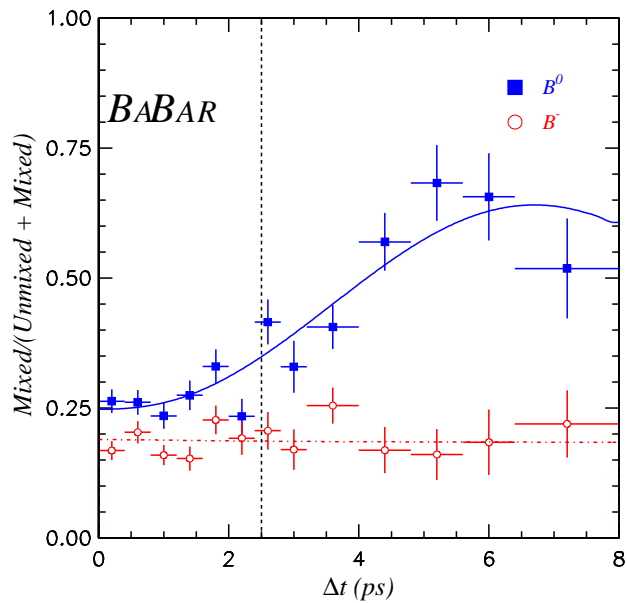,width=9cm}
\end{center}
\caption{The fraction of mixed events as a function of $|\Delta t|$ for
data events in the hadronic sample for neutral $B$ mesons (full squares) and
charged $B$ mesons (open circles).  The dot-dashed line at $t_{cut}=2.5$ ps
indicates the bin boundary for the time-integrated single-bin method.}
\label{fig:mix}
\end{figure}

A more accurate estimate of $w$ comes from a time-dependent analysis of the 
fraction of mixed events.  The mixing probability is smallest at low 
$\Delta t$ so that this region is governed by the mistag fraction.  
Figure~\ref{fig:mix} shows the fraction of mixed events versus $\Delta t$.
The resultant tagging performances are shown in table~\ref{tab:tag}.  

\begin{table}
\vspace{0.3cm}
\begin{center}
\begin{tabular}{|l|c|c|c|} \hline
Tagging category & $\epsilon$ (\%) & $w$ (\%) & $Q$ (\%) \\ \hline \hline
{\tt Lepton}  & $11.2\pm0.5$ & $9.6\pm1.7\pm1.3$  & $7.3\pm0.3$  \\
{\tt Kaon}    & $36.7\pm0.9$ & $19.7\pm1.3\pm1.1$ & $13.5\pm0.3$ \\
{\tt NT1}     & $11.7\pm0.5$ & $16.7\pm2.2\pm2.0$ & $5.2\pm0.2$  \\
{\tt NT2}     & $16.6\pm0.6$ & $33.1\pm2.1\pm2.1$ & $1.9\pm0.1$  \\ \hline
\hline
all           & $76.7\pm0.5$ &                    & $27.9\pm0.5$  \\
\hline
\end{tabular}
\end{center}
\caption{Tagging performance as measured from data.} 
\label{tab:tag}
\end{table}

\paragraph{\bf Extracting \sb}
A blind analysis technique was adopted for the extraction of \sb\ to 
eliminate possible experimenter bias.  The technique hides both the result of
the likelihood fit and the visual \CP\ asymmetry in the $\Delta t$
distribution.  This method allows systematic studies to be performed while 
keeping the numerical value of \sb\ hidden.

Possible systematic effects due to uncertainty in the input parameters to the
fit, incomplete knowledge of the time resolution function, uncertainties in the
mistag fractions and possible limitations in the analysis procedure were all
studied.  Details can be found in~\cite{Hitlin:2000tm}.  The systematic 
errors are summarized in table~\ref{tab:syst}.

\begin{table}
\vspace{0.3cm}
\begin{center}
\begin{tabular}{|l|c|} \hline 
Source of uncertainty & Uncertainty on \sb\ \\ \hline \hline
uncertainty on $\tau^0_B$ & 0.002 \\
uncertainty on $\Delta m_d$ & 0.015 \\
uncertainty on $\Delta z$ resolution for \CP\ sample & 0.019 \\
uncertainty on time-resolution bias for \CP\ sample & 0.047 \\
uncertainty on measurement of mistag fractions & 0.053 \\
different mistag fractions for \CP\ and non-\CP\ samples & 0.050 \\
different mistag fractions for \Bz\ and \Bzb\ & 0.005 \\
background in \CP\ sample & 0.015 \\ \hline \hline
total systematic error & {\bf 0.091} \\
\hline
\end{tabular}
\end{center}
\caption{Summary of systematic uncertainties.  The different contributions are
added in quadrature.}
\label{tab:syst}
\end{table}

\paragraph{\bf Results and checks}
The maximum likelihood fit for \sb, using the full tagged sample of 120
$\Bz \rightarrow \jpsi \KS$ and $\Bz \rightarrow \psip \KS$ events yields:

\begin{equation}
\sb\ = 0.12 \pm 0.37\, (stat) \pm 0.09\, (syst)\ (preliminary).
\end{equation}

\begin{figure}
\begin{center}
\epsfig{file=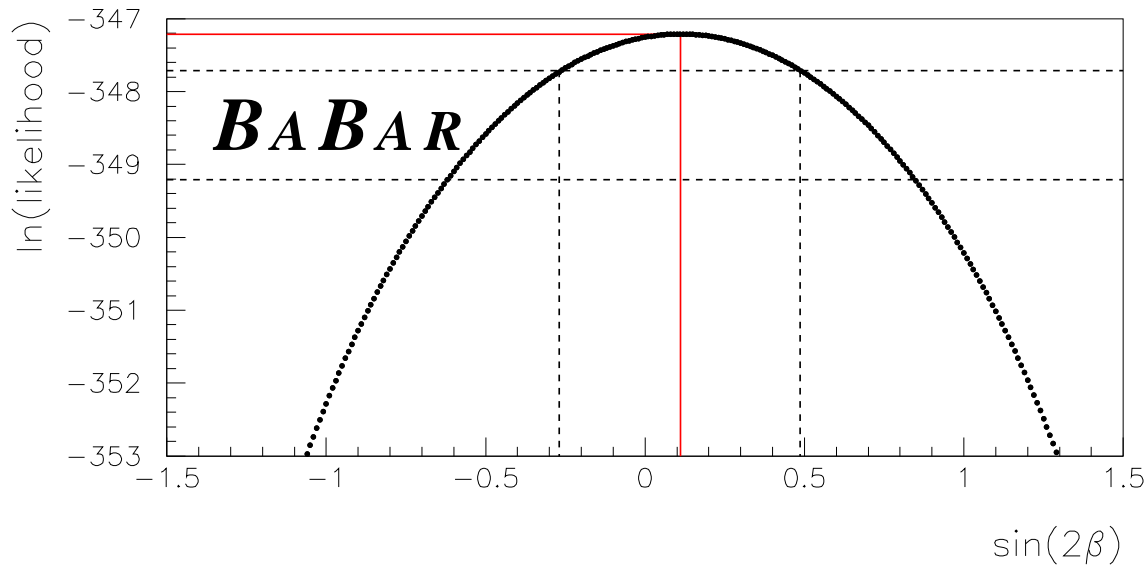,height=9cm}
\caption{Variation of the log likelihood as a function of \sb.  The 
two horizontal dashed lines indicate changes in the log-likelihood 
corresponding to one and two statistical standard deviations.}
\label{fig:like}
\end{center}
\end{figure}

\begin{figure}
\begin{center}
\epsfig{file=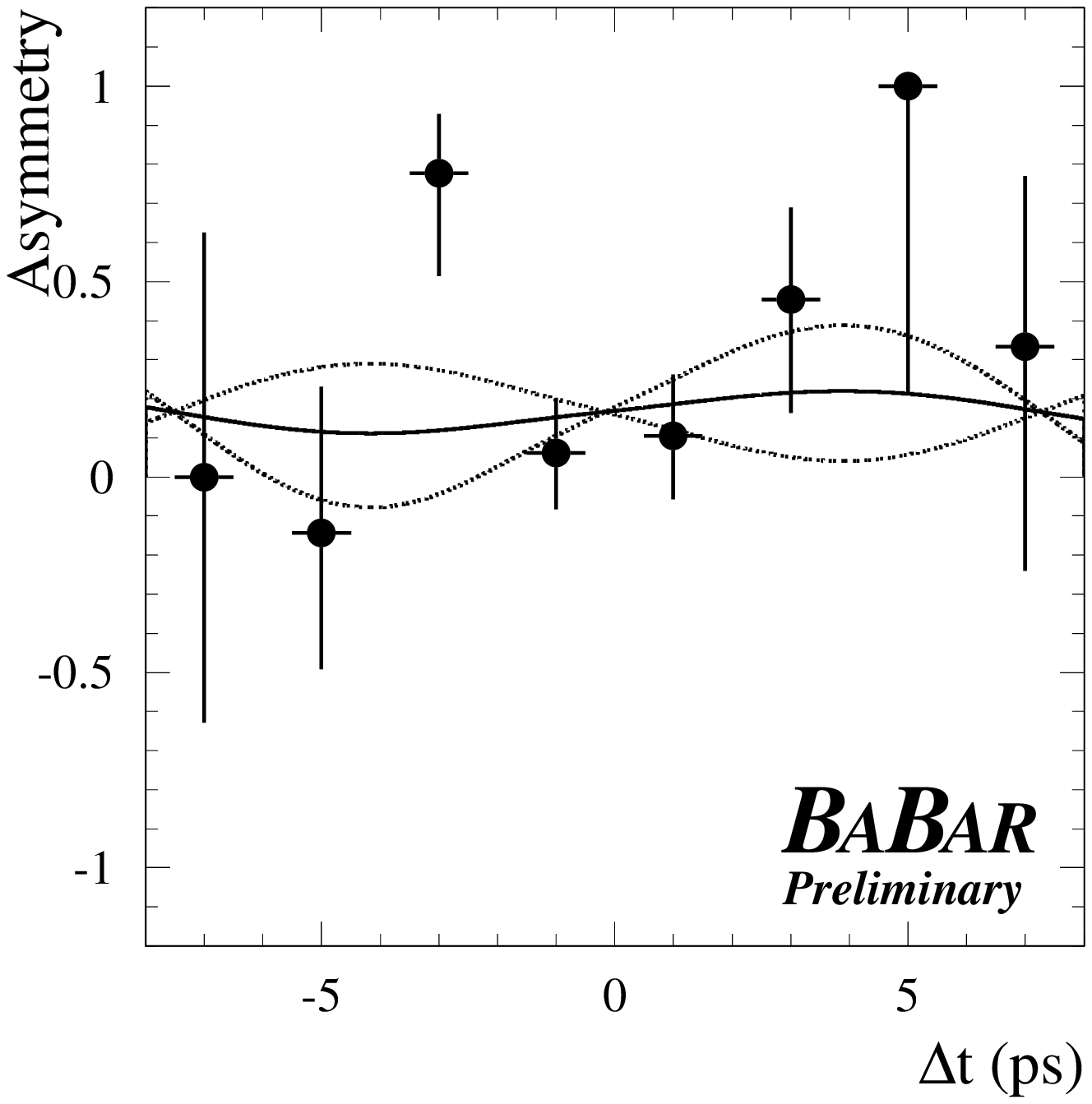,width=9cm}
\caption{The raw $B^0-\Bzb$ asymmetry $(N_{B^0}-N_{\Bzb})/
(N_{B^0}+N_{\Bzb})$.  The time-dependent asymmetry is represented by a solid
curve for the central value of \sb, and by two dotted curves for the values at
plus and minus one statistical standard deviation from the central value.  The
curves are not centered at (0,0) because the \CP\ sample contains an unequal 
number of \Bz\ and \Bzb\ events (70 \Bz\ versus 50 \Bzb).  The $\chi^2$ 
between the binned asymmetry and the result of the maximum likelihood fit is
9.2 for 7 degrees of freedom.}
\label{fig:raw}
\end{center}
\end{figure}

\noindent
The log likelihood is shown as a function of \sb\ in figure~\ref{fig:like}.
The raw asymmetry as a function of $\Delta t$ is shown in figure~\ref{fig:raw}

The probability of obtaining a statistical uncertainty of 0.37 is estimated
by generating a large number of toy Monte Carlo experiments with the same 
number of tagged \CP\ events as in the data sample.  The errors are
distributed around 0.32 with a standard deviation of 0.03, meaning that the
probability of obtaining a larger statistical error that the one observed is
5\%.  From a large number of full Monte Carlo simulated experiments, we 
estimate that the probability of finding a lower value of the likelihood than
the one observed is 20\%.

Several cross-checks are performed to validate the main analysis.  The 
charmonium and fully-reconstructed hadronic control samples are composed of 
events that should exhibit no time-dependent asymmetry.  These events are 
fitted in the same way as the signal \CP\ events to extract an ``apparent
\CP\ asymmetry''.  The results are shown in table~\ref{tab:chk}.  

\begin{table}
\vspace{0.3cm}
\begin{center}
\begin{tabular}{|l|c|} \hline 
Sample & Apparent \CP\ asymmetry \\ \hline \hline
Hadronic charged $B$ decays & $0.03\pm0.07$ \\ \hline
Hadronic neutral $B$ decays & $-0.01\pm0.08$ \\ \hline
$J/\psi K^+$ & $0.13\pm0.14$ \\ \hline
$J/\psi K^{*0} (K^{*0} \to K^+\pi^-)$ & $0.49\pm0.26$ \\ \hline
\end{tabular}
\end{center}
\caption{Summary of systematic uncertainties.  The different contributions are
added in quadrature.}
\label{tab:chk}
\end{table}    

\begin{figure}
\begin{center}
\epsfig{file=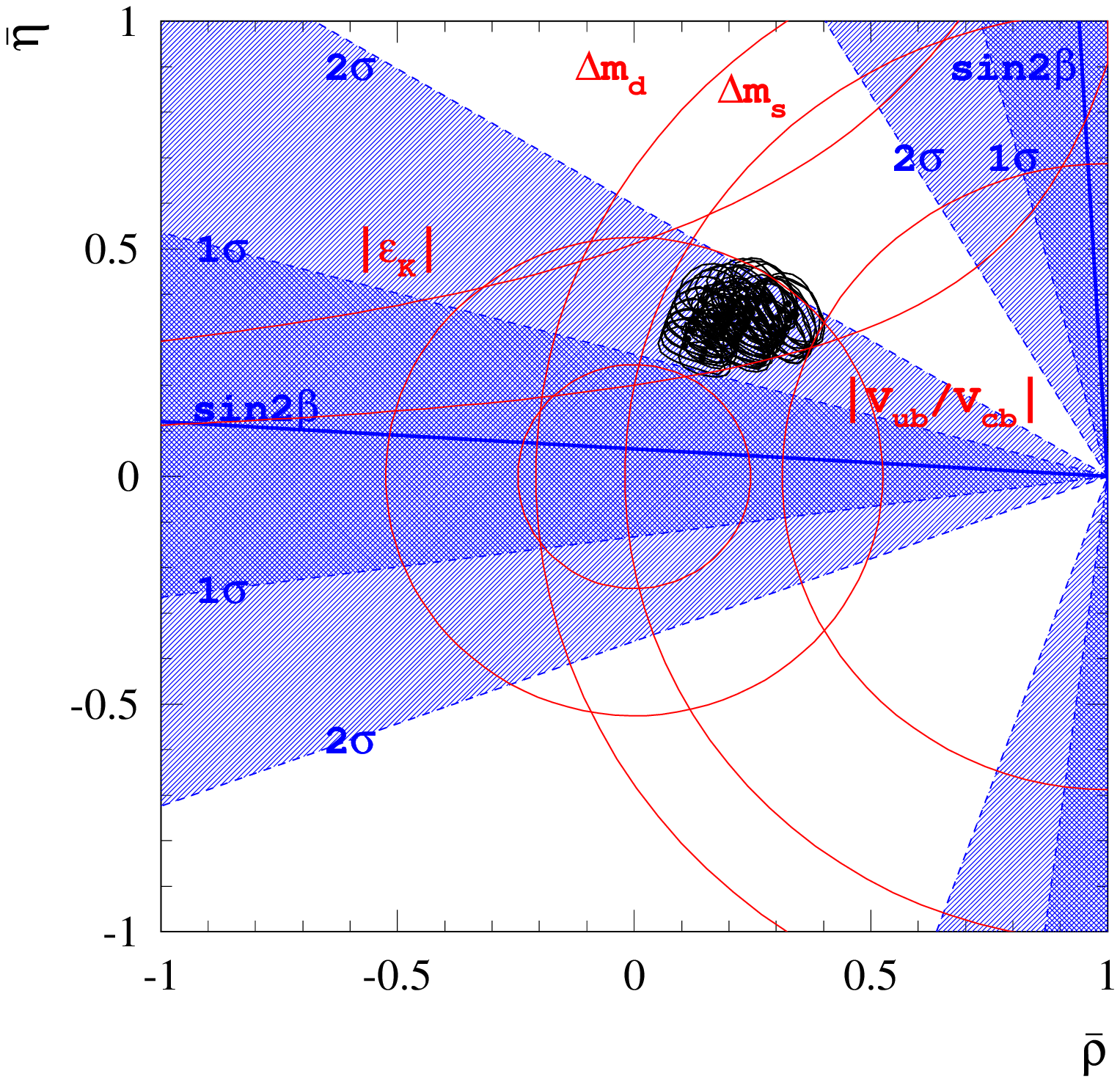,width=10cm}
\caption{Present constraints on the position of the apex of the Unitarity
Triangle with the \babar\ result indicated by the cross-hatched regions.}
\label{fig:ut}
\end{center}
\end{figure}

\paragraph{\bf Constraints on the unitarity triangle}
The Unitarity Triangle in the ($\overline{\rho},\overline{\eta}$) plane is 
shown in figure~\ref{fig:ut}.  The two solutions corresponding to the 
measured central value are shown as 
straight lines.  The cross-hatched regions correspond to one and two times 
the one-standard-deviation experimental uncertainty.  The ellipses represent
regions allowed by all other measurements that constrain the triangle.  They
are shown for a variety of choices of theoretical parameters.  More details
can be found in~\cite{Harrison:1998yr}.

\subsubsection{\bf Future prospects}
The preceding pages describe only a preliminary measurement of \sb\ by 
the \babar\ experiment.  More data will allow extra channels to be included in
the final fit as well as providing more events for the currently used decay 
modes.  The new channels will bring extra experimental and theoretical 
challenges with them.  Such present and future issues are discussed in the 
sections that follow.

\paragraph{\bf Available modes}
The B decay modes that have been used to measure \sb\ up to now are clean
in that they are vector-scalar, $b \to ccs$ transitions which have no 
significant pollution from penguin diagrams.  The next step is to add 
vector-vector modes such as $B^0 \to J/\psi K^*$.  These modes require an 
angular analysis of the
vector meson decay products, due to the different partial waves and 
therefore admixture of \CP\ odd and \CP\ even that is present in the final state.
Such an angular analysis has already yielded preliminary results for the 
$J/\psi K^*$ modes.  Once one has measured the polarizations in these modes, 
they are as clean, theoretically, as the vector-scalar modes.
Another obvious addition is $B^0 \to J/\psi K_L$ decays where the challenge
here is to understand the background well enough to make the channel feasible.
Work is ongoing in this area.

A different kind of difficulty is presented by channels with a significant
degree of penguin contamination, such as $b \to ccd$ scalar-scalar modes
(e.g. $B^0 \to D^+ D^-$).  Here the fit must take into account the fact that
the true value of \sb\ is shifted by an amount proportional to the ratio
of tree to penguin contributions.  This ratio is model dependent and subject
to large theoretical uncertainties.

Finally, modes such as $B^0 \to D^* D^*$ and $B \to J/\psi \rho^0$ which are
vector-vector, $b \to ccd$ transitions face the theoretical challenges of the
penguin contaminated modes described above, as well as requiring an angular
analysis to solve the vector-vector \CP\ admixture problem.

\paragraph{\bf Experimental considerations}
There are also experimental analysis issues which need to be resolved or 
studied in greater
depth in the future.  The tagging algorithms that \babar\ uses should be
developed and extended to include extra tagging categories such as the using
the soft pion from $D^*$ decays and incorporating leptons at an intermediate 
momentum (i.e. from a cascade).  It would be useful to take account of 
correlations within an event, such as when two different tagging categories 
report an answer.  This can give more information about the event if the 
correlations are well understood.  There is also an open question when it 
comes to measuring the tagging performance from the hadronic or semileptonic 
$B$ decay samples.  One then needs to be absolutely sure that using exactly
the same numbers for the \CP\ signal event sample is a valid thing to do.

The measurement of $\Delta z$ is another crucial part of the analysis and it is
important that the errors and biases to this distribution are understood.
The distribution tends to be biased by the decays of particles which fly 
significantly 
from the original $B$ decay vertex, such as $D^0$s.  These can be rejected
by looking explicitly for cascade decays.  The parameterization of the 
resolution function incorporates detector effects such as misalignments and 
electronics readout effects.  All contributions to the width should be 
studied in order to fully understand the error on $\Delta z$.

Backgrounds to the various \CP\ modes can also be a problem.  The channels vary
in terms of how much background they experience and this background can be
particularly dangerous if it has a significant structure in $\Delta z$.
For charmonium channels, much of the background comes from events containing
a real $J/\psi$.  In that case, one needs to study exactly which modes 
contribute and what their shape is in the final distributions (if they cannot
be removed otherwise).  Non-resonant backgrounds to vector-vector modes such
as the $J/\psi K^0 \pi^0$ contribution to $J/\psi K^{*0}(K^{*0} \to K^0\pi^0)$
are in principle dangerous since they can have \CP\
violating properties but no angular structure.  However, the branching ratios 
for these non-resonant modes are typically poorly known and consistent with 
zero making it difficult to simulate them in the correct proportions.

\paragraph{\bf Study of statistical error}
It seems anomalous that both \babar\ and Belle record higher statistical errors
than one would expect.  The fitting procedure is, and continues to be a
vigorously studied part of the analysis as we need to be certain that the
likelihood function is of exactly the correct form for the final fit.

\subsubsection{\bf Conclusions}
A preliminary measurement of \sb\ by \babar\ has been presented.  The errors
on the final result make it difficult to express its significance in terms
of constraints on the Unitarity Triangle.  However, results based on a much 
larger data sample ($\sim$20 fb$^{-1}$) will soon be available.  Combined
with a better understanding of systematic effects, this should make the next
measurement of \sb\ even more interesting than the current one.  It is also
expected that other \CP\ modes will soon be available for
analysis including $B^0 \to J/\psi K^{*0} (K^{*0} \to K_S \pi^0)$ and
$B^0 \to J/\psi K_L$.  The larger data sample with additional \CP\ modes 
should yield a value of \sb\ for which the statistical and systematic errors
are about one-half of their current values.

\subsection{\bf The first results from Belle}

{\it Yoshihito Iwasaki, KEK-IPNS (representing the Belle Collaboration)}

\subsubsection{\bf Introduction}

Observation of \CP\,  violation in the $B$ meson system is one of the most 
exciting physics targets at a $B$ factory experiment. In the standard model,
\CP \, violation is a natural consequence of the complex phase of quark mixing 
in the weak interaction as described by the Kobayashi-Maskawa matrix\cite{ckm}
. This phase can be detected in physics processes where amplitudes with 
different KM phase interfere.

In the decay of the neutral $B$ meson to a \CP \, eigenstate, at least two 
amplitudes ($A(B^0 \to f_{CP})$ and $A(B^0 \to \bar{B}^0 \to f_{CP})$) exist
due to \BBbar mixing. These amplitudes interfere. The time dependent \CP 
asymmetry in the decay rate of $B^0$ and $\bar{B}^0$, $a_{CP}(t)$, can be 
written as\cite{bigi_sanda}
\begin{eqnarray}
a_{CP}(t) & = &{{\Gamma (B^0(t)\to f_{CP})-\Gamma (\overline B^0(t)\to 
f_{CP})} \over {\Gamma (B^0(t)\to f_{CP})+\Gamma (\overline B^0(t)\to 
f_{CP})}} \\
 & = & \eta _{CP}\sin 2\phi _1\cdot \sin (\Delta m t),
\end{eqnarray}
where $t$ is the proper decay time of the $B$, $\eta_{CP}$ is the \CP \, 
eigenvalue of the final state $f_{CP}$, $\Delta m$ is the mass difference 
between the two $B^0$ mass eigenstates, and $\phi_1$ (also known as $\beta$)
is one of the three angles of the CKM unitarity triangle formed by the $d$ and
$b$ quark,
\begin{equation}
\phi_1 \equiv \pi - \arg(\frac{-V^*_{tb}V_{td}}{-V^*_{cb}V_{cd}}).
\end{equation}

In a $B$ factory experiment, a $B \bar {B}$ pair is created from $\Upsilon_
{4S}$ decay in a coherent quantum state. At the decay of one $B$, the other
$B$ oscillates starting with the opposite flavour of the $B$. Experimentally
we measure $\Delta t = t_{CP} - t_{tag}$ instead of $t$, where $t_{CP} $ is
the decay time of the neutral $B$ meson decaying to a \CP \, 
eigenstate($B_ {CP}
$), and $t_ {tag}$ is the decay time of the other $B$($B_{tag}$). The flavour
of $B_{tag} $ specifies the flavour of $B_{CP}$ at the start of the \BBbar 
mixing. To extract \PhiOne, we measure the proper time difference 
distributions instead of $a_{CP}(t)$ :
\begin{equation}
\frac{dN}{dt} = \frac{1}{2 \tau_{B^0}} \exp(- \frac{|\Delta t|}{\tau_{B^0}})(1 - 
\eta_{CP} \sin 2\phi_1 \sin(\Delta m_d \Delta t))
\end{equation}

\subsubsection{\bf KEKB accelerator and Belle detector}

KEKB is an asymmetric energy $e^+ e^-$ collider that produces a boosted 
$\Upsilon_{4S}$ in the laboratory frame. The beam energies of $e^+$ and $e^-$
are $3.5$ and $8.0$ GeV, respectively. The boost factor of the $\Upsilon_{4S}
$ is $0.425$. The beam size at the interaction point is $2 \mu m$ and $100 \mu
m$ in the vertical and horizontal direction, respectively.

The Belle detector is located at the interaction point of KEKB in the Tsukuba
experimental hall. Construction was started in 1994 and completed in 1998. 
Belle is a general-purpose detector with a Silicon Vertex Detector (SVD), a 
Central Drift Chamber (CDC), an Aerogel Cherenkov Counter (ACC), a Time Of 
Flight scintillation counter (TOF), an Electro-magnetic Calorimeter (ECL), a
solenoid magnet, and a $K_L$ muon catcher (KLM). Because the beam energies are
asymmetric, the detector shape is also asymmetric in the beam direction in 
order to cover a large solid angle in the $\Upsilon_{4S}$ rest frame.

Charged tracks are reconstructed from CDC and SVD hits as the  particle 
spirals in the solenoidal 1.5 Tesla magnetic field. The transverse momentum 
resolution is $\sigma_{p_t} / p_t = 0.0019 p_t \oplus 0.0034$. The impact 
parameter resolution in the plane perpendicular to the beam axis is $\sigma_
{r\phi} = 21 \oplus \frac{69}{p \beta \sin^{3/2} \theta}$ $\mu$m where $p$ is
the momentum in GeV and $\beta$ is the velocity. The impact parameter 
resolution along the beam direction is $\sigma_z = 39 \oplus \frac{51}{p\beta
\sin^{5/2} \theta}$ $\mu$m.

Photons are reconstructed from the energy deposited in the ECL with a 
resolution of $ \sigma_E / E = 0.013 \oplus 0.0007 / E \oplus 0.008 / E^{1/4}
$, where $E$ is the measured energy in GeV. Kaons are identified using 
probabilities calculated from $dE/dx$ measured by CDC, TOF, and hits in ACC.
The time resolution of the TOF is $95$ ps. The refractive index used by the 
ACC is chosen to provide good $\pi$/K separation for $1.5< p < 3.5$ GeV. The
efficiency is $\sim 80$\% and the fake rate is $\sim 10$\% for momentum up
to $3.5$ GeV. Electron identification is done using $dE/dx$, ACC hits, and 
energy deposited in the ECL. The efficiency is above 90\% for $p > 1.0$ GeV.
The fake rate is below $0.5$\%. Muon identification is done with hits in the
KLM. The efficiency is above $90$\% and the fake rate is below $2$\%.

\subsubsection{\bf Event selection of \CP \, eigenstates}

The \CP \, eigenstates we search for are charmonium plus $K^0$ as listed in 
table~\ref{table_decays}. All decay modes include $b \to c \bar{c} s$ 
transitions except for $B^0\to J / \psi (l^+l^-)+\pi ^0$, which is a $b \to
c \bar{c} d$ transition.

Candidate $J/\psi$ mesons are formed from pairs of oppositely charged tracks
where at least one track is positively identified as a lepton ($\mu$ or $e$)
and the other is consistent with a lepton. For the $J/\psi \to e^+e^-$ channel
we also include $\gamma$'s within $0.05$ radians of the electron direction to
recover events with initial state radiation from the $e$. We require the 
invariant mass of a lepton pair to be $ - 0.006$ GeV $< M_{\mu^+ \mu^-} - M_
{J/\psi} < 0.036 $ GeV and $- 0.150 $ GeV $< M_{e^+e^-} - M_{J/\psi} < 0.036$
GeV for $\mu^+\mu^-$ and $e^+e^-$ pairs, respectively. For $\psi ' \to l^+ 
l^-$ we require the invariant mass of the lepton pair to be $- 60$ MeV $< M_
{\mu^+ \mu^-} - M_{\psi '} < 36$ MeV and $- 150$ MeV $< M_{e^+e^-} - M_ {\psi
'} < 36$ MeV for $\mu^+\mu^-$ and $e^+e^-$ pairs, respectively. For $\psi ' 
\to J/\psi \pi^+ \pi^-$ we combine a $J/\psi$ candidate and a $\pi^+\pi^-$, 
where the invariant mass of the pion pair is required to be greater than $400$
MeV. We require the mass difference of $\psi '(l^+l^-\pi^+\pi^-)$ and $J/\psi$
to be between $0.58$ GeV and $0.60$ GeV. For $\chi_{c1} \to J/\psi \gamma$, 
we combine a $\gamma$ and $J/\psi$ where the $\gamma$ is not consistent with
forming $\pi^0$ when combined with any other $\gamma$.

For $K_S \to \pi^+ \pi^-$ we select a pair of oppositely charged tracks where
the closest distance of two tracks in the $z$ coordinate is consistent with 
zero. The invariant mass of the candidate is required to be within $\pm 3 
\sigma$ ($\pm 16$ MeV) of the $K_S$ mass peak. For $K_S \to \pi^0 \pi^0$ we 
use pairs of $\pi^0$'s where $\pi^0$ is reconstructed from two $\gamma$'s. We
require the invariant mass of $K_S$ and $\pi^0$ to be within the range $0.3$
GeV to $1.0$ GeV and $118$ MeV to $150$ MeV, respectively.

For $J/\psi \pi^0$ we select a high momentum $\pi^0$ reconstructed by two 
$\gamma$'s where the energy of each $\gamma$ is required to be greater than 
$100$ MeV. The invariant mass requirement is identical to the $\pi^0$ 
requirements in $K_S$ reconstruction.

\begin{table}
\label{table_decays}
\lineup
\begin{tabular}{@{}*{6}{l}}
\br
Decay mode & \CP & $\Delta E$ (MeV) & $N$ & $N_{bg}$ & $N_{tagged}$ \\
\mr
$B^0\to J / \psi (l^+l^-)+K_S(\pi ^+\pi ^-)$ & $-1$ & $\pm 40$ & 70 & 3.4 & 40\\

$B^0\to J / \psi (l^+l^-)+K_S(\pi ^0\pi ^0)$ & $-1$ & $\pm 100$ & 4 & 0.3 & 4\\

$B^0\to \psi '(l^+l^-)+K_S(\pi ^+\pi ^-)$     & $-1$ & $\pm 40$ & 5 & 0.2 & 2\\

$B^0\to \psi '(J / \psi(l^+l^-) \pi ^+\pi ^-)+K_S(\pi ^+\pi ^-)$ & $-1$ & $\pm 40$ & 
8 & 0.6 & 3 \\

$B^0\to \chi _{C1}(J / \psi(l^+l^-) \gamma )+K_S(\pi ^+\pi ^-)$ & $-1$ & 
$^{+30}_{-40}$ & 5 & 0.75 & 3 \\

$B^0\to J / \psi (l^+l^-)+K_L$ & $+1$ & - & 102 & 56 & 42 \\

$B^0\to J / \psi (l^+l^-)+\pi ^0$ & $+1$ & $^{+50}_{-100}$ & 10 & 1 & 4 \\

\br
\end{tabular}
\caption{The decay modes of $B^0$ going to \CP \, eigenstate. $l$ represents $e$ or 
$\mu$.}
\end{table}

\begin{figure}
\begin{center}
\epsfxsize=14cm
\epsfbox{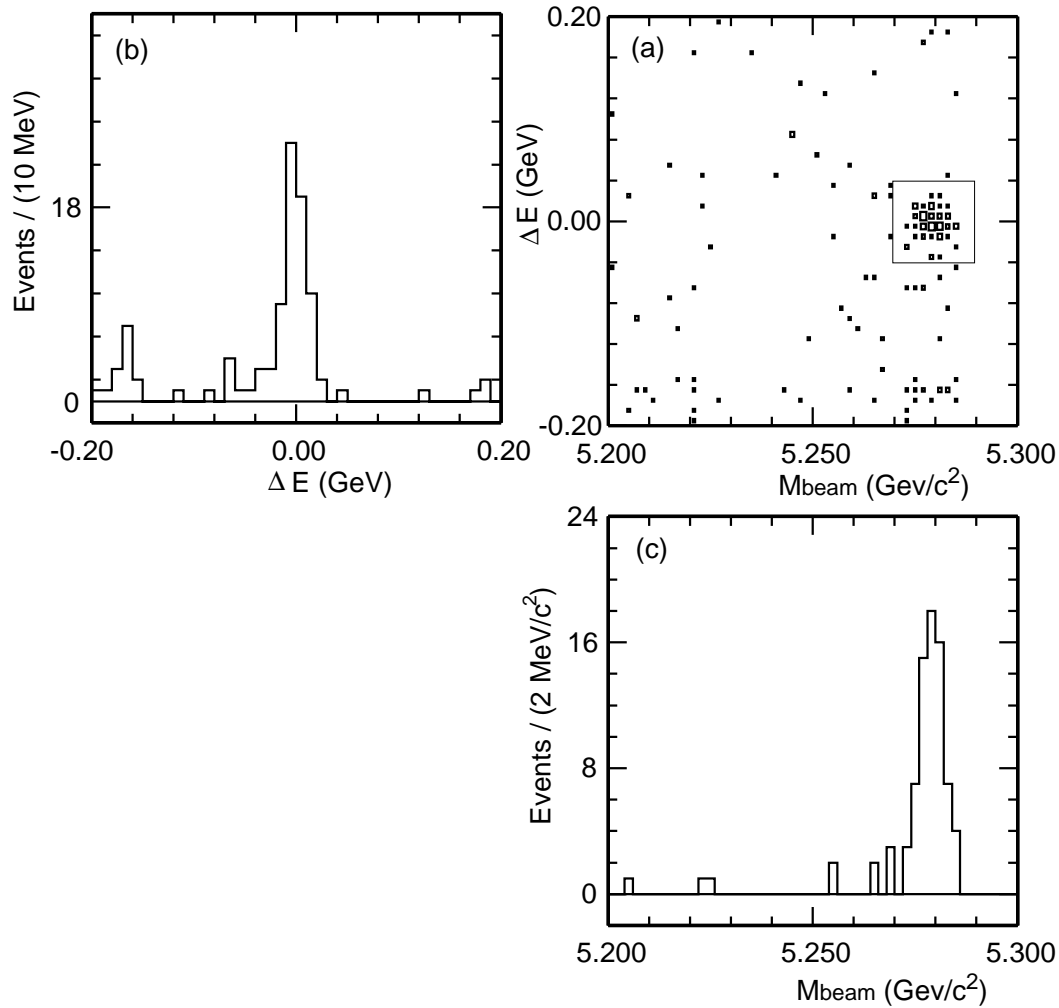}
\end{center}
\caption{(a) The scatter plot of $\Delta E$ versus $M_{beam}$ for $B^0 \to 
J/\psi K_S$ candidates. (b) The projection in $\Delta E$ with the cut $|M_
{beam} - M_{B^0}|<0.01$ GeV. (c) The projection in $M_{beam}$ with the cut $|\Delta
E| < 0.04$.}
\label{fig_psiks_candidates}
\end{figure}

For selection of $B$ candidates in all modes except for $B \to J/\psi K_L$, 
we use the energy difference $\Delta E \equiv E_B - E_{cm} / 2$ and the 
beam-constrained mass $M_{beam} \equiv \sqrt{(E_{cm} / 2)^2 - P_B^2}$. In 
figure \ref {fig_psiks_candidates}, the scatter plot of $\Delta E$ versus $M_
{beam}$ is shown for $B \to J/\psi K_S(\pi^+\pi^-)$. We define the signal 
region to be $|\Delta E| < 40$ MeV and $|M_{beam} - \left\langle M_{beam} 
\right\rangle | < 10$ MeV where $\left\langle M_{beam} \right\rangle$ is the
mean of the observed $M_{beam}$. The signal region in $\Delta E$ is varied 
depending on the decay mode (see table \ref{table_decays}). However, the 
signal region in $M_{beam}$ is the same as that for $B \to J/\psi K_S$ because
the error on $M_{beam}$ is dominated by the beam energy spread.

\begin{figure}
\begin{center}
\epsfxsize=8cm
\epsfbox{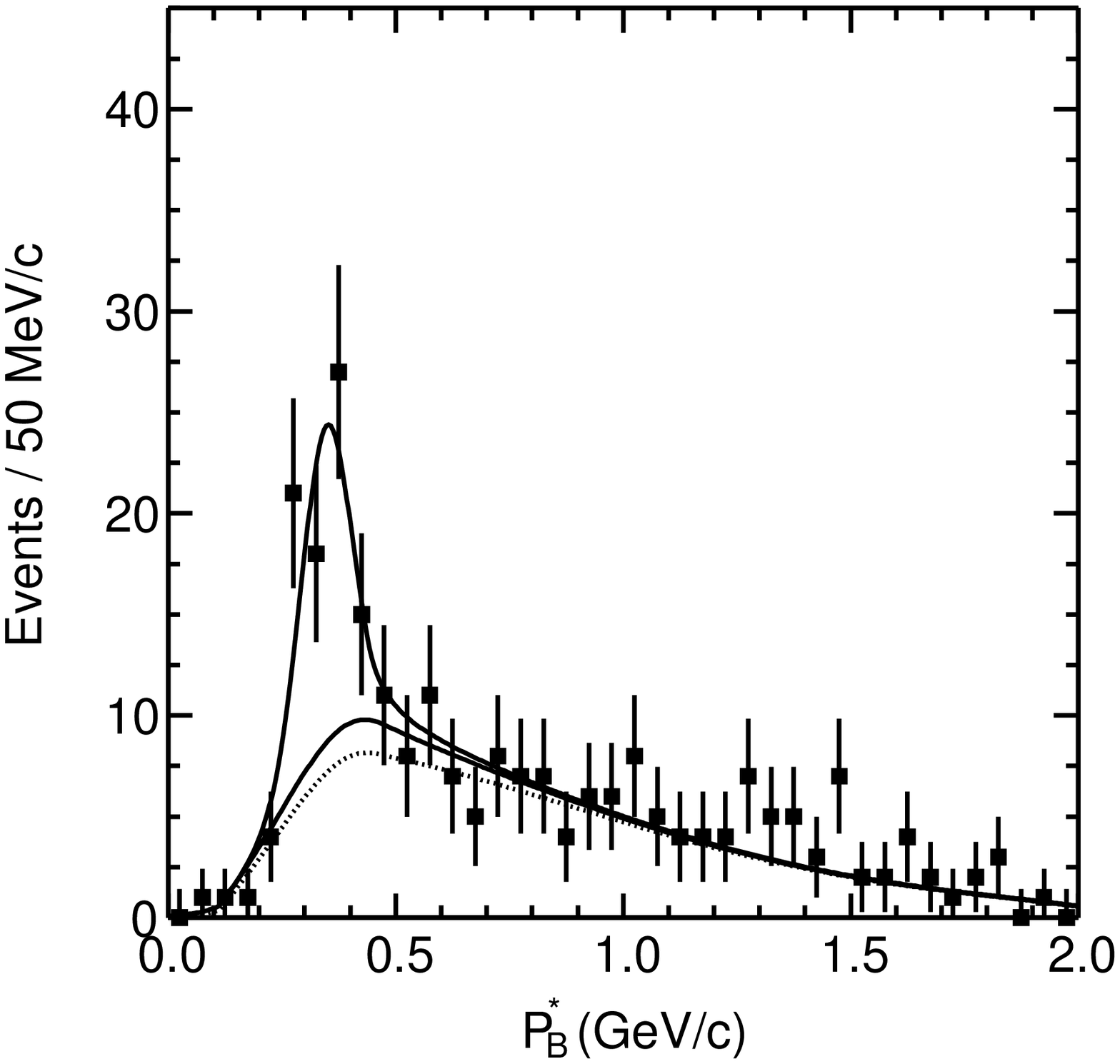}
\end{center}
\caption{The $P^*_B$ distribution for $B \to J/\psi K_L$ candidates. The upper
solid line is the fit of a sum of the signal and background expected from MC
simulation studies. The lower solid line is the total background. The dotted
line is the background component from sources other than $B \to J/\psi K_L 
\pi^0$.}
\label{fig_psikl_candidates}
\end{figure}

For $B \to J/\psi K_L$ we use tighter cuts for $J/\psi$ reconstruction by 
requiring positive identification for both leptons and the momentum of the 
$J/\psi$($P^*_{J/\psi}$) in the CMS frame to be $1.42 < P^*_{J/\psi} < 2.00$
GeV. We reconstruct the momentum of the $K_L$ from the momentum of the 
$J/\psi$ with the assumption of a two body decay $B \to J/\psi K_L$. We also
require associated KLM hits in the direction of the $K_L$ momentum. To select
signal events we require the momentum of the $B$ in the CMS, $P^*_B$, to be 
in the range $200$ MeV $\le P_B \le 450$ MeV. A true candidate should peak 
around $340$ MeV corresponding to the initial momentum of the $B$ from the 
$\Upsilon_{4S}$. In figure \ref {fig_psikl_candidates}, the $P^*_B$ 
distribution for $J/\psi K_L$ candidates is shown with the expectation 
obtained by a full MC simulation study. In the signal region, we have 102 
$J/\psi K_L$ candidates where we expect $8$ background events from $B \to 
J/\psi K_L \pi^0$ (a mixture of $\CP+1$ and $\CP-1$) and $48$ background 
events from other sources.

\subsubsection{\bf Flavour tagging}

To determine the flavour ($B^0$ or $\bar{B}^0$) of the $B_{CP}$ candidates, 
we partially reconstruct the other $B_{tag}$ in the event. The flavour of the
$B_{CP}$ should be opposite to that of the $B_ {tag}$ at the decay time of the
$B_{tag}$($\Delta t = 0$). We apply four methods sequentially. \begin
{enumerate}

\item{the charge of a high momentum lepton ($P^*_l > 1.1$ GeV)} : This tags the flavour 
via its primary leptonic decay. We assign $B_{CP} 
= B^0(\bar{B}^0)$ if the charge is positive(negative).

\item{sum of the charge for positively identified kaons} : This relies on $s$ 
quark pop-up in cascade decays. We assign $B_{CP} = B^0(\bar{B}^0)$ if the
sum of the charge is positive(negative).

\item{the charge of a medium momentum lepton} : This is similar to method (1)
but for the case where there is large missing momentum in the CMS($P^*_{miss}
$). We require the lepton momentum to be in the range $0.6 < P^*_l < 1.1$ GeV
and the missing momentum in the CMS should satisfy $P^*_l + P^*_{miss} \ge 
2.0$ GeV. The flavour assignment is identical to method (1).

\item{the charge of slow pion} : This tags the flavour of $D^*$ coming from 
$B_{tag}$. We require the momentum of the slow $\pi$ to be less than $200$ 
MeV. We assign $B_{CP} = B^0(\bar{B}^0)$ if the pion charge is negative
(positive).

\end{enumerate}

The flavour tagging efficiency, $\epsilon$, and the wrong tag fraction, 
$\omega$, are measured from data with self-tagging decay modes. We 
exclusively reconstruct $B^0 \to D^{(*)-} l^+ \nu$, and apply the flavour 
tagging methods for the rest of the event. Because of \BBbar mixing, the 
probability to find the opposite or same flavour for the exclusively reconstructed
$B$ and the result of the flavour tagging is
\begin{eqnarray}
P_{opposite}(\Delta t) & \propto & 1 + (1 - 2 \omega) \cos (\Delta m_d \Delta t) \\
P_{same}(\Delta t)     & \propto & 1 - (1 - 2 \omega) \cos (\Delta m_d \Delta t).
\end{eqnarray}
Then, $\omega$ can be extracted from the amplitude of the \BBbar mixing
\begin{equation}
A \equiv \frac{P_{opposite} - P_{same}}{P_{opposite} + P_{same}} = (1 - 2 \omega) 
\cos (\Delta m_d \Delta t).
\end{equation}

The vertex position of the $D^{(*)-} l^+ \nu$ is determined by requiring the
$l$ and $D^{(*)}$ to form a common vertex. The determination of the vertex 
position of the $B_{tag}$ is described in the next section. $\Delta t$ is 
calculated from the difference of the two vertices in the $z$ direction. We 
perform an unbinned maximum likelihood fit to the amplitude of the \BBbar 
mixing to obtain $\omega$ with $m_d$ allowed to be free. In table \ref
{table_flavour}, $\epsilon$ and $\omega$ are summarized. We obtain $\Delta m_d
= 0.488 \pm 0.026$ ps$^{-1}$, which is consistent with the world average\cite
{pdg5}. In table \ref{table_decays}, the number of tagged events for \CP 
\, eigenstates are listed. We find 98 tagged events in total.

\begin{table}
\label{table_flavour}
\lineup
\begin{tabular}{@{}*{3}{l}}
\br
Method & $\epsilon$ & $\omega$ \\
\mr
High momentum lepton & $0.014 \pm 0.021$ & $0.071 \pm 0.045$ \\
Kaon                 & $0.279 \pm 0.042$ & $0.199 \pm 0.070$ \\
Medium momentum lepton & $0.029 \pm 0.015$ & $0.29 \pm 0.15$ \\
Slow pion              & $0.070 \pm 0.035$ & $0.34 \pm 0.15$ \\
\br
\end{tabular}
\caption{The flavour tagging efficiency($\epsilon$) and the wrong tag fraction
($\omega$)}
\end{table}

\subsubsection{\bf Proper-time difference}

The proper-time difference is estimated from the difference of $z$ coordinate 
of the vertices of $B_{CP}$ and $B_{tag}$ in good approximation,
\begin{equation}
\Delta t = \frac{\Delta z}{\gamma \beta c}
\end{equation}
where $\gamma \beta$(=0.425) is the boost factor of $\Upsilon_{4S}$. The $B_
{CP}$ vertex is determined from the two lepton tracks in the $J/\psi$ decay.
The $B_ {tag}$ vertex is determined by the tracks used for the flavour tagging
after poorly measured tracks are removed. The expected vertex resolutions are
$\sim 40$ $\mu$m and $\sim 85$ $\mu$m for the $B_{CP}$ and $B_{tag}$ vertices,
respectively.

The resolution of $\Delta t$, $R(\Delta t)$, is parametrized by two Gaussian
distributions, where the first Gaussian is for the intrinsic vertex resolution
and the effect of the secondary charmed mesons in the $B_{tag}$ side, and the 
second Gaussian accounts for the tail due to poorly measured tracks:
\begin{equation}
R(\Delta t) = \frac{f}{\sigma \sqrt{2 \pi}} \exp{(- \frac{(\Delta t - \mu)^2}
{2 \sigma^2} )} + \frac{f_{tail}}{\sigma_{tail} \sqrt{2 \pi}} \exp{(- \frac
{(\Delta t - \mu_{tail})^2}{2 \sigma^2_{tail}})}.
\end{equation}
where $f_{tail} = 1 - f$. The means ($\mu$ and $\mu_{tail}$) and widths 
($\sigma$ and $\sigma_{tail}$) of the two Gaussians are calculated 
event-by-event from the errors on the two vertices. The fraction of the first
Gaussian, $f$, is $0.96 \pm 0.04$, determined from full MC simulation studies
and $B \to D^* l \nu$ data.

\subsubsection{\bf Extraction of \PhiOne}

The probability density function with a \CP\, eigenvalue of $\eta_f$ is
\begin{equation}
S(\Delta t, \eta_f, q) \equiv \frac{1}{\tau_{B^0}} \exp(- \Delta t / \tau_
{B^0}) \times \{1 - q(1 - 2 \omega) \eta_f \sin 2 \phi_1 \sin(\Delta m_d 
\Delta t)\}
\end{equation}
where $q = 1$$(-1)$ if $B_{tag} = B^0(\bar{B}^0)$. The wrong tag fraction, 
$\omega$, is a function of the tagging method as listed in table \ref 
{table_flavour}. We fix $\tau_{B^0}$ and $\Delta m_d$ to the world 
averages\cite{pdg5}, $1.548 \pm 0.032$ and $0.472 \pm 0.017$ ps$^{-1}$, 
respectively. The probability density function for backgrounds is $B(\Delta 
t) = \frac{1}{\tau_{bkg}} exp(- \Delta t / \tau_{bkg})$ where the lifetime of
the backgrounds, $0.73 \pm 0.12$ ps, is obtained from the side bands of the 
signal in $\Delta E$ and $M_{beam}$.

To extract \PhiOne, we define the likelihood of an event:
\begin{equation}
p = f_s \int^{+\infty}_{-\infty} R(\Delta t - s) S(s, \eta_f, q) ds
+ (1 - f_s) \int^{+\infty}_{-\infty} R(\Delta t - s) B(s) ds
\end{equation} 
where $f_s$ is the signal fraction. The extraction of \PhiOne is done by 
minimizing the log-likelihood, $- \sum_{i} \ln p_i$, as a function of \PhiOne.
The results are summarized in table \ref{table_cp}.

To verify our analysis, we analyze control data sample which should not have
any $CP$\,  asymmetry. The control data samples are $B^0 \to J/\psi K^{*0} (K^
{*0} \to K^+ \pi^-)$, $B^- \to J/\psi K^-$, $B^- \to D^0 \pi^-$, and $B^0 \to
D^{*-} l^+ \nu$. All results obtained from the control samples are consistent
with zero asymmetry.

In table \ref{table_syst}, the systematic errors on \PhiOne are listed. The 
largest error is due to $\omega$, which is obtained from data. The total 
systematic error on \PhiOne is the quadratic sum of all sources, $+0.07$ for 
positive side and $-0.09$ for negative side.

\begin{figure}
\begin{center}
\epsfxsize=16cm
\epsfbox[-39 160 1117 695]{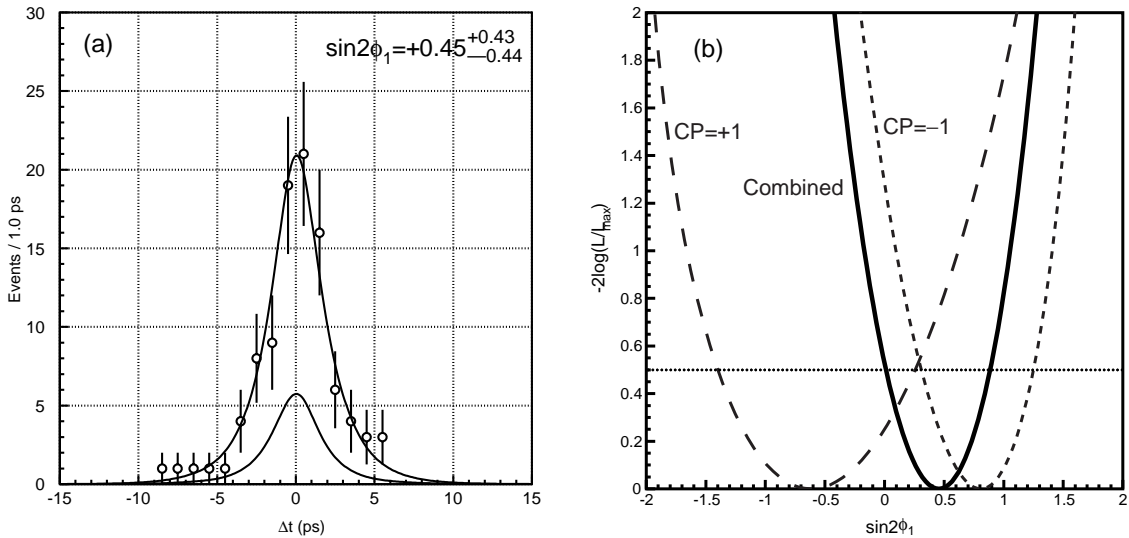}
\end{center}
\caption{(a) The fitted proper-time difference distribution for all candidates
: $\sum_{\eta_f=-1,+1}(dN/d(\eta_f \Delta t)|_{B^0} + dN/d(- \eta_f \Delta t)
|_{\bar{B}^0})$. (b) The log-likelihoods for $CP + 1$, $CP - 1$, and both $CP
\pm 1$.}
\label{f_cp_fit}
\end{figure}

\begin{table}
\label{table_cp}
\lineup
\begin{tabular}{@{}*{2}{l}}
\br
Decay mode & \PhiOne \\
\mr
All $CP -1$ modes & $+0.81^{+0.44}_{-0.50}$ \\
All $CP +1$ modes & $-0.61^{+0.87}_{-0.78}$ \\
All $CP \pm 1$ modes & $+0.45^{+0.43}_{-0.44}$ \\
\br
\end{tabular}
\caption{Results of the \CP \, fit}
\end{table}

\begin{table}
\label{table_syst}
\lineup
\begin{tabular}{@{}*{3}{l}}
\br
Source & $+ \sigma$ & $- \sigma$ \\
\mr
Wrong tag fraction($\omega$) & $+0.050$ & $-0.066$ \\
Resolution function & $+0.026$ & $-0.025$ \\
Background shape & $+0.029$ & $-0.042$ \\
Background fraction & $+0.029$ & $-0.032$ \\
$\tau_{B^0}$, $\Delta m_d$ & $+0.005$ & $-0.006$ \\
IP profile & $+0.004$ & $-0.000$ \\
\mr
Total & $+0.07$ & $-0.09$ \\
\br
\end{tabular}
\caption{Summary of the systematic errors on \PhiOne}
\end{table}

\subsubsection{\bf Conclusion}

We collected $6.2 fb^{-1}$ of data on the $\Upsilon_{4S}$ by the end
of the summer 2000. Using this data sample, we made a preliminary 
measurement of \PhiOne using $B \to J/\psi K_S (\pi^+ \pi^-)$, $B \to J/\psi
K_S (\pi^0 \pi^0)$, $B \to \psi' K_S(\pi^+ \pi^-)$, $B \to \chi_{C1} K_S 
(\pi^+ \pi^-)$, $B \to J\psi \pi^0$, and $B \to J/\psi K_L$. We found \PhiOne
$=$ \PhiOneValue. Our result is consistent with the standard model prediction.
We expect to improve the statistical errors as more data become available.

\subsubsection*{\bf References}


\subsection{\bf CP Violation in $B$-Decays at CDF: Results and Prospects.}

{\it Farrukh Azfar, Oxford University}


\subsubsection{\bf Introduction}

Charge-Parity (CP) violation in particle decays is necessary to explain the 
matter-anti-matter asymmetry observed in the universe today.
The amount of Standard Model (SM) CP violation is too small to account for the 
observed asymmetry. A detailed study of CP violation provides us an excellent opportunity to 
search for physics beyond the Standard Model.

\paragraph{\bf The CKM matrix}
CP violation in the Standard model has its origins in the complex couplings of the 
Cabibbo-Kobayashi-Maskawa quark mixing matrix. The interference of various decay amplitudes
are expected to give rise to large CP violating effects in the $B$ system.

The CKM matrix can be expressed in terms of 3 real parameters ($\rho$, $\lambda$, $A$) and one imaginary 
($\eta$) parameter in a representation known as the Wolfenstein parameterization. This matrix is unitary and 
several orthogonality relations between its rows and columns can be derived. We represent one of these
$V_{ud}V^*_{ub}+V_{cd}V^*_{cb}+V_{td}V^*_{tb}=0$, as a triangle in the complex plane. 

\begin{figure}
\begin{center}
\epsfig{file=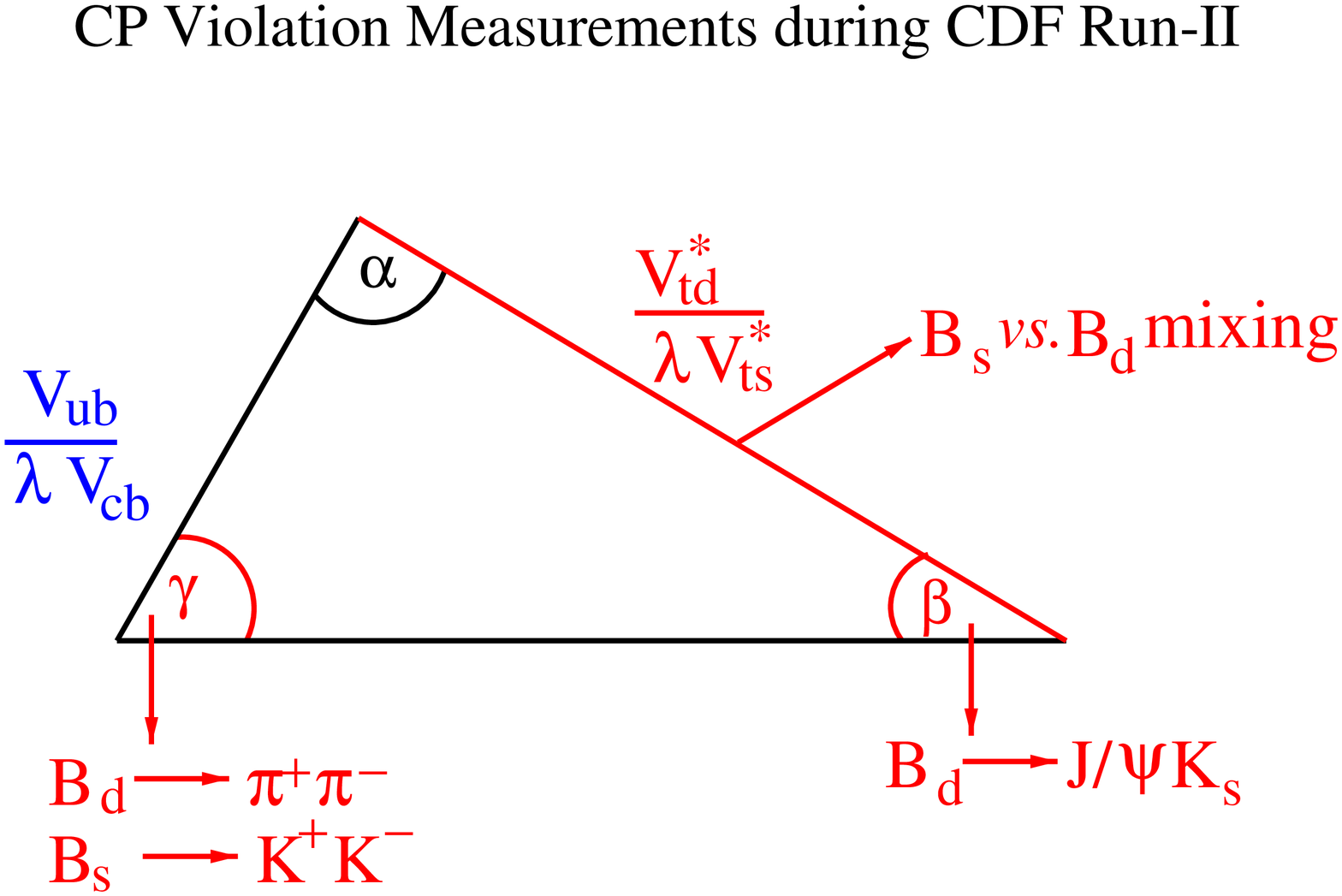,width=0.45\textwidth,angle=0}
\end{center}
\caption{CDF plans for measuring the unitarity triangle (red).}
\label{tri}
\end{figure}

At CDF we intend to measure one side and two angles of the unitarity triangle, by measuring the time 
dependent CP asymmetry in the modes $B_d^0 \to \pi^+ \pi^-$, $B_s^0 \to K^+ K^-$, $B_d^0 \to J/\psi K_S^0$ 
and the $B_s^0 - \bar{B}_s^0$ mixing parameter $x_s$ as illustrated in Fig.~\ref{tri}. A measurement of 
the time dependent CP asymmetry utilizing $B_d^0 \to J/\psi K_S^0$ decays with 100 pb$^{-1}$ of data 
collected during Run-I has already been done at CDF~\cite{sin2b}.

The time dependent CP asymmetry will also be measured in modes where the SM asymmetry is expected to be 
small {\it e.g.} $B_s^0 \to J/\psi \phi$ and $B_s^0 \to D_s^+ D_s^-$. A measurement of the width difference 
between the weak eigenstates of the $B_s^0$ is complimentary to such measurements and would utilize the same 
two modes. Comparisons of the width difference to the mixing parameter are also sensitive to non-SM Physics.

\paragraph{\bf Mixing and CP violation in the neutral $B$ meson system}

In $p-\bar{p}$ collisions $B$ and $\bar{B}$ mesons are created as strong interaction eigenstates which then
``mix'' into each other due to second order weak interactions represented by the box-diagram. 
The heavy and light weak interaction eigenstates are very nearly CP eigenstates 
each with its distinct mass and width, the quantities used to describe
this two state system are: $M_H$, $M_L$, $\Gamma_H$, $\Gamma_L$, $\Delta M = M_H-M_L$, 
$\Delta \Gamma = \Gamma_H-\Gamma_L$, $\Gamma = \frac{\Gamma_H+\Gamma_L}{2}$, 
$\tau_{H,L}=\frac{1}{\Gamma_{H,L}}$ and $x_{d,s} = \frac{\Delta M_{d,s}} {\Gamma_{B_d^0,B_s^0}}$.

If $B$ and $\bar{B}$ can decay to a CP eigenstate $\mid f >$, CP violation can occur if 
there is more than one amplitude contributing to the decay. If the complex CKM phases 
in both amplitudes are different then they will interfere causing an asymmetry in the rates $B \to \mid f >$
{\it vs.} $\bar{B} \to \mid f >$. The interference is caused by second order weak-interactions that are 
represented by the box or penguin diagrams.

Depending on whether we know (tag) the initial and final flavours of the $B$ and if the decay is
to a CP eigenstate or a flavour specific final state, we will get a particular time evolution, these are 
summarized below:
\begin{itemize}
\item [1.] The flavour of the $B$ is unknown and the decay is to a CP eigenstate:
          $P(t) = \frac{e^{\frac{-t}{\tau_{L,H}}}}{\tau_{L,H}}$. Examples are $B_s^0 \to D_s^+ D_s^-$,
$B_d^0 \to \pi^+ \pi^-$ (untagged).
\item [2.] Flavour at birth is known, but not at decay, and the final state is a CP eigenstate:
$P(t) \approx \frac{e^{(-\frac{t}{\tau_{L,H}})}}{\tau}(1 \pm A_{CP}^{Mix}\sin \Delta M t \pm A_{CP}^{Dir}\cos \Delta M t)$ Examples: $B_d \to J/\psi K_S^0$, $A_{CP}^{Mix} = \sin 2\beta$, where $A_{CP}^{Dir}=0$, and $B_d^0 \to \pi^+ \pi^-$ $A_{CP}^{Mix} \neq 0$ and $A_{CP}^{Dir} \neq 0$, (tagged).
\item  [3.] Flavour at birth and at decay is known (mixing):      
$P(t) \approx \frac{e^{(-\frac{t}{\tau})}}{\tau}(1 \pm \cos (\Delta M t))$ Examples $B_s^0 \to D_s^{\pm} l^{\mp} \nu(\bar{\nu})$ $B_s^0 \to D_s^{\pm} \pi^{\mp}$, $B_d^0 \to D^{\pm} l^{\mp} \nu(\bar{\nu})$.
\end{itemize}
Experimentally the path length before decay of the $B$ meson, $ct$ is measured.

\subsubsection{\bf The CDF detector, run-I and the run-II upgrade.}

In this section, we briefly describe components of the CDF detector relevant to $B$-physics. 
A partially instrumented detector with several upgraded components, was tested with $p-\bar{p}$
collisions at a centre-of-mass energy of 2 TeV in October 2000. We expect the upgrade described to be 
completed by March 2001~\cite{tdr}. The detectors present in Run-I are described along with their Run-II 
successors.

\paragraph{\bf Tracking in the central region}

Tracking in the Central Region is provided by wire drift chambers. In Run-I the Central Tracking Chamber (CTC)
 was used, this had $\approx$ 6000 axial and stereo sense wires with a transverse momentum resolution of
$\frac{\delta P_T}{P_T^2}$ = 0.3 \%, covering the region $-1 \le \eta \le 1$. For Run-II the Central Outer 
Tracker (COT), with the same coverage and $P_T$ resolution will be used, this is already installed, and has 
$\approx$ 30,000 axial and stereo sense wires. The increased number of wires provide better $\frac{dE}{dx}$, 
somewhat improving particle identification.

\paragraph{\bf Silicon microstrip detectors}

Silicon Microstrip Detectors for precise vertex determination are needed for $B$ lifetime 
measurements. The Run-I Silicon Vertex detector (SVX), had four axial layers and a 
coverage of $-1 \le \eta \le 1$. The proper decay length resolution was $\sigma_{c\tau}$: 35 $\mu$m = 0.12 ps.
This detector provided measurements in the $r-\phi$ plane only. The Run-II Silicon Vertex Detector 
is known as the SVX-II, and has seven stereo and axial layers providing coverage in $-2 \le \eta \le 2$ with 
an additional layer (Layer 00) at a distance of 1.4 cm from beam-pipe. Axial and Stereo strips provide the 
capability to do 3 dimensional tracking using only the SVX-II. The SVX-II $\sigma_{c\tau}$ is expected to be 
15 $\mu$m = 0.045 ps.

\paragraph{\bf Particle identification}

A Time Of Flight (TOF) detector built from scintillators has been mounted on the outer surface of the COT, 
this provides us with some particle identification capability at low momenta. Differentiating kaons 
from pions is crucial for tagging the flavour of $B$s. In addition to the TOF, the coverage of the Muon 
Chambers has been extended from $\eta=$1.0 in Run-I to $\eta=$ 1.4. High momentum muons are easily 
identified, and allow us to select events with $b \to \mu \nu_{\mu} X$ and $b \to c \bar{c}$ with $c \bar{c} \to J/\psi \to \mu^+ \mu^-$.

\begin{figure}
\begin{center}
\epsfig{file=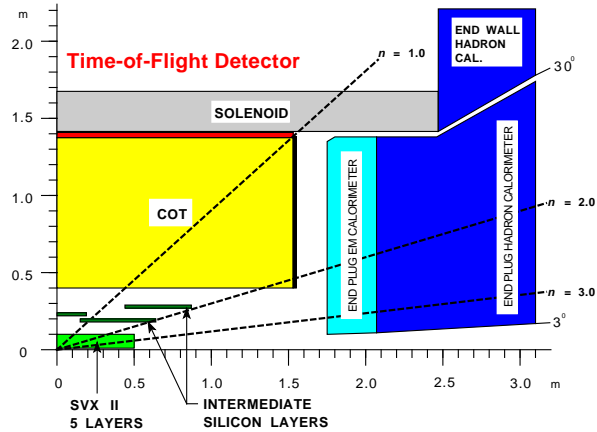,width=0.5\textwidth,angle=0}
\end{center}
\caption{A side view of the CDF detector.}
\label{det}
\end{figure}

\subsubsection{\bf Triggers and data selection }

At the Tevatron with $p-\bar{p}$ collisions at a centre-of-mass energy of $1.8$TeV the $b$ production 
cross-section is $\sigma_{b\bar{b}}=$100$\mu$b, however the inelastic cross section is three orders of
magnitude higher. Although $\sigma_{b\bar{b}}$ is much higher than for $e^+-e^-$ 
at the $\Upsilon(4S)$ and $Z$ resonances (1 and 7 nb), the high inelastic cross section requires 
specialized triggers for the selection of $b\bar{b}$ events. 

\paragraph{\bf CDF run-I triggers}

The selection of $b\bar{b}$ events during Run-I at CDF relied on high transverse momentum ($P_T$) leptons.
The  inclusive lepton $\ell=(\mu,e)$ trigger selected events with $P_T(B) > $8 GeV, 
the quark level decays being $b \to \ell \nu cX$ or $b\to cX$ $c \to \ell \nu X$.
Dilepton triggers selected $(\mu e)$ and $(\mu \mu)$ events with $P_T(B)\approx$10 GeV were also in use.
These were crucial in selecting $b \to J/\psi X$ with $J/\psi \to \mu \mu$ and $b \to \mu^- X$, 
$\bar{b} \to e^+ X$. The $\sin 2\beta$ analysis requires the reconstruction of the mode 
$B_d^0 \to J/\psi K_S^0$, for which the data sample was selected using the $J/\psi \to \mu^+ \mu^-$ trigger.
The long-lifetime of $B$ mesons was not utilized in any trigger during Run-I.

In Run-I the single lepton trigger data samples were used to measure $B_d^0$ mixing. The presence of a 
neutrino in the final state introduces uncertainties in the $B$ decay vertex determination. Despite this the 
CDF measurement of the $B_d^0$ mixing parameter $x_d=$ 0.77 $\pm$ 0.040 compares well with
world average $x_d$= 0.739 $\pm$ 0.023~\cite{mixing}. However this uncertainty will degrade the $B_s^0$ 
mixing parameter $x_s$ measurement significantly, since the oscillation period is much smaller. Clearly it is 
crucial to be able to trigger on fully reconstructible modes.

\paragraph{\bf CDF run-II triggers}

The Run-II CDF triggers used to select $B$ decays will include all Run-I triggers, in addition a new high 
impact parameter track trigger using SVX-II hit information will be used in order to select displaced 
hadronic tracks from $B$ decays. Therefore in Run-II CDF it will be possible to select $B$ decays with 
fully reconstructible purely hadronic decay states, such as $B_s^0 \to D_s^+ D_s^-$, $B_d^0 \to \pi^+ \pi^-$ 
and $B_s^0 \to D_s^{\pm} \pi^{\mp}$. These decays allow us to measure the $B_s^0$ width difference and mixing.

\subsubsection{\bf The CP asymmetry, $\sin 2\beta$ in $B_d^0 \to J/\psi K_S^0$: run-I}

To measure the CP asymmetry in $B_d^0 \to J/\psi K_S^0$, the flavour of the 
$B_d^0$ at production has to be tagged, and the decay fully reconstructed. 
The tagging of the flavour is neither fully efficient nor correct each time, we define the dilution variable 
$D = \frac{N_R-N_W}{N_R+N_W}$ where $N_R$ and $N_W$ are the number of correct and incorrect tags. If the 
tagging efficiency is given by $\epsilon$ then a CP asymmetry measured using $N$ tagged events will have the 
statistical power of $\epsilon D^2 N$  fully efficiently and correctly tagged events.

\paragraph{\bf Flavour tagging: opposite and same side}

Opposite side tagging algorithms use decay products from the $b$ quark on the opposite side of 
the reconstructed decay of interest. The decay products used are the jet associated with 
the hadronization of the $b$ into a $B$ or the sign of the lepton in case the $B$ decays leptonically. In 
case of non-leptonic decays a weighted sum of charges of all tracks is used to tag the $B$ flavour. 
The lepton tagging is known as the Soft Lepton Tag, has high dilution but low efficiency, the figure of 
merit, $\epsilon D^2$ is 0.91 $\pm$ 0.1 \%. The Jet Charge technique is more efficient but has worse 
dilution, with $\epsilon D^2$= 0.78 $\pm$ 0.12 \%.

There are two disadvantages of the opposite side tagging techniques: the opposite $B$ is within detector 
acceptance only 40 \% of the time, and if the opposite $B$ is neutral then it may have mixed. These factors 
degrade efficiency and dilution.

The same side tagging (SST) algorithm uses the hadronization products on the same side as the reconstructed 
decay to determine the flavour of the $B$. This has a figure of merit $\epsilon D^2$=  1.8 $\pm$ 0.4\%, as 
measured in Run-I.

\paragraph{\bf Results from run-I}
The candidate events for reconstructing the decay $B_d^0 \to J/\psi K_S^0$ with $J/\psi \to \mu^+ \mu^-$ and $K_S^0 \to \pi^+ \pi^-$ were selected using the Run-I $J/\psi \to \mu^+ \mu^-$ trigger. Roughly 400 such events 
were reconstructed with some 200 within the acceptance of the SVX. The tagging algorithms described were 
tested and tuned on a sample of $B_u^{\pm} \to J/\psi K^{\pm}$ decays, and the dilutions determined. The 
flavour of the $B$ is already known in this decay mode, so it can be used for tuning the algorithms.

All 400 events were used in an unbinned likelihood multi-parameter fit of the mass and decay length 
distributions, and tag information of the reconstructed $B_d^0 \to J/\psi K_S^0$ decays, each tagging 
algorithm was used for each event with an appropriate weighting accounting for agreement or disagreement 
between the taggers. 

\begin{figure}
\begin{center}
\epsfig{file=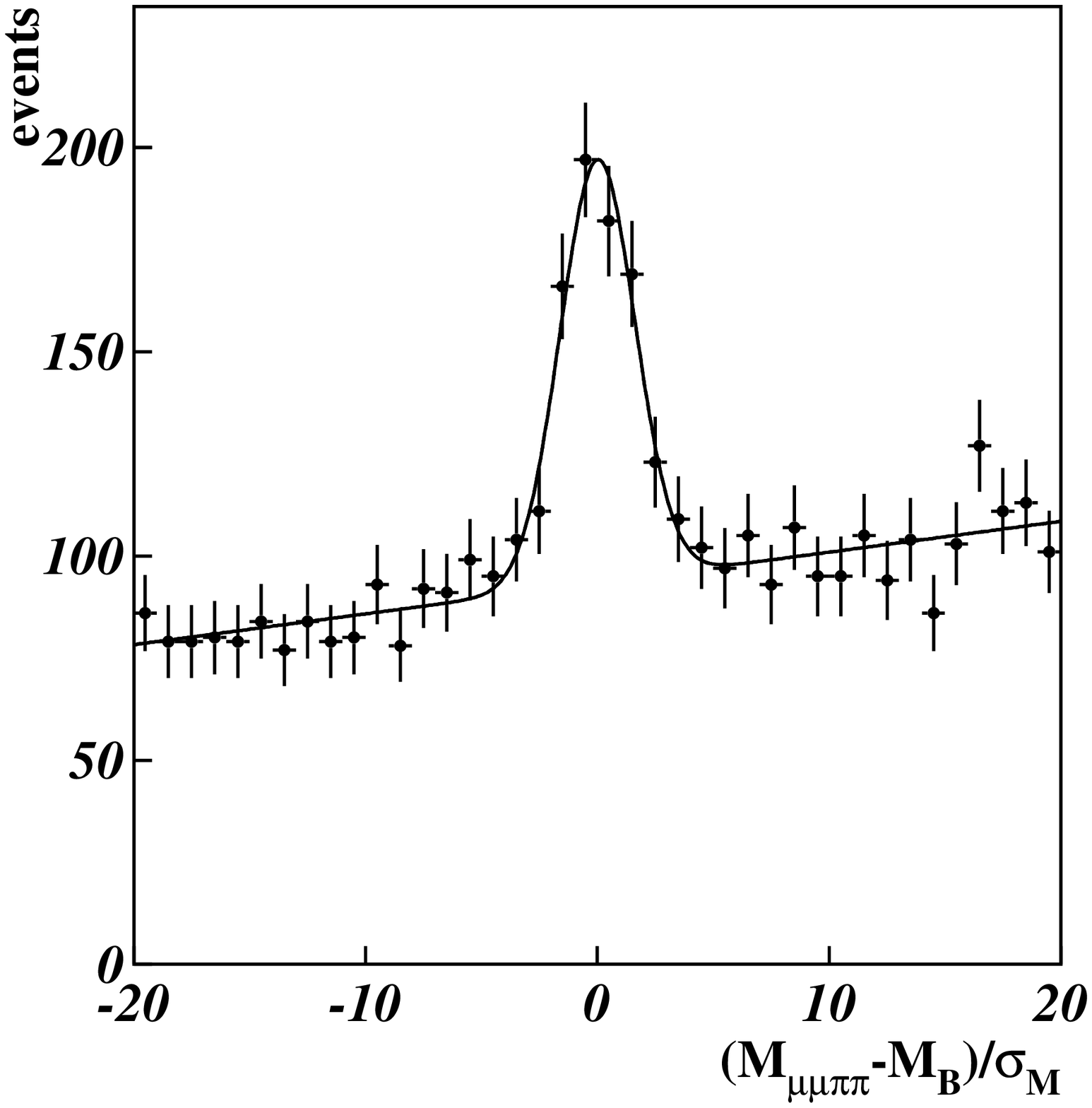,width=0.402\textwidth,angle=0}
\epsfig{file=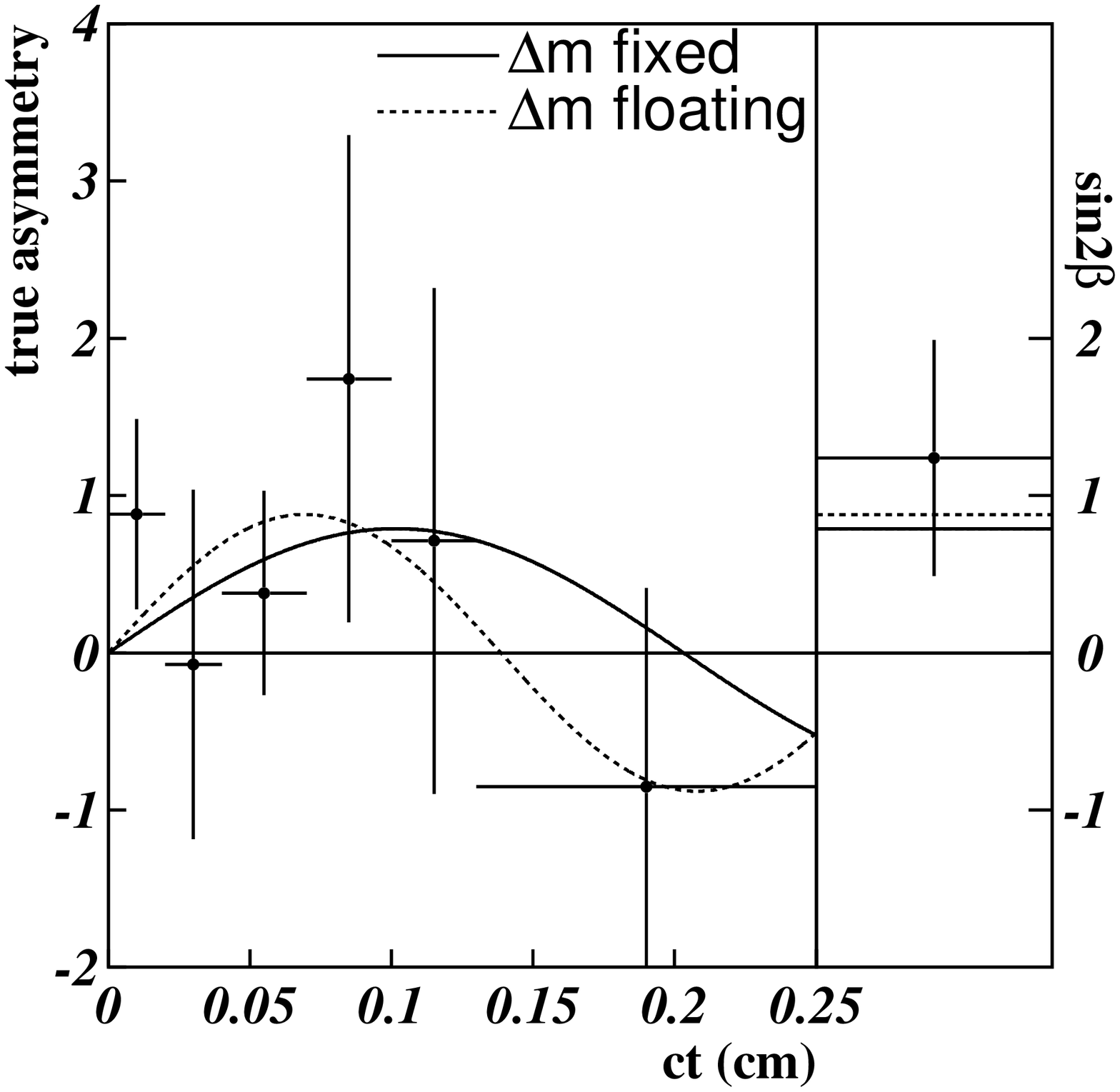,width=0.424\textwidth,angle=0}
\end{center}
\caption{Mass spectrum of $B_d^0 \to J/\psi K_S^0$ decays, signal: 395 $\pm$ 31 (left) and the 
CP asymmetry as a function of time (right).}

\label{s2beta}
\end{figure}

The reconstructed mass spectrum and time dependent asymmetry for $B_d^0 \to J/\psi K_S^0$ are shown 
in Fig.~\ref{s2beta}. The best fit value of the parameter 
$\sin 2\beta$ is 0.79 $\pm$ 0.39 (stat) $\pm$ 0.16(sys) where the systematic error is 
dominated by uncertainties in the dilution.

\subsubsection{\bf Prospects for run-II}

The plan for Run-II is to constrain the unitarity triangle by measuring one side using $B_s^0$ mixing, 
two angles $\gamma$ using CP asymmetries in $B_d^0 \to \pi^+ \pi^-$ and $B_s^0 \to K^+ K^-$ and $\beta$ 
($\sin 2\beta$), as in Run-I, by measuring the asymmetry in $B_d^0 \to J/\psi K_S^0$, this is shown in
Fig.~\ref{tri}. 

\paragraph{\bf Preparing the ground for projections} 

We expect at least 2fb$^{-1}$ of data in Run-II which is a factor of 20 higher than Run-I, 
this figure has been used as the basis for all projections in this article, it is important to note
that this is a very conservative estimate, the laboratory director's recent stated goal being 15 fb$^{-1}$.

In addition to the increase in luminosity, the increased coverage of the SVX-II and
Muon chambers will increase our statistics for all Run-I modes by a factor of 50-60.

All Run-I analyses such as the CP asymmetry in $B_d^0 \to J/\psi K_S^0$ will be repeated with much higher
statistics and several new modes, selected using the hadronic displaced track trigger will be used to search 
for CP violation and $B_s^0$ mixing.

\paragraph{\bf $\sin 2\beta$ in run-II}

We expect at least 10,000 events in the channel $B_d^0 \to J/\psi K_S^0$, 
therefore the statistical error  $\sigma_{\sin 2\beta}$ will be $\approx$ 0.067.
We expect a similar decrease in the systematic error, since it is dominated by errors in measuring 
dilution, due to the larger  $B_u^{\pm} \to J/\psi K^{\pm}$ calibration sample.
We have not included the contribution of other decays that the hadronic displaced track trigger will select:
such as $B_d^0 \to D^{*+} D^{*-}$, $B_d^0 \to D^+ D^-$ and $B_d^0 \to \phi K_S^0$, where the time 
dependent CP asymmetry is also proportional to $\sin 2\beta$.

\paragraph{\bf $\gamma$ using $B_d^0 \to \pi^+ \pi^-$ and $B_s^0 \to K^+ K^-$}

In the absence of penguin-graphs the asymmetry in the decay $B_d^0 \to \pi^+ \pi^-$ will measure 
$\sin 2(\gamma + \beta)$ and similarly the asymmetry in the decay $B_s^0 \to K^+ K^-$ should only measure 
$\sin 2\gamma$. However penguins are present in both decays, the $\pi^+\pi^-$ mode is tree dominated 
the $K^+K^-$ mode is penguin dominated. The CP asymmetry for $B_d^0 \to \pi^+ \pi^-$ is
$A_{CP, \pi^+ \pi^-}^{dir} \cos \Delta M_d t + A_{CP, \pi^+ \pi^-}^{mix} \sin \Delta M_d t$ and for the mode
$B_s^0 \to K^+ K^-$ it is $A_{CP, K^+ K^-}^{dir} \cos \Delta M_s t + A_{CP, K^+ K^-}^{mix} \sin \Delta M_s t$.
 The four $A_{CP}$ terms can be expressed in terms of 4 parameters, $d$ the ratio of the 
penguin to tree amplitudes (assuming SU(3) symmetry), $\theta$ the ratio of the phases of the penguin and 
tree amplitudes, and the CKM angles $\gamma$ and $\beta$. These can be solved for by using the 4 asymmetries 
and the measurement of $\sin 2\beta$~\cite{fleischer}. We expect the hadronic displaced track trigger to
select 5000 $B_d^0 \to \pi^+ \pi^-$ and 10000 $B_s^0 \to K^+ K^-$, and a simulation based on these
estimates was performed. An estimated SU(3) breaking of 20 \% was put into the simulation and the resulting 
inaccuracy is part of the combined systematic and statistical error of 10 degrees in measuring $\gamma$.
All estimates of background were based on Run-I data.

\paragraph{\bf Determination of $x_s$ the $B_s^0$ mixing parameter}

The ratio of the mixing parameters $\frac{x_d}{x_s}$ of the $B_d^0$ and $B_s^0$ neutral mesons
is proportional to $\frac{\mid V_{ts} \mid^2}{ \mid V_{td} \mid^2}$. Since  $V_{ts} >>$ $V_{td}$, the 
$B_s^0$ oscillates much faster than the $B_d^0$, and we expect $x_s \approx 25$ ($x_d \approx$=0.75)

The measurement of $x_s$ oscillation is challenging. In Run-I we measured the 
oscillation parameter of the $B_d^0$ meson, this means measuring an oscillation of period 2.12 ps
with a $ct$ resolution of 0.1 ps. To measure the $x_s$ we will try to measure a period of 0.07 ps with the 
SVX-II resolution of 0.045ps.

We expect 20,000 $B_s^0 \to D_s^{\pm} \pi^{\mp}$ and $B_s^0 \to D_s^{\pm} \pi^{\mp}\pi^{\mp}\pi^{\pm}$ with
$D_s^{\pm} \to \phi \pi^{\pm}$ which the  hadronic trigger will select. We have estimated our accuracy for 
measuring $x_s$  with a signal to noise of 2:1 and 1:2 based on Run-I estimates. We expect to measure a $x_s$ 
of up to 63 with a significance of 5$\sigma$. This accuracy in measuring $x_s$ corresponds to an accuracy of
7 \% in measuring $\mid \frac{V_{td}}{\lambda V_{ts}} \mid$. In addition to measuring this side of the
unitarity triangle the value of $x_s$ can be used as a constraint in CP asymmetry measurements in CP violating $B_s^0$ decays.

\subsubsection{\bf Non-SM surprises from the $B_s^0$ width difference and CP violation in $B_s^0$ decays}

In the standard model the $B_s^0$ width difference $\Delta \Gamma$ can be written in a CKM independent form
in terms of the top, bottom, W and charm masses and the $B_s^0$ mixing parameter $x_s$~\cite{isi}.
The fractional width splitting between the light and heavy $B_s^0$ $\frac{\Delta \Gamma}{\Gamma}$ 
is expected to be large $\approx$ 0.15. If we measure both $x_s$ and $(\frac{\Delta \Gamma}{\Gamma})_{B_s^0}$ 
we can test the SM prediction. In particular, we are interested in cases where $x_s$ is too large to 
measure but $\frac{\Delta \Gamma}{\Gamma}$ is measurable or conversely, if $\frac{\Delta \Gamma}{\Gamma}$ is 
too small to measure and $x_s$ is measurable. In the presence of new Physics a non-SM CP violating phase
$\phi_{BSM}$ can appear, and will reduce the width difference: $\Delta \Gamma_{Measured} = \Delta \Gamma_{SM} \cos \phi_{BSM}$. Complimentary to this effect the CP asymmetry in modes such as $B_s^0 \to J/\psi \phi$ will 
be proportional to $\sin 2\phi_{BSM}$.

In Run-I approximately 60 $B_s^0 \to J/\psi \phi$ decays were reconstructed from a data sample selected 
using the $J/\psi \to \mu^+ \mu^-$ trigger and a single lifetime was measured~\cite{bslife}.
 The final state $J/\psi \phi$ is however to a mixture of CP even and odd final states, it therefore contains
two lifetimes, for the heavy ($\approx$ CP odd) and light ($\approx$ CP even) $B_s^0$ states. Using angular 
variables to disentangle the CP content of the final state we can fit for two lifetimes utilizing a 
likelihood function normalized over invariant mass, lifetime and the angular variable~\cite{isi2}. We expect 
a signal of 4000 $B_s^0 \to J/\psi \phi$ in Run-II. A simulation based on 4000 signal events and Run-I
observations of signal and background shapes has been used to predict that we can measure
a $(\frac{\Delta \Gamma}{\Gamma})_{B_s^0}$ of 0.15 with a precision of 0.05. This does not include any 
contribution from the purely CP even mode $B_s^0 \to D_s^+ D_s^-$ which will be selected using the 
hadronic displaced track trigger, the lifetime measured in this mode can be compared with the results of 
a single-lifetime fit to semi-leptonic $B_s^0$ decays and $(\frac{\Delta \Gamma}{\Gamma})_{B_s^0}$ can be 
extracted.

The SM prediction for CP asymmetry in the $B_s^0 \to D_s^+ D_s^-$ and $B_s^0 \to J/\psi \phi$ modes
is of order 3\%, we do not expect to be able to see this at CDF, however if there are non-SM sources
of CP violation, we may be able to observe this, by doing a conventional tagging analysis in 
$B_s^0 \to D_s^+ D_s^-$ and a tagging analysis of the disentangled CP states of $B_s^0 \to J/\psi \phi$.

\subsubsection{\bf Conclusion}
During Run-II at CDF we have been able to measure of $\sin 2\beta =$0.79 $\pm$ 0.39 
(stat) $\pm$ 0.16(sys) and the $B_d^0$ mixing parameter $x_d$ $=$ 0.77 $\pm$ 0.040(stat) $\pm$0.039 
(syst), and have fully reconstructed the largest sample of $B_s^0$ decays.

Thus the viability of a rich CP violation program for Run-II, during which we expect a factor of 50 
increase in data, has been established. The introduction of the displaced track trigger will allow us to 
select fully reconstructible hadronic modes of all $B$ mesons. We expect to be able to constrain one side 
and two angles of the unitarity triangle. We expect to measure $\mid \frac{V_{td}}{\lambda V_{ts}} \mid$ to 
$\approx$ 7\% accuracy, using $B_s^0$ mixing. We emphasize that for the next few years CDF is unique in its 
ability to analyse $B_s^0$ decays.
We expect to use $B_d^0 \to \pi^+ \pi^-$, $B_s^0 \to K^+ K^-$ 
to measure $\gamma$ with a precision $\sigma_{\gamma} \approx 10^{o}$, and using $B_d^0 \to J/\psi K_S^0$ 
to measure $\sin 2\beta$ to a precision of 0.067.

\subsection{\bf B physics potential of 
future experiments at hadron machines}
{\it V Gibson, Cambridge}

%
%
\def\ra{\rightarrow}
\def\ba{\begin{array}}
\def\ea{\end{array}}
\def\iby2{\frac{i}{2}}
\def\bi{\begin{itemize}}
\def\ei{\end{itemize}}
\def\cp{\mathrm{CP}}
\def\re{\Re e}
\def\im{\Im m}
%
%
\def\mev{\mathrm{MeV}}
\def\gev{\mathrm{GeV}}
\def\mevc{\mev / \mathrm c^2}
\def\ns{\mathrm{ns}}
\def\pps{\mathrm{ps}^{-1}}
%
%
\def\u{\mathrm u}
\def\d{\mathrm d}
\def\c{\mathrm c}
\def\s{\mathrm s}
\def\t{\mathrm t}
\def\b{\mathrm b}
\def\ub{\bar{\u}}
\def\db{\bar{\d}}
\def\cb{\bar{\c}}
\def\sb{\bar{\s}}
\def\tb{\bar{\t}}
\def\bb{\bar{\b}}
%
%
\def\ee{e^+e^-}
\def\ppb{p\bar p}
\def\pp{pp}
%
%
\def\delm{\Delta M}
\def\delg{\Delta \Gamma}
\def\avg{\overline{\Gamma}}
\def\magr{|r|}
\def\magrb{|\overline r|}
%
%
\def\kz{\mathrm K^0}
\def\kzb{\overline{\kz}}
\def\ks{\kz_{\mathrm S}}
\def\kl{\kz_{\mathrm L}}
\def\taus{\tau_{\mathrm S}}
\def\taul{\tau_{\mathrm L}}
\def\mks{M_{\mathrm S}}
\def\mkl{M_{\mathrm L}}
\def\delk{\delta_{\mathrm K}}
\def\eps{\varepsilon}
\def\epsp{\eps^{\prime}}
%
%
\def\bhad{\mathrm B}
\def\bbhad{\overline{\bhad}}
\def\bz{\bhad^0}
\def\bzb{\overline{\bz}}
\def\bd{\bz_{\mathrm d}}
\def\bdb{\overline{\bd}}
\def\bs{\bz_{\mathrm s}}
\def\bsb{\overline{\bs}}
\def\bp{\bhad^+}
\def\bm{\bhad^-}
\def\bpm{\bhad^{\pm}}
\def\bstst{\bhad^{**}}
\def\bc{\bpm_{\mathrm c}}
\def\xd{x_{\mathrm d}}
\def\xs{x_{\mathrm s}}
\def\delb{\delta_{\bhad}}
%
%
\def\ckm{\mathrm{CKM}}
\def\vud{V_{ud}}
\def\vcd{V_{cd}}
\def\vtd{V_{td}}
\def\vus{V_{us}}
\def\vcs{V_{cs}}
\def\vts{V_{ts}}
\def\vub{V_{ub}}
\def\vcb{V_{cb}}
\def\vtb{V_{tb}}
\def\asym{A}
\def\tasym{\asym (t)}
\def\al{\alpha}
\def\be{\beta}
\def\ga{\gamma}
\def\dga{\delta\ga}
%
%
\def\kb{\overline{\mathrm K}}
\def\kp{\mathrm K^+}
\def\km{\mathrm K^-}
\def\kpm{\mathrm K^{\pm}}
\def\kmp{\mathrm K^{\mp}}
\def\kst{\mathrm K^{*0}}
\def\jpsi{J/\psi}
\def\dz{\mathrm D^{\mathrm 0}}
\def\dzb{\overline{\dz}}
\def\ds{\mathrm D_{\mathrm s}}
\def\done{\mathrm D_1}
\def\dtwo{\mathrm D_2}
\def\dst{\mathrm D^{(*)\pm}}
%
%
\def\lb{\Lambda_{\b}}
%
%
\def\jk{\jpsi\ks}
\def\jkl{\jpsi\kl}
\def\jphi{\jpsi\phi}
\def\jeta{\jpsi\eta^{(\prime)}}
\def\pipi{\pi^+\pi^-}
\def\popo{\pi^0\pi^0}
\def\pipipo{\pi^+\pi^-\pi^0}
\def\kk{\mathrm K^+ \mathrm K^-}
\def\dstpi{\dst \pi^{\mp}}
\def\mumu{\mu^+ \mu^-}
%
%
\def\bdtojk{\bd \ra \jk}   
\def\bdtopipi{\bd \ra \pipi}
\def\bdtopik{\bd \ra \pi^+ \mathrm{K}^-}
\def\bdtorhopi{\bd \ra \rho^{\pm}\pi^{\mp}}
\def\bdtorhopo{\bd \ra \rho^0 \pi^0}
\def\bdtopipipo{\bd \ra \pipipo}
\def\bdtodkst{\bd \ra \mathrm D \kst}
\def\bdtodzkst{\bd \ra \dz \kst}
\def\bdtodzbkst{\bd \ra \dzb \kst}
\def\bdtodonekst{\bd \ra \done \kst}
\def\bdtodstpi{\bd \ra \dstpi}
\def\bdtojkst{\bd \ra \jpsi \kst}   
\def\bdtodd{\bd \ra \mathrm D^+ \mathrm D^-}
\def\bdtomumu{\bd \ra \mumu}
\def\bdtokstga{\bd \ra \kst \gamma}
\def\bdtokstmumu{\bd \ra \kst \mumu}
\def\bdtorhomumu{\bd \ra \rho \mumu}
%
%
\def\bstojphi{\bs \ra \jphi}
\def\bstojeta{\bs \ra \jeta}
\def\bstokk{\bs \ra \mathrm K^+ \mathrm K^-}
\def\bstopik{\bs \ra \pi^+ \mathrm K^-}
\def\bstodsk{\bs \ra \ds^{\pm} \kmp}
\def\bstodskp{\bs \ra \ds^- \mathrm K^+}
\def\bstodskm{\bs \ra \ds^+ \mathrm K^-}
\def\bstodspi{\bs \ra \ds^{\pm} \pi^{\mp}}
\def\bstodspip{\bs \ra \ds^- \pi^+}
\def\bstodspipipi{\bs \ra \ds^- \pi^+ \pi^+ \pi^-}
\def\bstomumu{\bs \ra \mu^+ \mu^-}
\def\bstomumukst{\bstomumu \mathrm K^{*0}}
\def\bstojk{\bs \ra \jk}
\def\bstodd{\bs \ra \mathrm D^+_{\s} \mathrm D^-_{\s}}
\def\bstophimumu{\bs \ra \phi \mumu}
%
%
\def\bpmtodkpm{\bpm \ra \mathrm D \kpm}
\def\bpmtodzkpm{\bpm \ra \dz \kpm}
\def\bpmtodzbkpm{\bpm \ra \dzb \kpm}
\def\bpmtodonekpm{\bpm \ra \done \kpm}
\def\bptodzkp{\bp \ra \dz \kp}
\def\bptodzbkp{\bp \ra \dzb \kp}
\def\bptodonekp{\bp \ra \done \kp}
\def\bmtodzkm{\bm \ra \dz \km}
\def\bmtodzbkm{\bm \ra \dzb \km}
\def\bmtodonekm{\bm \ra \done \km}
%
%
\def\btodonek{\bhad \ra \done \mathrm K}
\def\btodzk{\bhad \ra \dz \mathrm K}
\def\btodzbk{\bhad B \ra \dzb \mathrm K}
\def\bbtodonekb{\bbhad \ra \done \kb}
\def\bbtodzkb{\bbhad \ra \dz \kb}
\def\bbtodzbkb{\bbhad \ra \dzb \kb}
\def\btopik{\bhad \ra \pi \mathrm K}
%
%
\def\lbtoppi{\lb \ra \mathrm p \pi^-}
\def\lbtopk{\lb \ra \mathrm p \mathrm K^-}
%
%
\def\lhcb{LHCb}
\def\btev{BTeV}
\def\atlas{ATLAS}
\def\cms{CMS}
\def\babar{BaBar}
\def\belle{BELLE}
\def\herab{HERA-B}
\def\cdf{CDF}
\def\d0{D0}
\def\lumi{1.5\times 10^{32} \, \mathrm{cm}^{-2}\mathrm{s}^{-1}}
\def\lowlumi{1\times 10^{33} \, \mathrm{cm}^{-2}\mathrm{s}^{-1}}
%
%
\def\tree{\mathrm T}
\def\peng{\mathrm P}
\def\magpt{\left|\frac{\peng}{\tree}\right|}
\def\dtree{\delta_{\tree}}
\def\dpeng{\delta_{\peng}}
\def\ptree{\phi_{\tree}}
\def\ppeng{\phi_{\peng}}
\def\br{\mathrm{Br}}
%
\def\CPbar{\hbox{{\rm CP}\hskip-1.80em{/}}}
%

\subsubsection{\bf Introduction}
\begin{center}
\begin{figure}[htb]
\epsfxsize=5.2cm
\epsfysize=5.2cm
\epsffile[-95 20 120 230]{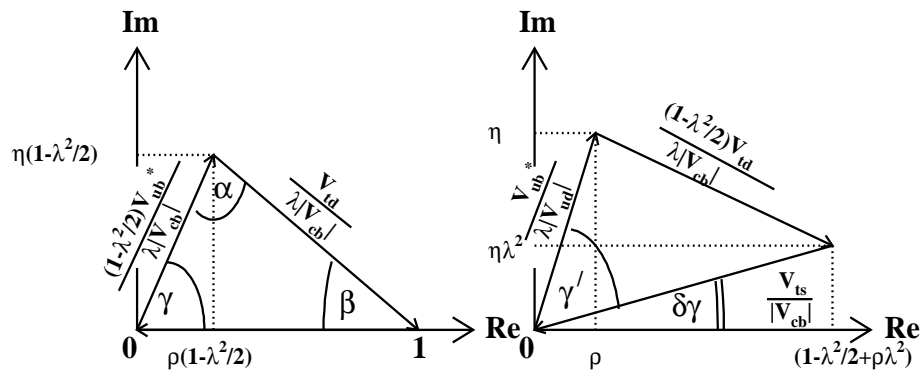}
\caption{\label{fig:triangles}
The unitarity triangles characterized by the relations
$\vud\vub^* + \vcd\vcb^* + \vtd\vtb^* =  0$ and 
$\vud\vtd^* + \vus\vts^* + \vub\vtb^* =  0$.  
The phase convention is chosen such that $\vcd\vcb^*$ 
is real and all sides are normalized to $|\vcd\vcb^*|$.}
\end{figure}
\end{center}

CP violation remains one of the enigmas of particle physics today.
Experimentalists have just started a full programme of
research to study CP violation in B decays. 
The experiments can be divided into two phases.
The main goal of the current phase of experiments will be to observe 
CP violation in the B system in the exciting exploratory phase,
with the potential to establish a breakdown in the Standard Model~\cite{ref:current}.
This review concentrates on the next generation of experiments which 
become operational around 2006.
The experiments considered are the dedicated B physics experiments,
LHCb~\cite{ref:lhcb2} at the Large Hadron Collider (LHC) and the 
proposed BTeV experiment~\cite{ref:btev} at the Tevatron, as well as the 
general purpose experiments ATLAS~\cite{ref:atlas} and CMS~\cite{ref:cms} at the LHC. 
The main aims of these experiments will be to make
precision measurements of CP violating observables using many different decay
channels and species of $\bhad$ hadrons.
They will thoroughly test the internal consistency of the Standard Model 
description of CP violation and have the sensitivity to search 
for the necessary new physics beyond. 
%
\paragraph{\bf CP violation in the Standard Model and beyond}
CP violation arises naturally in the Standard Model 
through the presence of a single phase in the 
unitary Cabibbo-Kobayashi-Maskawa ($\ckm$) quark-mixing matrix~\cite{ref:ckm}.
The unitarity of the $\ckm$ matrix is clearly exposed when 
using an explicit parameterization.
A very popular parameterization is the perturbative form 
suggested by Wolfenstein~\cite{ref:wolf},
which can be expanded to order $\lambda^5$ 
($\lambda \equiv \sin\theta_{C} \sim 0.22$ where $\theta_{C}$ is the Cabibbo angle):
\begin{eqnarray}
V_{\ckm} 
& = & 
\left(
\ba{ccc}
1-\frac{1}{2}\lambda^2  & \lambda                & A\lambda^3(\rho - i\eta)  \\ \nonumber
-\lambda                & 1-\frac{1}{2}\lambda^2 & A\lambda^2 \\ \nonumber
A\lambda^3(1-\rho-i\eta)& -A\lambda^2            & 1                        
\ea
\right) \\
& + & 
\left(
\ba{ccc}
0                        & 0                          & 0  \\ \nonumber
-iA^2\lambda^5\eta       & 0                          & 0  \\
A\lambda^5(\rho+i\eta)/2 & A\lambda^4(1/2-\rho-i\eta) & 0
\ea
\right).
\end{eqnarray}
The parameter $\eta$ represents the CP violating phase 
in the Standard Model and appears in three of the matrix elements.
The unitarity of the CKM matrix implies that there are six 
orthogonality conditions, which can be represented geometrically 
as triangles in the complex plane.
Two such {\it unitarity triangles}, 
shown in Figure~\ref{fig:triangles}, 
are expected to have angles, which are all non-trivial.

The angles of the unitarity triangles are all related to 
the single CP-violating phase in the $\ckm$ matrix 
and are designated by $\al$, $\be$, $\ga$ and $\dga$;
\begin{eqnarray*}
\be  = \tan^{-1} \left[  \frac{\eta (1-\lambda^2/2)}{1-\rho(1-\lambda^2/2)} \right],\;\;\;
\ga  =  \tan^{-1} \frac{\eta}{\rho}, \;\;\;
\dga =  \eta \lambda^2,
\end{eqnarray*}
and $\al = \pi-\be-\ga$.
The angles $\be, \dga$ and $\ga$ are commonly referred 
to as {\it the $\bd$ mixing phase, the $\bs$ mixing phase} and 
{\it the weak decay phase} respectively.
By 2006,
it is expected that a measurement of $\sin 2\be$ will have 
been made with a precision of 
$\sim 0.02$~\cite{ref:betaexp}. 
There will be no good or direct measurement of 
$\ga$ and there will be no sensitivity to $\dga$.
It is expected that the single Standard Model $\ckm$ phase is 
insufficient to explain the observed baryon asymmetry in 
the universe~\cite{ref:gavela} and that new physics must intervene.
A large source of CP violation at the electroweak scale could 
be provided for in extensions to the Standard Model~\cite{ref:cohen}.
A summary of physics beyond the Standard Model and its effects on the $\ckm$ 
parameters can be found in references~\cite{ref:beyond,ref:yellow}.
In order to search for new physics it is essential to measure
and calculate as many processes as possible and to compare
the resulting $\ckm$ parameters with each other.
%
\paragraph{\bf The next generation experiments}
The LHC and Tevatron colliders will provide an 
intense source of the full spectrum 
of B hadrons $(\bpm, \, \bd, \, \bs, \, \bc$ and b-baryons).
The running parameters of the colliders and experiments are given 
in Table~\ref{tab:comp}. 
The expected cross-section for the production of $\b\bb$ 
pairs at the LHC is approximately five times that expected at the 
Tevatron, with a relatively smaller inelastic cross-section.
%
\begin{table*}[htb]
\centering
\begin{small}
\begin{tabular}{|l|c|c|c|}
\hline
                              & Tevatron        & \multicolumn{2}{|c|}{LHC}            \\ \hline
Energy/collision mode         & 2.0~TeV $\ppb$  & \multicolumn{2}{|c|}{14.0~TeV $\pp$} \\
$\b\bb$ cross-section         & $\sim 100\mu$b  & \multicolumn{2}{|c|}{$\sim 500\mu$b} \\
Inelastic cross-section       & $\sim 50$mb     & \multicolumn{2}{|c|}{$\sim 80$mb}    \\ 
Ratio $\b\bb$/inelastic       & 0.2\%           & \multicolumn{2}{|c|}{0.6\%}          \\
Bunch spacing                 & 132 ns          & \multicolumn{2}{|c|}{25 ns}          \\
Bunch length                  & $\sim 30$cm     & \multicolumn{2}{|c|}{$\sim 5$cm}     \\
\hline
                              & \btev           & \lhcb              & \atlas / \cms   \\
\hline
Detector configuration        & Two-arm forward & Single-arm forward & Central detector \\
Running luminosity            & $\lumi$         & $\lumi$            & $\lowlumi$       \\
$\b\bb$ events per $10^7$ sec & $2\times 10^{11}\times accep$ & $1\times 10^{12}\times accep$
& $5\times 10^{12}\times accep$ \\
$\langle$Interactions/crossing$\rangle$ 
                              & $\sim 2.0$      & $\sim 0.5$ (30\% single int.) & $\sim 2.3$ \\
Mass resolution $\bdtojk$     & 9.3 $\mevc$     & 7 $\mevc$          & 18/16 $\mevc$     \\
Proper time res. $\bstodspi$ & 43 fs           & 43 fs              & 50/65 fs         \\
\hline 
\end{tabular}
\end{small}
\caption{
Comparison of the LHC and Tevatron collider running parameters and experiments.
}
\label{tab:comp}
\end{table*}

The forward detector geometries of the LHCb and BTeV experiments 
exploit the expected forward peaked and strongly correlated production of 
$\b$ and $\bb$ hadrons. 
LHCb is instrumented on one arm with a dipole spectrometer~\cite{ref:lhcb2} and
is designed to run at a low LHC luminosity of $\lumi$.
It employs a precision ministrip silicon detector and has an efficient
multi-level trigger, which includes a vertex, trigger at the second level.
The experiment employs two RICH detectors for particle identification and 
has hadronic and electromagnetic calorimetry. 
The key design features of the BTeV detector include a forward 
two-arm spectrometer, 
a precision silicon pixel vertex detector, 
a vertex trigger at the first level, 
a RICH detector and a lead-tungstate calorimeter 
for neutral particle reconstruction~\cite{ref:btev}. 
The geometrical acceptance, the use of a vertex trigger at the first 
level and of multi-bunch interactions mostly compensate for the smaller $\b\bb$ 
cross-section at the Tevatron.
ATLAS and CMS are central detectors designed for general-purpose use 
at the LHC~\cite{ref:atlas,ref:cms}.
During the first three years of running, 
the LHC will operate at a low luminosity,
$\lowlumi$, thereby enabling ATLAS and CMS to pursue a B physics programme.
ATLAS and CMS have tracking capabilities in the central region 
and employ precision silicon pixel and microstrip detectors.
Specialist $\bhad$ triggers are achieved by 
reducing the lepton $p_T$ thresholds to a minimum.

The requirements of the dedicated experiments are governed by the need to measure
time-dependent CP asymmetries for $\bhad$ and $\bbhad$ hadrons 
decaying to the same final state.
The experimental acceptance and trigger efficiencies mainly cancel.
However, precise measurements require good decay time and mass resolutions, 
efficient triggers for low and high multiplicity 
$\bhad$ final states, 
particle identification for $\pi / \mathrm K$ separation
and photon detection for neutral final states. 
An example of the $\bdtojk$ mass resolution is given in Table~\ref{tab:comp}.
The flavour of the $\bhad$ hadron at production needs to be determined 
either using the signal $\bhad$ or the other $\bhad$ in the event and
good control of systematic uncertainties is crucial~\cite{ref:yellow}.
The experimental need for good flavour tagging and $\bhad$ proper time 
resolution is demonstrated through the measurement of $\bs$ oscillations.
The bench-mark decay mode for this measurement is $\bstodspi$ in which the
flavour of the $\bs$ is given by the charge of the $\ds$.
The proper time resolutions obtained by the experiments is summarised in Table~\ref{tab:comp}.
The event yields and expected measurable values for the $\bs$ mixing parameter
are given in Table~\ref{tab:summary}.
%
\subsubsection{\bf Direct measurements of CKM angles and search for new physics}
\begin{figure}[htb]
  \begin{minipage}[b]{.46\linewidth}
    \epsfxsize=5.5cm
    \epsfysize=5.5cm
    \begin{center}
    \epsffile{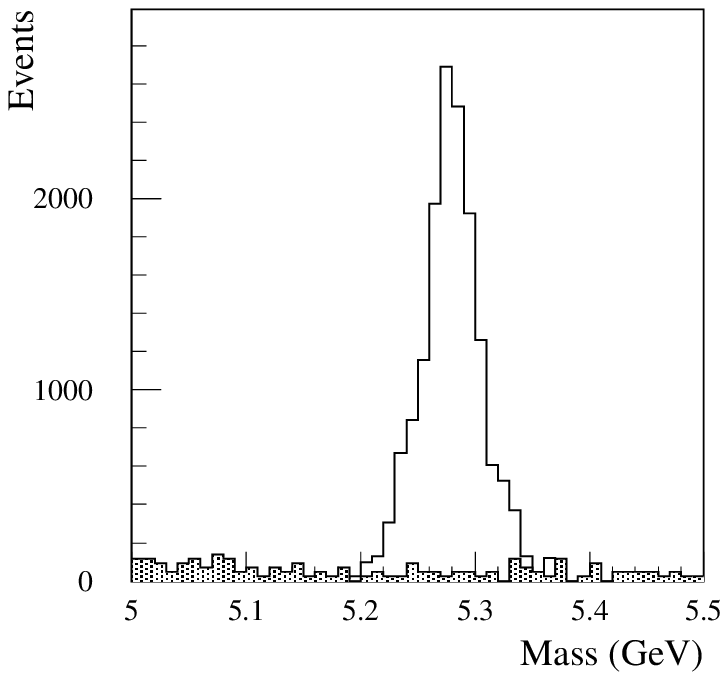}
    \end{center}
   \end{minipage}\hfill
  \begin{minipage}[b]{.46\linewidth}
    \begin{center}
    \epsfxsize=5.5cm
    \epsfysize=5.5cm
    \epsffile{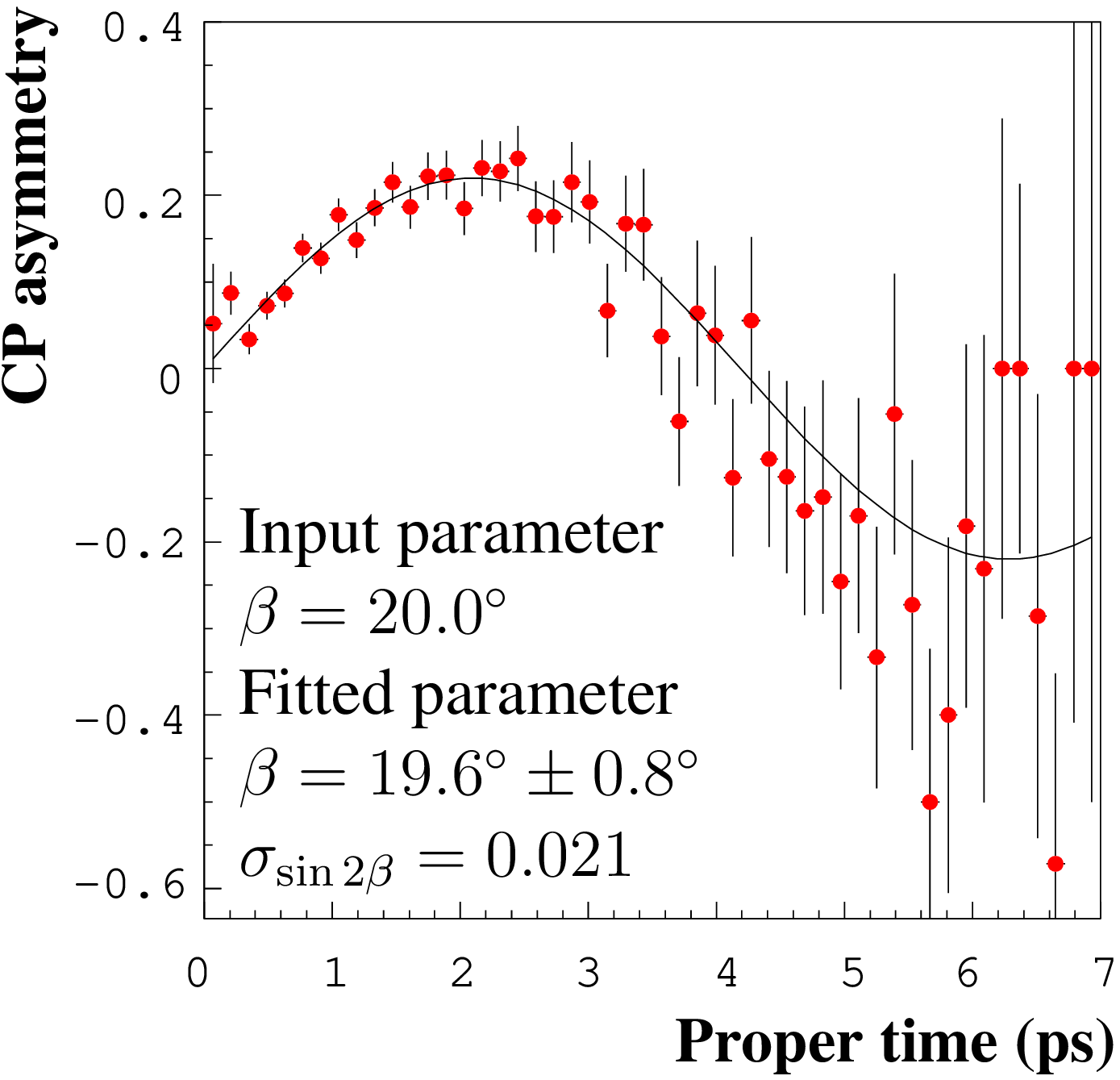}
    \end{center}
  \end{minipage}\hfill
\caption{\label{fig:bdtojkfig}
The ATLAS $\bdtojk, \; \jpsi \ra \ee$ reconstructed mass distribution 
after 3 years of data and the LHCb CP asymmetry with one year's data.
}
\end{figure}

\begin{figure}[htb]
  \begin{minipage}[b]{.46\linewidth}
    \epsfxsize=5.5cm
    \epsfysize=5.5cm
    \begin{center}
    \epsffile{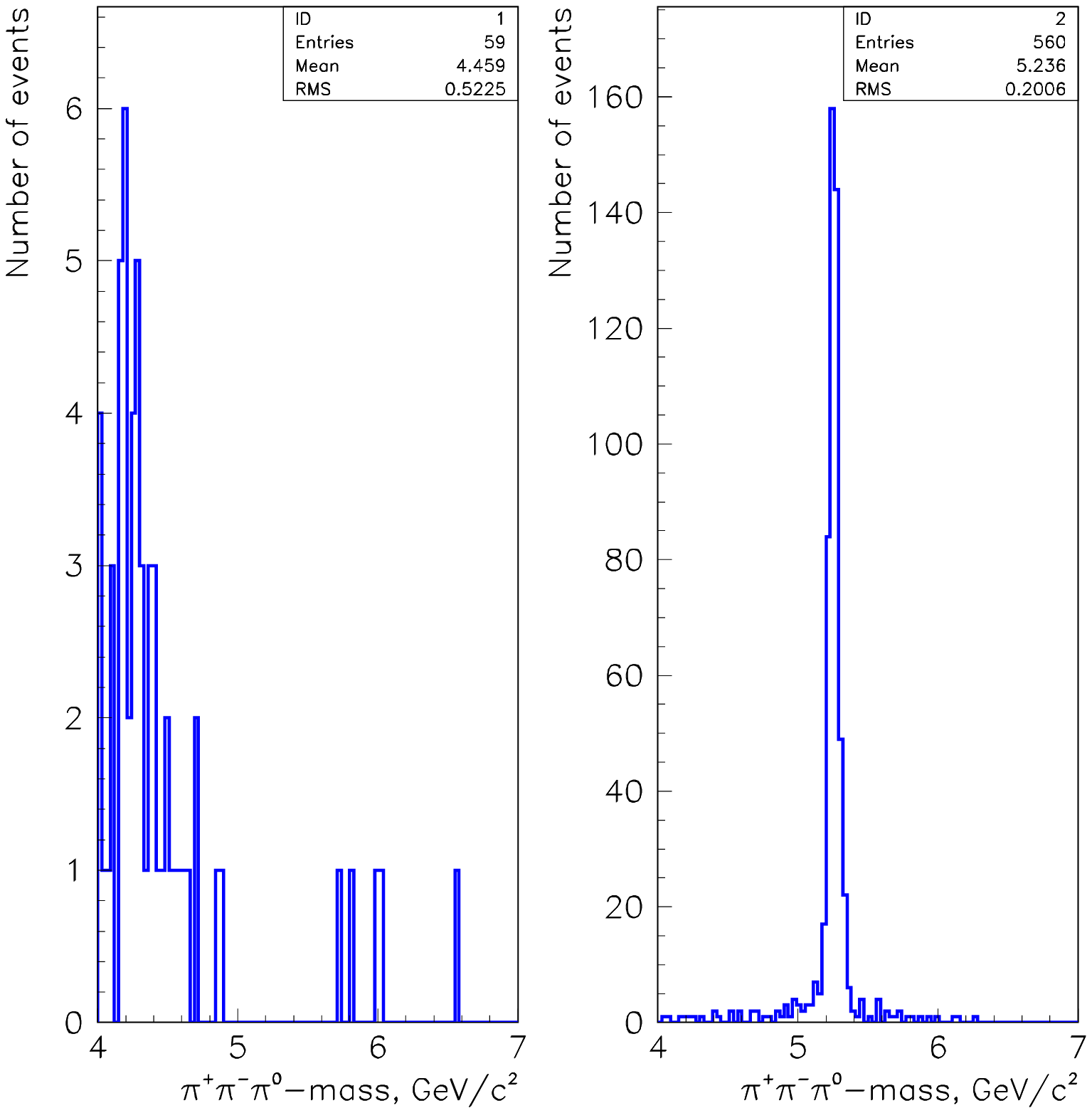}
    \end{center}
   \end{minipage}\hfill
  \begin{minipage}[b]{.46\linewidth}
    \begin{center}
    \epsfxsize=5.5cm
    \epsfysize=5.5cm
    \epsffile{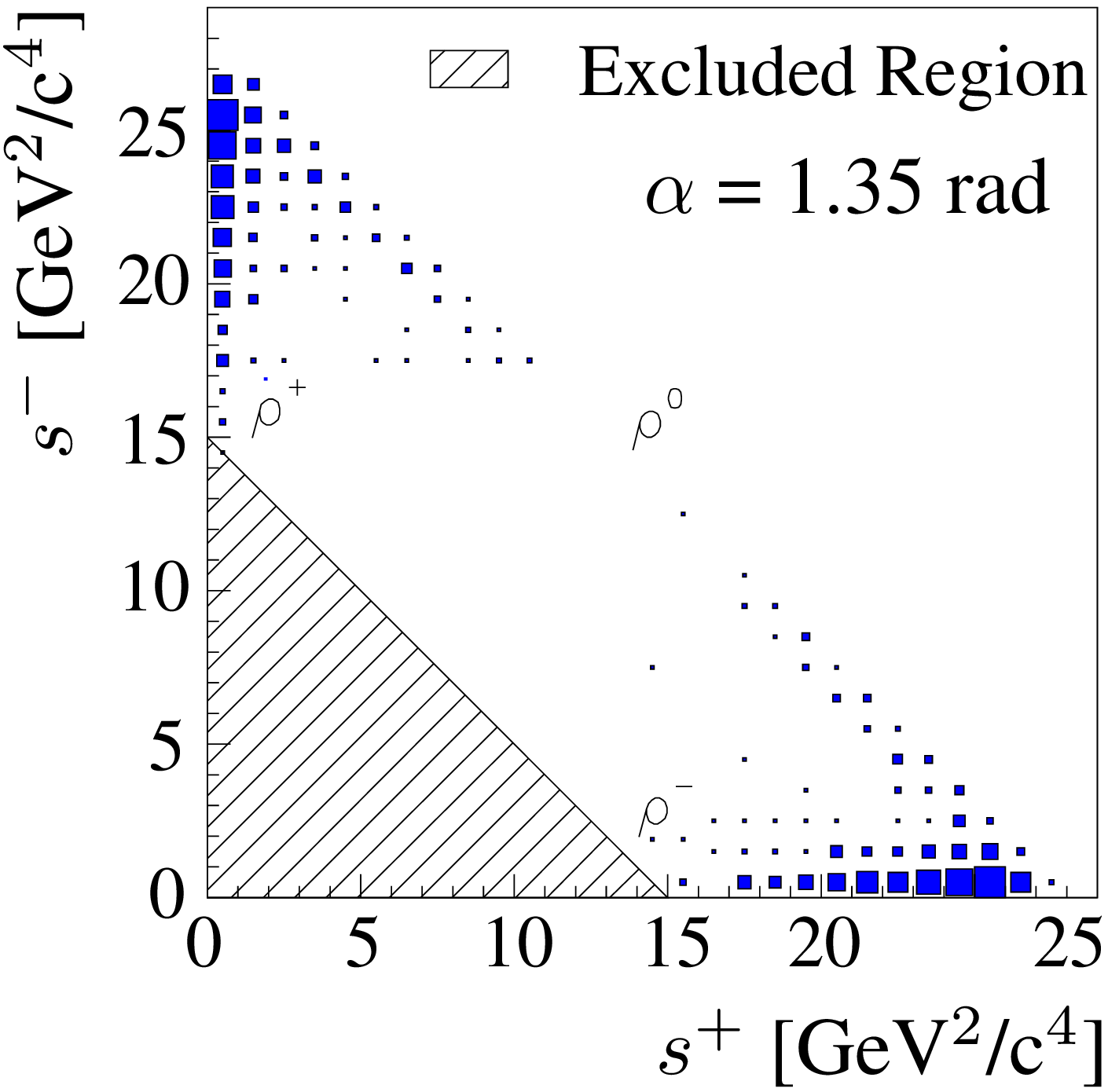}
    \end{center}
  \end{minipage}\hfill
\caption{\label{fig:bdrhopi}
The $\pipipo$ invariant mass distribution for background and signal 
events for $\bdtorhopo$ decays from BTeV and the LHCb Dalitz plot.
}
\end{figure}

\begin{figure}[htb]
  \begin{minipage}[b]{0.3\linewidth}
    \epsfxsize=5.1cm
    \epsfysize=5.1cm
    \begin{center}
    \epsffile{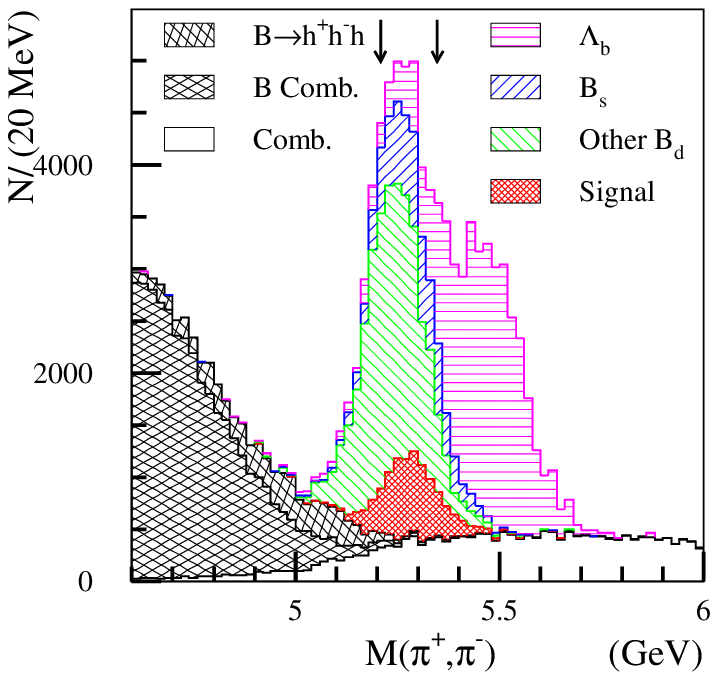}
    \end{center}
   \end{minipage}\hfill
  \begin{minipage}[b]{0.3\linewidth}
    \begin{center}
    \epsfxsize=5.1cm
    \epsfysize=5.1cm
    \epsffile{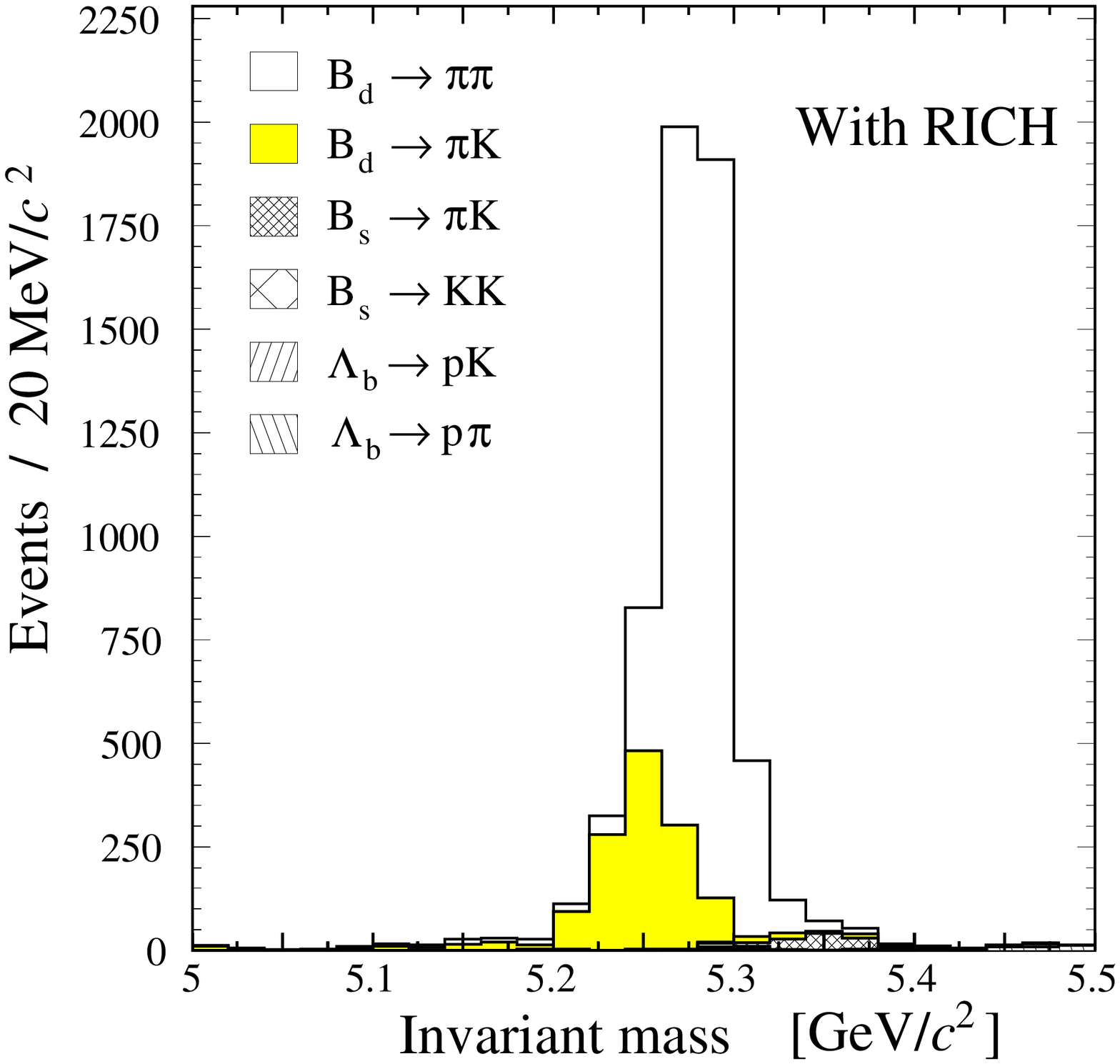}
    \end{center}
  \end{minipage}\hfill
  \begin{minipage}[b]{0.3\linewidth}
    \begin{center}
    \epsfxsize=5.1cm
    \epsfysize=5.1cm
    \epsffile{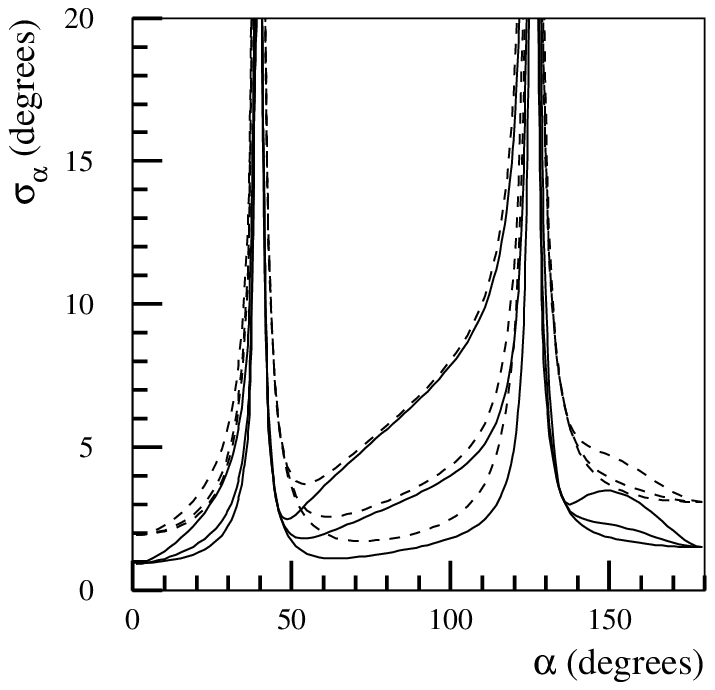}
    \end{center}
  \end{minipage}\hfill
\caption{\label{fig:alpexp}
The ATLAS $\pipi$ invariant mass distribution (left),
the LHCb $\pipi$ invariant mass distribution after the 
application of RICH information (centre) and 
the expected LHC combined sensitivity to $\al$ as a function of $\al$ (right),
for $\delta = 30^{\circ}$, $\left|\peng / \tree \right| = 0.2$
after one year (dashed lines) and 5 years (solid lines).
}
\end{figure}

The internal consistency of the Standard Model description
of quark mixing in weak interactions can be thoroughly tested
by measuring CP violating observables in the decays of B mesons.
If the Standard Model is correct then all such measurements will be describable
with a single set of CKM parameters.
New physics outside the Standard Model could lead to additional phases in the
CKM matrix and an inconsistency between the measurements.
This review discusses the current status of studies performed by the 
new generation experiments to extract the CKM angles 
$\al , \be, \ga$ and $\dga$.
The event yields and sensitivities presented are mainly taken from 
references~\cite{ref:btev} and~\cite{ref:yellow} where further details can be found.
Since the experiments are at different stages in their
preparation for physics,
the results quoted should only be considered as a current snapshot. 
%
\paragraph{\bf $\bd$ mixing phase}
The decay $\bdtojk$ is a transition into a CP eigenstate and
is dominated by only {\it one} CKM amplitude. 
Hence, the time-dependent CP asymmetry is governed by the 
$\bd$ mixing phase, $\be$.
The decay is also experimentally clean and can be reconstructed with
relatively low backgrounds.
Examples of the $\bdtojk$ reconstructed mass distribution and 
CP asymmetry are shown in Figure~\ref{fig:bdtojkfig}. 
The expected event yields and sensitivities for the next generation experiments
are given in Table~\ref{tab:summary}.
Ultimately,
a combined statistical precision of 0.005 should be achievable at the LHC. 
%
%
\paragraph{\bf The $\bs$ mixing phase}
The decays $\bstojphi$ and $\bstojeta$ are $\bs$ 
counterparts to the decay mode $\bdtojk$.
Once again these decays are dominated by only one CKM amplitude
and are sensitive to the $\bs$ mixing phase, $\dga$.
Experimentally, the decay mode $\bstojphi$ is very clean,
but requires a full angular analysis to disentangle the mixture of CP 
eigenstates in the final state.
The reconstruction of the decay $\bstojeta$ 
benefits from good electromagnetic calorimetry and, 
because it is a transition into a CP eigenstate,
the $\dga$ can be extracted directly from the
decay asymmetry.
The expected sensitivities of the experiments for the extraction
of $\dga$ are summarised in Table~\ref{tab:summary} and depend strongly
on the proper time resolution and $\bs$ mixing parameter. 
It is interesting to note that the expected sensitivity for the extraction of
$\dga$ from $\bstojeta$ decays 
in one year of running (BTeV) is comparable to 5 years of 
running for the full angular analysis of $\bstojphi$ (LHCb).
It has been suggested that some of this difference can be recuperated 
using a transversity angle analysis~\cite{ref:dighe}. 
A particularly interesting feature of these decays is that
they exhibit very small CP violating effects within the Standard Model,
$\dga = \lambda^2\eta \approx 0.03$,
and hence are very sensitive to new physics 
(see~\cite{ref:yellow} and references therein).
%
\paragraph{\bf The angle $\al$}
%
\subparagraph{\bf $\bdtorhopi$: }
%
A full three-body analysis of the decay $\bdtopipipo$ in the $\rho$ resonance
region, 
taking into account interference effects between vector mesons of different charges, 
has been proposed to 
extract all parameters that describe both the tree and
penguin contributions to $\bdtorhopi$,
including the CKM angle $\al$~\cite{ref:snyder}.
Although the method is theoretically clean, 
it is experimentally difficult due to the need to reconstruct $\pi^0$'s and
the presence of a large combinatorial background.
The invariant mass for $\bdtorhopo$ candidates reconstructed in BTeV and
the $\pipipo$ Dalitz plot from LHCb are shown in Figure~\ref{fig:bdrhopi}.
The event yields,
given in Table~\ref{tab:summary},
are expected to be sufficient, so that an unambiguous value for $\al$ can be extracted. 
%
\subparagraph{\bf $\bdtopipi$: }
In principle, the decay mode $\bdtopipi$ 
allows the $\ckm$ angle $\al$ to be probed.
However, penguin contributions to the transition amplitude 
introduce a direct CP violation contribution,
$A^{dir}_{\cp}$,
into the decay asymmetry, 
\begin{eqnarray*}
\tasym = A^{dir}_{\cp}\cos(\delm t) + A^{mix}_{\cp}\sin (\delm t). 
\end{eqnarray*} 
Experimentally, the analysis of $\bdtopipi$ decays is complicated by the existence
of backgrounds with similar topologies and with unknown CP asymmetries,
such as $\bdtopik$, $\bstokk$, $\bstopik$, $\lbtoppi$ and $\lbtopk$. 
This is illustrated in Figure~\ref{fig:alpexp} (left) which shows the  
reconstructed two pion invariant mass distribution obtained in ATLAS.
In order to reject the non $\pipi$ background,
LHCb exploits the powerful RICH particle identification 
detectors,
shown in Figure~\ref{fig:alpexp} (centre).
The expected $\bdtopipi$ event yields
and the experimental sensitivities for the CP observables 
are given in Table~\ref{tab:summary}.
The observables can be written to leading order in the ratio 
of penguin to tree amplitudes, $\left|\peng / \tree \right|$,
\begin{eqnarray*}
A^{mix}_{\cp} =  -\sin 2\al - 2 \magpt\sin\al\cos 2\al\cos(\delta), \;\;\;
A^{dir}_{\cp} = 2 \magpt\sin\al \sin(\delta),            \nonumber
\end{eqnarray*} 
where $\delta$ is the CP conserving strong phase, $\delta = arg(PT^*)$.
Unfortunately, a theoretically reliable prediction for 
$ \left|\peng / \tree \right| $ 
which would allow the extraction of $\al$, 
albeit with a four-fold discrete ambiguity,
is very challenging. 
Figure~\ref{fig:alpexp} (right) shows the expected sensitivity 
to $\al$ for different values of the uncertainty on $ \left|\peng / \tree \right| $
and integrated luminosities.
It can be seen that for values of $\al$ around $90^{\circ}$,
the sensitivity to $\al$ is already limited after one year of running,
if the uncertainty on $\left|\peng / \tree \right| $ is not better than $10\%$.
%
\paragraph{\bf The weak decay phase $\ga$}
\subparagraph{\bf $\bdtopipi$ and $\bstokk$: }
A strategy has been proposed to combine the CP observables 
from the decay $\bdtopipi$ with those from $\bstokk$ to provide a 
simultaneous determination of the angles $\be$ and $\ga$~\cite{ref:fleischer1}.
The decays $\bdtopipi$ and $\bstokk$ are related to each other by
interchanging all down and strange quarks (U-spin symmetry).
Assuming the $\bs$ mixing phase $\dga$ is known, 
the four observables,
depend on four unknowns: two hadronic parameters, 
the $\bd$ mixing phase $\be$ and the weak phase $\ga$.
If $\be$ is also fixed from external measurements then the
weak phase, $\ga$,
can be extracted in a theoretically clean way. 
This approach is unaffected by penguin topologies and 
final state interaction effects and is only limited theoretically 
by U-spin breaking effects.
Moreover, since penguin processes play an important role in the decays
of $\bdtopipi$ and $\bstokk$,
this strategy is very promising with regard to the search for new physics. 
Experimentally, the experiments expect large statistics for both
$\bdtopipi$ and $\bstokk$ decays. 
The expected event yields and sensitivity for $\ga$ are given in Table~\ref{tab:summary},
where the range of sensitivity quoted reflects the dependence of the result on the $\bs$ mixing frequency.
%
\subparagraph{\bf $\bpmtodzkpm$ and $\bdtodzkst$: }
The decays $\bpmtodzkpm$ and $\bdtodzkst$ are pure tree decays and 
can be used to give a theoretically clean determination of the 
weak decay phase $\ga$.
The approach is through the measurement of six exclusive decay rates,
either $\bptodonekp$, $\bptodzbkp$ and $\bptodzkp + c.c.$;
or $\bdtodzbkst$, $\bdtodzkst$ and $\bdtodonekst + c.c.$
Here, $\done = ( \dz + \dzb )/\sqrt{2}$ is the CP even state and
leads to relationships between the decay amplitudes 
that can be used to extract $2\ga$ with a two-fold ambiguity.
Experimentally, the separation between the decay modes $\bptodzkp$ and $\bptodzbkp$
is extremely difficult as the $\dz$ and $\dzb$ decay to the 
same hadronic final state.
Also the decays of neutral D mesons into CP eigenstates,
such as $\done \ra \pipi, \kk$,
require excellent particle identification.
A method has been suggested to 
make use of the large interference effects by measuring at least 
two different final states of the $\dz$ and $\dzb$~\cite{ref:atwood}.
BTeV have investigated this approach and have studied the two decays
modes; 
$\dz \ra \mathrm K^+ \pi^-$ and $\dz \ra \kk$.
The annual event yield for this type of decay
assuming branching ratios of $10^{-7}-10^{-6}$ 
is given in Table~\ref{tab:summary}.
The expected sensitivity for $\ga$,
quoted in Table~\ref{tab:summary},
depends strongly on the strong phase and the
value of $\ga$ itself. 
LHCb,
benefiting from the hadron trigger and RICH particle identification,
have investigated the possibility to determine $\ga$
in the family of $\bdtodzkst$ decays and
have demonstrated that it will be possible to reconstruct samples
of such events. 
However, the visible branching ratios are very small,
$10^{-8}-10^{-7}$, 
leading to low event yields.
%
\subparagraph{\bf $\bdtodstpi$: }
The decays $\bdtodstpi$ are transitions into non-CP eigenstates and 
receive contributions only
from tree diagrams which lead to interference effects between the $\bd-\bdb$ mixing
and decay processes.
Measurements of the time-dependent asymmetries for the final states
$\mathrm D^- \pi^+$ and $\mathrm D^+ \pi^-$ lead to a measurement of $2\be+\ga$. 
LHCb have investigated the potential of measuring $\ga$ using $\bdtodstpi$ decays
where the $\dst$ decays strongly.
Two methods have been studied: a conventional exclusive reconstruction with
$\dzb \ra K^+ \pi^-$ and an inclusive approach.
For the inclusive approach,
the momentum of the $\bhad$ is reconstructed using the momenta of the pion 
coming directly from the $\bhad$, 
the momenta of the pion from the $\dst$ decay and the direction of the $\bhad$.
The expected error on $2\be+\ga$ depends strongly on the value of $2\be+\ga$
and is $>11^{\circ}$ for one year's data.
%
\subparagraph{\bf $\bstodsk$: }
The decays $\bstodsk$ are the $\bs$ counterparts to the $\bdtodstpi$ decay modes
and likewise receive only tree diagram contributions,
thereby probing the CKM angle combination $-2\dga+\ga$ in a theoretically clean manner.
The interference effects in $\bstodsk$ are much larger as
they are not doubly-Cabibbo suppressed as in the case of $\bdtodstpi$.
Experimentally, the event selection is challenging as $\bstodspi$ events,
which come with a 20 times larger branching ratio,
need to be rejected.
The LHCb and BTeV RICH detectors are therefore crucial to the analysis of this decay.
The precision with which the $\ckm$ phase combination can be measured after one year's
operation, 
given in Table~\ref{tab:summary},
depends strongly on the $\bs$ decay width difference and mixing frequency.  
%
\subparagraph{\bf $\btopik$: }
Due to the dominant role of
QCD penguins,
$\btopik$ decays potentially offer a determination of
the weak decay phase $\ga$ which is sensitive to new physics~\cite{ref:fleischer2}.
Experimentally, the strategy involving $\mathrm K^+\pi^-$ and $\kz \pi^+$ final states 
provides the cleanest channel.
However, the measurements require the knowledge of the trigger
and reconstruction efficiencies for the different final states and hence will incur 
an additional source of systematic error in contrast to most CP asymmetry measurements.
A value for $\ga$, with a four-fold ambiguity, 
can only be extracted once the ratio of tree to penguin decay amplitudes 
is determined from theory.
This is limited by rescattering effects and colour suppressed electroweak penguins.
A preliminary study by LHCb shows that a precision 
$\sim 2^{\circ}$ and $\sim 7^{\circ}$ for two of the solutions can be obtained 
if the ratio of tree to penguin amplitudes is $0.18 \pm 10\%$.
%
\subparagraph{\bf $\bhad_{d(s)} \ra \jk$,
$\bhad_{d(s)} \ra \mathrm D^+_{(s)} \mathrm D^-_{(s)}$: }
A strategy to extract the weak decay phase from 
$\bhad_{d(s)} \ra \jk$ and $\bhad_{d(s)} \ra \mathrm D^+_{(s)} \mathrm D^-_{(s)}$
decays has recently been proposed~\cite{ref:fleischer3}.
The method makes use of the U-spin symmetry of the decays and is 
sensitive to new physics due to the presence of penguins.
For example,
the decay $\bstojk$ is used to extract three observables
from the time-dependent CP asymmetry.
The ratio of the $\bstojk$ to $\bdtojk$ untagged time-integrated decay rates
provides a fourth observable.
These are then used to extract the four unknowns : two hadronic parameters,
$\ga$ and $\dga$.
Experimentally, although the $\jk$ final state is clean, 
the isolation of $\bstojk$ events is challenging due to a low event yield,
a large combinatorial background and the close $\bd$ mass peak.  
A study by CMS using an event selection tailored to $\bstojk$ decays
indicates that a measurement of the CP asymmetry in $\bstojk$ is 
feasible at the LHC and that $\ga$ can be measured with a precision
of $\sim 9^{\circ}$ in 3 years operation.
The strategy for extracting $\ga$ from 
$\bhad_{d(s)} \ra \mathrm D^+_{(s)} \mathrm D^-_{(s)}$
decays is analogous to the strategy for $\bhad_{d(s)} \ra \jk$ decays.
The time-dependent asymmetries are measured using $\bdtodd$ decays and
the overall $\bdtodd$ normalization is fixed through the CP-averaged $\bstodd$ decay rate.
The method benefits experimentally as it is neither necessary to resolve
the rapid $\bs$ oscillations, 
nor does flavour tagging reduce the already suppressed yield in $\bs$ events.
However, as the final state consists of six hadrons, 
a hadron trigger and RICH particle identification are vital for this analysis.
A very preliminary study by LHCb suggests that a precision on $\ga$ of a few
degrees should be achievable.
%
\paragraph{\bf Rare B decays}
Flavour-changing neutral current decays involving $b \ra s$ or
$b \ra d$ transitions occur only at a loop-level in the Standard Model
with very small branching ratios and 
therefore provide an excellent probe of indirect effects of new physics. 
The study of rare B decays at the next generation experiments is at a preliminary
stage and only a few remarks will be made here. 
A first assessment of the physics potential of these experiments shows
that it will be possible to:
\bi
\item observe $\bstomumu$ and measure its branching ratio ($\cal O$$(10^{-9})$ in the SM),
\item perform a high sensitivity search for $\bdtomumu$ (branching ratio $\cal O$$(10^{-10})$),
\item measure the branching ratio and decay characteristics of $\bdtokstga$,
\item measure the branching ratios of $\bdtokstmumu$, $\bdtorhomumu$ and $\bstophimumu$ and
study their decay kinematics; and
\item measure the forward-backward asymmetry in $\bdtokstmumu$ decays,
allowing the distinction between the Standard Model and a large class
of SUSY models.
\ei
The expected number of signal and background events 
at the LHC is given in Table~\ref{tab:summary}.
\begin{table}
\begin{footnotesize}
\begin{tabular}{|c|c||c|c|c|c||c|c|c|c|} \hline
Measurement            &Channel                 &\multicolumn{4}{|c||}{Event Yield}&\multicolumn{4}{|c|}{Sensitivity}\\
                       &                        &\btev &\lhcb &\atlas &\cms &\btev &\lhcb &\atlas &\cms  \\ \hline
\multicolumn{10}{|l|}{$\bd$ Mixing Phase}                                                      \\ \hline
$\sin 2\be$            &$\bdtojk$               &80.5k &88k   &165k   &433k &0.025 &0.021 &0.017  &0.015 \\ \hline
\multicolumn{10}{|l|}{$\bs$ Mixing Phase}                                                      \\ \hline
$\sin 2\dga$           &$\bstojphi (\dagger)$   &-     &370k  &300k   &600k &-     &0.03  &0.05   &0.03  \\ 
$\sin 2\dga$           &$\bstojeta$             &9.2k  & -    & -     & -   &0.033 & -    & -     & -    \\ \hline
\multicolumn{10}{|l|}{$\ckm$ Angle $\al$}                                                      \\ \hline
$\tasym$               &$\bdtopipi$             &23.7k &12.3k &2.3k   &0.9k &0.024 & -    & -     & -    \\
$A^{mix}_{\cp}$        &$\bdtopipi$             & -    & -    & -     & -   &-     &0.07  &0.21   &0.14  \\
$A^{dir}_{\cp}$        &$\bdtopipi$             & -    & -    & -     & -   &-     &0.09  &0.16   &0.11  \\ 
$\sin 2\al$            &$\bdtorhopi$            &10.8k &3.3k  & -     & -   &-     &$2.5^\circ$ & - & -  \\ \hline
\multicolumn{10}{|l|}{Weak Decay Phase}                                                        \\ \hline
$\ga$                  &$\bpmtodzkpm$           &2.1k  &-     & -     & -   &$2-10^\circ$ &-     &-      & -    \\
$\ga$                  &$\bdtodzkst$            &-     &0.4k  & -     & -   & -    & $\sim 10^{\circ}$ & - &- \\
$\ga$                  &$\bstokk$               & -    &4.6k  &0.54k  &0.96k &-  &\multicolumn{3}{|c|}{$4-8^\circ$}\\
                       &($\bdtopipi$)         &      &      &       &      &     &      &       &     \\ 
$2 \be +\ga$           &$\bdtodstpi$            & -    &340k  &-      &-     &-    & $>11^{\circ}$   & -     &      \\ 
$-2\dga+\ga$           &$\bstodsk$              &13.1k &6k    &-      &-     &$6-15^{\circ}$ & $2-12^{\circ}$ &- &- \\ \hline
\multicolumn{10}{|l|}{$\bs$ mixing}                                                            \\ \hline
$\xs (5\sigma)$        &$\bstodspi$             &103k  &86k        &3.5k   &4.5k  &75   &75    & 46    & 42   \\ \hline
\multicolumn{10}{|l|}{Rare B Decays $(\dagger)$}                                                           \\ \hline
S/B                    &$\bstomumu $            & -    &33/10      &27/93  &21/3  &     &      &       &      \\
S/B                    &$\bdtokstmumu $         & -    &22.4k/1.4k &2k/0.3k & - &  &      &       &      \\ \hline
\end{tabular}
\end{footnotesize}
\caption{
A comparison of the expected physics performance for one year of running 
for all experiments proposing to measure CP violating observables in the
B system.
The numbers quoted for channels marked with a $\dagger$ are for 5 years 
of data for LHCb and 3 years of data for ATLAS and CMS.
}
\label{tab:summary}
\end{table}
\subsubsection{\bf Concluding remarks}
%
The experimental search for the origin of CP violation has just 
entered an exciting new era.
If the Standard Model description of CP violation is correct,
then it is expected that CP violation in B decays will be discovered 
around the year 2001.
However, 
in order to thoroughly test the internal consistency
of the Standard Model and search for the necessary physics beyond,
CP violating observables in the B system must be ultimately
measured with very high statistics and in many different decay channels. 
The next generation of experiments will become operational around 2006 and
will be the ultimate source of CP violation studies 
in the B system for many years to come.   
%
%

%

\section{ {\bf CKM phenomenology and new physics}\\
 {\bf and what can we learn from $sin 2 \beta$?}}

\subsection{\bf Determination of the CP violating weak phase $\gamma$}~\footnote{The work  was supported in part by Seo-Am (SBS)
Foundation, in part by  BK21 Program, SRC Program and Grant No. 
2000-1-11100-003-1
of the KOSEF, and in part by the KRF Grants, £ Project No. 2000-015-DP0077.}

{\it C.S. Kim, Yonsei Univ.}

\def\beq{\begin{equation}}
\def\eeq{\end{equation}}
\def\ol{\overline}
\def\ul{\underline}
\def\o{\over}
\def\bmat{\begin{array}}
\def\emat{\end{array}}
\def\barr{\begin{eqnarray}}
\def\earr{\end{eqnarray}}
\def\h{\hbox}
\def\l{\left}
\def\r{\right}
\def\non{\nonumber}
\def\wt{\widetilde}
\def\lsim{\raise0.3ex\hbox{$\;<$\kern-0.75em\raise-1.1ex\hbox{$\sim\;$}}}
\def\gsim{\raise0.3ex\hbox{$\;>$\kern-0.75em\raise-1.1ex\hbox{$\sim\;$}}}
\def\mf{{\cal M}_F}
\def\md{{\cal M}_D}
\def\irad{{\cal I}_\tau}
\def\dmsq{\Delta m^2}
\def\gh{\Gamma^d_H}
\def\gl{\Gamma^d_L}
\def\dg{\Delta \Gamma_d}

\def\np#1#2#3{           {\it Nucl. Phys. }{\bf #1} (#2) #3}  
\def\pl#1#2#3{           {\it Phys. Lett. }{\bf #1} (#2) #3}  
\def\pr#1#2#3{           {\it Phys. Rev. }{\bf #1} (#2) #3} 
\def\prl#1#2#3{          {\it Phys. Rev. Lett. }{\bf #1} (#2) #3}


\subsubsection{\bf Introduction}
The source for CP violation in the Standard Model (SM) with \emph{three}
generations is a phase in the
Cabibbo-Kobayashi-Maskawa (CKM) matrix.  One of the goals of
$B$ factories is to test the SM through
measurements of the unitarity triangle of the CKM matrix.
An important way of verifying the CKM paradigm is to
measure the three angles \cite{2},
\begin{eqnarray}
\alpha &\equiv& \rm{Arg}[-(V_{td} V^*_{tb})/(V_{ud} V^*_{ub})], \nonumber\\
\beta &\equiv& \rm{Arg}[-(V_{cd}
V^*_{cb})/(V_{td} V^*_{tb})], \nonumber\\
{\rm and}~~~~ \gamma &\equiv& \rm{Arg}[-(V_{ud} V^*_{ub})/(V_{cd}
V^*_{cb})],
\end{eqnarray}
of the unitarity triangle independently of many experimental observables and
to check whether the sum of these three angles is equal\footnote{ The sum of
those three angles, defined as the intersections of three lines, would be
always equal to 180$^0$, even though the three lines may not be closed to make
a triangle.} to 180$^0$, as it should be in the paradigm.    It is well known that
among the three angles, $\gamma$ would be the most difficult to determine in
experiment.  There have been a lot of works to propose methods measuring
$\gamma$ using $B$ decays, but at present there is no gold-plated way to
determine this angle. In particular, a class of methods using $B \rightarrow D
K$ decays have been proposed \cite{7,9,10,11}. 

\subsubsection{\bf Methods to extract $\gamma$ }
In Ref. \cite{7}, Gronau, London, and Wyler (GLW) suggested a method
for extracting $\gamma$ from measurements of
the branching ratios of decays $B^{\pm} \rightarrow D^0 K^{\pm}$,
$B^{\pm} \rightarrow \bar D^0 K^{\pm}$ and
$B^{\pm} \rightarrow D_{CP} K^{\pm}$, where $D_{CP}$ is a CP eigenstate.
However, the GLW method suffers from
serious experimental difficulties, mainly because the process
$B^- \rightarrow \bar D^0 K^-$ (and its CP conjugate
process $B^+ \rightarrow D^0 K^+$) is difficult to measure in experiment.
That is, the rate for the CKM-- and
color--suppressed process $B^- \rightarrow \bar D^0 K^-$ is suppressed
by about two orders of magnitudes relative
to that for the CKM-- and color--allowed process $B^- \rightarrow D^0 K^-$,
and it causes experimental
difficulties in identifying $\bar D^0$ through
$\bar D^0 \rightarrow K^+ \pi^-$ since doubly Cabibbo--suppressed
$D^0 \rightarrow K^+ \pi^-$ following $B^- \rightarrow D^0 K^-$
strongly interferes with $\bar D^0 \rightarrow K^+
\pi^-$ following the rare process $B^- \rightarrow \bar D^0 K^-$.

To overcome these difficulties in the GLW method,
a few variant methods have been proposed.  The
Atwood-Dunietz-Soni (ADS) method \cite{9} uses the processes
$B^- \rightarrow D^0 (\bar D^0) K^- \rightarrow f K^-$
with two neutral $D$ decaying into final states $f$
that are not CP eigenstates, such as
$f= K^+ \pi^-, ~K \pi\pi$, {\it etc.}  In this method, large CP asymmetries
are possible  since magnitudes of the two interfering
amplitudes are comparable; {\it i.e.}, the process
$B^- \rightarrow D^0 K^- \rightarrow f K^-$ is CKM-- and
color--suppressed in $B$ decay, while the process
$B^- \rightarrow \bar D^0 K^- \rightarrow f K^-$ is doubly
Cabibbo--suppressed in $D$ decay.
The extraction of $\gamma$ can be allowed without  measurement of the
branching ratio for $B^- \rightarrow \bar D^0 K^-$.
Note that the decay amplitudes of
$B^- \rightarrow D^0 K^-$ and $D^0 \rightarrow f$
contain the CKM factors $V^*_{us} V_{cb}$ and $V^*_{cd} V_{us}$,
respectively, while the amplitudes of
$B^- \rightarrow \bar D^0 K^-$ and $\bar D^0 \rightarrow f$
contain the CKM factors $V^*_{cs} V_{ub} = |V^*_{cs} V_{ub}| e^{-i \gamma}$
and $V^*_{ud} V_{cs} = |V^*_{ud} V_{cs}|$, respectively.
We define the following quantities: $(i = 1,2)$
\begin{eqnarray}
a &=& A(B^- \rightarrow D^0 K^-) = |A(B^- \rightarrow D^0 K^-)|
   e^{i \delta_a},   \nonumber \\
b &=& A(B^- \rightarrow \bar D^0 K^-) = |A(B^- \rightarrow \bar D^0 K^-)|
   e^{-i \gamma }  e^{i \delta_b},  \nonumber \\
c_i &=& A(D^0 \rightarrow f_i) = |A(D^0 \rightarrow f_i)|
   e^{i \delta_{c_i}}, \nonumber \\
c_i^{\prime} &=& A(D^0 \rightarrow \bar f_i) = |A(D^0 \rightarrow \bar f_i)|
     e^{i \delta_{c^\prime_i}}, \nonumber \\
d_i &=& A(B^- \rightarrow  D^0 (\rightarrow f_i) K^-),
\end{eqnarray}
where $A$ denotes the relevant decay amplitude
and $\delta$'s are the relevant strong rescattering phases.  
Similarly, we also define
$\bar a$, $\bar b$, $\bar c_i$, $\bar c_i^{\prime}$ and $\bar d_i$
as the CP-conjugate decay amplitudes
corresponding to $a$, $b$, $c_i$, $c_i^{\prime}$ and $d_i$, respectively,
such as $\bar d_i = A(B^+ \rightarrow [\bar f_i] K^+)$,
{\it etc.} Here
$[f_i]$ in $d_i$ denotes that $f_i$ originates from a $D^0$ or
$\bar D^0$ decay.
Note that $|x| = |\bar x|$
with $x = a,b,c_i,c_i^{\prime}$, but in general
$|d_i| \neq |\bar d_i|$, as shown below.
Then, the amplitude $d_i$ can be written as
\begin{eqnarray}
d_i = |a c_i| e^{(\delta_a +\delta_{c_i})} +|b \bar c_i^{\prime}|
  e^{-i \gamma} e^{i (\delta_b +\delta_{c_i^{\prime}})},
\end{eqnarray}
which leads to
\begin{eqnarray}
|d_i|^2 = |a c_i|^2 +|b \bar c_i^{\prime}|^2 +2 |a b c_i \bar c_i^{\prime}|
   \cos(\gamma +\Delta_i), \nonumber \\
|\bar d_i|^2 = |a c_i|^2 +|b \bar c_i^{\prime}|^2
+2 |a b c_i \bar c_i^{\prime}| \cos(\gamma -\Delta_i),
\label{ddd}
\end{eqnarray}
where $\Delta_i = \delta_a -\delta_b +\delta_{c_i} -\delta_{c_i^{\prime}}$.
We see that $|d_i| \neq |\bar d_i|$,
unless $\Delta_i = n \pi$ ($n =0, 1, ...$).
Now the four equations for $i=1,2$ in Eq. (\ref{ddd})
contain the four unknowns $|b|$, $\gamma$, $\Delta_1$,
$\Delta_2$, assuming that the quantities $|a|$, $|c_i|$, $|c_i^{\prime}|$,
$|d_i|$ and $|\bar d_i|$ are known, but
$|b|$ is unknown. By solving the equations
one can determine $\gamma $, as well as the other unknowns such as
$|b|=|A(B^- \rightarrow \bar D^0 K^-)|$. 

In Ref. \cite{10}, Gronau
suggested a method to determine $\gamma$ using only the color--allowed
processes, $B^- \rightarrow D^0 K^-$ and $B^- \rightarrow D_{CP} K^-$, and
their CP-conjugate processes.

In Ref. \cite{11} two groups, Gronau and Rosner (GR), Jang and Ko (JK),
proposed a method to extract $\gamma$ by exploiting
Cabibbo--allowed decays $B \rightarrow D^{(*)} K$ and using the isospin
relations. 
In GR/JK method \cite{11}, the decay modes $B \rightarrow DK$
with the quark process $b \rightarrow u \bar c s$
contain the CKM factor $|V_{ub} V^*_{cs}| e^{-i \gamma}$
and their amplitudes can be written as
\begin{eqnarray}
A(B^- \rightarrow \bar D^0 K^-) &=& \left( {1 \over 2} A_1 e^{i \delta_1}
+{1 \over 2} A_0 e^{i \delta_0} \right) e^{-i \gamma }, \nonumber \\
A(B^- \rightarrow D^- \bar K^0) &=& \left( {1 \over 2} A_1 e^{i \delta_1}
-{1 \over 2} A_0 e^{i \delta_0} \right) e^{-i \gamma }, \nonumber \\
A(\bar B^0 \rightarrow \bar D^0 \bar K^0) &=& A_1 e^{i \delta_1}
     e^{-i \gamma },
\label{AAA}
\end{eqnarray}
where $A_i$ and $\delta_i$ denote the amplitude and the strong
re-scattering phase for the isospin $i$ state.
In this method, three triangles are drawn to extract
$2 \gamma$, using the isospin relation
\begin{eqnarray}
A(B^- \rightarrow \bar D^0 K^-) +A(B^- \rightarrow D^- \bar K^0)
 =A(\bar B^0 \rightarrow \bar D^0 \bar K^0)
\end{eqnarray}
and the following relations
\begin{eqnarray}
A(B^- \rightarrow D_1 K^-) &=& A(\bar B^0
\rightarrow D_1 \bar K^0) +{1 \over \sqrt{2}}
   A(\bar B^0 \rightarrow D^+ K^-),  \nonumber \\
A(B^+ \rightarrow D_1 K^+) &=& A(B^0 \rightarrow
D_1 K^0) +{1 \over \sqrt{2}} A(B^0 \rightarrow D^- K^+),
\end{eqnarray}
where $D_1$ is a CP eigenstate of $D$ meson, defined by
$D_1 = {1 \over \sqrt{2}} (D^0 + \bar D^0)$.

Recently  a new method has been presented by Kim and Oh (KO) \cite{12}, 
which is similar to the ADS method, but uses
$B \rightarrow D \pi$ decays instead of $B \rightarrow D K$ decays used in
the ADS method.   In fact, CLEO
Collaboration have observed \cite{17} that the branching ratio for
$B^- \rightarrow D^0 \pi^-$ is much larger than that for
$B^- \rightarrow D^0 K^-$,
\begin{eqnarray}
{{\cal B}(B^- \rightarrow D^0 K^-) \over {\cal
B}(B^- \rightarrow D^0 \pi^-)} = \rm{0.055} \pm \rm{0.014} \pm \rm{0.005} \; .
\end{eqnarray}
This new KO (Kim and Oh) method considers the decay processes
$B^- \rightarrow D^0 \pi^- \rightarrow f \pi^-$,
$B^- \rightarrow \bar D^0 \pi^- \rightarrow f \pi^-$
and their CP-conjugate processes, where $D^0$ and $\bar D^0$ decay into
common final states $f = K^+ \pi^-$, $K^+ \rho^-$, $K \pi\pi$, and so forth.
The mode $B^- \rightarrow \bar D^0 \pi^-$ is much suppressed relative to the
mode $B^- \rightarrow D^0 \pi^-$, and this fact causes serious experimental
difficulties in using $B^- \rightarrow \bar D^0 \pi^-$ decays for the
 GLW-type
method.  However, in KO method one needs not to perform the difficult task of
measuring the branching ratio for $B^- \rightarrow \bar D^0 \pi^-$,
similar to the case of  the ADS method.
Note that the decay amplitudes of
$B^- \rightarrow D^0 \pi^-$ and $D^0 \rightarrow f$
contain the CKM factors $V^*_{ud} V_{cb}$ and $V^*_{cd} V_{us}$,
respectively, while the amplitudes of
$B^- \rightarrow \bar D^0 \pi^-$ and $\bar D^0 \rightarrow f$
contain the CKM factors $V^*_{cd} V_{ub} = |V^*_{cd} V_{ub}| e^{-i \gamma}$
and $V^*_{ud} V_{cs} = |V^*_{ud} V_{cs}|$, respectively: $(i = 1,2)$
\begin{eqnarray}
a &=& A(B^-\rightarrow D^0 \pi^-) =
   |A(B^- \rightarrow D^0 \pi^-)| e^{i \delta_a},   \nonumber \\
b &=& A(B^- \rightarrow \bar D^0 \pi^-) = |A(B^- \rightarrow \bar D^0 \pi^-)|
   e^{-i \gamma } e^{i \delta_b}, \nonumber \\
d_i &=& A(B^- \rightarrow  D^0 (\rightarrow f_i)\pi^-). \label{abcd}
\end{eqnarray}
Then, the amplitude $d_i$ can be written as
\begin{eqnarray}
d_i &=& A(B^- \rightarrow D^0 \pi^-) A(D^0 \rightarrow f_i)
  +A(B^- \rightarrow \bar D^0 \pi^-) A(\bar D^0 \rightarrow f_i) \nonumber\\
&=& a c_i + b \bar c_i^{\prime}  \nonumber \\
&=& |a c_i| e^{(\delta_a +\delta_{c_i})} +|b \bar c_i^{\prime}|
e^{-i \gamma}  e^{i (\delta_b +\delta_{c_i^{\prime}})}.
\label{di}
\end{eqnarray}
Thus, $|d_i|^2$ and $|\bar d_i|^2$ are given by
\begin{eqnarray}
|d_i|^2 = |a c_i|^2 +|b \bar c_i^{\prime}|^2
+2 |a b c_i \bar c_i^{\prime}| \cos(\gamma  +\Delta_i),
\nonumber \\ |\bar d_i|^2 = |a c_i|^2 +|b \bar c_i^{\prime}|^2
+2 |a b c_i \bar c_i^{\prime}|  \cos(\gamma -\Delta_i).
\label{dd}
\end{eqnarray}
The expressions in Eq. (\ref{dd}) represent four
equations for $i =1,2$. Now let us assume that the quantities $|a|$, $|c_i|$,
$|c_i^{\prime}|$, $|d_i|$ and $|\bar d_i|$ are measured by experiment, but
$|b|$ is unknown. Then there are the four unknowns $|b|$, $\gamma $,
$\Delta_1$, $\Delta_2$ in the above four equations.  By solving the equations
one can determine $\gamma $, as well as the other unknowns such as
$|b|=|A(B^- \rightarrow \bar D^0 \pi^-)|$. 

\subsubsection{\bf Experimental considerations}
Now we study the experimental feasibility of the ADS and KO method, 
by solving Eqs. (4,11) analytically,
\begin{equation}
cos(\gamma \pm \Delta_i)={{|d_i|^2 - |a c_i|^2 - |b \bar c_i^\prime|^2} \over
 {2 |a c_i b \bar c_i^\prime|}} . \nonumber
\end{equation}
To make a rough numerical estimate of possible statistical error on $\gamma$,
we use the following experimental result:
\begin{eqnarray}
BR(B^- \to D^0 \pi^-) &>& BR(~\to D^0 K^-) > BR(~\to \bar D^0 K^-)
 > BR(~\to \bar D^0 \pi^-) \nonumber \\
 &\propto& ~~{\cal O}(100) ~:~ {\cal O}(10) ~:~ {\cal O}(1) ~:~ {\cal O}(0.1) .
\nonumber 
\end{eqnarray} 
Therefore, if we assume the precision of 1 \% level experimental determination
for the product of branching  ratios,
$BR(B^- \to D^0 \pi^-) \times BR(D^0 \to K^- \pi^+)$, then we roughly get
\begin{eqnarray}
|a c_i|^2(\pi)=100 \pm 1, &{}&~|b \bar c_i^\prime|^2(\pi)=0.1\pm 0.03 \nonumber\\
|a c_i|^2(K)=10 \pm 0.3,  &{}&~|b \bar c_i^\prime|^2(K)=1\pm 0.1 . \nonumber
\end{eqnarray} 
Then,  we can estimate the statistical error as
\begin{eqnarray}
\Delta[cos(\gamma \pm \Delta)(B^\pm \to D(\to f)\pi^\pm)] &\sim& 0.3 ,
\nonumber \\ 
\Delta[cos(\gamma \pm \Delta)(B^\pm \to D(\to f)K^\pm)] &\sim& 0.15 .
\nonumber   
\end{eqnarray} 
Even though the ADS method can give approximately twice better precision
statistically for determination of $\gamma$ than KO method, however,
KO method can have other advantages:
\begin{itemize}
\item
The value of $|d_i|^2 \propto BR(B^- \to [f_i]\pi^-)$ is an order of magnitude
bigger than $|d_i|^2 \propto BR(B^- \to [f_i]K^-)$. Therefore, if B-factories
can produce only handful of such events, KO method can be only the possible
option. 
\item
Systematic errors could be much smaller for KO method due to final state
particle identification.
\end{itemize}

We can extend the GLW method to $B_c$ decay \cite{18} from the relations,
\begin{eqnarray}
\sqrt{2} A(B_c^+ \to D_s^+ D^0_1) &=& 
A(B_c^+ \to D_s^+ D^0) + A(B_c^+ \to D_s^+ \bar D^0),  \nonumber \\
\sqrt{2} A(B_c^- \to D_s^- D^0_1) &=& 
A(B_c^- \to D_s^- \bar D^0) + A(B_c^- \to D_s^-  D^0),
\end{eqnarray}
and we can obtain $\gamma$ from
$A(B_c^+ \to D_s^+ D^0) = e^{2 i \gamma} A(B_c^- \to D_s^- \bar D^0)$.
Here the advantage is that the amplitude with the rather small CKM element
$V_{ub}$ is not color suppressed, while the larger element $V_{cb}$
comes with a color suppression factor. Therefore, the two amplitudes are
similar in size,  as
$$
A(B_c^+ \to D_s^+ D^0)~:~ A(B_c^+ \to D_s^+ \bar D^0)
\propto \lambda^3/2 ~:~ \lambda^3/N_c,
$$
because of $|V_{ub}/V_{cb}| \approx \lambda/2$.

\subsubsection{\bf $\gamma$ and new physics}
Finally, I would to make a short note on new physics effects on determination
of weak phase $\gamma$. There can be two independent approach to find out
new physics beyond the Standard Model, if it exists.
\begin{itemize}
\item
We can assume the unitarity of CKM matrix. In this case, new physics effects 
can only come out from new virtual particles  or through new interactions
in penguin or box diagrams in $B$ meson decays. If this is the case,
all the methods, which I described, will give the exactly same $\gamma$.
\item
We can generalize CKM matrix to the non-unitary matrix.
In this case, new physics effects can appear even in tree diagram decays.
And the values of $\gamma$ extracted from each method can be different.
Therefore, I will describe in more detail for this second case.
\end{itemize}
In fact, in models beyond
the SM, the CKM matrix may not be unitary; for instance, in a model
with an extra down quark singlet (or more than
one), or an extra up quark singlet, the CKM matrix is no longer
unitary \cite{16}.
If the unitarity constraint of the CKM
matrix is removed, the generalized CKM matrix possesses 13
independent parameters (after absorbing 5 phases to quark fields) -- it
consists of 9 real parameters and  \emph{4 independent phase angles}.
The generalized CKM matrix can be parametrized as \cite{19}
\begin{eqnarray}
\left( \matrix{ |V_{ud}| & |V_{us}| & |V_{ub}|e^{i \delta_{13}} \cr
                |V_{cd}| & |V_{cs}|e^{i \delta_{22}} & |V_{cb}| \cr
|V_{td}|e^{i \delta_{31}} & |V_{ts}| & |V_{tb}|e^{i \delta_{33}} \cr } \right).
\label{ckm}
\end{eqnarray}
Then, the GLW method would measure the angle $(\gamma - \theta)$, where
\begin{eqnarray}
\gamma \equiv -\delta_{13}~~~~{\rm and}~~~~ \theta \equiv \delta_{22},
\end{eqnarray}
instead of $\gamma$.
The ADS method would still measure $\gamma$, but GR/JK method would measure
$(\gamma -\theta)$ rather than $\gamma$, and KO method would measure
$(\gamma +\theta)$

\subsection{\bf The unitarity triangle and new physics}

{\it Paolo Gambino, CERN}


\subsubsection{\bf Introduction}
The determination of Cabibbo--Kobayashi--Maskawa matrix has  
enormously improved 
in the last few years \cite{stocchi}. All data agree remarkably
with the Standard Model (SM) within experimental
and theoretical errors. Only the recent direct measurement of $\sin
2\beta$ from $B_d^0\to \psi K_S$ at BaBar and Belle
\cite{babar,belle} seem to show a mild discrepancy with SM
expectations and to hint to new physics. In the case these  results are
confirmed, the main theoretical interest will be in discriminating
among different new physics scenarios. It is therefore important to
have a clear picture of the ways the experimental determination of the
CKM matrix would be affected by physics beyond the SM.

Let us briefly recall the standard way of determining the CKM matrix
in the Wolfenstein parameterization.  $\lambda$ and $A$ 
are determined from semileptonic K and B
decays sensitive to the elements $|V_{us}|$ and $|V_{cb}|$
respectively; as these decays are tree level processes, 
this determination is to an excellent approximation
independent of new physics. 
$\bar\varrho$ and $\bar\eta$ are determined by constructing
from various decays the unitarity triangle.

The standard construction of this triangle involves the ratio
$|V_{ub}/V_{cb}|$ extracted from inclusive and exclusive tree level B decays
and flavour changing neutral current processes such as $B^0_d-\bar B^0_d$
mixing (the mass difference $(\Delta M)_d$) and indirect CP violation 
in $K_L$ decays (the parameter $\varepsilon$), both sensitive to the
CKM element $V_{td}$. There is also a constraint coming from the
lower bound on the mass difference $(\Delta M)_s$ describing 
$B^0_s-\bar B^0_s$ mixing. In the case of 
$B^0_{d,s}-\bar B^0_{d,s}$ mixings $(\Delta M)_{d,s}$ 
is given by \cite{lectures}
\begin{equation}
(\Delta M)_{d,s} = \frac{G_F^2}{6 \pi^2} \eta_B m_{B_{d,s}} 
\hat B_{B_{d,s}} F_{B_{d,s}}^2 
 M_W^2 F_{tt} |V_{t(d,s)}|^2
\label{eq:xds}
\end{equation}
Here $F_{tt}$ is a function of $m_t$ and $M_W$
 resulting from box diagrams with top
quark exchanges,
$\hat B_B$ is a non-perturbative parameter, $F_B$ is
the B meson decay constant and 
$\eta_B$ the short distance QCD factor  common
to $(\Delta M)_{d}$ and $(\Delta M)_{s}$.
Similarly, the experimental value for $\varepsilon$ combined
with the theoretical calculation of box diagrams describing 
$K^0-\bar K^0$ mixing gives the constraint for
$(\bar\varrho,\bar\eta)$ in the form of the following 
hyperbola \cite{lectures}:
\begin{equation}\label{100}
\bar\eta \left[(1-\bar\varrho) A^2 \eta_2 F_{tt}
+ P_c(\varepsilon) \right] A^2 \hat B_K = 0.226~.
\end{equation}
Here $\hat B_K$ is a  non-perturbative parameter analogous to
$\hat B_{B_{d,s}}$, $\eta_2$ is a short distance
QCD correction, $F_{tt}$ is the same function
present in (\ref{eq:xds}), and
$P_c(\varepsilon) =0.31\pm0.05$   summarizes 
charm--charm and charm--top contributions.

Combining  these and other observables which will become available in 
the future in a global fit,  one determines the
range of values of $(\bar\varrho,\bar\eta)$ consistent with the
data \cite{stocchi}.  If the SM is correct 
all these measurements will result in a unique value of
$(\bar\varrho,\bar\eta)$.

This procedure of testing the SM can be applied to its extensions
as well.  The determination of the unitarity triangle
 is affected by new physics in the following ways:
\begin{enumerate}

\item New contributions to $F_{tt}$ and to similar
short distance functions entering rare decays
that in the SM depend only on $m_t$ and $M_W$. These new contributions
depend on masses and couplings of the new particles.

\item New contributions  that are {\it
    not}  proportional to the same combination of CKM  elements as the
  SM top contribution (disruption of the GIM cancellations). This
  means, e.g., new contributions to $P_c$ in
  eq.~(\ref{100}) or new contributions to $(\Delta M)_{d,s}$
  proportional to $|V_{cd(s)}^*V_{cb}|^2$.  

\item New phases beyond the CKM one.
 For instance, the CP asymmetry in $B\to \psi K_S$ will no longer measure
$\beta$ but $\beta+\theta_{NP}$ where $\theta_{NP}$ is a new phase.

\item New local operators contributing to the relevant amplitudes
beyond those present in the SM, e.g., with different chirality. 
This would introduce additional
non-perturbative factors $B_i$ and new box and penguin functions.
\end{enumerate}
 
It is clear from (\ref{eq:xds}) and (\ref{100}) that any
modification of the function $F_{tt}$ will change the values of the
extracted $(\bar\varrho,\bar\eta)$. A recent analysis of this type in
the MSSM can be found in \cite{ALI00}.  
  The presence of new
physics and of new phases will be signaled by inconsistencies in the
$(\bar\varrho,\bar\eta)$ plane when observables are calculated in the SM.

\subsubsection{\bf The universal unitarity triangle}
While in principle a global fit of all experimental data
can be used to test the SM and its
extensions it is desirable to develop strategies which allow to make
these tests in a transparent manner.
In order to sort out which kind of new physics is responsible for the 
possible inconsistencies between different observables, 
it is useful to introduce \cite{noi} a broad class of
models very similar to the SM, 
 that do not have any new operators beyond those present in the SM
 and in which all flavour changing transitions are governed
 by the CKM matrix and there is no phase other than the CKM phase.
 Furthermore, we ask that 
in these models the only sizable new contributions be
 proportional to the same CKM parameters as the SM top contributions.

This class of models represent the slightest modification of the
SM, or the most drastic simplification among new physics scenarios. 
It includes the  Two Higgs Doublet models  I and II and the 
MSSM with minimal flavour and CP
violation (MFV) --- see \cite{mfv}.
The kind of new physics it characterizes affects the determination of the
unitarity triangle only through (i) above, namely through the function
$F_{tt}$. Therefore, the CKM parameters 
extracted  from a set of data independent of the 
loop functions like $F_{tt}$ are universal in this class of models. 
Correspondingly,   there exists a {\it universal unitarity triangle}
(UUT) \cite{noi}. In fact, there are as many UUT as 
are the sets of observables meeting the above requirement. 

For example, from (\ref{eq:xds}) one finds that the ratio
\begin{equation}\label{107x}
\frac{|V_{td}|}{|V_{ts}|}= 
\xi\sqrt{\frac{m_{B_s}}{m_{B_d}}}
\sqrt{\frac{(\Delta M)_d}{(\Delta M)_s}}\equiv\kappa,
\ \ \ \xi = 
\frac{F_{B_s} \sqrt{\hat B_{B_s}}}{F_{B_d} \sqrt{\hat B_{B_d}}}.
\end{equation}
depends only on the measurable quantities 
$(\Delta M)_{d,s}$, $m_{B_{d,s}}$ and the non-perturbative parameter
$\xi$. To good  accuracy this fixes one of the two unconstrained sides of the
unitarity triangle,  
$R_t={\kappa}/{\lambda}$,
independently of new parameters characteristic for a given model. 
There are other quantities which may allow
a clean measurement of $R_t$ within our class of extensions of the SM,
like  the ratios of branching ratios for $B\to X_d\nu\bar\nu$ 
over $B\to X_s\nu\bar\nu$ \cite{noi}.
Notice that the hadronic uncertainties in (\ref{107x}) are only due to 
$SU(3)$ breaking effects, which should be eventually known well from
lattice simulations. 

Having measured $R_t$ from (\ref{107x}), in
 order to complete the determination of  the UUT one can use 
$\sin2\beta$ extracted either from the CP asymmetry
in $B_d\to\psi K_S$ \cite{babar,belle} or from $K\to\pi\nu\bar\nu$ decays 
\cite{bb}. 
Both extractions of $\sin 2\beta$  are  to an excellent accuracy
independent of the  new physics parameters and 
hadronic uncertainties  have been found
to be negligibly small \cite{RS}. 
An alternative to $\sin 2 \beta $ is represented by a measurement of
$\gamma$ from tree level decays  (see \cite{ROB00} and refs.\ therein). 
Another possibility is to use the measurement
of $\sqrt{\bar\varrho^2+\bar\eta^2}$ by means of $|V_{ub}|/|V_{cb}|$ but this
strategy suffers from much larger hadronic uncertainties.

In our class of models all these different methods
determine the   values of 
$\bar\eta$ and $\bar\varrho$ independently of the parameters of new
physics. Using them,
one can calculate $\varepsilon$, $\varepsilon'/\varepsilon$,
 $(\Delta M)_d$, $(\Delta M)_s$ and the BRs for
rare decays \cite{noieps}.
 As these quantities depend on the parameters characteristic
for a given model, the results for the SM, the MSSM and other models
of this class will generally differ from
each other. Comparison with data will then single out models within
our class or exclude them all, in the case new physics goes beyond the
simple paradigm we have considered.

\subsubsection{\bf Lower bound on $\sin 2\beta$}
An interesting application of the strategy described so far is the 
derivation of a lower bound on   $\sin 2\beta$
in the models of our class \cite{burasburas}.
Direct measurements of $\sin 2\beta$ from $B_d^0\to \psi K_S$
lead to $\sin 2\beta=0.42\pm 0.24$ \cite{babar,belle}, which is 
only marginally consistent with the SM fit \cite{stocchi}. 
Although it is certainly too soon to conclude that a signal of new physics
has been detected, it is intriguing to speculate what kind of new
physics could be responsible for the low value of $\sin 2\beta$
\cite{new}. Two obvious possibility would be $(a)$ a new CP violating phase
and/or $(b)$ a modification of the function $F_{tt}$ entering $\varepsilon$ and
$\Delta M_{d,s}$. 
In our  class of models only the second possibility is open and we can
parametrize the new physics contribution by 
$F_{tt}=F_{tt}^{SM} (1+f)$, where $f$ represents the
variation relative to the SM. 
From $\Delta M_{d}$ and $\varepsilon$ given in (\ref{eq:xds},\ref{100})
we can determine $\sin 2\beta$ indirectly and in model dependent way
\cite{burasburas}: 
\begin{equation}
\sin 2\beta=\frac{1.26}{R_0^2\,\eta_2}\left[\frac{0.226}{A^2 \,\hat{B}_K}-
\bar{\eta}\, P_c(\varepsilon)\right],
\end{equation}
where $R_0$ depends only on $\Delta M_{d}$ and relative hadronic
parameters, $\eta_2$ is a short distance QCD correction factor which
depends on new physics only at NLO.  A mild dependence on new
physics comes mostly in an indirect way 
 through $\bar{\eta}$. The unitarity of the CKM matrix alone places a
 constraint on $f$, $-0.4<f<5.5$. However, in a minimal
supergravity scenario with universal soft terms (which belongs to
our class of models) $0<f<0.75$ \cite{ALI00}, while in a generic MSSM model
with MFV one finds $0<f<1.13$ \cite{noieps}. Using this information together
with the present ranges for the hadronic parameters, it is possible to
derive lower bounds on 
$\sin 2\beta$ for the  various scenarios \cite{burasburas}:
in minimal supergravity $\sin 2\beta>0.53$, in the MSSM
with MFV $\sin 2\beta>0.40$ and --- for a generic  model 
of our class --- $\sin 2\beta>0.34$. Assuming a reduction by a factor
2 of all theoretical errors, the latter absolute bound becomes much more
stringent, $\sin 2\beta>048$.

\subsubsection{\bf Conclusions}%

In summary, the UUT provides 
a transparent strategy to distinguish between models belonging to
the class we have considered  and to  search for physics
beyond the SM. It  usefully decouples the determination of the CKM
matrix from that of  new physics parameters, in a
way  essentially free from hadronic uncertainties.
A very simple and clear application  of this strategy leads to the
derivation of a stringent lower bound on  $\sin 2 \beta$ in a broad
class of models. If the preliminary result of BaBar, 
$\sin 2 \beta=0.12 \pm 0.38$, were confirmed
with smaller errors, the whole class of models with MFV would be ruled out.

\subsection{\bf $CP$-violation, the CKM matrix and new physics}

{\it D. Wyler, University of Zurich}

\subsubsection{\bf Introduction}

Observation of novel phenomena often paves the way to new physics. For 
instance, $\beta$
decays, parity and flavor violation required
the existence of a new force, the weak interactions. 
At present, it is often thought that $CP$-violation could signal new
physics beyond the standard model. Although the latter can indeed
account for the observed effects \footnote{a notable exception is the
baryon asymmetry in the universe} (even $\epsilon'/\epsilon$
may be described by the standard model) 
its predictions are not well tested (compared to physics
at LEP) and therefore a comprehensive study of $CP$-violation experiments
is important. As sketched in figure~1, $CP$-violation manifests itself in
many areas; only a comparison between them 
can determine the correct description.
\begin{figure}[thb]
\vspace{0.8cm}
\leavevmode
\begin{center}
\includegraphics[width=10cm]{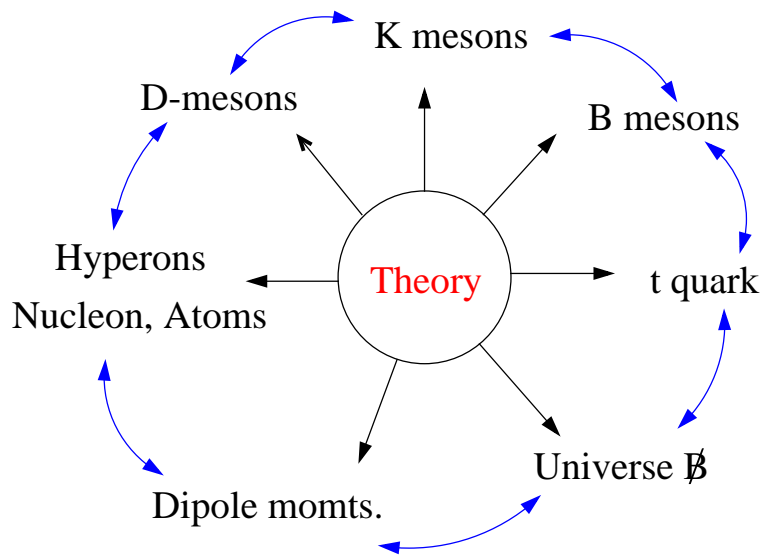}
\end{center}
\caption{\label{f1}CP-Violation}
\end{figure}   
In the standard model, all $CP$-violation resides in the CKM matrix
\footnote{I do not discuss the so-called $\theta$ term} which
describes the couplings of the W-bosons to the quarks of different
charges. Thus all appreciable $CP$-violation occurs
within flavor physics. Thus, one obvious strategy to search for new
forces and particles would be to look for non-zero $CP$-violating
effects in electric dipole
moments or asymmetries in nuclear reactions. Unfortunately, the effects of
new physics are judged to be quite small (apart from the dipole
moments). Therefore more chance is given to the flavor sector instead, 
that is the physics of
Kaons and mostly B-mesons. For a recent extensive review of $CP$-violation,
see ref. (\cite{nir}).

The unitarity of the CKM matrix 
\begin{equation}\label{2.67}
V =
\left(\begin{array}{ccc}
V_{ud}&V_{us}&V_{ub}\\
V_{cd}&V_{cs}&V_{cb}\\
V_{td}&V_{ts}&V_{tb}
\end{array}\right)
\end{equation}
implies among others the triangle relation
\begin{equation}\label{2.87h}
V_{ud}^{}V_{ub}^* + V_{cd}^{}V_{cb}^* + V_{td}^{}V_{tb}^* =0
\end{equation}
which relates observable products of matrix elements and gives stringent
tests of the validity of the standard model. Using the Wolfenstein
parameterization, and scaling as usual the bottom side to one, 
we can write for the other sides of the scaled triangle
\begin{equation}\label{2.88b}
R_b = \frac{1}{A\lambda^3}V_{ud}^{}V_{ub}^*
=\bar\varrho+i\bar\eta
\qquad,
\qquad
R_t = \frac{1}{A\lambda^3}V_{td}^{}V_{tb}^*
=1-(\bar\varrho+i\bar\eta).
\end{equation}
Here, following ref. \cite{BLO}, the quantities
\begin{equation}
\bar\varrho = \varrho(1 - \frac{\lambda^2}{2})\,\,\,
\bar\eta = \eta(1 - \frac{\lambda^2}{2})
\end{equation}
are introduced to take into account even higher powers of $\lambda$.
\begin{figure}[thb]
\vspace{0.8cm}
\leavevmode
\begin{center}
\includegraphics[width=10cm]{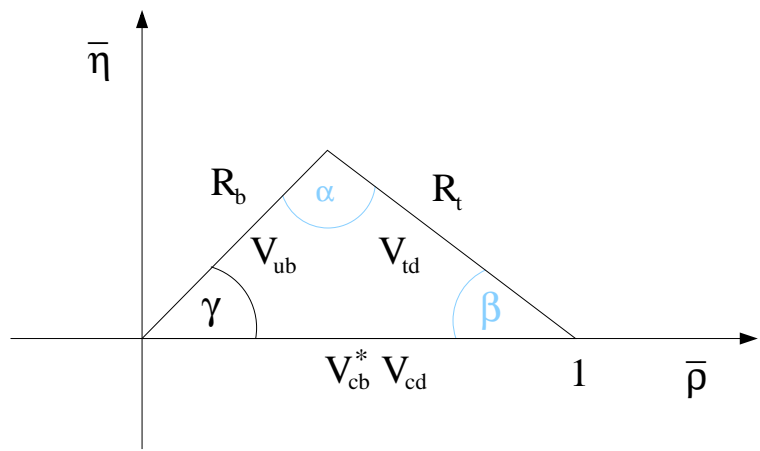}
\end{center}
\caption{\label{f2}Unitarity triangle in the complex $(\bar\varrho,\bar\eta)$ plane.  }
\end{figure}     

An elaborate analysis of superallowed $\beta$ decay, semileptonic
Kaon and $D$-meson decays and decays of $B$ mesons into charmed and charmless
final states yields \cite{pdg2}
\begin{equation}
\begin{array}{cc}
V_{ud}= 0.9736 \pm 0.001&V_{cs}= 1.010 \pm 0.16\\
V_{us}= 0.2205 \pm 0.0018&V_{cd}= 0.224\pm 0.016\\
V_{ub}= 0.04  \pm 0.002&V_{cb}= 0.0036\pm 0.006\\
\end{array}
\end{equation}

These are (apart from corrections) all tree-level processes and therefore
thought to be governed by the standard model \footnote{of course, the small
$b \to u$ transition could be due to new physics}. They are however not
sufficient to check unitarity (unless very precise data from $t$ decays
would be available, or if the sum of the squares would be significantly
away from $1$). 

Further input comes from loop-induced observables. They can be
calculated within perturbation theory and input from hadronic
physics. While the former are rather reliable and usually give
results accurate to 10 percent or so, the latter are generally
difficult to estimate. One usually considers the Kaon-mixing
quantity $\epsilon_K$, the mass difference of the $B$ and the
$\bar B$ mesons (and also of the $B_s$ and $\bar B_s$ mesons).
This analysis has resulted in the range of values for the three
angles $\alpha$, $\beta$ and $\gamma$ of the unitary
triangle and its sides.
The hadronic uncertainties are summarized in \cite{hadro} and are
reflected by
\begin{equation}
|R_b| = 0.39 \pm 0.07~~~~~~|R_t| = 0.98 +0.04 -0.22
\end{equation}
and by \cite{Stocchi,ALI002,SCHUNE}
\begin{equation}
\label{sinth}
(\sin 2\beta)_{\rm SM} =
0.75\pm0.20.
\end{equation}
The new results this summer concern the angle $\beta$. It was found that
the coefficient a of $sin(\Delta M_{B_d})$ in the asymmetry for $B \to
J/\Psi K_S$ is 
\begin{eqnarray}
a = 0.79 \pm 0.4 (CDF)\cite{CDFB}\\
a = 0.45 \pm 0.4 (Belle)\cite{Belle}\\
a = 0.12 \pm 0.4 (BaBar)\cite{BaBar}
\end{eqnarray}

In the standard model, one has $a=sin(2\beta)$; comparing
eqs. (\ref{sinth}) and (10) we see a surprising inconsistency.
Of course, this is a preliminary
result, and may disappear as experiments collect more statistics.
However, it makes it mandatory to investigate $CP$-violation in
a (standard) model independent way. Unless $CP$-violation within the 
standard model is grossly wrong, this program essentially amounts
to making many measurements and extracting discrepancies between
quantities thought to be the same in the standard model. Many
authors have discussed this situation; see e.g. \cite{NIR00, SW, NK00, XING,
bubu}.

\subsubsection{\bf A more general framework}

New physics may affect every process. Because the standard model
describes the most important weak decays, we will assume that it 
accounts for semileptonic and tree-level quark
decays, at least to the required accuracy. This assumption can
be tested, by investigating the consistency of different semileptonic
decays, bounds from LEP etc. As an example consider the strengths of the
effective Hamiltonians
\begin{eqnarray}
\label{forf}
{\cal H}_{eff} = 
G_F (\bar c_L \gamma_{\mu}b_L)(\bar s_L \gamma^{\mu}c_L)\\
{\cal H}_{eff} =
G_F (\bar u_L \gamma_{\mu}b_L)(\bar s_L \gamma^{\mu}u_L).
\end{eqnarray}
In the standard model,  they are 
difficult to estimate. One usually considers the Kaon-mixing
quantity $\epsilon_K$, the mass difference of the $B$ and the  
$\bar B$ mesons (and also of the $B_s$ and $\bar B_s$ mesons).
On the other hand, a new
neutral intermediate boson, say $Z'$, may exist, coupled to the current
$(\bar s_L \gamma_{\mu}b_L)$. and $(\bar c_L \gamma_{\mu}c_L)$. 
If it also couples to quark and lepton pairs, such as
$(\bar u_L \gamma_{\mu}u_L)$ and $(\bar c_L \gamma_{\mu}c_L)$, it would 
contribute to the above interaction, to $B_s$ mixing,
to $B_s \to l^+ l^-$ etc. If the couplings are the same for all
these pairs, the effective strength would be the same for the two
terms in eqs. (\ref{forf}) and (12).
Therefore a new $Z'$-mediated interaction would induce a deviation
from the standard model result that the couplings of the two 
interactions have a relative strength of $\lambda^2$. 
Thus detailed studies could in principle also test the first assumption.
But of course, there are various experimental and theoretical
difficulties to overcome before one will obtain accurate enough results.

\begin{figure}[thb]
\vspace{0.8cm}
\leavevmode
\begin{center}
\includegraphics[width=10cm]{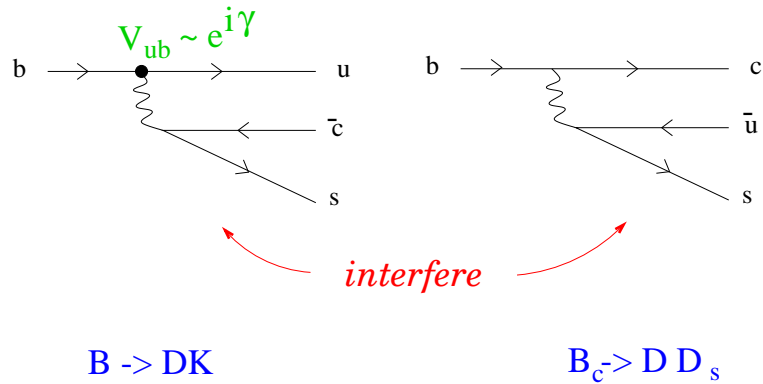}
\end{center}
\caption{\label{}two quark diagrams whose interference gives $\gamma$}
\end{figure}      

From fig. 3 we see that the determination of the angle $\gamma$ from
tree level processes involves the interference of amplitudes proportional
to $V_{ub}$ and $V_{ub}$ respectively. This is achieved in processes
where the two diagrams of fig. 3 contribute. A well known example are
the decays $B \to D K$ \cite{gw, ads}; more recently the advantage 
of $B_c \to D D_s$
was stressed \cite{flwy}. The idea is the same as in the previous
papers on $B \to D K$ : One needs to measure the six
amplitudes shown in Fig. 4.                                                                One usually considers the Kaon-mixing
quantity $\epsilon_K$, the mass difference of the $B$ and the
$\bar B$ mesons (and also of the $B_s$ and $\bar B_s$ mesons).
the sides of the triangles in Fig. 3 are of similar length and an extraction of
$\gamma$ seems possible with the $10^{10}$ or so $B_c$-mesons expected
at LHC. This method does not suffer from hadronic uncertainties.
\begin{figure}
\vspace*{-1.8truecm}
\centerline{
\epsfysize=25.5truecm
\epsffile{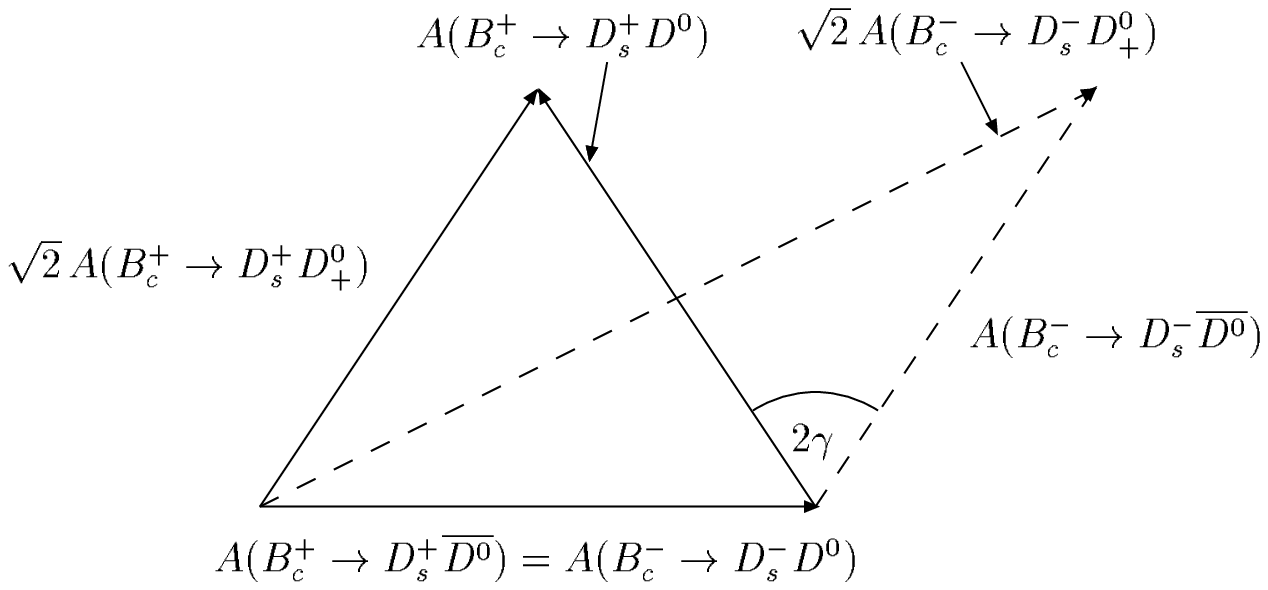}
}
\vspace*{-17.7truecm}
\caption{The extraction of $\gamma$ from
$B_c^\pm\to D^\pm_s\{D^0,\overline{D^0},D^0_+\}$ decays.}
\end{figure}
           
The experimental difficulties associated with these decays 
have lead to other possibilities. The decays $B \to K \pi$ are
sensitive to the interference
of the tree level diagram (with $V_{ub}$) and the penguin diagram. This
also yields
the angle $\gamma$ if the penguin graph has no extra phase.
This decay has been discussed by many people \cite{mf}.

A third possibility that was discussed are the decays $B^0 \to D^{\pm}
\pi^{\mp}$ \cite{fleisch}. The usual mixing-decay formalism yields
for the time dependent asymmetries the coefficients 
\begin{eqnarray}
a \sim Im(e^{-i(2 \phi_{mix} +\gamma)})const\\
\bar a ~ \sim Im(e^{-i(2 \phi_{mix} +\gamma)})/const.
\end{eqnarray}
where const is an unknown hadronic number. It cancels in the product
which then yields the combination 
\begin{equation}
2 \phi_{mix} +\gamma
\end{equation}
The $B \bar B$ mixing angle $\phi_{mix}$ can be determined as usual from
the decay $B \to J/\Psi K_s$.

The other angles of the triangle cannot be determined by a tree
level analysis independently. But we see, that the tree level analysis allows
difficult to estimate. One usually considers the Kaon-mixing
quantity $\epsilon_K$, the mass difference of the $B$ and the
$\bar B$ mesons (and also of the $B_s$ and $\bar B_s$ mesons). 
to determine the unitary triangle of the standard model. It yields, 
in principle, also the unknown 
side $R_t$
and the angle $\beta$. Any further independent measurement of these
quantities checks the standard model with high accuracy, but it requires
loop effects.

\subsubsection{\bf New physics: phenomenology}

Among the $CP$-violating observables, the mixing-decay asymmetry is
the cleanest theoretically \cite{bs}. It is therefore reasonable to start an
investigation of new physics with this quantity. Denoting the coefficient
of $sin(\Delta m t)$ by $a$, one has in general
\begin{equation}
a_{M \to F} = Im((\frac{p}{q})_M\frac{a}{\bar a}(\frac{p}{q})_F)
\end{equation}
where ($\frac{p}{q}$) are the mixing parameters and $a$, $\bar a$ the
amplitudes for $M \to F$ and $M \to \bar F$, respectively. 

Setting for the $B$-meson mixing element $M_{12}$
\begin{equation}
M_{12} = r^2 e^{2 i \phi^{NP}}e^{2 i \beta}|M_{12}^{SM}|
\end{equation}
to account for a possible new phase and magnitude of the mixing, 
the asymmetry coefficient is given in the table below:

\begin{center}
\begin{tabular}{|c|c|c|c|c|} \hline
$quarks$&$B_d$&$a$&$B_s$&$a$\\ \hline
$b \to c \bar c s$&$\Psi K_s$&$\beta + \phi^{NP}_d$&$D D_s$&$\phi^{NP}_s$\\
$b \to s \bar s s$&$\Phi K_s$&$\beta + \phi^{NP}_d + \phi^{A}$&$\Phi \Phi$&$\phi^{NP}_s + \phi^{A}$\\
$b \to u \bar u s$&$\pi \pi$&&&\\
$b \to c \bar c d$&$D^+ D^-$&&&\\
$b \to u \bar u s$&$\pi^0 K_s$&&&\\
$b\to s \bar s s$&$\Phi \pi$&&&\\ \hline
\end{tabular}
\end{center}
The phase $\phi^A$ takes into account a possible new phase in the decay.
The entries left out receive possibly sizeable contributions from
penguin diagrams and cannot be brought to the simple form. This result
tells us that comparing the different asymmetries, we can check the
consistency of the standard model and determine the phases of new physics. 

New physics will also influence other $CP$-violating observables, such
as the  direct asymmetries of, say charged B-meson decays. In cases 
such as $B \to K \pi$, where the asymmetry is small in the standard model 
new physics may give rise to sizeable asymmetries. Of course,
one need to continue the search for these, but because of the
difficulty of calculating direct asymmetries, only quantitative
statements are possible.

\subsubsection{\bf New physics: analysis}

If new physics is associated with a scale $\Lambda$ much above
the weak scale ($\sim M_W$), the total Lagrangian density may be
written in the form \cite{bw}
\begin{equation}
{\cal L} = {\cal L}^{SM} + \sum {d_i {\cal O}_i^{NP}}
\end{equation}

where the $O_i$ are operators of dimension six induced by new physics 
and their coefficients $d_i$ are of order($1/{\Lambda^2}$). 
This 'effective'
Lagrangian is not renormalizable, and therefore one usually uses the
new operators only at tree level (see a discussion by). The $CP$-
violation induced by effective operators ${\cal O}_i^NP$ can in most 
cases only be seen
when they are in loops, because the imaginary part (discontinuity)
of the corresponding Feynman graph is responsible for $CP$-asymmetry.
\footnote{an exception is the electric dipole moment} At low energies 
we then have an effective Hamiltonian
\begin{equation}
{\cal H} =\sum {c_i{\cal O}_i^{SM}} + \sum {d_i {\cal O}_i^{NP}}
\end{equation}

The amplitudes for a process $I \to F$ and the CP conjugated
one $\bar I \to \bar F$ then are
\begin{equation}
A(I \to F) =\sum {c_j (R_j + iI_j)^{SM}}  + \sum {d_j^* (R_j + iI_j)^{NP}}
\end{equation}
where $R$ and $I$ are the dispersive and absorptive parts of the
matrix elements. For the charge-conjugated process we have similarly
\begin{equation}
A(\bar I \to \bar F) =\sum {c_j^* (R_j + iI_j)^{SM}}  + \sum {d_j (R_j + iI_j)^{NP}}
\end{equation}
When we calculate the $CP$-violating asymmetry $\alpha \sim (|A(I \to F)|^2-
|A(\bar I \to \bar F)|^2)$, we obtain in leading order in QCD and in NP
\begin{equation}
\alpha \sim Im(cd^*)(R^{SM}I^{NP}-R^{NP}I^{SM}).
\end{equation}
$R^{NP}$ is a (finite) tree level amplitude, however also the loop 
$ I^{NP}$ is finite.
Therefore the problems associated with a the non-renormalizable theory
$\sum {d_i {\cal O}_i^{NP}}$ disappear and exact predictions are indeed
possible for the the $CP$-violating asymmetry. Therefore, an analysis
of the effects of new operators is possible also at for CP-violating
asymmetries, and not just at tree level!

\subsubsection{\bf New physics: models}

Virtually any model beyond the standard one carries new sources for
flavour and $CP$-violations. It is therefore more economical to
look at them in increasing complexity.

The simplest one are the minimal flavour violating ones (MFV) where
all sources of flavour violation reside in the CKM matrix. This
results in many cases in a simple modification of the 
coefficients in the usual loop expressions. However, there still
is a unitary triangle, but its sizes and angles may change. It was analyzed by
Ali and London \cite{ALI002}; recently Buras and Buras \cite{bubu} found a
clever lower bound on $sin(2\beta)$. The idea is simple. For both
$\epsilon$ and the $B$-meson mass difference, the standard model
contribution consists mostly of a $W-W-t-t$ box diagram; its value
might be denoted by $F_{tt}$. The $MFV$ modify this to
\begin{equation}
F_{tt}=S_0(m_t)~ (1+f)~.
\end{equation}
Then we can write for $\epsilon_K$
\begin{equation}\label{100W}
\epsilon_K  \sim \bar\eta \left[(1-\bar\varrho) A^2 \eta_2 F_{tt}
+ P_c(\varepsilon) \right] A^2 \hat B_K
\end{equation}
while the $B$-meson mass difference yields the relation
\begin{equation}\label{RT}
R_t= 1.26~ \frac{ R_0}{A}\frac{1}{\sqrt{F_{tt}}}~,
\end{equation}
where
\begin{equation}\label{R0}
 R_0= \sqrt{\frac{(\Delta M)_d}{0.47/{\rm ps}}}
          \left[\frac{200~mev}{F_{B_d} \sqrt{\hat B_d}}\right]
          \sqrt{\frac{0.55}{\eta_B}}~.
\end{equation}

With
\begin{equation}
\label{ss}
\sin 2\beta=\frac{2\bar\eta(1-\bar\varrho)}{R^2_t}
\end{equation}
one gets \cite{BLO}
\begin{equation}
\label{main}
\sin 2\beta=\frac{1.26}{ R^2_0\eta_2}
\left[\frac{0.226}{A^2 \hat B_K}-\bar\eta P_c(\varepsilon)\right].
\end{equation}
Since unitarity implies $\bar\eta \le R_b~$, there exists a lower bound
on $\sin 2\beta$. A careful numerical analysis implies
\begin{equation}
\bar \eta \ge 0.34.
\end{equation}
The lower bound in fact corresponds to a $F_{tt}$ which is three times
larger than the standard model value.

Supersymmetry is a attractive candidate for new physics. In general, there
are many new $CP$-violating phases. Since they can directly affect
observables such a the electric dipole moment, it is natural
to take them to be small (approximate $CP$-violation, \cite{nir}). In
this situation, also $CP$-violating effects in the $B$-system are small.
This implies a small angle $\beta$. 
This is in contrast to the standard model, where the flavour structure
suppresses CP-violation.

The problem with this scheme is that it is hard to get $\epsilon_K$ right
and that $\epsilon'/\epsilon$ tends to be to small. 

Similarly, models with left-right symmetry tend to have small $CP$-violating
phases, thus the effects tend to be small also.

\subsubsection{\bf $CP$-violation in $D$-mesons}

In the standard model, $CP$-violation is small in the $D$-System. This
is partly due to the rather large tree-level decay rates and small
coupling of the third generation. Therefore one would expect new
physics $CP$-violation mostly in the mixing (see \cite{nir} for a
more detailed discussion).
Recent studies of time-dependent decay rates of $D^0\rightarrow K^+\pi^-$ 
by the CLEO collaboration \cite{cleo} and measurements of the combination of 
$D^0\rightarrow K^+K^-$ and $D^0\rightarrow K^-\pi^+$ rates by the FOCUS
collaboration \cite{focus} gave first information on the mixing.

As usual, one define the mixing quantities
\begin{equation}\label{DelMG}
x\equiv{m_2-m_1\over\Gamma},\ \ \ 
y\equiv{\Gamma_2-\Gamma_1\over2\Gamma}.
\end{equation}
CP-violation in the mixing is defined by the angle $\phi$. The experiments
find that the quantity $y\cos\phi$ is significantly larger than the
expectation in the standard model. The errors being large, this result
is not yet significant, but it shows the potential of $D$-meson physics.

\subsubsection{\bf $K$-physics}

Finally let me mention $K$-physics. Of course, efforts continue in
calculating $\epsilon'/\epsilon$.
However, the rare decays $K^+ \to \pi^+ \nu \bar \nu$ and
$K^0 \to \pi^0 \nu \bar \nu$ provide a theoretically
clean way to measure (in the standard model) $|V_{td}|$ and $Im V_{td}$
\cite{kburas}.
Clearly, this can be used as a test of the unitary triangle, however
the measurement of the neutral decays is not easy and probably many years
away.

\subsubsection{\bf Conclusions}

The new results on $\sin 2\beta$ are surprising; they may indicate
a failure of the standard model. Several parameters have to be stretched
beyond their reasonable values to account for them. One can
modify the standard model to accommodate the small value of
 $\sin 2\beta$,
but it is not clear that these modifications are consistent. 

Nevertheless, the result brings back the (old) view, that a (standard) 
model independent and broad analysis of $CP$-violation
is required in order to fully understand this phenomenon and the
need for new interaction. this implies in particular measurements of 
many decay channels. 

I have sketched strategies to determine the source of $CP$-violation for
the case that the standard model accounts for tree level processes
and given a phenomenological framework to calculate the effects
of new operators. Needless to say that all of this will take many years
of hard work on both the experimental and the theoretical side and
that also less perfect measurements have to be pursued.

\section{\bf What can we learn from $\epsilon'/\epsilon$?}

\subsection{\bf CP violation in the kaon system}

{\it  V. Martin, University of Edinburgh}

CP violation was first observed in the neutral kaon system by Christenson, Cronin, 
Fitch and Turlay in 1964 ~\cite{ccft64} when they observed the long-lived neutral kaon decay 
into a $\pi^+\pi^-$ state.
To date, the neutral kaon system is the only place where CP violation has been conclusively 
observed.

In the neutral kaon system CP violation is classified into two types:  indirect 
and direct, parameterised by $\epsilon$ and $\epsilon'$ respectively.
Indirect CP violation may be observed through the asymmetry in the mixing of the neutral kaons.
Direct CP violation may be observed through the asymmetry in the decays of the neutral kaons.
As direct CP violation is a much smaller effect than indirect CP violation, the ratio 
$\mathcal R e ({\epsilon'/\epsilon}) \approx {\epsilon'/\epsilon}$ 
is usually considered as the measurement of direct CP violation.
The last generation of  $\mathcal R e (\epsilon'/\epsilon)$ experiments announced final results in 1993.
The NA31 experiment at CERN had a result which was compatible with direct CP violation:
$\mathcal R e (\epsilon'/\epsilon) = (23.0\pm6.5)\times 10^{-4}$~\cite{na31}, whereas the E731 experiment 
at FNAL found a result compatible with no direct CP violation: 
$\mathcal R e (\epsilon'/\epsilon) = (7.4\pm5.9)\times 10^{-4}$~\cite{e731}.

This disagreement prompted a new generation of direct CP violation experiments 
with the aim of measuring $\mathcal R e (\epsilon'/\epsilon)$ at the $10^{-4}$ level.

\begin{figure}[t]
\begin{center}
\includegraphics[width=11cm]{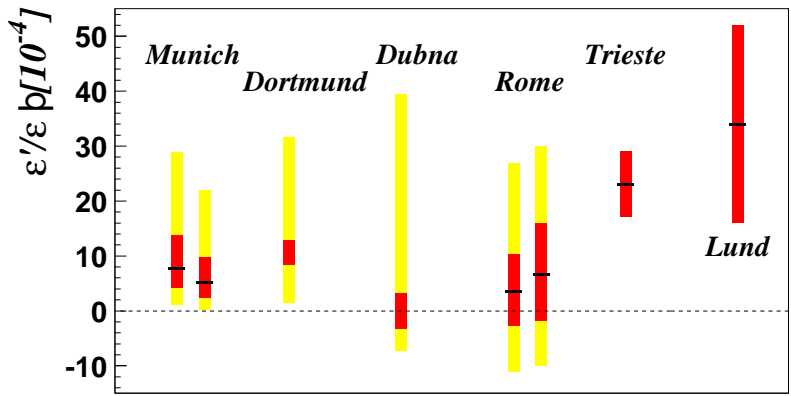}
\caption{\label{fig:eprimetheory}
Recent theoretical estimates of $\mathcal R e (\epsilon'/\epsilon)$ in chronological order 
\cite{buras99,hambye99,belkov99,ciuchini99,bertolini00}.
 The red/dark-grey error bars
correspond to a Gaussian treatment of the input parameters.  
The yellow/light-grey error bars correspond to a 
flat scanning of all the input parameters.
The cross lines corresponds to the most favoured value.
Where there are two different results from one group, these correspond to different 
renormalisation schemes.}
\end{center}
\end{figure}

Theoretical estimates for the value of $\mathcal R e (\epsilon'/\epsilon)$ are based 
on QCD calculations.
The short distance contributions have been reliably calculated using perturbative-QCD~\cite{buchalla96}.
Various methods have been used to estimate the long distance effects, 
such as chiral perturbation theory and lattice QCD. 
Figure~\ref{fig:eprimetheory} shows some recent theoretical predictions of $\mathcal R e (\epsilon'/\epsilon)$.

Experimentally, the parameter $\mathcal R e (\epsilon'/\epsilon)$ is obtained by measuring
the double ratio of decay rates ($\mathbf R$) of the long-lived and short-lived neutral kaons 
$\mathrm{K_L}$ and $\mathrm{K_S}$ into two pion final states:
\begin{equation} 
\mathbf R \equiv \frac{\Gamma(\mathrm{K_L}\rightarrow \pi^0 \pi^0)/
                  \Gamma(\mathrm{K_S}\rightarrow \pi^0 \pi^0)}
                 {\Gamma(\mathrm{K_L}\rightarrow \pi^+ \pi^-)/
                  \Gamma(\mathrm{K_S}\rightarrow \pi^+ \pi^-)} = 1 - 6\, \mathcal R e (\epsilon'/\epsilon)
\end{equation}

Currently there are three dedicated experiments analysing or taking data with the aim of measuring 
$\mathcal R e (\epsilon'/\epsilon)$:  KTEV at FNAL, NA48 at CERN and KLOE at the 
DA$\Phi$NE $\Phi$ factory at Frascati. 
Each of these experiments was designed specifically to measure $\mathcal R e (\epsilon'/\epsilon)$.
In particular, they were designed to record high statistics and to minimise systematic 
errors in the measurement of $\mathcal R e (\epsilon'/\epsilon)$.
The following discussion here will focus on the measurements from NA48 and KTeV, which were presented
at this workshop.

In the measurement of the double ratio {\bf R}, many systematic effects cancel, such as the number of
$\mathrm{K_S}$ and $\mathrm{K_L}$ particles produced, and the efficiency of detecting the
charged and neutral final states.  

The NA48 and KTeV experiments use a similar experimental set-up, aiming to exploit many of the 
cancellations in the measurement of $\mathbf{R}$.
In order to be able to detect all four $\mathcal R e (\epsilon'/\epsilon)$ decay modes simultaneously,
both experiments employ a double beam technique:  
two neutral kaon beams are produced at different distances from the detector.
Decays from the beam with the longer path length are predominantly $\mathrm{K_L}$ decays 
whereas decays from beam with the shorter path length are predominantly $\mathrm{K_S}$.

KTeV uses two identical, parallel, neutral kaon beams. A plastic scintillator regenerator is placed in one
of the beams to regenerate $\mathrm{K_S}$ particles.  The position of the decay products in the detectors 
can be used to determine whether the decay came from the vacuum or regenerator beam.
The two NA48 beams are created by colliding the SPS protons on identical beryllium targets. 
The so-called $\mathrm{K_L}$ and $\mathrm{K_S}$ targets are placed respectively 
$240\;\mathrm{m}$ and $122\;\mathrm{m}$ upstream of the detector. The targets are arranged
such that the two beams converge, meeting at the centre of the electromagnetic calor\-i\-meter. 
The times of the protons in transport to the $\mathrm{K_S}$ target are recorded by the 
proton tagging detector.  The proton times can be compared offline to the time of the decay 
products in the detectors to determine which of the beams the decay came from.

To detect the decay products of the neutral kaons both the NA48 and KTeV experiments have:
a magnetic spectrometer to detect charged tracks and reconstruct $\mathrm K \rightarrow \pi^+ \pi^-$ decays;
a high-resolution electromagnetic calorimeter to detect photons from $\pi^0 \rightarrow 2\,\gamma$ decays
and a muon detector to reject background from 
$\mathrm {K_L} \rightarrow \pi^{\pm} \mu^{\mp} \nu_{\mu}$ decays.

NA48 uses a quasi-homogeneous Liquid Krypton calorimeter which contains $9\;\mathrm{m^3}$ of liquid krypton 
kept at $121\;\mathrm{K}$.
KTeV has a Cesium-Iodide crystal calorimeter, made of 3,100 blocks of pure Cesium-Iodide.
The quality of these two calorimeters is reflected in the resolution in measuring reconstructed 
masses from the $\mathrm K \rightarrow \pi^0 \pi^0 \rightarrow 4\,\gamma$ decays.
NA48 achieves a resolution on the reconstructed $\gamma \gamma$ mass of $1.1 \;\mathrm{MeV/c^2}$, and
KTeV achieves a resolution on the reconstructed kaon mass of $1.7 - 1.9 \;\mathrm{MeV/c^2}$.

The analysis of the data from the $\mathcal R e (\epsilon'/\epsilon)$ experiments is, in principle, straightforward.
The number of events in each of the four decay modes in counted and a initial value of {\bf R} is obtained.
Corrections must be made to this value of {\bf R} for backgrounds, trigger efficiencies, 
detector energy scale, beam scattering, accidental effects and detector acceptance.  
In addition both experiments make corrections for effects
due to mis-identifying which of the beams the decay came from.
Most of these corrections are small due to the cancellations present in {\bf R}.

The largest potential correction to {\bf R} is due to the differential acceptance of the four decay modes.
To reduce this correction NA48 uses the technique of {\it weighting} the $\mathrm{K_L}$ events
so that the decay distribution of weighted $\mathrm{K_L}$ events is the same as the (unweighted) 
$\mathrm{K_S}$ events.
This reduces the correction to {\bf R} due to the acceptance from around $13\%$ to around $3\times 10^{-3}$,
but leads to a reduction of statistical precision of $35-40\%$.
KTeV instead relies on a detailed Monte Carlo simulation of the experiment to calculate the acceptance.
They calculate a correction to {\bf R} of around $4\%$, which they measure with a precision 
of $13.4\times10^{-3}$~\cite{ktev}.

KTeV collected data for the $\mathcal R e (\epsilon'/\epsilon)$ measurement during 1996, 1997 and 1999.
A preliminary result based on some of the data taken during 1996 and 1997 is described in~\cite{ktev}.
The result of that analysis is $\mathcal R e (\epsilon'/\epsilon) 
= \left(28.0 \pm  3.0 (\mathrm{stat}) \pm 2.6 (\mathrm{syst}) \pm 1.0 (\mathrm{MC\; stat})\right)
\times 10^{-4}$.

NA48 has collected data for the $\mathcal R e (\epsilon'/\epsilon)$ measurement
in the years 1997--99 and will take more data in 2001.  
Preliminary results based on the data taken in 1997 and 1998 have been announced~\cite{na48,ceccucci00}.
The combined result from these data sets is:
$\mathcal R e (\epsilon'/\epsilon) = (14.0 \pm 4.3) \times 10^{-4}$,
where the statistical and systematic errors have been added in quadrature,
and the small correlation between systematic effects has been taken into account.
The error on this result is currently dominated by systematic effects, but as the systematic error 
includes some statistical component, the error will improve significantly when more data is analysed.

Table~\ref{table:eprimestats} shows the number decays in each $\mathcal R e (\epsilon'/\epsilon)$ 
mode analyzed by NA48 (in 1998) and KTeV for these preliminary results.

\begin{figure}
\begin{center}
\includegraphics[width=8.5cm]{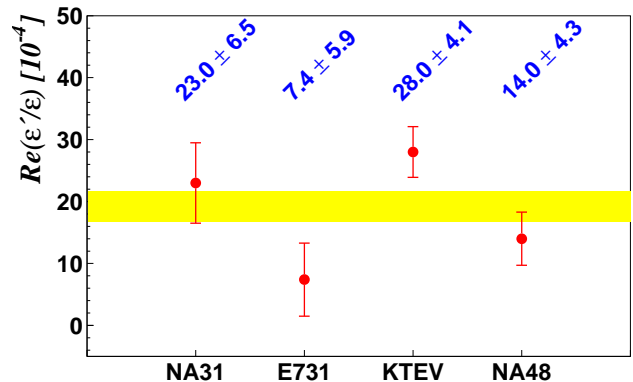}
\caption{\label{fig:eprimeexperiments}
Recent measurements of $\mathcal R e (\epsilon'/\epsilon)$~\cite{na31,e731,ktev,ceccucci00}.
Where both a statistical and systematic error is given, these have been added in quadrature.
The yellow/light-grey band shows the $\pm 1 \sigma$ allowed region of the average of these results.}
\end{center}
\end{figure}

Figure~\ref{fig:eprimeexperiments} shows the preliminary NA48 and KTeV results along with the results
from E731 and NA31.
A weighted average of these four results gives $\mathcal R e (\epsilon'/\epsilon)_{\mathrm{global}} = (19.2 \pm 2.5) \times 10^{-4}$
with a $\chi^2/\mathrm{ndf} = 10.4/3$.
The yellow/light-grey band on figure~\ref{fig:eprimeexperiments} shows the $\pm 1 \sigma$ allowed region of 
this average.
As the $\chi^2$ per degree of freedom on the average is larger than 1, 
the approach proposed by the PDG can be used to rescale the error by 
$(\chi^2/\mathrm{ndf})^{1/2}=1.8$, resulting in 
$\mathcal R e (\epsilon'/\epsilon)_{\mathrm{global}} = (19.5 \pm 4.6) \times 10^{-4}$.

\begin{table}
\begin{center}
\begin{tabular}{|c||c|c|c|c|}
\hline
experiment & $\mathrm{K_L}\rightarrow \pi^0\pi^0$ & $\mathrm{K_S}\rightarrow \pi^0\pi^0$ & 
             $\mathrm{K_L}\rightarrow \pi^+\pi^-$ & $\mathrm{K_S}\rightarrow \pi^+\pi^-$\\
\hline
NA48       &  $1.80\times 10^{6}$ & $1.14\times 10^{6}$ & $7.46\times 10^{6}$ & $4.87\times 10^{6}$ \\
KTeV       &  $0.86\times 10^{6}$ & $1.42\times 10^{6}$ & $2.61\times 10^{6}$ & $4.51\times 10^{6}$\\
\hline
\end{tabular}
\caption{\label{table:eprimestats}
Numbers of decays for each $\mathcal R e (\epsilon'/\epsilon)$ decay mode 
collected by NA48 in 1998 and by KTeV for their preliminary result.
The numbers have been corrected for background decays.
The KTeV numbers actually refer to the numbers of decays counted from the vacuum beam 
($\mathrm{K_L}$) and the regenerator beam ($\mathrm{K_S}$).}
\end{center}
\end{table}

The theoretical calculations shown in figure~\ref{fig:eprimetheory} generally predict smaller values
for $\mathcal R e (\epsilon'/\epsilon)$ than the experimental results, 
leading to some speculation that the measurement of $\mathcal R e (\epsilon'/\epsilon)$ may be a
sign of new physics.  However due to the relatively large uncertainties  
reported by both theoretical calculations and experiments it is obviously too early to draw any conclusions
on this issue.
 
Concluding on CP violation in kaon physics:
the experimental picture is now much clearer than it was in 1993.  
Direct CP violation in the neutral kaon system is now firmly established.
However, due to the large variation in the measurements, it is too early to settle on 
an exact value for $\mathcal R e (\epsilon'/\epsilon)$.
New, more precise experimental results will be produced by NA48, KTeV and KLOE in the future,
allowing $\mathcal R e (\epsilon'/\epsilon)$ to be known with an uncertainty of around $1 \times 10^{-4}$.

\subsection{\bf Supersymmetry predictions for $\varepsilon'/\varepsilon$}

{\it Shaaban Khalil, University of  Sussex}

\newcommand{\no}{\nonumber\\}
\def\tgb{\mbox{$\tan{\beta}~$}}
\def\bsg{$b\to s \gamma$~}
\def\eps{$\varepsilon$~}
\def\epspeps{$\varepsilon^{\prime}/\varepsilon$~}
\def\Lsoft{${\cal L}_{SB}$~}
\def\mch{$m_{\chi^{\pm}}$~}
\def\mneu{$m_{\chi^{0}}$~}
\def\mglu{$m_{\tilde{g}}$~}
\def\stop{$m_{\tilde{t}}$~}
\def\mgrav{$m_{3/2}$~}
\def\Ibanez{Iba\~{n}ez~}
\def\Munoz{Mu\~{n}oz~}
\newcommand{\BXcenu}{B\rightarrow X_c e \nu}
\newcommand{\mub}{\mu_b}
\newcommand{\mb}{m_b}
\newcommand{\alphas}{\alpha_s}
\newcommand{\alphae}{\alpha_e}
\newcommand{\BRg}{{\rm BR}(B\to X_s \gamma)}
\newcommand{\BR}{{\rm BR}}
\newcommand{\Bsg}{B\to X_s \gamma}

\def\be{\begin{equation}}
\def\ee{\end{equation}}
\def\bea{\begin{eqnarray}}  
\def\eea{\end{eqnarray}}   
\def\etal{{\it et al.}}   
\def\eg{{\it e.g.}}
\def\ie{{\it i.e.}}
\def\Frac#1#2{\frac{\displaystyle{#1}}{\displaystyle{#2}}}
\def\lsim{\raise0.3ex\hbox{$\;<$\kern-0.75em\raise-1.1ex\hbox{$\sim\;$}}}
\def\gsim{\raise0.3ex\hbox{$\;>$\kern-0.75em\raise-1.1ex\hbox{$\sim\;$}}}
\renewcommand{\O}{{\cal O}}
\def\ap#1#2#3{     {\it Ann. Phys. (NY) }{\bf #1} (#2) #3}
\def\arnps#1#2#3{  {\it Ann. Rev. Nucl. Part. Sci. }{\bf #1} (#2) #3}
\def\npb#1#2#3{    {\it Nucl. Phys. }{\bf B #1} (#2) #3}
\def\npbps#1#2#3{    {\it Nucl. Phys. }(Proc. Suppl.){\bf B #1} (#2) #3}
\def\plb#1#2#3{    {\it Phys. Lett. }{\bf B #1} (#2) #3}
\def\prd#1#2#3{    {\it Phys. Rev. }{\bf D #1} (#2) #3}
\def\prep#1#2#3{   {\it Phys. Rep. }{\bf #1} (#2) #3}
\def\prl#1#2#3{    {\it Phys. Rev. Lett. }{\bf #1} (#2) #3}
\def\ptp#1#2#3{    {\it Prog. Theor. Phys. }{\bf #1} (#2) #3}
\def\rmp#1#2#3{    {\it Rev. Mod. Phys. }{\bf #1} (#2) #3}
\def\zpc#1#2#3{    {\it Zeit. f\"ur Physik }{\bf C #1} (#2) #3}
\def\mpla#1#2#3{   {\it Mod. Phys. Lett. }{\bf A #1} (#2) #3}
\def\sjnp#1#2#3{   {\it Sov. J. Nucl. Phys. }{\bf #1} (#2) #3}
\def\yf#1#2#3{     {\it Yad. Fiz. }{\bf #1} (#2) #3}
\def\nc#1#2#3{     {\it Nuovo Cim. }{\bf #1} (#2) #3}
\def\jetpl#1#2#3{  {\it JETP Lett. }{\bf #1} (#2) #3}
\def\ibid#1#2#3{   {\it ibid. }{\bf #1} (#2) #3}



\subsubsection{\bf Introduction}

The most recent results of \epspeps, which measures the size of the direct 
CP violation in $K_L \to \pi \pi$, reported by KTeV
\cite{CP1} and NA48 \cite{CP2}
lead to a world average
of Re~\epspeps =$(21.4\pm 4.0)\times 10^{-4}$~\cite{CP3}.
This result is higher than the Standard Model (SM) predictions \cite{epsp1},
opening the way to the interpretation that it may be a signal of
new physics beyond the SM. The SM predictions for \epspeps suffer from  large
theoretical uncertainties \cite{epsp2} such that one can not draw a definite conclusion
if this observed high value of \epspeps can be accommodated in the SM. In any case, one
may wonder if the supersymmetry (SUSY) can be responsible for enhancing \epspeps. 

In the minimal supersymmetric extension of the SM (MSSM) there is no way of generating
a sizable SUSY contribution to \epspeps even if one assume that the SUSY CP violating
phases are large and the electric dipole moments (EDM) of the electron and neutron are
less than the experimental bounds due to the cancellation between the different
contributions. This is mainly due to the assumption of universal boundary conditions
of the soft-breaking terms \cite{GG,abel,khalil1,demir}. It has been shown that, 
without new flavor structure beyond the usual Yukawa couplings, general SUSY models with 
phases of the soft terms of order $\O(1)$ (but with a vanishing CKM phase 
$\delta_{\mathrm{CKM}}=0$)
can not give a sizeable contribution to the CP violating processes
\cite{abel,khalil1,demir,barr}. This means that the presence of non--universal soft breaking
terms besides large SUSY phases is crucial to enhance these CP violation effects.
In agreement with this, it has been explicitly shown
that contributions to $\varepsilon_K$ are small within the
dilaton--dominated SUSY breaking of the weakly coupled heterotic string
model~\cite{barr}, where $A$--terms as well as gaugino masses are universal. 
On the other hand, it is well--known that the strict universality in the soft
breaking sector is a strong assumption not realized in many supergravity and
string inspired models~\cite{ibanez1,ibanez2}.
All these arguments indicate not only that the presence of non--universal
soft terms can solve the problem of too large contributions to EDMs but also
that it allows for large SUSY contributions in CP violation experiments.
Hence, in this work we will follow this avenue and analyze the effects of
non--universal soft terms in CP violation in the $K$--system~\cite{khalil2,vives,emidio}.

\subsubsection{\bf CP violation in minimal supergravity model}

It is well known that in SUSY models there are new possibilities for CP violation. 
In particular, the soft SUSY breaking terms contain several parameters that  may be 
complex, as can also be the $\mu$-parameter. 
In the minimal supergravity model there are
only two new CP-violating phases. This can be seen as  
follows. The parameters $M, A$ and $B$ and $\mu$ can be complex.
But of these four phases only two are physical. First, by an
R-rotation with R-charge $Q_R=1$ for lepton and quark superfields
and $Q_R=0$ for the vector and the Higgs superfields, the gaugino
mass parameter $M$ can be made real. Second, $B \mu$ can be made
real by a change of phase of the Higgs superfield. This ensures   
that the Higgs vacuum expectation values are real. The remaining
phases cannot be defined away and violate CP. One is in $A=A_0  
e^{i \phi_A}$ and the other in $B=B_0 e^{i\phi_B}$. The $\mu$
parameter then has a fixed phase $\mu=\mu_0 e^{-i\phi_B}$.
In any phase convention 
\be 
\phi_A= \mathrm{arg}(AM^*),  \hspace{3cm}
\phi_B= \mathrm{arg}(BM^*).
\ee
These phases can cause at one
loop level an electric dipole moment (EDM) for the quarks and
leptons, and therefore also for the neutron. It has been known for
a long time that in the SUSY models the contributions to the
neutron electric dipole moment are larger than the experimental
limit $6.3\times 10^{-26}$ e cm unless either the new `SUSY
phases' are tuned to be of order $10^{-3}$, or the SUSY masses are
of order a TeV. Such small phases can not generate sizable CP violation.
Also they constitute a fine tuning. This is known as `` SUSY CP problem".
However, in the last few it has been suggested that a natural   
cancellation mechanism exists whereby the electric dipole moment
of the neutron may be made small without such fine-tuning. In this case 
large SUSY phases are expected and still satisfy experimental bounds
on the values of EDM of the electron and neutron.

In this section  we will study the effect of these phases in CP violation 
observables as \eps and \epspeps. We assume that $\delta_{CKM}=0$ to maximize 
this effect. The value   
of the indirect CP violation in the Kaon decays, $\varepsilon$, is 
defined as $ \varepsilon = e^{i\frac{\pi}{4}}
{\rm Im} M_{12}/\sqrt{2} \Delta m_K,$
where $\Delta m_K= 2 {\rm Re} \langle K^0 \vert H_{eff} \vert \bar{K}^0
\rangle = 3.52 \times 10^{-15}$ GeV.
The amplitude
$M_{12}=\langle K^0 \vert H_{eff} \vert \bar{K}^0 \rangle$.
The relevant supersymmetric contributions to $K^0-\bar{K}^0$ are 
the gluino and the chargino contributions, (\ie, the transition
proceeds through box diagrams exchanging gluino-squarks and
chargino-squarks). It is usually expected that the gluino is the   
dominant contribution. However, as we will show, it is impossible
in the case of degenerate $A$-terms that the gluino gives any significant
contribution to $\varepsilon$ when the CKM matrix is taken to be real even with
large phase of $A$.
The amplitude of the gluino contribution is given in terms of the mass insertion 
$\delta_{AB}$ defined by $\delta_{AB} = \Delta_{AB}/\tilde{m}^2$ where
$\tilde{m}$ is an average sfermion mass and $\Delta$ is off-diagonal
terms in the sfermion mass matrices. The mass insertion to accomplish
the transition from $\tilde{d}_{iL}$ to $\tilde{d}_{jL}$ ($i,j$ are
flavor indices) is given by
\begin{eqnarray}
(\Delta^d_{LL})_{ij}&\simeq&-\frac{1}{8\pi^2}\left[\frac{K^{\dag}
(M_u^{diag})^2 K}{v^2 \sin^2\beta} \ln(\frac{M_{GUT}}{M_W})
\right](3\tilde{m}^2+\vert X \vert^2),
\\
(\Delta^d_{LR})_{ij}&\simeq&-\frac{1}{8\pi^2}\left[\frac{K^{\dag}
(M_u^{diag})^2 K\ M_d}{v^2 \sin^2\beta \cos\beta}   
\right] \ln(\frac{M_{GUT}}{M_W}) X,
\\
(\Delta^d_{RL})_{ij}&\simeq&-\frac{1}{8\pi^2}\left[\frac{M_d\ K^{\dag}
(M_u^{diag})^2 K }{v^2 \sin^2\beta \cos\beta}
\right] \ln(\frac{M_{GUT}}{M_W}) X,
\\
(\Delta^d_{RR})_{ij}&=&0,
\end{eqnarray}
where $X= A_d -\mu\ \tan\beta $. It is clear that $\Delta_{ij}$ in
general are complex due to the complexity of the CKM matrix, the
trilinear coupling $A$ and $\mu$ parameter. Here we assume the
vanishing of $\delta_{CKM}$ to analyze the effect of the SUSY
phases. We notice that $(\Delta^d_{LL})_{12}$ is proportional to
$\vert X \vert^2$ \ie, it is real and does not contribute to
$\varepsilon$ whatever the phase of $A$ is. Moreover, the values
of the $(\Delta^d_{LR})_{12}$ and $(\Delta^d_{RL})_{12}$ are
proportional to $m_s$ and $m_d$, hence they are quite small.
Indeed in this case we find the gluino contribution to
$\varepsilon$ is of order $10^{-6}$. 

For the chargino contribution the amplitude is given
by~\cite{branco}
\begin{eqnarray}
\hspace{-1cm}\langle K^0 \vert H_{eff} \vert \bar{K}^0 \rangle &=& -\Frac{G_F^2
M_W^2}{(2\pi)^2} (V_{td}^* V_{ts})^2 f_K^2 M_k \biggl[ \frac{1}{3} 
C_1(\mu) B_1(\mu)\no
&+& \bigl(\frac{M_k}{m_s(\mu) +m_d(\mu)}\bigr)^2 
\bigl(-\frac{5}{24} C_2(\mu) B_2(\mu) + \frac{1}{24} C_3(\mu)
B_3(\mu)\bigr)\biggr].
\end{eqnarray}
The complete expression for these function can be found in
Ref.~\cite{branco}. For low and moderate values of $\tan \beta$ the value
of $C_3$ is much smaller than $C_1$ since it is suppressed by the
ratio of $m_s$ to $M_W$. However, by neglecting the flavor mixing
in the squark mass matrix $C_1$ turned out to be exactly real~\cite{demir}.
The imaginary part of $C_1$ is associated to the size of the
intergenerational sfermion mixings, thus it is maximal for large
$\tan \beta$. In low $\tan \beta$ case, that we consider, the imaginary part
of $C_1$ is very small, and the gluino contribution is still the dominant
contribution $\varepsilon$. In particular, we found that \eps is of order 
$10^{-6}$, which is less than the experimental value $2.26\times 10^{-3}$.

Now we consider the effect of these two phases ($\phi_A$ and $\phi_{\mu}$)
on the direct CP violation parameter \epspeps. Similar to the case of indirect CP 
violation parameter \eps, in the gluino contribution the $L$--$L$ transitions are almost real
and the $L$--$R$ transitions are suppressed by two up Yukawa couplings and a down quark 
mass. Moreover, the analysis of chargino contribution is also the same as in the indirect CP 
violation. Even the experimental bounds on the branching ratio of $b \to s \gamma$ decay 
impose sever constraint on the LR transition. Therefore we do not find any significant SUSY
CP violation effect in \epspeps too.

\subsubsection{\bf Non--universal soft terms and SUSY CP violation}

In the previous section, we have shown that CP violation effects are always very small 
in SUSY models with universal soft SUSY breaking terms. Recently, it has been shown that
the non--universality of $A$--terms is very
effective to generate large CP violation effects
\cite{abel,khalil1,khalil2,vives,masieromur,non-u}. In fact, the presence of
non--degenerate $A$--terms is essential for enhancing the gluino contributions
to $\varepsilon'/\varepsilon$ through large imaginary parts of the $L$--$R$
mass insertions, $\mathrm{Im}(\delta_{LR})_{12}$ and
$\mathrm{Im}(\delta_{RL})_{12}$.
These SUSY contributions can, indeed, account for a sizeable part of the
recently measured experimental value of $\varepsilon'/\varepsilon$
\cite{CP1,CP2}. In the following, we will present an explicit realization
of such mechanism in the framework of a type I superstring inspired SUSY
model. Within this model, it is possible to obtain non--universal soft
breaking terms, i.e. scalar masses, gaugino masses and trilinear
couplings. 

Type I string models contain nine--branes and three types of
five--branes ($5_a$, $a=1,2,3$).    
If we assume that the gauge group $SU(3) \times U(1)$ is on one of the branes (9--brane)
and the gauge group $SU(2)$ is on another brane ($5_1$--brane).
Chiral matter fields correspond to open strings spanning between
branes. Thus, they have non--vanishing
quantum numbers only for the gauge groups corresponding to
the branes between which the open string spans. For example, the chiral field corresponding 
to the open string between the $SU(3)$ and $SU(2)$ branes can have
non--trivial representations under both $SU(3)$ and $SU(2)$,
while the chiral field corresponding to the open string,
which starts and ends on the $SU(3)$--brane, should be
an $SU(2)$--singlet.
There is only one type of the open string which spans between the 9 and $5_1$--branes,
which we denote it as $C^{95_1}$.  
However, there are three types of open strings which could start and end on the
9--brane, that is, the $C^9_i$ sectors (i=1,2,3), which corresponding to the $i$-th 
complex compact dimension among the three complex dimensions. If we 
assign the three families to the different $C^9_i$ sectors we obtain non--universality
in the right--handed sector. In particular, we assign the $C^{9}_1$ sector
to the third family and $C^{9}_3$ and $C^{9}_2$ to the first and second families 
respectively. Under these assumption the soft SUSY breaking terms are obtained,
following the formulae in Ref.~\cite{ibanez1,ibanez2}.
The gaugino masses are obtained
\begin{eqnarray}
\label{gaugino}
M_3 & = & M_1 = \sqrt 3 m_{3/2} \sin \theta\  e^{-i\alpha_S}, \\
M_2 & = &  \sqrt 3 m_{3/2} \cos \theta\ \Theta_1 e^{-i\alpha_1}. 
\end{eqnarray}
While the $A$-terms are obtained as
\begin{equation}
A_{C_1^9}= -\sqrt 3 m_{3/2} \sin \theta\ e^{-i\alpha_S}=-M_3,
\label{A-C1}
\end{equation}
for the coupling including $C_1^{9}$, i.e. the third family,
\begin{equation}
A_{C_2^9}= -\sqrt 3 m_{3/2}(\sin \theta\ e^{-i\alpha_S}+
\cos \theta\ (\Theta_1 e^{-i\alpha_1}- \Theta_2 e^{-i\alpha_2})),
\label{A-C2}
\end{equation}
for the coupling including $C_2^{9}$, i.e. the second
family  and
\begin{equation}
\label{A-C3}   
A_{C_3^9}= -\sqrt 3 m_{3/2}(\sin \theta\ e^{-i\alpha_S}+
\cos \theta\ (\Theta_1 e^{-i\alpha_1}- \Theta_3 e^{-i\alpha_3})),
\end{equation}
for the coupling including $C_3^{9}$, i.e. the first family.
Here $m_{3/2}$ is the gravitino mass, $\alpha_S$ and $\alpha_i$ are
the CP phases of the F-terms of the dilaton field $S$ and
the three moduli fields $T_i$, and $\theta$ and $\Theta_i$ are
goldstino angles, and we have the constraint, $\sum \Theta_i^2=1$.
Thus, if quark fields correspond to different
$C_i^9$ sectors, we have non--universal A--terms.
Then we obtain the following A--matrix for both of the
up and down sectors,
\begin{eqnarray}
A= \left(  
\begin{array}{ccc}
A_{C^9_3}  & A_{C^9_2} & A_{C^9_1} \\ A_{C^9_3} & A_{C^9_2} &
A_{C^9_1} \\ A_{C^9_3} & A_{C^9_2} & A_{C^9_1}
\end{array}
\right) \label{A-1}.
\end{eqnarray}
Note that the non--universality appears
only for the right--handed sector.
The trilinear SUSY breaking matrix, $(Y^A)_{ij}=(Y)_{ij}(A)_{ij}$,
itself is obtained
\begin{equation}
\label{trilinear}
Y^A = \left(\begin{array}{ccc}
 &  &  \\  & Y_{ij} &  \\  &  & \end{array}
\right) \cdot
\left(\begin{array}{ccc}
A_{C^9_3} & 0 & 0 \\ 0 & A_{C^9_2} & 0 \\ 0 & 0 & A_{C^9_1} \end{array}
\right),   
\end{equation}
in matrix notation. In addition, soft scalar masses for quark doublets and
the Higgs fields are obtained,
\begin{equation}
\label{doublets}
m_{C^{95_1}}^2=m_{3/2}^2(1-\Frac{3}{2} \cos^2 \theta\ (1-
\Theta_1^2)).
\end{equation}  
The soft scalar masses for quark singlets are obtained as
\begin{equation}
\label{singlets}
m_{C_i^9}^2=m_{3/2}^2(1-3\cos^2 \theta\ \Theta^2_i),
\end{equation}
if it corresponds to  the $C_i^{9}$ sector.

In models with non-degenerate $A$--terms we have to fix the Yukawa
matrices to completely specify the model. In fact, with universal
$A$--terms the textures of the Yukawa matrices at GUT scale affect
the physics at EW scale only through the quark masses and usual
CKM matrix, since the extra parameters contained in the Yukawa
matrices can be eliminated by unitary fields transformations. This
is no longer true with non-degenerate $A$--terms. 
Here, we choose our Yukawa texture to be
\be
\hspace{-1.5cm}Y^u=\frac{1}{v\cos{\beta}} {\rm diag}\left(
m_u,m_c,m_t\right)~,~~
Y^d=\frac{1}{v\sin{\beta}}  K^{\dagger} \cdot {\rm diag }
\left(m_d, m_s, m_b\right) \cdot  K
\ee
where $K$ is the CKM matrix. In this case one find that the mass insertion $\delta^d_{LR}$
can be written as~\cite{vives}
\begin{eqnarray}
\label{DLR}
\hspace{-1.5cm}(\delta_{LR}^{(d)})_{i j}&=& \frac{1}{m^2_{\tilde{q}}}\ m_i\ \Big(
\delta_{ij}\ (c_{A} A_{C^9_3}^*\ +\ c_{\tilde{g}}\ m_{\tilde{g}}^* -\ 
\mu e^{i\varphi_{\mu}} \tan\beta ) \nonumber \\
&+&K_{i 2}\ K^*_{j 2}\ c_{A}\ ( A_{C^9_2}^* - A_{C^9_3}^* ) +
K_{i 3}\ K^*_{j 3}\ c_{A}\ ( A_{C^9_1}^* - A_{C^9_3}^* ) \Big)
\end{eqnarray}
where $m^2_{\tilde{q}}$ is an average squark mass and $m_i$ the quark mass.
This expression shows the main effects of the non--universal $A$--terms.
In the first place, we can see that the diagonal elements are still very
similar to the universal $A$--terms situation. Apart of the usual scaling with the
quark mass, these flavor--diagonal mass insertions receive
dominant contributions from the corresponding $A_{C^9_i}$ terms
plus an approximately equal contribution from gluino to all three generations
and an identical $\mu$ term contribution. Hence, given that the gluino
RG effects are dominant, also the phases of 
these terms tend to align with the gluino phase, as in the minimal supergravity. 
Therefore, EDM bounds constrain mainly the relative phase between $\mu$ and gluino
(or chargino) and give a relatively weaker constraint to the relative
phase between $A_{C^9_3}$ (the first generation $A$--term) and the relevant
gaugino. Effects of different $A_{C^9_i}$ in these elements are suppressed by squared CKM
mixing angles. However, flavor--off--diagonal elements are completely new
in this model. They do not receive significant contributions from gluino
nor from $\mu$ and so their phases are still determined by the $A_{C^9_i}$
phases and, in principle, they do not directly contribute to EDMs .

\begin{figure}
\begin{center}
\epsfxsize = 11cm
\epsffile{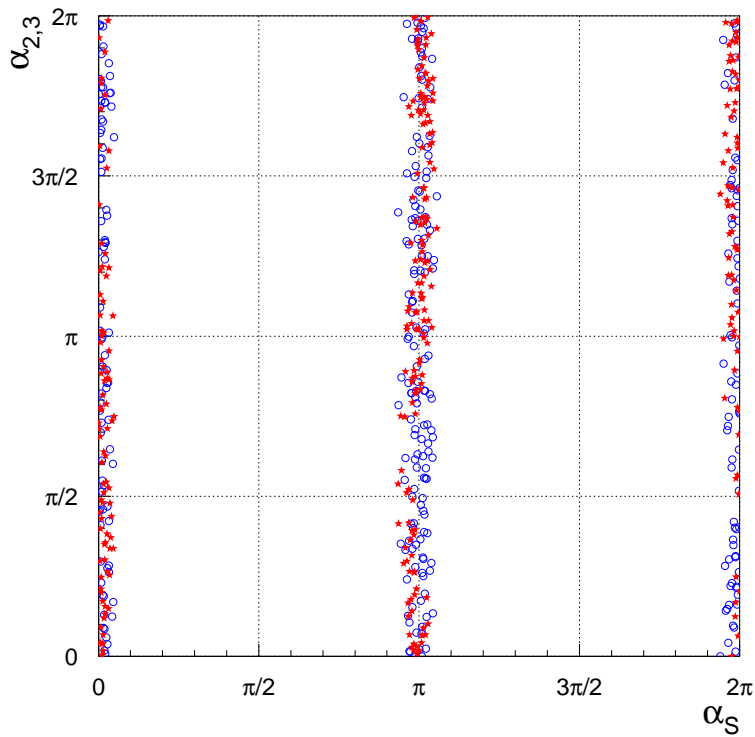}
\leavevmode
\end{center}
\caption{Allowed values for $\alpha_2$--$\alpha_S$ (open blue circles) and
$\alpha_3$--$\alpha_S$ (red stars)}
\label{scat}
\end{figure}

In figure \ref{scat} we show the allowed values for $\alpha_S$, $\alpha_2$
and $\alpha_3$ assuming $\alpha_1=\varphi_\mu=0$. We 
have imposed the EDM, \eps and $b \to s \gamma$ bounds with the usual bounds on SUSY 
masses. We can see that, similarly to the minimal supergravity, $\varphi_\mu$ is
constrained to be very close to the gluino and chargino phases
(in the plot $\alpha_S \simeq 0, \pi$), but $\alpha_2$ and
$\alpha_3$ are completely unconstrained.

\begin{figure}
\begin{center}
\epsfxsize = 11cm
\epsffile{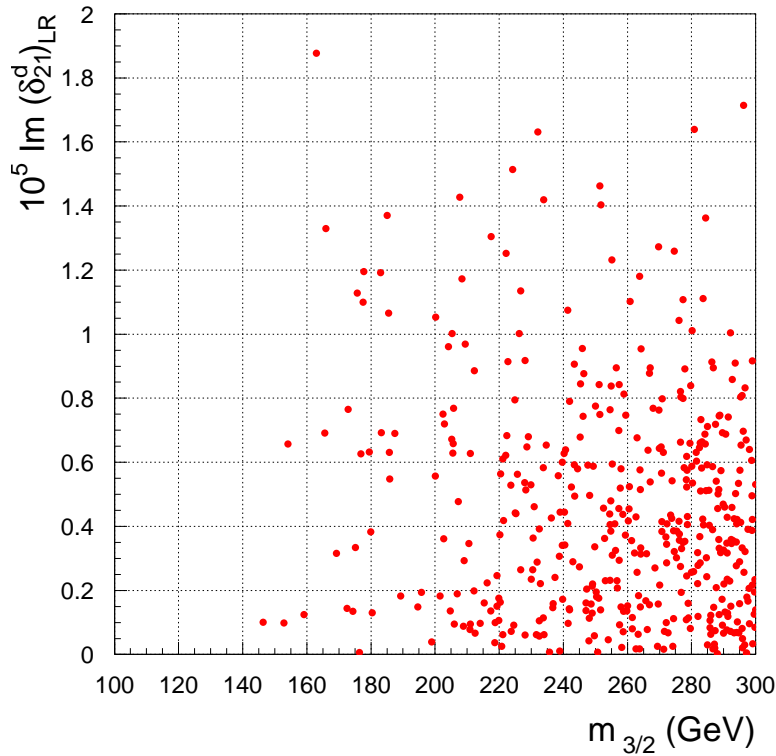}
\leavevmode
\end{center}
\caption{$(\delta_{LR}^{(d)})_{2 1}$ versus $m_{\tilde{g}}$ for experimentally
allowed regions of the SUSY parameter space}
\label{eps'}
\end{figure}

Finally, in figure \ref{eps'} we show the values of
$Im (\delta^{(d)}_{LR})_{2 1}$ versus the gluino mass in the same regions of
parameter space and with the same constraints as in figure \ref{scat}. 
As we can see due to the effect of the off-diagonal phases a large percentage of points 
are above or close to $1 \times 10^{-5}$, hence, sizeable supersymmetric contribution to
$\varepsilon^\prime/\varepsilon$ can be expected in the presence of
non-universal $A$--terms. 

\subsubsection{\bf Conclusions}

Non--universal Supersymmetry soft breaking terms are a natural consequence
in many supergravity or string inspired SUSY models. Moreover, the
non--universality of the $A$--terms has a significant effect in the CP violation.
We have shown that in these models a sizeable supersymmetric contribution to CP 
observables \epspeps and  \eps can be easily obtained. 




\section{\bf $B_s \to J/\psi \phi$ for $\delta \gamma$ and the                 
width difference in the $B_d$ - $\overline{B}_d$\, system}


\def\beq{\begin{equation}}
\def\eeq{\end{equation}}
\def\barr{\begin{eqnarray}}
\def\earr{\end{eqnarray}}
\def\aa{{\cal A}}
\def\rp{{\cal R}_P}
\def\rcp{{\cal R}_{CP}}
\def\dg{\delta \gamma}
\def\bb{{\cal B}}
\def\pp{|\vec{p}_{K^+} \vec{p}_{K^-} \rangle}
\def\pj{|\vec{p}_{\psi} J_{z\psi} \rangle}
\def\pppj{|\vec{p}_{K^+} \vec{p}_{K^-} \vec{p}_{\psi} J_{z\psi}
  \rangle _\phi}

\subsection{\bf The decays $B_s \to J/\psi \eta$ vs.
$B_s \to J/\psi \phi$ for $\delta \gamma$}

{\it Amol S. Dighe, CERN}

\subsubsection{\bf Disentangling the CP eigenstates}

The decay rate of the mode  $B_s \to J/\psi \eta$              as
a function of time is approximately given as
\beq
\frac{d\Gamma(B_s \to J/\psi \eta)}{dt} \approx
|\aa(0)|^2 ~ \left[ e^{-\Gamma_L t} + 2 e^{-\bar{\Gamma} t}  
\sin (\Delta m_s t) \delta \gamma \right]~~,
\label{tagged-eta}
\eeq
where $\bar{\Gamma} \equiv (\Gamma_L+\Gamma_H)/2$.
The non-oscillatory part of (\ref{tagged-eta}) gives
the value of $\Gamma_L$ whereas the oscillatory part
gives $\dg$.

The decay mode $B_s \to J/\psi(\ell^+ \ell^-) \phi(K^+ K^-)$
is another candidate for the determination of 
$\delta \gamma$. It has the advantages of having a larger
branching ratio and a higher efficiency of detection
(since all the four final state particles are charged).
However, the final state $J/\psi \phi$ is not a pure
$CP$ eigenstate, but an admixture of $CP$ odd and
$CP$ even components. 
Here, we argue that in spite of this, the mode 
$J/\psi \phi$ is a better candidate than $J/\psi \eta$
for determining $\delta \gamma$.

The most general amplitude for
this decay can be written in terms of the polarizations
$\epsilon_{J/\psi}, \epsilon_\phi$ of the
two vector mesons as \cite{ddlr,ddf}:
\beq
A(B_s \to J/\psi \phi) =  A_0 
\left(\frac{m_\phi}{E_\phi} \right) 
\epsilon^{*L}_{J/\psi} \epsilon^{*L}_\phi - 
\frac{A_\parallel}{\sqrt{2}}~ 
\epsilon^{*T}_{J/\psi} \cdot \epsilon^{*T}_\phi
- i \frac{A_\perp}{\sqrt{2}}~
\epsilon^*_{J/\psi} \times \epsilon^*_\phi 
\cdot {\bf \hat p}~~,
\label{general2}
\eeq
where $E_\phi$ is the energy of  $\phi$ and
${\bf \hat p}$ is the unit vector in the direction of $\phi$ 
in the $J/\psi$ rest frame.
The superscripts $L$ and $T$ represent the longitudinal
and transverse components respectively.
Since the direct $CP$ violation in this mode is negligible, 
the amplitudes $A_0$ and $A_\parallel$ are
$CP$ even whereas $A_\perp$ is $CP$ odd.

The final states with different $CP$ parities can be
separated through their different angular distributions.
The decay $B_s \to J/\psi(\ell^+ \ell^-) \phi(K^+ K^-)$,
however, has four final state particles, which implies
that the angular distributions will, in general, be in
terms of three physical angles \cite{ddlr,ddf}. 
Such a three angle analysis has been performed \cite{lhc-work},
where preliminary studies indicate that 
that the values of $\dg$ as low as 0.03 may be accessible,
depending on the value of $x_s \equiv \Delta m_s/\bar{\Gamma}$.

However, more the number of angular terms, more formidable 
the task of disentangling their coefficients, which are the
physical quantities of interest. 
The method of angular moments \cite{ddf,senf} may be of
some help, but with as many as 6 angular 
terms in the complete three angle distribution, the errors
in the determination of the coefficients are large. Moreover,
it involves a fit to as many as 8 independent parameters,
some of which seem to have strong correlations. Getting rid of
these problems would increase the sensitivity on the value of
$\dg$, and might bring its SM value within the domain of
measurability.

The angular analysis can be simplified a lot if the
angles are chosen such that the angular distribution 
that separates the $CP$ odd and even terms can be written 
in terms of a single angle. 
Here we show that such a angle may be defined 
for this decay. It follows on the lines of the 
arguments given in \cite{dqstl}.
The {\it transversity} angle distribution reduces
the problem to disentangling only two angular terms
(as opposed to six earlier) and only five independent 
parameters (as opposed to eight).

\subsubsection{\bf The transversity angle}
\label{transv}

In the rest frame of $\phi$, the decay $B \to J/\psi K^+ K^-$
is planar. Let us define the {\it transversity} axis as the one
perpendicular to this decay plane. Let the decay plane be the
$x-y$ plane, so that the $z$ axis is the transversity axis.
The transversity angle is the angle made by the spin of $J/\psi$
with this axis.

The final state $\{J/\psi K^+ K^- \}$ in the rest frame of $\phi$
may be represented in the basis of $\pppj$,
where $J_{z\psi}$ is the $z$ component of the spin of $J/\psi$.
Clearly, $\vec{p}_{K^+} = -\vec{p}_{K^-}$. 
Consider the operator $\rcp$, which combines charge conjugation
with the reflection in the $x-y$ plane:
\beq
\rcp \equiv {\cal C} \rp = {\cal CP} e^{i \pi J_z}~~
\label{rcp-def}
\eeq
where ${\cal C}$ and ${\cal P}$ are the charge conjugation and 
parity transformation operators respectively. Since the total
angular momentum of the final state is zero, from
(\ref{rcp-def}), we have 
\beq
\rcp \pppj = (CP)_f \pppj~~,
\label{rcp-whole}
\eeq
where $(CP)_f$ is the $CP$ parity of the final state.

On the other hand, both $\pp_\phi$ and $\pj_\phi$ are
separately eigenstates of  $\rcp$, and $\rcp$ commutes
with any boost in the decay plane. Then, if we denote the
boost from $\phi$ rest frame to $J/\psi$ rest frame by
$\bb$, so that $\bb \pj_\phi = \pj_{J/\psi}$, then
\barr
\rcp \pppj & = & \rcp \pp_\phi~ \bb^{-1} \rcp \bb \pj_\phi 
\\
 & = & (CP)_\phi \pp_\phi~ \bb^{-1} (CP)_{J/\psi}
e^{i \pi \tau} \pj_{J/\psi}~,
\earr
where $\tau$ is the {\it transversity} of $J/\psi$, i.e.
the projection of its spin along the transversity axis.
Using $(CP)_\phi = (CP)_{J/\psi} = +1$ and the fact that
$\bb$ commutes with the other operators, we can write
\barr
\rcp \pppj & = & \pp_\phi ~e^{i \pi \tau} \pj_\phi \\
 & = & (-1)^{\tau} ~\pppj~~.
\label{rcp-combined}
\earr
From (\ref{rcp-whole}) and (\ref{rcp-combined}), we get
\beq
(CP)_f \pppj = (-1)^{\tau}~ \pppj~~,
\eeq
which shows that the final states with even and odd $CP$ 
parities correspond to the states with the transversity of
$J/\psi$ equal to 0 and $\pm 1$ respectively.
There two states can easily be separated by using 
the parity conserving decay $J/\psi \to \ell^+ \ell^-$
as an analyzer, and studying the angular distribution 
in the transversity angle, the angle made by the decay  
products with the transversity axis in the rest frame of 
$J/\psi$.

The above argument may be generalized and shown to be applicable
in the case of all decays of the form $B \to A C(\to C_1 C_2)$,
where 
(a) $A$ and $C$ are self-conjugate particles with
$C_1$ and $C_2$ spinless and charge conjugates of each other
(as in the case under discussion), 
(b) $A, C_1, C_2$ all are self-conjugate particles, or
(c) $A$ and $C$ are charge conjugates of each other, with
$C_1$ and $C_2$ spinless particles. 
In all such cases, a transversity angle can be defined such that
the angular distribution separates $CP$ odd and even final states
\cite{dqstl}.

\subsubsection{\bf Transversity angle distribution in $J/\psi \phi$}
\label{sec-distr}

\begin{figure}
\begin{center}
\epsfxsize=5in
\epsfbox{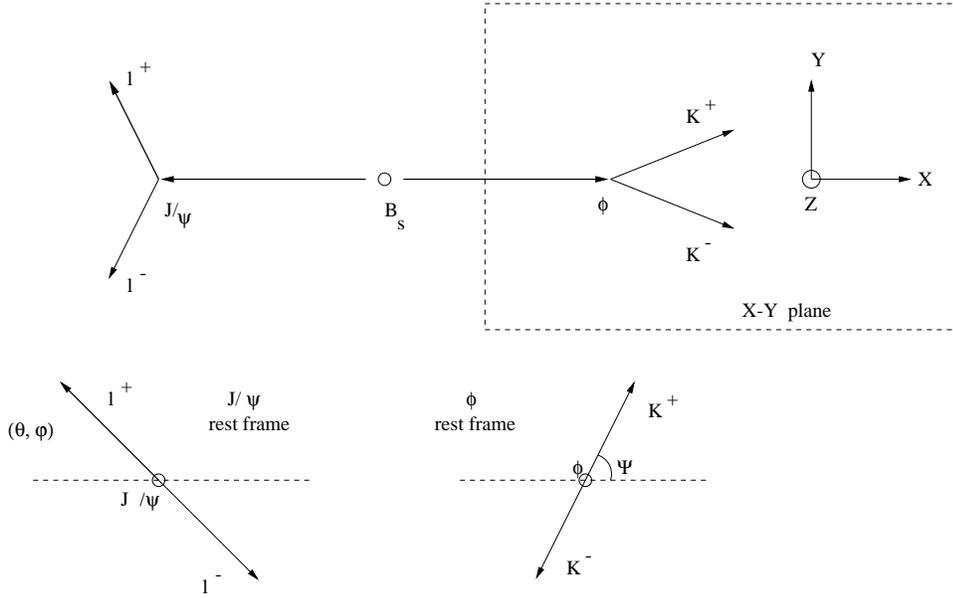}
\end{center}
\caption{The definitions of angles $\theta, \varphi, \psi$.
Here $\theta$ is the transversity angle.}
\label{ang-def}
\end{figure}

Let us define the angles as shown in Fig.~\ref{ang-def}.
Let $x$ axis be the direction of $\phi$ in the $J/\psi$
rest frame, and let $z$ axis be perpendicular to the decay plane of
$\phi \to K^+ K^-$, with positive $y$ direction
chosen such that $p_y(K^+) \geq 0$.
Then we define $(\theta, \varphi)$ as the decay direction of
$l^+$ in $J/ \psi$ rest frame and $\psi$ as the angle made by
$K^+$ with $x$ axis in $\phi$ rest frame.

The angular distribution is then given by
\beq
\frac{d\Gamma[B_s \to J/\psi \phi]}{d\cos \theta dt} 
\approx \frac{3}{8} |\aa_+(t)|^2 (1 + \cos^2 \theta) +
\frac{3}{4} |\aa_-(t)|^2 \sin^2 \theta~~,
\label{angular}
\eeq
where $ |\aa_+(t)|^2 \equiv |A_0(t)|^2 + |A_\parallel(t)|^2$
and $ |\aa_-(t)|^2 \equiv |A_\perp(t)|^2$ are the $CP$
even and $CP$ odd components respectively.
The time evolution of these components is given by
\barr
|\aa_+(t)|^2  & & = |\aa_+(0)|^2 \times \nonumber \\
 & & \left[ \cos^2 (\dg) e^{-\Gamma_L t} +
\sin^2 (\dg) e^{-\Gamma_H t} +
e^{-\bar{\Gamma} t}  
\sin (\Delta m_s t) \sin (2\dg) \right],
\label{aplus-t} \\
|\aa_-(t)|^2  & & =   |\aa_-(0)|^2 \times \nonumber \\
& & \left[ \sin^2 (\dg) e^{-\Gamma_L t} +
\cos^2 (\dg) e^{-\Gamma_H t} -
e^{-\bar{\Gamma} t}  
\sin (\Delta m_s t) \sin (2\dg) \right].
\label{aminus-t}
\earr
The time evolution of either one (or both of) the
components lets us determine the values of
$\Gamma_H, \Gamma_L$ and $\delta \gamma$.

The value of $\delta \gamma$ is small ($\approx 0.015$)
in the standard model. To  first order in $\dg$, eqs.
(\ref{aplus-t}) and (\ref{aminus-t}) may be written as
\barr
|\aa_+(t)|^2  & \approx &  |\aa_+(0)|^2 ~
\left[ e^{-\Gamma_L t} +
2 e^{-\bar{\Gamma} t}  
\sin (\Delta m_s t) \dg \right]~~, 
\label{aplus-t1} \\
|\aa_-(t)|^2  & \approx &  |\aa_-(0)|^2 ~ 
\left[ e^{-\Gamma_H t} -
2 e^{-\bar{\Gamma} t}  
\sin (\Delta m_s t) \dg \right]~~.
\label{aminus-t1}
\earr
The time evolution (\ref{aplus-t1}) is exactly the one
used for the determination of $\dg$ through
$J/\psi \eta$ in (\ref{tagged-eta}). 
Since only two angular terms are present in (\ref{angular}),
the separation of their coefficients is not hard.
Therefore, the accuracy obtained through
these two modes would be similar if the number of 
events were the same. However, the mode $J/\psi \phi $
is expected to have almost an order of magnitude 
more number of events than $J/\psi \eta$.

An interesting situation arises when the measured value of
$\dg$ is large. This is clearly a signal of physics beyond
the SM, in particular of an extra phase appearing in the 
$B_s - \bar{B}_s$ mixing. Using the exact expressions 
for time evolution (eqs. \ref{aplus-t} and \ref{aminus-t})
will give a measurement of this physically interesting 
phase, whereas The approximate time evolution 
(eqs. \ref{tagged-eta}, \ref{aplus-t1} and \ref{aminus-t1})
will fail to do so. In this case, any advantage $J/\psi \eta$
would have had due to its being close to a $CP$ eigenstate
is lost. The non-oscillatory part of (\ref{tagged-eta})
can no longer be used to give a clean measurement of
$\Gamma_L$, since the time evolution now consists of 
two exponential decays with similar lifetimes.
In \cite{senf}, the information content {\it per event}
in such decays has been quantified, and it has been shown that
the mode with angular information may have orders
of magnitude more information {\it per event} than the
mode with time information alone. Adding this to the
larger number of expected events in $J/\psi \phi$,
the case for this mode becomes even stronger.

\subsection{\bf Width Difference in the $B_d$ - $\overline{B}_d$\, system}

{\it A. Dighe, CERN, T. Hurth,  CERN, C.S. Kim, Yonsei University}

\def\beq{\begin{equation}}
\def\eeq{\end{equation}}
\def\ol{\overline}
\def\ul{\underline}
\def\o{\over}
\def\bmat{\begin{array}}
\def\emat{\end{array}}
\def\barr{\begin{eqnarray}}
\def\earr{\end{eqnarray}}
\def\h{\hbox}
\def\l{\left}
\def\r{\right}
\def\non{\nonumber}
\def\wt{\widetilde}
\def\lsim{\raise0.3ex\hbox{$\;<$\kern-0.75em\raise-1.1ex\hbox{$\sim\;$}}}
\def\gsim{\raise0.3ex\hbox{$\;>$\kern-0.75em\raise-1.1ex\hbox{$\sim\;$}}}
\def\mf{{\cal M}_F}
\def\md{{\cal M}_D}
\def\irad{{\cal I}_\tau}
\def\dmsq{\Delta m^2}
\def\gh{\Gamma^d_H}
\def\gl{\Gamma^d_L}
\def\dg{\Delta \Gamma_d}
\def\aa{{\cal A}}


\newcommand{\asl}{$a_{SL}$}
\newcommand{\apks}{$a_{\psi K_S}$}
\newcommand{\app}{$a_{\pi\pi}$}
\newcommand{\arp}{$a_{\rho\pi}$}
\newcommand{\bbar}{$B^0-\bar B^0$}
\newcommand{\goff}{$\Gamma_{12}$}
\newcommand{\moff}{$M_{12}$}
\def\sm{Standard Model}
\newcommand{\mot}{$M_{12}$}
\newcommand{\msmot}{$M^0_{12}$}
\newcommand{\dmot}{$\delta M_{12}$}
\def\epem{$e^+e^-$\ }
\newcommand{\betam}{$\tilde\beta$}
\newcommand{\fb}{$\sqrt{B_B}f_B$}

\def\npb#1{Nucl.\ Phys.\ {\bf B #1}}
\def\plb#1{Phys.\ Lett.\ {\bf B #1}}
\def\prd#1{Phys.\ Rev.\ {\bf D #1}}
\def\prl#1{Phys.\ Rev.\ Lett. {\bf #1}}
\def\ncim#1{Nuo.\ Cim.\ {\bf #1}}
\def\zpc#1{Z.~Phys.\ {\bf C #1}}
\def\prep#1{Phys.\ Rep.\ {\bf #1}}
\def\rmp{Rev.\ Mod.\ Phys.\ }
\def\ijmpa#1{Int.\ J.\ Mod.\ Phys.\ {\bf A #1}}
\def\progtp#1{Prog.\ Th.\ Phys.\ {\bf #1}}
\def\hep#1{hep-ph\ {#1}}
\def\epj#1{Eur.\ Phys.\ J.\ {\bf #1}}

Within the standard model, the difference in lifetimes of
$B_d$ mesons is CKM suppressed  compared with the ones of
$B_s$ mesons. A rough estimate leads to:
\beq
\frac{\Delta \Gamma_d}{\Gamma_d} \approx
\frac{\Delta \Gamma_s}{\Gamma_s}
\left| \frac{V_{cs}}{V_{cd}} \right| ^2
\approx  0.5 \% ~~,
\label{estimate}
\eeq
where $\Gamma_d$ is the average lifetime of the
light and heavy $B_d$ mesons ($B_d^L$ and $B_d^H$ respectively).
We denote these lifetimes by $\Gamma^d_L,
\Gamma^d_H$ respectively, and define $\dg \equiv \gl - \gh$.

More precisely, we get for $\Gamma_{12}$
at leading order using the standard notations \cite{SM,SM2}

\begin{equation}
\Gamma_{12} = {\displaystyle{
-\frac{G_F^2m_b^2m_B}{24\pi}[
\frac{5}{3}\frac{m_B^2}{(m_b+m_d)^2}
     (K_2-K_1)f_B^2B_S(V_{tb}V_{td}^{*})^2}} \nonumber
\end{equation}
\begin{equation}
{\displaystyle{
+ \frac{8}{3}(K_1+\frac{K_2}{2})f_B^2B_B(V_{tb}V_{td}^{*})^2 +
8(K_1+K_2)f_B^2B_B\frac{m_c^2}{m_b^2}V_{cb}V_{cd}^{*}V_{tb}V_{td}^{*}}}].
\label{leading}
\end{equation}

$K_1=-0.39$ and $K_2=1.25$  are combinations of Wilson coefficients.
$B_S$ and $B_B$ are the
bag factors corresponding to the matrix elements of the operators
$Q_S \equiv (\bar b d)_{S-P}(\bar b d)_{S-P}$ and
$Q \equiv (\bar b d)_{V-A}(\bar b d)_{V-A}$. Using the known
expression for $M_{12}$ (see i.e. \cite{SM})
and $m_b=4.5$ GeV, we have

\beq
\frac{\Gamma_{12}^0}{M_{12}^0} = -5.0\times 10^{-3}
                                 \left (1.4\frac{B_S}{B_B} + 0.24 +
                                 2.5 \frac{m_c^2}{m_b^2}
                                 \frac{V_{cb}V_{cd}^*}{V_{tb}V_{td}^*}
                                 \right ).
\label{leading2}
\eeq
In the vacuum saturation
approximation one has $B_S/B_B=1$ at some typical hadronic scale.

With ${\rm Re}(\Gamma_{12}/M_{12}) \simeq \Delta\Gamma/\Delta
m$, and the measured value $x_d = \Delta m_{d} / \Gamma_d = 0.73
\pm 0.05$ we confirm the rough estimate~(\ref{estimate}) taking
into account the large hadronic uncertaintities.

No experimental measurement of $\Delta \Gamma_d$ is
currently available.
Indeed, any new physics contribution
can only decrease $\dg$ \cite{yuval}, thus
taking it still out of the range of sensitivity of
the $B$ factories or at the present hadronic machines
that concentrate on $B$ mesons. Moreover, no motivation for
its measurement (other than just measuring another number to
check with the SM prediction) has been discussed, and hence
the study of the lifetime difference between $B_d$ mesons
has hitherto been neglected as compared to the $B_s$ mesons.
The corresponding lifetime difference in the case of $B_s$
is expected to be $\Delta \Gamma_s / \Gamma_s \approx 15 \%$.

The time resolution and the high statistics expected at the
LHC \cite{lhc} might make it possible to measure
$\Delta \Gamma_d$.
Here we shall discuss the possible measurement of
$\Delta \Gamma_d$ at the LHC, the effect of
$\Delta \Gamma_d$ on the measurement of $\sin(2\beta)$
through $B_d \to J/\psi K_s$ and
the possible resolution of the discrete ambiguity in $\beta$
through the decay $B_d \to J/\psi K^\ast$.

The lifetime of $B_d$ is $\Gamma_d \approx 1.5$ ps \cite{pdg3}.
It is conventionally measured through the self-tagging
semileptonic decays,
and in terms of the lifetimes $B_d^L$ and $B_d^H$, is
given by
\beq
\Gamma_d(sl) = \frac{(\Gamma^d_L+ \Gamma^d_H) \Gamma^d_L \Gamma^d_H }
{(\Gamma^d_L)^2 + (\Gamma^d_H)^2}
= \bar{\Gamma}_d
\frac{1 - \frac{1}{4} \left( \frac{\dg}{\Gamma_d} \right) ^2}
{1 + \frac{1}{4} \left( \frac{\dg}{\Gamma_d} \right) ^2}
\label{gam-sl}
\eeq
Since the difference between $\Gamma_d(sl)$ and
$\bar{\Gamma}_d \equiv (\Gamma^d_L+ \Gamma^d_H)/2$ is
quadratic in the small quantity
$\Delta \Gamma_d / \bar{\Gamma}_d$, we
shall neglect it and shall take
$\Gamma_d \equiv \Gamma_d(sl) =  \bar{\Gamma}_d$
in our analysis.

The lifetime difference $\dg$ affects the measurement
of $\beta$ through $B_d \to J/\psi K_s$.
The $CP$ asymmetry measured in order to determine $\beta$
through this mode is
\barr
\aa_{CP} & = &\frac{\Gamma[B_d(t) \to J/\psi K_s] -
\Gamma[\bar{B}_d(t) \to J/\psi K_s]}
{\Gamma[B_d(t) \to J/\psi K_s] +
\Gamma[\bar{B}_d(t) \to J/\psi K_s]} \\
 & = & \frac{e^{-\Gamma_d t}\sin(\Delta M_d t) \sin(2\beta)}
{\cos^2 \beta~ e^{-\gl t} + \sin^2 \beta ~e^{-\gh t}}~~,
\label{new-exp}
\earr
where $\bar{\beta}$ is the experimentally measured value.
In the limit $\dg \to 0$, the expression
reduces to
$\aa_{CP} = \sin(\Delta M_d t) \sin(2\beta)$, which is the
approximation used in the absence of accurate enough
time measurements. The error introduced in the measurement
of $\beta$ due to this aproximation is
\barr
\frac{\sin(2\beta)}{\sin(2\tilde\beta)} & = &
\cos^2 \beta~ e^{-\dg t/2} + \sin^2 \beta ~e^{\dg t/2} -1 \\
& \approx & - \cos(2\beta) \dg t~~,
\earr
so that for $t \sim 1/\Gamma_d$, the error introduced
due to neglecting the lifetime difference is
of the order of $\dg/\Gamma_d \approx 0.5\%$.

At LHCb, the proper time resolution is expected to be
$\Delta \tau \approx 0.03$ ps \cite{lhc}, at least when
the decay vertex lies within the
silicon strips and its position can be accurately determined.
This would imply that, if the number of relevant events
with the proper time of decay measured with the
precision $\Delta \tau$ above is $N$, then the value
of $\Delta \Gamma_d / \Gamma_d$ is measured with an
accuracy of $2 \cdot 10^{-2} / \sqrt{N}$. With a
sufficiently large number of events $N$, it would be
possible to reach the accuracy of $0.5 \cdot 10^{-2}$
or better.

The above naive argument works if there is a final state
that decays with a given lifetime $\gl$ or $\gh$.
However, such a state does not exist (e.g. decays
into  $CP$ eigenstates does not help
since the $B_d - \bar{B}_d$ mixing phase ($2\beta$)
is large and the $CP$ eigenstates are far away from the
lifetime eigenstates). So we have to find a way to
separate a mixture of two exponential decays with very
similar lifetimes.

The time evolution of the decay rate of $B_d$
into $CP$ eigenstates is given by
\barr
\hspace{-2cm} |\aa_+(t)|^2 & = & |\aa_+(0)|^2 \left[ \cos^2 \beta~ e^{-\gl t}
+ \sin^2 \beta ~e^{-\gh t} +
e^{-\Gamma_d t}\sin(\Delta M_d t) \sin(2\beta)\right]~~,
\label{aplussq} \\
\hspace{-2cm} |\aa_-(t)|^2 & = & |\aa_-(0)|^2 \left[
\sin^2 \beta ~ e^{-\gl t} +
\cos^2 \beta ~e^{-\gh t} -
e^{-\Gamma_d t}\sin(\Delta M_d t)\sin(2\beta)\right]
\label{aminussq}
\earr
for $CP$ even and odd final states respectively.
Separating the two lifetimes from the non-oscillating
part of the time evolutions above is a formidable
task. However, if instead we use a decay to two
vector mesons ($B \to VV$),
the angular distribution of the events
can help us in separating the two lifetimes. It has
been shown in \cite{sen} that the additional information
due to the measurement of the angles can be an order of
magnitude more than the information in the decay time
alone.



Let us have a close look at the decay $B_d \to J/\psi K^\ast
(\to K_s \pi^0)$, which will be useful in the determination
of $\dg$.
The most general amplitude for this decay
is given in terms of the polarizations $\epsilon_{J/\psi},
\epsilon_{K^\ast}$ of the two vector mesons:
\beq
\hspace{-2cm} A(B_s \to J/\psi K^\ast) =  A_0
\left(\frac{m_{K^\ast}}{E_{K^\ast}} \right)
\epsilon^{*L}_{J/\psi} \epsilon^{*L}_{K^\ast} -
\frac{A_\parallel}{\sqrt{2}}~
\epsilon^{*T}_{J/\psi} \cdot \epsilon^{*T}_{K^\ast}
- i \frac{A_\perp}{\sqrt{2}}~
\epsilon^*_{J/\psi} \times \epsilon^*_{K^\ast}
\cdot {\bf \hat p}
\label{general}
\eeq
where $E_{K^\ast}$ is the energy of the $K^\ast$
and ${\bf \hat p}$ the unit vector in the direction of
$K^\ast$ in the $J/\psi$ rest frame.
The superscripts $L$ and $T$ represent the longitudinal
and transverse components respectively.
Since the direct $CP$ violation in this mode is negligible,
the amplitudes $A_0$ and $A_\parallel$ are $CP$ even
whereas $A_\perp$ is $CP$ odd.
Let us define the angles as follows.
Let $x$ axis be the direction of $K^\ast$ in the $J/\psi$
rest frame, and $z$ axis be perpendicular to the decay plane of
$K^\ast \to K_s \pi^0$, with positive $y$ direction
chosen such that $p_y(K_s) \geq 0$.
Then we define $(\theta, \varphi)$ as the decay direction of
$l^+$ in $J/ \psi$ rest frame and $\psi$ as the angle made by
$K_s$ with $x$ axis in $K^\ast$ rest frame.

The transversity angle distribution, which is sufficient
to separate the $CP$ odd and even final states,
is given by
\beq
\hspace{-2cm} \frac{d\Gamma[ B \to J/\psi(\ell^+ \ell^-)K^*(K_S \pi^0)]}
{d\cos \theta dt} \approx
\frac{3}{8} |\aa_+(t)|^2 (1 + \cos^2 \theta) +
\frac{3}{4} |\aa_-(t)|^2 \sin^2 \theta~~,
\label{angularT}
\eeq
where $ |\aa_+(t)|^2 \equiv |A_0(t)|^2 + |A_\parallel(t)|^2$
and $ |\aa_-(t)|^2 \equiv |A_\perp(t)|^2$ are the $CP$
even and odd components respectively.

The complete solution to the time evolutions involves a
maximal likelihood fit to the five parameters $\gl, \gh,
\beta, \Delta M_d, |\aa_-(0)/\aa_+(0)|$. The value of
$\Delta M_d$ can also be taken from other experiments.
This not only would give the values of $\gl$ and $\gh$
separately, but it will also determine the value of
$\sin^2 \beta$, so that $\beta$ is measured without
the discrete ambiguity $\beta \to \pi/2 -\beta$ present
in its determination through
the gold-plated decay $B_d \to J/\psi K_s$.

More ways for the determination of $\dg$ as well as $\beta$
are available through
the angular distribution in the angles $\theta$ and $\varphi$:
\barr
\hspace{-2cm} \frac{d^3 \Gamma [B_d \to (\ell^+\ell^-)_{J/\psi} (K_s \pi^0)_{K^\ast}]}
{d \cos \theta~d \varphi~dt}
& = & \frac{3}{8 \pi} [|A_0|^2 (1 - \sin^2 \theta \cos^2 \varphi)
\nonumber \\
\hspace{-2cm} + |A_\parallel|^2 (1 - \sin^2 \theta \sin^2 \varphi)
& + &  |A_\perp|^2 \sin^2 \theta  - {\rm Im}~(A_\parallel^* A_\perp)
\sin 2 \theta \sin \varphi ]~~~.
\label{twoangle}
\earr
The time evolution of the coefficients of the four angular
terms is
\barr
|A_0(t)|^2 & &=  |A_0(0)|^2 \times \nonumber \\
& &  \left[ \cos^2 \beta~ e^{-\gl t}
+  \sin^2 \beta ~e^{-\gh t} +
e^{-\Gamma_d t}\sin(\Delta M_d t) \sin(2\beta)\right]
\label{a0sq} \nonumber \\
|A_{\|}(t)|^2 & & =  |A_{\|}(0)|^2 \times \nonumber \\
& &\left[ \cos^2 \beta~ e^{-\gl t}
+  \sin^2 \beta ~e^{-\gh t} +
e^{-\Gamma_d t}\sin(\Delta M_d t)\sin(2\beta)\right]
\label{aparpsq} \nonumber \\
|A_{\perp}(t)|^2 & & =  |A_{\perp}(0)|^2 \times \nonumber \\
& & \left[\sin^2 \beta ~ e^{-\gl t}
+
\cos^2 \beta ~e^{-\gh t} -
e^{-\Gamma_d t}\sin(\Delta M_d t)\sin(2\beta)\right]
\label{aperpsq} \nonumber \\
\mbox{Im}\{A_{\|}^*(t)A_{\perp}(t)\}& & =|A_{\|}(0)||A_{\perp}(0)|
\times \nonumber \\
& &\Bigl[
e^{-\Gamma_d t}\left\{\sin\delta_1\cos(\Delta M_d t)
 - \cos\delta_1
\sin(\Delta M_d t)\cos(2\beta) \right\} \nonumber \\
& &-\frac{1}{2}\left(e^{-\gh t}-
e^{-\gl t}\right)\cos\delta_1\sin(2\beta)\Bigr]~~,
\label{interf}
\earr
where $\delta_1 = \rm{Arg}(A_{\|}^*(0)A_{\perp}(0))$.
Note that even before reaching the precision to able to
separate $\gh$ and $\gl$, the above can
already measure the value of $\sin(2\beta)$
through $\mbox{Im}\{A_{\|}^*(t)A_{\perp}(t)\}$.

The discrete ambiguity $\beta \to \pi/2 -\beta$ would
remain unresolved in the absence of the lifetime separation,
since the sign of $\cos \delta_1$, and hence the sign of
$\cos(2\beta)$ is undetermined. One way to resolve this
ambiguity by using $B_s \to J/\psi \phi$ angular distribution
and $U$-spin symmetry is suggested in \cite{ambig}. But
the time evolution (\ref{interf}) above offers a way
without having to take recourse to any other decay
or flavour symmetry.

The non-oscillating part of (\ref{interf}) is
\beq
\hspace{-2cm} {\cal C} \equiv
[\mbox{Im}\{A_{\|}^*(t)A_{\perp}(t)\}]_{NO} =
-\frac{1}{2} |A_{\|}(0)||A_{\perp}(0)|
\left(e^{-\gh t}-e^{-\gl t}\right)
\cos\delta_1\sin(2\beta)~~,
\label{no-osc}
\eeq
which is also the non-oscillating part of the
corresponding charge conjugate decay
$\bar{B}_d \to J/\psi K^\ast (\to K_s \pi^0)$.
The sign of this quantity is the same as the sign of
$\cos \delta_1$, since $\gl > \gh$.
This in turn establishes the sign of $\cos (2\beta)$
through (\ref{interf}).
Note that, in the absence of any $B_d - \bar{B}_d$
production asymmetry, the non-oscillating part of
(\ref{interf}) is exactly the quantity measured if
the initial $B$ meson was not tagged.
Then determination of the sign of
${\cal C}$ would need neither tagging nor time
measurements.

The $B_d - \bar{B}_d$ production asymmetry may be
measured through the asymmetries in the
pairs of self-tagging decays
(a) $B_d \to D_s^+ D^-$ vs.  $\bar{B}_d \to D_s^- D^+$
or
(b) $B_d \to J/\psi K^*(K^+ \pi^-)$ vs.
$\bar{B}_d \to J/\psi K^*(K^- \pi^+)$.
The asymmetry in both the pairs of decays is a combination of
the production asymmetry and a very small amount of direct $CP$
violation. However, though the production asymmetry is the same
in both the pairs, the direct $CP$ violation is expected to be
smaller for the pair (a). This can give us a handle on the
$B_d - \bar{B}_d$ production asymmetry at the hadronic machines.
In the case of $B$ factories that run at the $\upsilon(4S)$
resonance, the production asymmetry is absent and the sign of
${\cal C}$ can be cleanly measured.
For further issues we refer the reader to a forthcoming paper
\cite{newt}.

\section{\bf Trigger and Event Selection Studies}

\subsection{\bf Overview of the CDF B-physics trigger strategy}

{\it M.S.Martin, Oxford University}

\subsubsection{\bf Trigger hardware}
The CDF Run2 Trigger is a significant improvement over its Run1
counterpart. The luminosity expected in Run2 is expected to go up to
$L=2 \times 10^{32} cm^{-2}s^{-1}$\cite{Run2Handbook}. This will require the trigger to
be able to take the interaction rate (about the $7.6 MHz$ bunch
crossing rate) and convert this into the rate to tape (about 50 $ev/s$).
\par The physics capabilities of the Run2 trigger are greatly
increased over Run1. We will be able to examine high $pt$ tracks at
Level 1 (hardware). We will be able to reconstruct (in 2D) high impact
parameter tracks at Level 2 (hardware), and we hope to be able to
perform offline quality reconstruction at Level 3 (software).
\paragraph{\bf Level 1}
At Level 1, we will seek to find physics objects based on a subset of
detector information. These include high energy muons and electrons,
but also high $pt$ tracks using the Extremely Fast Tracker\cite{XFT} (XFT). The
XFT takes a subsection of the main tracking chamber\cite{COT} (Central Outer
Tracker) data and applies a two stage algorithm to locate high $pt$
tracks. The COT is a large open cell drift chamber consisting of 96
layers which are subdivided into 8 superlayers. For each one of four
layers of the COT, the first stage of the XFT algorithm applies a mask
of possible hit configurations for $pt \geq 1.5 GeV$. Then the second
stage of the algorithm then examines $1.25^{\circ}$ angular bins to
put the high $pt$ hits in these layers together to form high $pt$
track candidates. This trigger forms the basis for all of the hadronic
B-Physics triggers at CDF Run2.
\paragraph{\bf Level 2}
Level 2 uses programmable hardware processors\cite{L2} to perform more
sophisticated cuts on the physics objects obtained from Level 1. It
will have an input rate of about 40 $kHz$. A significant Level 2
improvement over Run1 for B-Physics is the Secondary Vertex Trigger\cite{SVT}
(SVT) which uses the main silicon detector (SVX) to find high impact
parameter tracks.  \par The SVT operates on a subset of the data from
the SVX and the COT. It takes the high $pt$ track candidates from
Level 1 and forms {\em superstrips} which are used in conjunction with
SVX data in the pattern recognition stage. The digitised data from the
SVX is fed into the Hit Finder which clusters the strips into hits.
The combination of hits and superstrips are then fed in parallel to a
number of Associative Memory chips which compare the data to a list of
previously computed legitimate combinations.  Finally, these
legitimate combinations (roads) are sent to the Track Fitter Farm
where they are fed into a linear approximation fit to yield 2D tracks
with impact parameter information.

\paragraph{\bf Level 3} \label{l3}
Level 3 consists of a fast switch serving a farm of Linux PC's. The
output rate of Level 2 is going to be about $300 Hz$, and the rate to
tape is going to be about $50 ev/s$. The necessary rejection will be
obtained with a combination of confirmation of Level 2 measurements
and more sophisticated cuts. We will only track in the parts of the
detector which are necessary since there are too many channels (eg in
the silicon) to do global tracking.

\subsubsection{\bf B-physics triggers}
There are two types of B-Physics trigger to be implemented in Run2,
Leptonic and Hadronic. The following sections describe the specific
trigger strategies we intend to pursue for B-Physics. The modes
mentioned are by no means exhaustive, but are rather the modes more
relevant for CKM element measurements and CP-Violation.
\paragraph{\bf Leptonic triggers}
Firstly, building on extensive experience in Run1 we will be
triggering on B-decays with leptons in the final state. The $J/ \psi
\rightarrow \mu \mu$ trigger\cite{jpsimumu} will be the most powerful
version of this trigger and will make use of Level 2 muon objects to
form $J/ \psi$ candidates. We will also implement a $J/ \psi
\rightarrow ee$\cite{jpsiee} trigger. \par Secondly, we will implement
a lepton plus displaced track trigger\cite{semi} to select more
generic leptonic B-decays. The lepton will come from the Level 2 muon
and electron objects, and the displaced track will come from the SVT.
\par Finally we are going to try to implement a Radiative B-Decay
Conversion trigger\cite{conv1} \cite{conv2}, where we attempt to trigger on an electron from
the conversion, and a nearby displaced track from the B to bring the
trigger cross-section down. This has a large overlap with the Lepton
plus displaced track trigger.

\subparagraph{\bf $J/ \psi \rightarrow \mu \mu$ triggers\cite{jpsimumu}}
There are several interesting physics modes which are accessible via a
$J/ \psi$ trigger. The two I will highlight here are the $B
\rightarrow J/ \psi K_{s}$ and $B_{s} \rightarrow J/ \psi \phi$. \par
The baseline requirements at Level 2 (Level 1 mirrors Level 2) for a
$J/ \psi$ candidate are two central muons with $pt>2.0 GeV$. A central
muon in the trigger is defined as a stub in the CMU\cite{muon}
(Central Muon) or CMX (Central Muon Extension) which matches a track
in the COT. The muons must also pass a transverse mass requirement
$1.5 < m_{T} < 3.25 GeV$. We are also investigating lowering the CMU
$pt$ requirement to $1.5 GeV$ to improve this trigger.
\par At Level 3 we will require an invariant mass cut $2.85 < m < 3.25 GeV$. It is
expected that the cross-section out of Level 3 for this trigger will
be $20 nb$ leading to 40 million $J/ \psi$ signal plus sideband events
in the first $2 fb^{-1}$ of Run2. The number of signal events we
expect for $2 fb^{-1}$ are given in table \ref{table:Jpsi}. These estimates
are obtained from scaling observed Run1 samples of these two modes.

\begin{table}[htbp] 
\caption{\label{table:Jpsi} Flagship $J/ \psi$ mode sample estimates}
\indent
\begin{tabular}{|l|l|} \hline
Mode & Expected sample size \\ \hline
 $B\rightarrow J/ \psi K_{s}$    &  20,000 \\ \hline
  $B_{s} \rightarrow J/ \psi \phi$   & 4,000 \\ \hline
\end{tabular}
\end{table}

\subparagraph{\bf $J/ \psi \rightarrow ee$ triggers\cite{jpsiee}}
Although the $J/ \psi \rightarrow \mu \mu$ trigger is by far the most
powerful, we will seek to use electrons to trigger on $J/ \psi$
candidates. \par At Level 1 we require 2 electron candidates with
$E_{T}> 2 GeV$ and $pt > 2 GeV$. We also require $Had/Em <0.125$ where
$Had$ is the deposit of energy in the hadronic calorimeter, and $Em$
is the same for the electromagnetic calorimeter. \par At Level 2 we
make positional requirements on the shower using the Shower-max
detector\cite{cal} (a proportional wire chamber), and require opposite
sign tracks with $\Delta \phi < 90^{\circ}$ (transverse angle). There
is also a full invariant mass cut (in contrast to the transverse mass
for muons) imposed $2.5 < m < 3.5 GeV$. This strategy could be changed
to match the muons though. \par At Level 3, we will impose the same
mass requirement as for the muon version.  In addition we will require
further cuts based upon soft lepton tagging\cite{jpsiee}. \par After
Level 1, the cross-section is 7 to 18 $\mu b$ depending on the
luminosity. This could be reduced by additionally imposing opposite
charge and an angle requirement at Level 1. After Level 2, the
cross-section (without the extra Level 1 cuts) is 100 $nb$, with
negligible luminosity dependence. The Level 3 cross-section is
estimated to be 6 nb leading to an estimate 12 million of these events
in the first $2 fb^{-1}$ of Run2.

\subparagraph{\bf Lepton plus displaced track triggers\cite{semi}}
Triggers of this type gain access primarily to semi-leptonic B-Decays.
Therefore this trigger is relevant for collecting samples to constrain
the CKM elements. This trigger subdivides into Muon, and Electron plus
displaced track. \par For the muon path at level one, we require a
central muon matched to an XFT track (the Level 1 track trigger) with
$pt > 4 Gev$. For the electron path, we require a $4 GeV$ Central
Electromagnetic calorimeter tower with $Had/Em<0.125$.  Also, as for
the muon we require a $4 GeV$ XFT track matched to the calorimeter
tower. These Level 1 triggers fully overlap with other triggers. \par
At Level 2 for the muon path we confirm Level 1 and require one SVT
track with $ 120 \mu m <|d0|< 1 mm$ which is correlated to the muon by
opening angle and transverse mass: $\Delta \phi \leq 90^{\circ}$ and
$m_{T} \leq 5 GeV$. For the electron path at Level 2, we require the
same correlation with the associated track, and some positional
information about the shower from the shower-max wire chamber. \par
For the muon path the cross-section after Level 2 is 39 $nb$ while for
the electron path it is 37 $nb$. The cross-section after Level 3 needs
more study, but the estimated data set size is 60-134 million for $2
fb^{-1}$.

\subparagraph{\bf Conversion trigger\cite{conv1}\cite{conv2}}
The physics goal of this trigger is to access the $B \rightarrow s
\gamma$ type of decay (for example $B \rightarrow K^{*} \gamma$, or $B
\rightarrow \phi \gamma$).\par This trigger is very similar to the
Lepton plus Displaced Track trigger, except that it doesn't have the
correlation requirements with the associated track. This means the
cross-section goes up, and so we may require a second SVT track at
Level 2. \par Without the second SVT track required at Level 2, the
cross-section out of Level 2 is $\sim 30 nb$. The Level 3 cross
section is estimated to be $6 nb$, leading to a 12 million event data
set for $2 fb^{-1}$.

\paragraph{\bf Hadronic triggers\cite{had1}\cite{had2}\cite{had3}\cite{had4}}
In contrast to leptonic B-decays, this class of decay will be
accessible solely via the SVT. The strategy is to perform a range of
cuts on combinations of two tracks. The generic term for this is the
{\em Two Track Trigger.} There are two flavours of this trigger: $B
\rightarrow \pi \pi$, and $B_{s}$ hadronic.

\subparagraph{\bf $B \rightarrow \pi \pi$}
The main physics goals of this trigger are to measure the CKM angles
$\alpha$ and $\gamma$. The measurement of $\gamma$ will be attempted in conjunctions with another mode \cite{fkw} which we hope to collect with this trigger, $B_{s} \rightarrow K^{+} K^{-}$. We also require $B \rightarrow J/ \psi K_{s}$ for this measurement. We also hope to collect samples of $B_{s/d} \rightarrow \pi K$ with this trigger with a view to a direct CP-Violation Limit/Measurement.
\par The cuts for this trigger are summarised in the table below.

\begin{table}[htbp]
\caption{\label{table:Bpipi} $B \rightarrow \pi \pi$ Two Track Trigger (d0 is impact parameter)}
\indent
\begin{tabular}{|l|l|l|} \hline
Trigger level & Cuts on each track & Cut on the pair \\ \hline
L1            & $pt>2 GeV$ & $\Sigma pt >5.5 Gev$, $\Delta \phi <135^{\circ}$ \\  \hline
L2            & $100 \mu <d0< 1mm$ & $20^{\circ} < \Delta \phi < 135^{\circ}$\\  \hline
L3            & Under discussion       & \\\hline
\end{tabular}
\end{table}

The $\Delta \phi$ cuts are made in 2D on the transverse angles. The
difference between this trigger and the next one is that the opening
angle of the two tracks is constrained to be quite large whereas the
opening angle in the $B_{s}$ hadronic trigger can be more restricted.
This effectively makes this trigger into a high Q-Value trigger, and
correspondingly the subsequent one into a low Q-Value trigger. \par At
Level 3 the base-line is verification of Level 2, which will require
tracking regionally in the silicon as mentioned in section \ref{l3}.
The cross-section of this trigger is considered with the subsequent
one below.

\subparagraph{\bf $B_{s}$ hadronic}
The main purpose for this trigger is to collect events for a
measurement of $B_{s}$ mixing. However, many other fully reconstructed
B decays will come in on this trigger. One example is $B_{s}
\rightarrow D_{s}^{(*)} D_{s}^{(*)}$ in which CP-Violation is predicted to be
zero.
\par The cuts for this trigger are summarised in the table below.

\begin{table}[htbp]
\caption{\label{table:Bs} $B_{s}$ hadronic Two Track Trigger (d0 is impact parameter)}
\indent
\begin{tabular}{|l|l|l|} \hline
Trigger level & Cuts on each track & Cut on the pair \\ \hline
L1            & $pt>2 GeV$ & $\Sigma pt >5.5 Gev$, $\Delta \phi <135^{\circ}$ \\  \hline
L2            & $100 \mu <d0< 1mm$ & $2^{\circ} < \Delta \phi < 90^{\circ}$, $(pt \cdot X_{\nu})>0$    \\  \hline
L3            & Under discussion &      \\  \hline
\end{tabular}
\end{table}

\begin{figure}[h!!!] \label{vert}
\begin{center}
\resizebox{5cm}{!}{\includegraphics{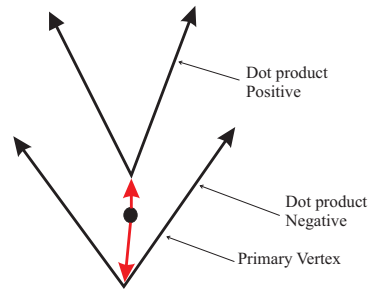}}
\caption{\label{figure:vert} Diagram explaining the effect of $(pt \cdot X_{\nu})>0$ cut}
\end{center}
\end{figure}

The $(pt \cdot X_{\nu})>0$ cut is defined as follows. $X_{\nu}$ is the
vector pointing from the primary vertex to the secondary vertex, and
so, the dot product $pt \cdot X_{\nu}$ will be positive for tracks
originating in front of the primary, and negative for tracks
originating behind it. See figure \ref{vert}.

\par At Level 3, as for $B \rightarrow \pi \pi$ we will at the very least verify Level 2, but we will probably need to do something more sophisticated than this to beat the cross-section down to an acceptable level. \par The trigger rates are estimated to be (for the initial configuration of the accelerator) $250 nb$ at level1, $560 \pm 125 nb$ at Level 2, and $200 nb$ at Level 3. This would lead to a 200 million event data set size.

\subsection{\bf A brief outline of the LHCb trigger}

{\it C. Shepherd-Themistocleous, Cambridge University}

\subsubsection{\bf Introduction}
B meson decay processes of interest to CP violation physics have small
branching ratios, typically of the order of 10$^{-5}$.  The production
of large quantities of B mesons is therefore required for a study of
these processes. Such large rates, of the order of 10$^5$ $\rm
b\bar{b}$ events/s will be available at the LHC collider. The price
that has to be paid for these large rates is a very high background
rate.  The rate of bunch crossings with at least 1 interaction at the
LHCb interaction point will be of the order of 12MHz.  One of the
greatest challenges of the experiment is therefore to be able to
trigger on interesting events. The event rate is so high that it is
not even possible to record all B events.  This therefore means that a
highly selective and efficient trigger is required. It also has the
important consequence that the designers of the trigger should be
aware of all potentially interesting final states while the trigger is
being designed.  If a particular channel is only found to be of
interest at a later date then it is likely that the trigger will be
inefficient for this particular channel. While the parts of the
trigger that are algorithms running on processors may easily be
changed, changing the hardware is a much more difficult task.  The
strategy that will be followed at the LHCb detector is briefly
outlined below.

\subsubsection{\bf Trigger strategy at the LHCb experiment}~\cite{ref:lhcb}

The beam crossing rate at the LHC is 40MHz and the LHCb trigger is 
designed to be ready to receive data at every beam crossing. 
This means that the detectors must be readout at a rate of 40MHz 
and a trigger decision must also be made a this rate. It is clear 
that a sophisticated trigger algorithm cannot be run in the 25ns available 
between beam crossings. The strategy followed therefore is to 
use a pipeline trigger and to use several different levels of trigger.
In a pipeline trigger data are stored for a fixed time while 
processing is performed to make a trigger decision. The processing takes
place in parallel for different events. Therefore, in the 
first level of the trigger for example,  while a decision is 
made at a frequency of 40MHz, 4$\mu$s are available for each 
event to be processed. 

The Level 0 trigger utilizes the fact that decay products of B decays
have on average higher momenta transverse to the beam (p$_t$)
direction than background particles. This trigger looks for high p$_t$
muons, electrons and hadrons as well as photons and possibly neutral
pions. These triggers take place in parallel and are then combined
into a final Level 0 decision. At this stage a veto trigger on beam
crossings that contain more than 1 interaction may also be used. This
uses part of the vertex detector to identify multiple primary
vertices.  This level of the trigger is designed to produced an output
of 1MHz.

The next trigger level utilizes the other major characteristic of 
B mesons, namely their long lifetimes. This leads to a mean decay 
length of 7mm at LHCb and hence to decay vertices clearly separated 
from the primary vertex. A silicon vertex detector placed close to the 
beam pipe is used to look for these secondary vertices. 

These first two levels of the trigger are ``hardware'' triggers in the
sense that it will be hard to change these triggers once the detector
is built. Any changes would have implications for aspects such as the
physical configuration in which the detectors are read out and the
speed of operation of specialized chips on electronics boards.  The
first two levels of the trigger have to operate a high rates and hence
essentially use one subdetector in each trigger for speed.

At this point all the data is made available to a processing farm and
all subsequent triggers are software algorithms. Data from several
subdetectors can now be combined to refine the earlier triggers. An
example is the refinement of the vertex trigger. The vertex detector
is not in a magnetic field and a source of fake secondary vertices is
from multiply scattered low momentum tracks. This background can be
removed by using information from tracking chambers. Complete tracking
information will also be available to selection algorithms in the
trigger.  After all generic B selection algorithms the event is rate
still too high to allow the recording of all events selected at this
point. Once this stage has been reached algorithms to select specific
B decays or classes of B decay are performed. Data will finally be
written to tape at rate of about 200Hz.\footnote{This number is under
revision and may well change in the near future.}  An important
difference between this trigger and the trigger used in CDF is that
analysis of data in the later stages of the trigger is not restricted
to regions of interest in the detector that have been defined by
earlier stages of trigger. The processing power available here is such
that this time saving device is not required.

The discussion about triggers that took place in the CP violation
working group proved to be a very useful one. In particular many
people where made aware of the difficulty of triggering in the hadron
collider environment and of the importance of knowing early in the
design phase what final states must be triggered on. Many useful
discussions on what ought to be included in the LHCb trigger took
place.

\subsection{\bf Separating $\Upsilon(4\mbox{S})$ decays from continuum events
using a neural network}

{\it Fabrizio Salvatore and Glen Cowan, Royal Holloway}

\subsubsection{\bf Introduction}

Artificial Neural Networks (NN) have been applied to a variety of problems 
in high energy physics in order to discriminate between different 
classes of events.  Here the technique is used to separate 
$b\bar{b}$ events from $\Upsilon(4\mbox{S})$ decays
(`signal') from the continuum background for 
$\mbox{e}^+\mbox{e}^-$ interactions at a centre-of-mass energy
$E_{\rm cm} = 10.58$ GeV, i.e., at the $\Upsilon(4\mbox{S})$ resonance. 
The analysis relies on the different topologies for
signal and background at a B Factory: while $b\bar{b}$ events are 
more `spherical', $uds$ and $c\bar{c}$ events are more `jet-like'. 
This is due to the fact that in the $\Upsilon(4\mbox{S})$ rest frame, the 
momenta of the produced B mesons are small and they decay
isotropically.  On the other hand, events where light quarks 
($u\overline{u}$, $d\overline{d}$ or $s\overline{s}$) 
are produced are characterized by a preferred direction with hadron
jets following roughly the quark and antiquark. A similar 
jet-like structure is also present in $c\overline{c}$ events, 
although there it is less pronounced.

For this study, two samples of 600k events of the type
$\mbox{e}^+\mbox{e}^- \rightarrow q\overline{q}$ have been generated using  
PYTHIA \cite{pythia} all at a centre-of-mass energy equal to 
the mass of the $\Upsilon(4\mbox{S})$, where the quark flavour $q$ is 
a mixture of $u$, $d$ and $s$ in the first set
and $c$ in the second.  A third data sample of the same size consisted
of $\Upsilon(4\mbox{S})$ decays.  The analysis has been performed at the 
level of the generated particles and no attempt has been made at this
stage to simulate detector effects.

\subsubsection{\bf The shape variables}
\label{sec:shape}

The variables used to discriminate jet-like from isotropic events are 
described below.  Further information on the variables can be
found in \cite{babar2, aleph}.

\begin{itemize}

\item[(a)] The ratio {\bf $R_2 = H_2 / H_0$} of the 2nd to the 0th 
Fox-Wolfram moments.  Neglecting particle 
masses, 4-momentum conservation requires that $H_0 = 1$. For a two-jet 
event, $H_1=0$ and $H_l \sim 1$ for $l$ even and $H_l \sim 0$ for $l$ odd. 
The $R_2$ distribution is shown in Fig.~\ref{fig:fig1}, 
where the different behaviour for signal and background events can be 
observed.

\item[(b)] The thrust {\bf $T = \sum_i |\hat{T}\cdot{\bf p_i}| / 
\sum_i |{\bf p_i}|$}, where $\hat{T}$ is the thrust axis of the event 
\cite{babar2}.  For an isotropic event one has {\bf $T=0.5$} while 
{\bf $T\sim 1$} for a highly directional one, as shown in Figs.~\ref{fig:fig2}
(a) and (c).

\end{itemize}

\setlength{\unitlength}{1.0 cm}
\renewcommand{\baselinestretch}{0.9}
\begin{figure}[htbp]
\begin{picture}(40.0,6.)
\put(3.5,-2.5)
{\includegraphics{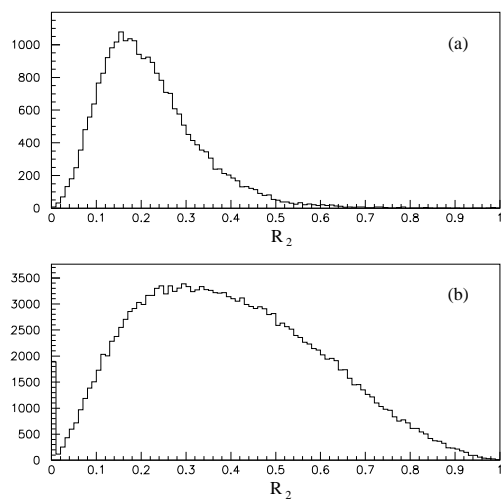}}
\end{picture}
\caption{\footnotesize $R_2$ distribution for (a) $\Upsilon(4\mbox{S})$ 
signal and (b) continuum 
events.}
\label{fig:fig1}
\end{figure}
\renewcommand{\baselinestretch}{1}
\footnotesize\normalsize

The variables defined above are quite powerful for identifying events 
where the two-jet topology is well defined and clearly distinguishable 
from the isotropic case.  The signal to background separation 
decreases, however, when one or more gluons are emitted, which is more 
likely to happen when light quarks are produced. In this case there 
are three or more jets, so that the event shape resembles more that
of the signal.

To discriminate further between signal and background, the Durham 
clustering algorithm \cite{durham_alg} is applied to group particles 
into jets.  For every pair of particles $i$ and $j$, the `distance' 
$y_{ij}$ is computed as

\begin{equation}
y_{ij} = \frac{2~\mbox{min}(E_i^2,E_j^2)~(1- \cos \theta_{ij})}{E_{\rm vis}^2}
\,,
\label{eqn:eq1}
\end{equation}

\noindent where $E_i$ and $E_j$ are the particles' energies, 
$\theta_{ij}$ is their opening angle and $E_{\rm vis}$ is the total 
(visible) energy in the event.  The pair with the smallest $y_{ij}$
is replaced by a pseudo-particle with four-momentum 
$p^{\mu} = p^{\mu}_{i} + p^{\mu}_{j}$. The procedure is repeated 
until $N$ jets  have been found.  In our case $N=3$ or 4 and the 
smallest $y_{ij}$ values are called $y_{3}$ and $y_{4}$, respectively.
Once the clustering has been done, the following discriminating 
variables can be added to the list:

\begin{itemize}

\item[(c)] {\bf $y_{3}$}, {\bf $y_{4}$};

\item[(d)] {\bf $\theta_{\rm BZ}$}, the angle between the highest and 
second highest energy jet;

\item[(e)] {\bf $\theta_{\rm KS}$}, the angle between the lowest and 
second lowest energy jet;

\item[(f)] {\bf $\theta_{\rm NR}$}, the angle between the planes defined 
by the highest and lowest energy jets and by the other two jets;

\item[(g)] {\bf $\theta_{34}$}, the angle between the planes defined 
by the highest and second highest jets and by the other two jets;

\item[(h)] the QCD four-jet matrix element squared, {\bf $|M_{QCD}|^2$}, 
averaged over all possible assignments of the jets to the final-state partons
\cite{ERT}.

\end{itemize}

As an example, the distribution of $\ln|M_{QCD}|^2$ is shown in 
Fig.~\ref{fig:fig2}(b) and (d) for signal and background events.

\setlength{\unitlength}{1.0 cm}
\renewcommand{\baselinestretch}{0.9}
\begin{figure}[htbp]
\begin{picture}(40.0,8.2)
\put(3.2,-3.5)
{\includegraphics{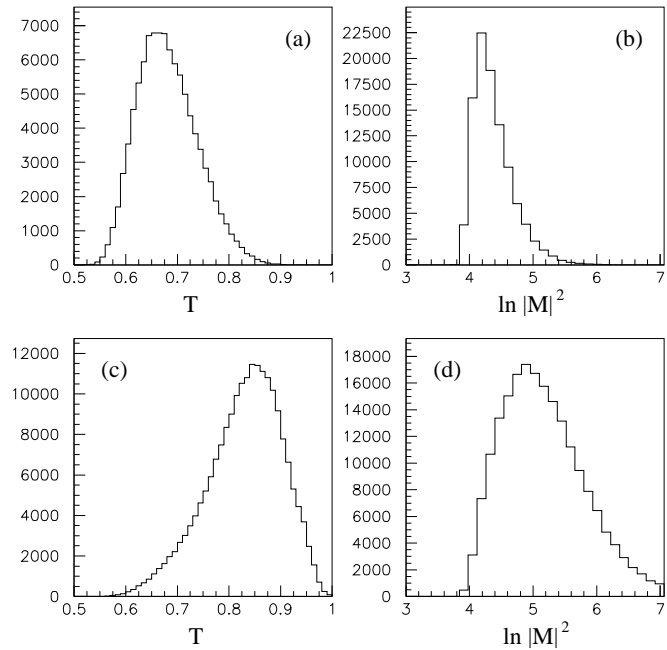}}
\end{picture}
\caption{\footnotesize Distribution of (a) thrust and (b) 
$\ln |M_{QCD}|^2$ (right) for signal events; (c) and (d) show the same
for continuum events.}
\label{fig:fig2}
\end{figure}
\renewcommand{\baselinestretch}{1}
\footnotesize\normalsize

\subsubsection{\bf The analysis results}
\label{sec:res}

All the variables defined in the previous section have been combined in a 
Neural Network using the package JETNET 3.51 \cite{jetnet}. The network is a
3 layered feed-forward one, with 9 input nodes, 12 hidden nodes and 
one output.  The network has been trained using 100k $b\bar{b}$, 
100k $uds$ and 100k $c\bar{c}$ events. Each input variable has been 
scaled linearly so that it lies in the range [-1,1];  in this way at 
the start of the training no single variable dominates the inputs 
to any neuron.  The training algorithm is `error back-propagation'.

After the training, the NN has been used with an independent sample of 
events, 500k for each type, and the output distributions are shown in 
Fig.~\ref{fig:fig3}. The signal and background curves are very well 
separated, with continuum background events accumulating around 0 and 
signal events around 1. 

\setlength{\unitlength}{1.0 cm}
\renewcommand{\baselinestretch}{0.9}
\begin{figure}[htbp]
\begin{picture}(40.0,9.1)
\put(2.,-3.4)
{\includegraphics{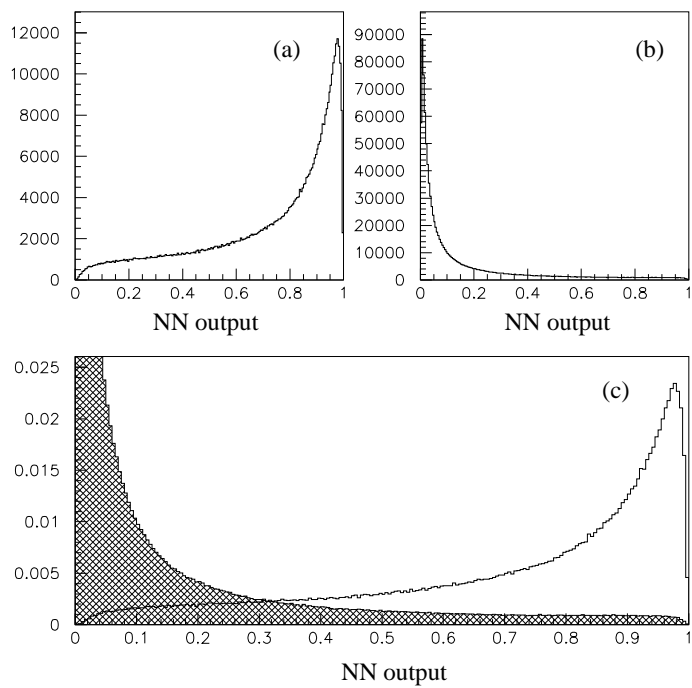}}
\end{picture}
\caption{\footnotesize Distributions of the NN output for 
(a) signal and (b) background events. The distributions 
are superimposed in (c) after being normalized to unity.}
\label{fig:fig3}
\end{figure}
\renewcommand{\baselinestretch}{1}
\footnotesize\normalsize

To evaluate the rejection power of the NN we can set a cut on the output
such that the efficiency for the signal is 90\%; this corresponds
to requiring an output greater than $0.35$. The fraction of background 
events surviving the cut is 14\% for $uds$ events and 18\% for 
$c\bar{c}$ events.

\subsubsection{\bf Conclusions}

A feed-forward neural network has been applied to separate $q\overline{q}$
continuum events from $\Upsilon(4\mbox{S})$ decays produced in 
simulated $\mbox{e}^+\mbox{e}^-$ interactions at the $\Upsilon(4S)$ 
resonance.  Preliminary results are quite encouraging. For a signal 
efficiency of 90\%, only $\sim 16\%$ of background events survive the 
NN cut.  More work needs to be done to verify that the results are still 
valid after a full detector simulation; nevertheless, the NN technique seems 
a promising option for this kind of analysis.

\section{\bf QCD Factorization}

\subsection{\bf Factorization in charmless B decays}

{\it W.N. Cottingham}

Apart from electro-weak terms and other small corrections and taking account 
of QCD corrections the effective Hamiltonian for b quark decay is a sum of 
current current terms $j \cdot j$.
Factorization approximates the quasi two body sum of matrix elements 
$\langle h_1 h_2\mid j \cdot j \mid B \rangle$
by a sum over products of single body matrix elements 
$\langle h_1 \mid j^{\mu} \mid B \rangle 
\langle h_2\mid j_{\mu} \mid 0 \rangle$.
So that the transition amplitude~\cite{noel1, noel2, noel3} becomes

\[
\langle h_1 h_2 \mid H_{eff} \mid B \rangle = 
G_F / \sqrt{2} \Sigma^6_i P_i [ Q_i(h_1,h_2) + Q_i(h_2,h_1)]
\]
\[
Q_1(h_1,h_2) = 
\langle h_1 \mid \bar u \gamma^{\mu} (1-\gamma_5) b \mid B \rangle
\langle h_2 \mid \bar q \gamma_{\mu} (1-\gamma_5) u \mid 0 \rangle 
\]
\[
Q_2(h_1,h_2) = 
\langle h_1 \mid \bar q \gamma^{\mu} (1-\gamma_5) b \mid B \rangle
\langle h_2 \mid \bar u \gamma_{\mu} (1-\gamma_5) u \mid 0 \rangle
\]
\[
Q_3(h_1,h_2) = 
\langle h_1 \mid \bar q \gamma^{\mu} (1-\gamma_5) b \mid B \rangle
\langle h_2 \mid \bar q^{'} \gamma_{\mu} (1-\gamma_5) q^{'} \mid 0 \rangle
\]
\[
Q_4(h_1,h_2) = 
\langle h_1 \mid \bar q^{'} \gamma^{\mu} (1-\gamma_5) b \mid B \rangle
\langle h_2 \mid \bar q \gamma_{\mu} (1-\gamma_5) q^{'} \mid 0 \rangle 
\]
\[
Q_5(h_1,h_2) = 
\langle h_1 \mid \bar q \gamma^{\mu} (1-\gamma_5) b \mid B \rangle
\langle h_2 \mid \bar q^{'} \gamma_{\mu} (1-\gamma_5) q^{'} \mid 0 \rangle
\]
\[
Q_6(h_1,h_2) = (-2)
\langle h_1 \mid \bar q^{'} (1-\gamma_5) b \mid B \rangle
\langle h_2 \mid \bar q (1+\gamma_5) q^{'} \mid 0 \rangle 
\]
\[
q =  d, s  \hspace{1cm}   q^{'}  =  u,  d, s 
\]

The coefficients $P_i$ are constructed from Wilson 
coefficients \cite{noel4, noel5} and CKM parameters. The Wilson coefficients 
have been quite reliably estimated from QCD.  The single particle matrix 
elements, for example 
$\langle \pi^+ \mid \bar u \gamma^{\mu} (1 - \gamma_5) d \mid 0 \rangle
= f_{\pi} p_{\pi}^{\mu}$, 
involve soft QCD parameters, in this case $f_{\pi}$ and many are quite 
precisely known for example from measured leptonic decay rates. The matrix 
elements $\langle h_1 \mid j^{\mu} \mid B \rangle$ 
are in principle measurable from B semi leptonic decays, they involve only 
a few parameters. For the moment they are only available as theoretical 
estimates.  

The physical justification for the factorisation approximation is the idea 
of colour transparency. It is argued that the b quark by virtue of its 
large mass imparts a large velocity to the light quark decay products, 
the pair which form the light particle $h_2$ move away but stay together 
as a colour singlet so escape from the gluonic environment of the B meson 
without interaction as does a lepton pair in the semi leptonic decay 
$B \rightarrow h_1 + l + \nu$
which is described by a matrix element 
$\langle h_1 \mid j^{\mu} \mid B \rangle$.
This rationale has recently been given theoretical 
justification~\cite{noel3}, it has been claimed that for decays into a 
pair of light mesons and in the heavy quark limit that the corrections 
to the factorisation approximation can be calculated from first principles 
of perturbative QCD.  The corrections involve, as you might expect, 
hard gluon interactions with the spectator quark but the only  long 
distance physics, other than the current matrix elements, involves 
the meson light cone distribution amplitudes.  The beauty of the results 
presented in~\cite{noel3} for $\pi\pi$
decay is that the coefficients $P_i$ acquire additional perturbative 
corrections. The difference between the decay rate formulae of the 
factorisation approximation and the formulae with the QCD corrections 
lies only in the coefficients $P_i$, the factorisation matrix elements 
are common to both. Generalisations to other decay channels can be 
found in~\cite{noel5, noel6, noel7}.

The corrections to the coefficients $P_i$ so far presented are to first order 
in $\alpha_s$. An encouraging feature is that they are not large, which 
leads one to hope that the precision with which the standard model can be 
tested will be determined by the proximity of the b quark mass to the heavy 
quark limit  and the precision of our knowledge of the soft QCD parameters, 
B meson semileptonic transition form factors, meson light cone distribution 
amplitudes etc. Because of the involvement of soft QCD parameters, some of 
which are only poorly known, and to be confident of the conclusions 
about the standard model, it will be important to have a consistent 
picture of as many channels of charmless B decay as possible.

Limiting the discussion to the light hadrons $h_1$ and $h_2$ being the 
lowest mass pseudo scaler and vector mesons there are only a few poorly 
known soft QCD parameters. Because the B has zero spin a spin one meson 
with a spin zero partner is confined to the helicity zero state.  Also 
in the heavy quark limit and to lowest order in $\left ( m_h / m_B \right )^2 $
only helicity zero states should be involved  in transitions to two spin 
one mesons . The poorly known B transition form factors should then involve 
only five parameters that describe the transitions 
$B \rightarrow \pi, B \rightarrow \rho, B \rightarrow \omega, B \rightarrow K,
B \rightarrow K^{*} $
Setting aside transitions involving the $\eta$ and $\eta^{'}$  there are 
about forty channels of $B^0$ and $B^+$ decay with branching ratios 
anticipated to be greater than  $10^{-6}$~\cite{noel2}. It can be expected 
that the basic validity of the improved factorisation approximation will 
be tested within the next year or two, and  the proximity of the 
b quark mass to the heavy quark limit will determine the sharpness of 
focus that this method provides on the standard model description 
of the weak interaction.

\subsection{{\bf Phenomenological impact of the QCD improved factorization 
approach}\\Content identical to \cite{proccc}}

{\it M.~Beneke, RWTH Aachen}

\subsubsection{\bf Introduction}

The observation of $B$ decays into $\pi K$ and $\pi \pi$ final 
states  has resulted in a large amount of theoretical and 
phenomenological work that attempts to interpret these 
observations in terms of the factorization approximation (FA), or in 
terms of general parameterizations of the decay amplitudes. 
A detailed understanding of these amplitudes would help us to pin down 
the value of the CKM angle $\gamma$ using only data on CP-averaged
branching fractions. Theoretical work on the heavy-quark limit has 
justified the FA as a useful starting 
point~\cite{BBNS99,BBNS00}, but predicts important and computable 
corrections. Here we discuss the most important consequences of 
this approach for the $\pi K$ and $\pi \pi$ final 
states.

To leading order in an expansion in powers of $\Lambda_{\rm QCD}/m_b$, 
the $B\to \pi K$ matrix elements obey the factorization 
formula
\begin{eqnarray}\label{fact}
   \langle\pi K|Q_i|B\rangle
   &=& f_+^{B\to\pi}(0)\,f_K\,T_{K,i}^{\rm I}*\Phi_K \nonumber\\
   &&\hspace*{-1.5cm}\mbox{}
    + f_+^{B\to K}(0)\,f_\pi\,T_{\pi,i}^{\rm I}*\Phi_\pi \\ 
   &&\hspace*{-1.5cm}\mbox{}
    + f_B f_K f_\pi\,T_i^{\rm II}*\Phi_B*\Phi_K*\Phi_\pi \,,
    \nonumber
\end{eqnarray}
where $Q_i$ is an operator in the weak effective Hamiltonian, 
$f_+^{B\to M}(0)$ are semi-leptonic form factors of a vector
current evaluated at $q^2=0$, $\Phi_M$ are leading-twist
light-cone distribution amplitudes, and the $*$-products imply an
integration over the light-cone momentum fractions of the 
constituent quarks inside the mesons. When the hard-scattering 
functions $T$ are evaluated to order $\alpha_s^0$, Eq.~(\ref{fact}) 
reduces to the conventional FA. The 
subsequent results are based on kernels including all corrections 
of order $\alpha_s$. A detailed justification of (\ref{fact}) is 
given in Ref.~\cite{BBNS00}. Compared to our previous discussion 
of $\pi\pi$ final states\cite{BBNS99} the present analysis 
incorporates three new ingredients: 

i) the matrix elements of electroweak (EW) penguin operators 
(for $\pi K$ modes); 

ii) hard-scattering kernels for general, asymmetric light-cone 
distributions; 

iii) the complete set of ``chirally enhanced'' $1/m_b$ 
corrections~\cite{BBNS99}.

\noindent
The second and third items have not been considered in 
other~\cite{DZ00} generalizations of Ref.~\cite{BBNS99} to the $\pi K$ 
final states. The third one, in particular, is essential for estimating 
some of the theoretical uncertainties of the approach. 


We now briefly present the input to our calculations. 
Following Ref.~\cite{BBNS99},
we obtained the coefficients $a_i(\pi K)$ of 
the effective factorized transition operator defined analogously to the 
case of $\pi\pi$ final states, but augmented by
coefficients $a_{7-10}(\pi K)$ related to EW penguin 
operators and electro-magnetic penguin contractions of current--current 
and QCD penguin operators. A sensible 
implementation of QCD corrections to EW penguin matrix elements
implies that one departs from the usual renormalization-group 
counting, in which the initial condition for EW penguin coefficients is 
treated as a next-to-leading order (NLO) 
effect. Our NLO initial condition hence includes 
the $\alpha_s$ corrections computed in Ref.~\cite{BGH00}. 

Chirally enhanced corrections arise from twist-3 two-particle 
light-cone distribution amplitudes, whose normalization involves the 
quark condensate. The relevant parameter, 
$2\mu_\pi/m_b = -4\langle \bar{q}q\rangle/(f_\pi^2 m_b)$, is formally 
of order $\Lambda_{\rm QCD}/m_b$, but large numerically. The
coefficients $a_6$ and $a_8$ are multiplied by this parameter.
There are also additional chirally enhanced corrections to the 
spectator-interaction term in (\ref{fact}), which turn out to be the 
more important effect. In both cases, these corrections involve 
logarithmically divergent integrals, which violate factorization. For
instance, for matrix elements of $V-A$ operators the hard spectator 
interaction is now proportional to ($\bar u\equiv 1-u$)
\[
   \int_0^1 \!\frac{du}{\bar{u}} \frac{dv}{\bar{v}}\, 
   \Phi_K(u)\left(\Phi_\pi(v)+\frac{2\mu_\pi}{m_b} \frac{\bar{u}}{u}
   \right) 
\]
when the spectator quark goes to the pion. (Here we used that the 
twist-3 distribution amplitudes can be taken to be the asymptotic
ones when one neglects twist-3 corrections without the chiral 
enhancement.) The divergence of the $v$-integral in the second term
as $\bar v\to 0$ implies that it is dominated by soft gluon
exchange between the spectator quark and the quarks that form the 
kaon. We therefore treat the divergent integral $X=\int_0^1(dv/\bar v)$
as an unknown parameter (different for the penguin and hard scattering
contributions), which may in principle be complex owing 
to soft rescattering in higher orders. In our numerical analysis we
set $X=\ln(m_B/0.35\,\mbox{GeV})+r$, where $r$ is chosen randomly 
inside a circle in the complex plane of radius 3 (``realistic'') or
6 (``conservative''). 
Our results also depend on the $B$-meson parameter\cite{BBNS99} $\lambda_B$,
which we vary between 0.2 and 0.5\,GeV. Finally, there is in some 
cases a non-negligible dependence of the coefficients $a_i(\pi K)$
on the renormalization scale, which we vary between $m_b/2$ and 
$2m_b$.

\subsubsection{\bf Results}

We take $|V_{ub}/V_{cb}|=0.085$ and 
$m_s(2\,\mbox{GeV})=110\,$MeV as fixed input, noting that ultimately
the ratio $|V_{ub}/V_{cb}|$, along with the CP-violating phase 
$\gamma=\mbox{arg}(V_{ub}^*)$, might be extracted from a simultaneous fit
to the $B\to\pi K$ and $B\to\pi\pi$ decay rates.

\paragraph{\bf $SU(3)$ breaking}

Bounds\cite{FM98,NR98} on the CKM angle $\gamma$ derived from ratios of 
$\pi K$ branching fractions, as well as the determination of $\gamma$ 
using the method of Ref.~\cite{NR}, rely on an estimate of $SU(3)$ 
flavour-symmetry violations. We find that ``non-factorizable'' 
$SU(3)$-breaking effects (i.e., effects not accounted for by the 
different decay constants and form factors of pions and kaons in the 
conventional FA) do not exceed a few percent at leading power.

\paragraph{\bf Amplitude parameters}

The approach discussed here allows us to obtain the decay amplitudes 
for the $\pi\pi$ and $\pi K$ final states in terms of the form factors 
and the light-cone distribution amplitudes. The $\pi^0\pi^0$ 
final state is very poorly predicted and will not be discussed here. 
We write 
\begin{displaymath}
{\cal A}(B^0\to\pi^+\pi^-) = T\,[e^{i\gamma}+(P/T)_{\pi\pi}] 
\end{displaymath}
and parametrize the $\pi K$ amplitudes by\cite{NR98}
\begin{eqnarray}\label{para}
   {\cal A}(B^+\to\pi^+ K^0) &=& P \left( 1 - \varepsilon_a\,
    e^{i\eta} e^{i\gamma} \right), \nonumber\\
   - \sqrt2\,{\cal A}(B^+\to\pi^0 K^+) &=& P \Big[ 1
    - \varepsilon_a\,e^{i\eta} e^{i\gamma} 
   - \varepsilon_{3/2}\,e^{i\phi} (e^{i\gamma} - q\,e^{i\omega})
    \Big],\\
   - {\cal A}(B^0\to\pi^- K^+) &=& P \Big[ 1
    - \varepsilon_a\,e^{i\eta} e^{i\gamma} 
   - \varepsilon_T\,e^{i\phi_T} (e^{i\gamma} - q_C\,e^{i\omega_C})
    \Big] , \nonumber 
\end{eqnarray}
and $\sqrt2\,{\cal A}(B^0\to\pi^0 K^0) = {\cal A}(B^+\to\pi^+ K^0)
+ \sqrt2\,{\cal A}(B^+\to\pi^0 K^+) - {\cal A}(B^0\to\pi^- K^+)$. 
Table~\ref{tab1} summarizes the numerical values for the  
amplitude parameters for the conservative variation of $X$,
and variation of the other parameters as explained above.
The LO results correspond to the conventional FA at the fixed
scale $\mu=m_b$. They are strongly scale dependent. In comparison, the 
scale-dependence of the NLO result is small, with the exception 
of $q_C\,e^{i \omega_C}$. 
One must keep in mind that the ranges may overestimate the 
true uncertainty, since the parameter $X$ may ultimately 
be constrained from a subset of branching fractions. This is true 
in particular for the quantity $\varepsilon_{3/2}$ in
Table~\ref{tab1}, which can be extracted from data.\cite{NR98}

\begin{table}
\begin{center}
\begin{tabular}{|c|c|c|} 
\hline 
 \raisebox{0pt}[12pt][6pt]{} & 
\raisebox{0pt}[12pt][6pt]{Range, NLO} & 
 \raisebox{0pt}[12pt][6pt]{LO}  \\
 \hline
 \raisebox{0pt}[12pt][6pt]{$-\varepsilon_a\,e^{i\eta}$} & 
 \raisebox{0pt}[12pt][6pt]{$(0.017\mbox{--}0.020)\,e^{i\,[13,21]^\circ}$}
 & \raisebox{0pt}[12pt][6pt]{$0.02$}\\
\hline
 \raisebox{0pt}[12pt][6pt]{$\varepsilon_{3/2}\,e^{i\phi}$} & 
 \raisebox{0pt}[12pt][6pt]{$(0.20\mbox{--}0.38)\,e^{i\,[-30,7]^\circ}$}
 & \raisebox{0pt}[12pt][6pt]{$0.36$}\\
\hline
 \raisebox{0pt}[12pt][6pt]{$q\,e^{i\omega}$} & 
 \raisebox{0pt}[12pt][6pt]{$(0.53\mbox{--}0.63)\,e^{i\,[-7,3]^\circ}$}
 & \raisebox{0pt}[12pt][6pt]{$0.64$}\\
\hline
 \raisebox{0pt}[12pt][6pt]{$\varepsilon_T\,e^{i\phi_T}$} & 
 \raisebox{0pt}[12pt][6pt]{$(0.20\mbox{--}0.29)\,e^{i\,[-19,3]^\circ}$}
 & \raisebox{0pt}[12pt][6pt]{$0.33$}\\
\hline
 \raisebox{0pt}[12pt][6pt]{$q_C\,e^{i\omega_C}$} & 
 \raisebox{0pt}[12pt][6pt]{$(0.00\mbox{--}0.22)\,e^{i\,[-180,180]^\circ}$}
 & \raisebox{0pt}[12pt][6pt]{$0.06$}\\
\hline
 \raisebox{0pt}[12pt][6pt]{$(P/T)_{\pi\pi}$} & 
 \raisebox{0pt}[12pt][6pt]{$(0.19\mbox{--}0.29)\,e^{i\,[-1,23]^\circ}$}
 & \raisebox{0pt}[12pt][6pt]{$0.16$}\\
\hline
\end{tabular}
\end{center}
\caption{Parameters for the $B\to\pi K$ amplitudes as defined in 
(\protect\ref{para}), for conservative variation of all input 
parameters (see text).} 
\label{tab1}
\end{table}

\paragraph{\bf Ratios of CP-averaged rates}

\begin{figure}
\centerline{\epsfig{figure=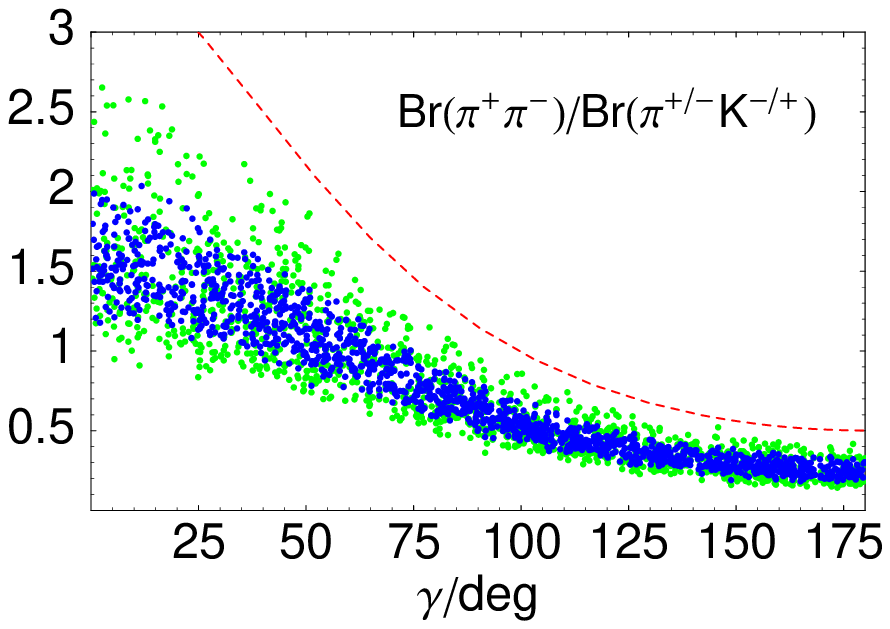,width=8.75cm,height=8.0cm}
\epsfig{figure=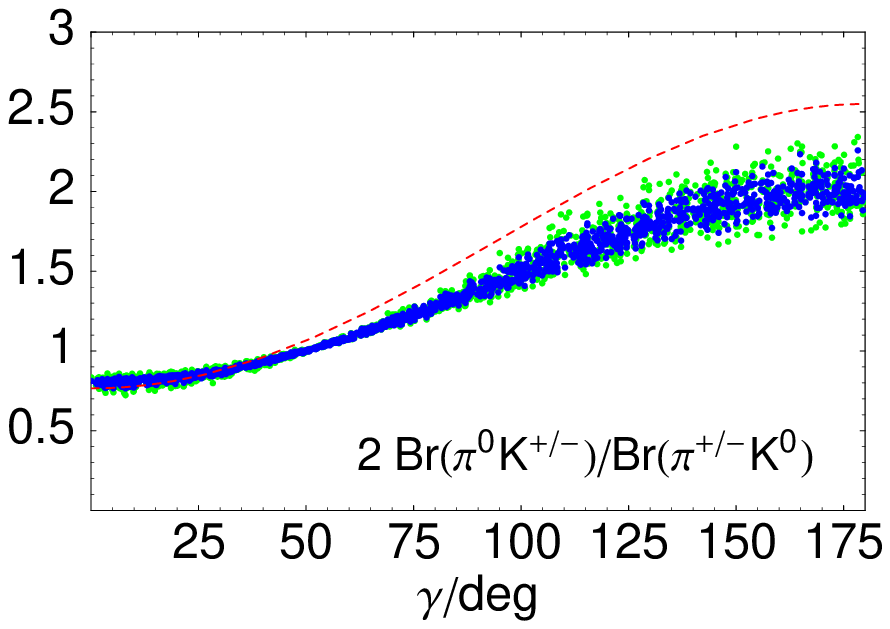,width=8.75cm,height=8.0cm}}
\vspace*{0.3cm}
\centerline{\epsfig{figure=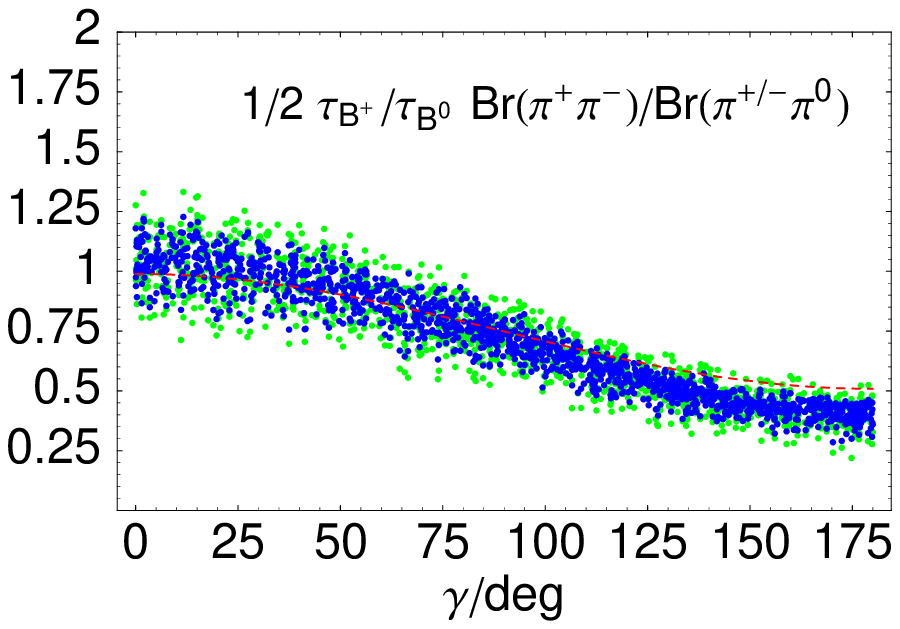,width=8.75cm,height=8.0cm}
\epsfig{figure=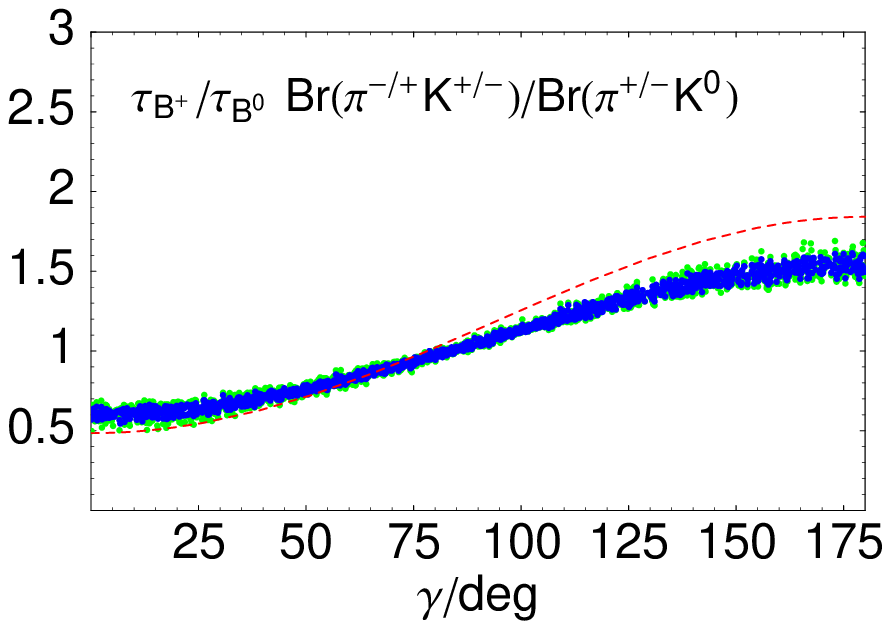,width=8.75cm,height=8.0cm}}
\vspace*{0.3cm}

\caption{Ratios of CP-averaged $B\to\pi K$ and $\pi\pi$ decay rates. 
The scattered points cover a realistic (dark) and conservative (light)
variation of input parameters. The dashed curve is the LO result, 
corresponding to conventional factorization.}
\label{fig1}
\end{figure}

Since the form factor $f_+(0)$ is not well known, we consider here
only ratios of CP-averaged branching ratios, discarding the 
$\pi^0\pi^0$ final state. We 
display these as functions of the CKM angle $\gamma$ in 
Fig.~\ref{fig1}. 


Table~\ref{tab1} shows that the corrections with respect to the 
conventional FA are significant (and 
important to reduce the re\-nor\-malization-scale dependence). Despite 
this fact, the {\em qualitative } pattern that emerges for the set of 
$\pi K$ and $\pi \pi$ decay modes is similar to that in conventional 
factorization. In particular, the penguin--tree interference is 
constructive (destructive) in $B\to\pi^+\pi^-$ ($B\to\pi^- K^+$)
decays if $\gamma<90^\circ$. Taking the currently favoured range 
$\gamma=(60\pm 20)^\circ$, we find the following robust predictions:
\begin{eqnarray}
   \frac{\mbox{Br}(\pi^+\pi^-)}{\mbox{Br}(\pi^\mp K^\pm)}
   &=& 0.5\mbox{--}1.9 \quad [0.25\pm 0.10] \nonumber\\
   \frac{\mbox{Br}(\pi^\mp K^\pm)}{2\mbox{Br}(\pi^0 K^0)}
   &=& 0.9\mbox{--}1.4 \quad [0.59\pm 0.27] \nonumber\\
   \frac{2\mbox{Br}(\pi^0 K^\pm)}{\mbox{Br}(\pi^\pm K^0)}
   &=& 0.9\mbox{--}1.3 \quad [1.27\pm 0.47] \nonumber\\
   \frac{\tau_{B^+}}{\tau_{B^0}}\,
   \frac{\mbox{Br}(\pi^\mp K^\pm)}{\mbox{Br}(\pi^\pm K^0)}
   &=& 0.6\mbox{--}1.0 \quad [1.00\pm 0.30] \nonumber
\end{eqnarray}
The first ratio is in striking disagreement with 
current CLEO data\cite{C00} (square brackets). 
The near equality of the second and third ratios is a 
result of isospin symmetry.\cite{NR98} We find
$\mbox{Br}(B\to\pi^0 K^0)=(4.5\pm 2.5)\times 10^{-6}\, 
(V_{cb}/0.039)^2 (f^{B\to\pi}_+(0)/0.3)^2$
almost 
independently of $\gamma$. This is three time smaller than the central
value reported by CLEO.

\paragraph{\bf CP asymmetry in $B\to \pi^+\pi^-$ decay}

\begin{figure}
\centerline{\epsfig{figure=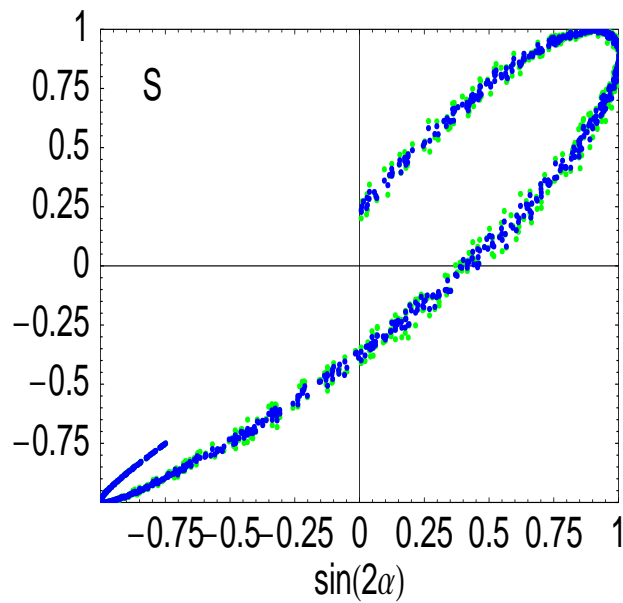,width=8.75cm,height=8.0cm}}
\caption{Mixing-induced CP asymmetry in $B\to \pi^+\pi^-$ decays. 
The lower band refers to values $45^\circ<\alpha<135^\circ$, the
upper one to $\alpha<45^\circ$ (right) or $\alpha>135^\circ$ (left). 
We assume $\alpha,\beta,\gamma\in [0,\pi]$.}
\label{fig2}
\end{figure}

The stability of the prediction for the $\pi^+\pi^-$ amplitude 
suggests that the CKM angle $\alpha$ can be extracted from 
the time-dependent mixing-induced CP asymmetry in this decay mode, 
without using isospin analysis. Fig.~\ref{fig2} displays the 
coefficient $S$ of $-\sin(\Delta M_{B_d} t)$ as a function of 
$\sin(2\alpha)$ for $\sin(2\beta)=0.75$, which may be compared with 
the result in Ref.~\cite{B99}. For some values of $S$ there
is a two-fold ambiguity (assuming all angles are between 
$0^\circ$ and $180^\circ$). A consistency check of the approach 
could be obtained, in principle, from the coefficient of the 
$\cos(\Delta m_{B_d} t)$ term.


\subsubsection{\bf Conclusions}

We have examined some of the consequences of the QCD factorization 
approach to $B$ decays into $\pi K$ and $\pi \pi$ final 
states, leaving a detailed discussion to a subsequent publication. 
Here we have focused on robust predictions for ratios of CP-averaged
decay rates.
Our result for the ratio of the $B\to\pi^+\pi^-$
and $B\to\pi^\mp K^\pm$ decay rates is in disagreement with the 
current experimental value, unless the weak phase $\gamma$ were 
significantly larger than $90^\circ$. 

\section{\bf Statistical Errors}

\subsection{\bf Feldman and Cousins for Pedestrians}

{\it R. Barlow, University of Manchester}

\rightline{\it ``What confidence is this wherein thou trustest?''}

\rightline {Isaiah Chap. 36 v. 4}

\subsubsection {\bf What is a confidence level?}

Confidence levels are the subject of much confusion and argument.
To bring out what they are, and what they are not, consider two statements which
might validly be made with 90\% confidence.
(For illustration this note uses 90\% confidence levels throughout, but of 
course any
value can be chosen.)

A grocer: ``Our potatoes weigh between
$100$ and $400 g$.'' 

A physicist:``The Higgs Boson has a mass between $100$ and $400 GeV/c ^2$.

The superficial similarity conceals a difference.
The grocer has weighed many potatoes and found
that 90\% of them have weights in the limits given.  The statement has 
a 90\% probability of being true, with probability defined 
in the conventional (frequentist) way as
$$P=\lim_{N\to\infty}{N_{In Range}^{objects} \over N_{Total}^{objects}}\eqno(1).$$

The physicist has not weighed any Higgs Bosons.  If they had, they would
discover either that all the bosons are within the range, or that all the 
bosons
are outside it.  Equation (1) will give a probability which is either 0 or 1.
What they mean by a 90\% confidence level statement is not

``A Higgs boson mass has a 90\% probability of lying between $100$ and $400 GeV/c^2$'' 

but

``The statement `The Higgs Boson Mass lies between  $100$ and $400 GeV/c^2$' has a 90\% probability of being true.'' 

If this physicist goes through life making 
statements of this type, they will be correct in  9 cases out of 10.
The `probability' of 90\% is still given as the limit of a frequency, but a rather different one. 
$$P=\lim_{N\to\infty}{N_{True}^{statements} \over N_{Total}^{statements}}\eqno(2)$$
As a technical detail, a confidence level of 90\% actually means that the
probability is
{\it at least} 90\% (this covers cases in which it is not calculable exactly.) So they are right at least 9 times out of 10.

\subsubsection{\bf Frequentist confidence levels}

The basic tool for constructing  confidence intervals is the {\it Confidence Diagram}\cite{rjb:kso}.
Suppose that a parameter of interest $a$ and an observed quantity 
$x$ are related by a pdf $P(x;a)$.
(For example: $x$ could be a number of events, and $a$ a 
branching ratio.)
 
Choose a value of the probability 
(e.g. 68\%, 90\%, 95\%) and a
strategy (e.g. one sided upper, one sided lower, two-sided symmetrical).
For any value of $a$, we can make statements about the probability of
$x$ lying in certain regions, and determine a range
$[x_-,x_+]$ within which we say (at the desired level of confidence)
that $x$ will lie.  

\begin{figure}[htbp]
\begin{center}
\epsfig{figure=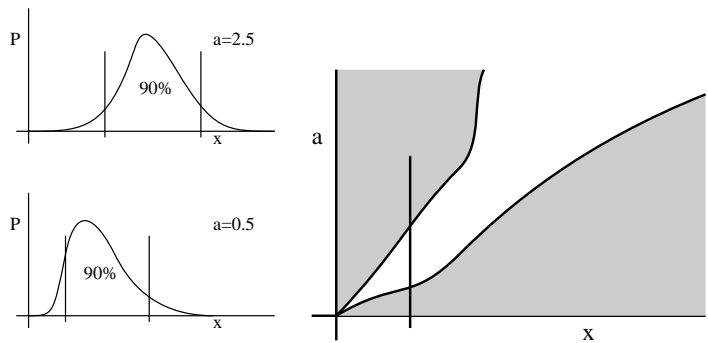,width=10cm,bbllx=0pt,bblly=170,bburx=611,bbury=470}
\end{center}
\caption{A two-sided confidence band.}
\label{fig:fig1B}
\end{figure}

We then plot $x_-$ and $x_+$ on the diagram as a function of $a$. 
Suppose we chose to quote 90\% central confidence limits. 
Any {\it horizontal} line across the plot represents a particular value of $a$ 
and thus a pdf with its limits, as shown in two particular cases. There
is a 5\% chance of a measurement lying below the lower limit, and a 5\% chance of it lying above the upper limit. 
The unshaded region between the limits is the {\it confidence band} (or {\it belt}).

The next step is a neat bit of logic.  For {\it any} value of $a$, the 
probability
that $x$ lies in the belt is, by construction, 90\%.  So if
you make an observation $x$ the probability of the pair $(x,a)$ lying
within the belt is $90\%$.  A {\it vertical} line
at the observed $x$ gives the limits for $a$ where it 
intersects the
edges of the confidence belt.


A similar technique gives one sided (upper) limits: for each value of $a$ the value
of $x$ is found for which the probability of a result this small or smaller
is only 10\% (or whatever),  and for a measured 
$x$ you read off the
upper limit $a_+$ for which the probability of such a low result is only 
10\% -- 
for a higher value of $a$ the probability is even smaller.

\begin{figure}[htbp]
\begin{center}
\epsfig{figure=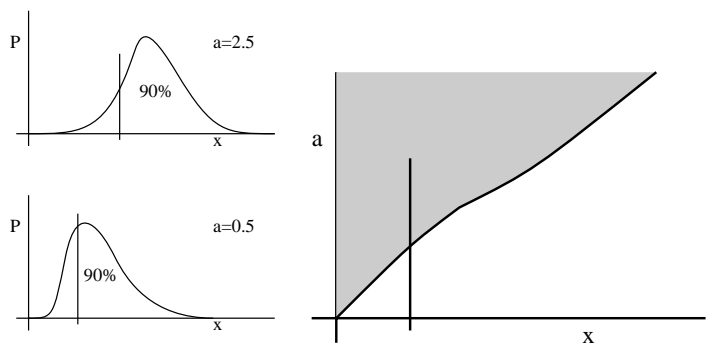,width=10cm,bbllx=0pt,bblly=170,bburx=611,bbury=470}
\end{center}
\caption{A one-sided confidence band.}
\label{fig:fig2B}
\end{figure}

\subsubsection{\bf Poisson statistics}

This theory of confidence limits applies to any distribution function, but 
in B physics its main use is in dealing with rare decays, where observed
numbers of events are governed by Poisson 
statistics.
This means that the variable $x$ is discrete (integer) -- denote it by $N$. 
Confidence diagrams have a continuous ordinate but a discrete abscissa and
the smooth curves become staircases.
Integrations are replaced by summations. For an observed number $N$
the limits are given by
$$\sum_0^N P(r;a_+)=0.05 \qquad \sum_0^{N-1} P(r;a_-)=0.95\eqno(3)$$
This means that in general
it won't be possible to find a range which satisfies $\sum P(r;a)=0.05$.
One has to use the conservative requirement $\sum P(r;a)\leq 0.05$.

$a$ is a branching ratio, lifetime, or similar quantity describing
a source of events. It trivially
translates into a number of events $S$.  The actual expected number is then
often given by $S+B$, where $B$ is some background which we will
take as being reliably and accurately estimated. The probability 
of observing $r$ events is the Poisson expression
$$e^{-(S+B)}(S+B)^r / r!\eqno(4)$$   

\subsubsection{\bf Problem: non-physical regions}

Often
this works perfectly well, but there are  
problems if the parameter 
estimates lie near some physical boundary. (For example
 Branching Ratios cannot be 
negative. Another relevant example is the interpretation of the
limit on the Higgs mass
from precision electroweak measurements when much of the 
range is already ruled out by direct searches.) 

For example, if $B$ is 5.1, and  
you observe 1 event, what can you conclude about $S$?
Following the above procedure, you calculate
$0.05=\sum_0^1 P(r;4.74)=\sum_1^\infty P(r;0.05)$ so $S+B$ lies between 
$0.05$ and $4.74$, so
you can say with 90\% confidence that $S$ lies
 between  
$-5.05$ and $-0.36$.
One would have to be an extreme purist to publish such a statement. 
It is technically correct:
you expect one in ten of your 90\% CL statements to be wrong.  In this case
you just have independent confirmation that this is one of them; you happen
to have had a large downward fluctuation in the background. There is no 
way of including the knowledge that $S$ must be non-negative.

\subsubsection{\bf Bayesian confidence levels}

Bayesian statistics\cite{rjb:bayes} handles these limits without problems using Bayes' 
theorem. 
$$P(a;x)={P(x;a) \over P(x)} P(a) $$
here $P(x;a)$ is the usual pdf.  $P(x)$ is a prior 
distribution for $x$, usually
ignored as being taken care of by normalisation. $P(a)$ is the prior
probability for $a$.   
A full discussion of Bayesian versus Frequentist methodologies can be found 
in \cite{rjb:morebayes}.

The  probabilities $P(a)$ and $P(a;x)$ do not have a frequentist definition
(except as a delta function) but can be defined as
subjective probabilities.
The form of $P(a)$ is often taken as uniform, 
under the  `principle of ignorance'; this title
is disingenuous as the assumption of a prior uniform in, say a lifetime
$\tau$ will lead to different limits than those obtained from a prior
uniform in, say, the width $\hbar / \tau$.
According to Jeffreys' prescription one should take a uniform distribution
for a {\it location} parameter, whereas for a scale parameter $\lambda$
the distribution should be proportional to ${1\over \lambda}$
(i.e. uniform in $\ln \lambda$)
\cite{rjb:dss}.  Jeffreys' criteria \cite{rjb:ch} suggests that for a Poisson source $S$
the appropriate prior should be taken as 
proportional to ${1 \over \sqrt S}$. 

With a step-function prior (zero for negative values and 
uniformly constant for positive ones), the requirement
for a one-sided confidence limit $S$ on the signal at the level $\alpha$
from $N$ observed events with expected background $B$ 
$$\alpha = \int_0^S P(S'|N)\, dS'$$
produces (after repeated integration by parts) the requirement
$$1-\alpha = {\sum_0^N P(r;B+S) \over \sum_0^N P(r;B)}\eqno(5)$$
which is found in early versions of the PDG \cite{rjb:pdg4}.
For large $N$, the denominator approaches the standard value of 1.
For small $N$ it expresses the 
fact that we have manifestly got a downward fluctuation of the background,
and one should temper the requirement on the signal accordingly.

\subsubsection{\bf What Feldman and Cousins do}

Feldman and Cousins \cite{rjb:fc} get us out of this dilemma by pointing out another one.

In real life, physicists (being human) do not decide beforehand
whether they will quote a one-sided or a two-sided limit.
If the observations are high, a two-tailed limit will be appropriate; 
if they are low then a single upper limit is the best information that 
can be given.

For example: given a background of 1.1, if you observed 10 events you would 
very possibly quote as a result the 
range of 4.3 to 15.1 (at 90\% confidence).  If you observed 2 events you 
would probably say (with
90\% confidence) that
any signal was less than 4.2. There is some intermediate value at which
you would `flip-flop' between the two states.

\begin{figure}[htbp]
\begin{center}
\epsfig{figure=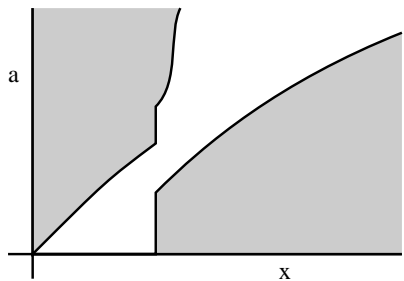,width=10cm,bbllx=0pt,bblly=170,bburx=611,bbury=430}
\end{center}
\caption{The effect of 'flip-flop'.}
\label{fig:fig3B}
\end{figure}

This may be sensible, but it destroys the confidence belt
logic.  For small $x$ values we have the one sided belt of Figure~\ref{fig:fig2B};
for large values the two-sided belt of Figure~\ref{fig:fig1B}.
The resulting hybrid (Figure~\ref{fig:fig3B}) violates the basis on which the
curves were constructed, as for the intermediate
values of $a$ the probability content of the belt is not the specified 90\%.

Feldman and Cousins thus argue that the strategy for deciding whether to quote 
a one-sided or two-sided result must be decided beforehand, and decided in
such a way that for all values of $a$ the probability content of the belt is
90\%. They propose a `unified approach' in which a single algorithm
decides whether to quote one-sided or two-sided limits, and gives the
value(s) in question.

There are an infinite number of solutions for limits $x_1$ and $x_2$
satisfying
$$\int_{x_1}^{x_2} P(x;a)\, dx = 0.90$$
(For discrete processes $\sum_{r_1}^{r_2} P(r;a) \geq 0.90$)

One initially plausible algorithm is to include the region where the 
pdf itself is greatest,
adjusting $C$ to satisfy
$$\int_{P(x;a)>C} P(x;a) \, dx = 0.90 \eqno(6)$$
This in fact gives the most compact region.

But consider an outcome with
1 events seen, with an expected signal and background of 0 and 15.1 
respectively.
This is very unlikely ($4 \, 10^{-6}$ probability) and this 
algorithm is bound to
exclude the possibility.  But given one observed event and a background of 15.1, 
any probabilities are going to be pretty small, and we would want to include
0 signal in the band, on the grounds that for any other values the probability
is even lower!
It would be fairer to include points by comparing their probability
to, say, the best that can be achieved.
$$\int_{P(x;a)/P(x;a_{best})>C} P(x;a) = 0.90\eqno(7)$$ 
 
For Poisson processes with a known background $B$ and unknown signal $S$,
$S_{best}$ is $x-B$, if the observed number is bigger than the expected 
background, and $0$ if it is smaller.

A Feldman-Cousins confidence interval for rare decays
 is thus constructed as  previously, with
a 90\% (or whatever) confidence band being constructed for each value of the
source strength $S$. For each $S$ the Poisson probability
is calculated for each number of observed events $r=0,1,2\dots$, 
and its ratio taken to the `best' probability (with
$S=r-B$ or $S=0$ as appropriate). $r$ values are included in the confidence
belt in order of their ranking in this ratio until the total probability
condition is fulfilled. The result is a diagram where
low values of $N$ will just give one limit whereas
higher ones give a range, and  the probability content of the belt is valid.

There is a loss of universality as such confidence diagrams, unlike those of
 the
previous scheme, are non-trivially different for different background values.

\subsubsection{\bf Some examples}

90\% confidence intervals are shown for various values for an
observed signal and expected background. Both 1-tailed and 2-tailed limits
are given for conventional frequentist confidence intervals, and for Bayesian
intervals, and Bayesian intervals are shown both for a step-function prior
(uniform for positive values) and a prior according to Jeffreys'
prescription, proportional to $1/\sqrt{s}$. (Taking a prior proportional to 
$1/s$ gives divergent integrals.)

\font\eightrm=cmr8
\font\eightmm=cmmi8
\begin{table}
{\let\rm=\eightrm
\textfont0=\eightrm
\textfont1=\eightmm
\halign{\strut \hfil $#$ \ &\hfil $#$\ \vrule &\ \hfil $#$ \hfil \quad &\hfil $ #$ \hfil \vrule 
&\hfil $#$\hfil &\quad \hfil $#$\hfil \vrule &\hfil $#$\hfil 
&\quad \hfil $#$\hfil\vrule &\ \hfil $#$\hfil\cr
N&B&\hbox{\rm Conventional}\span\omit 
&\hbox{\rm Uniform Bayesian}\span\omit
&\hbox{\rm Jeffreys' Bayesian}\span\omit 
&\hbox{\rm Feldman-}\ \cr
& & \hbox{\rm 1 Tailed} & \hbox{\rm 2 tailed} &\hbox{\rm 1 Tailed} &\hbox{\rm 2 Tailed} & \hbox{\rm 1 Tailed} & \hbox{\rm 2 Tailed} & \hbox{\rm\ Cousins}\cr
\noalign{\hrule}
10 & 0.0 & < 15.4 & 5.43 \dots 17.0 & < 15.4 & 6.17\dots 17.0 & < 14.8 & 5.80\dots 16.3 & 5.50\dots 16.5 \cr 
& 1.0 & < 14.4 & 4.43 \dots 16.0 & < 14.4 & 5.17\dots 16.0 & < 13.7 & 4.73\dots 15.3 & 4.50\dots 15.5 \cr 
& 5.0 & < 10.4 & 0.425 \dots 12.0 & < 10.4 & 1.43\dots 12.0 & < 9.24 & 0.345\dots 10.8 & 1.19\dots 11.5 \cr 
& 10. & < 5.41 & -4.57 \dots 6.96 & < 6.63 & 0.233\dots 8.08 & < 4.73 & 0.0106\dots 6.15 & < 6.50 \cr 
& 15. & < 0.407 & -9.57 \dots 1.96 & < 4.83 & 0.124\dots 6.10 & < 3.07 & 0.00498\dots 4.21 & < 2.65 \cr 
5 & 0.0 & < 9.27 & 1.97 \dots 10.5 & < 9.27 & 2.61\dots 10.5 & < 8.64 & 2.29\dots 9.84 & 1.84\dots 9.99 \cr 
& 0.50 & < 8.77 & 1.47 \dots 10.0 & < 8.77 & 2.11\dots 10.0 & < 8.07 & 1.71\dots 9.27 & 1.53\dots 9.49 \cr 
& 2.5 & < 6.77 & -0.530 \dots 8.01 & < 6.85 & 0.558\dots 8.09 & < 5.66 & 0.0580\dots 6.87 & < 7.49 \cr 
& 4.0 & < 5.27 & -2.03 \dots 6.51 & < 5.72 & 0.244\dots 6.92 & < 4.25 & 0.0127\dots 5.42 & < 5.99 \cr 
& 10. & < -0.725 & -8.03 \dots 0.513 & < 3.71 & 0.0907\dots 4.73 & < 2.30 & 0.00356\dots 3.19 & < 1.83 \cr 
1 & 0.0 & < 3.89 & 0.0513 \dots 4.74 & < 3.89 & 0.355\dots 4.74 & < 3.13 & 0.176\dots 3.91 & 0.104\dots 4.36 \cr 
& 0.50 & < 3.39 & -0.449 \dots 4.24 & < 3.51 & 0.141\dots 4.36 & < 2.48 & 0.00782\dots 3.25 & < 3.86 \cr 
& 1.0 & < 2.89 & -0.949 \dots 3.74 & < 3.27 & 0.100\dots 4.11 & < 2.18 & 0.00442\dots 2.93 & < 3.36 \cr 
& 5.0 & < -1.11 & -4.95 \dots -0.256 & < 2.67 & 0.0615\dots 3.45 & < 1.60 & 0.00237\dots 2.25 & < 1.20 \cr 
& 10. & < -6.11 & -9.95 \dots -5.26 & < 2.51 & 0.0564\dots 3.25 & < 1.48 & 0.00217\dots 2.10 & < 0.672 \cr 
0 & 0.0 & < 2.30 & None  & < 2.30 & 0.0513\dots 3.00 & < 1.35 & 0.00196\dots 1.92 & < 2.43 \cr 
& 0.10 & < 2.20 & None & < 2.30 & 0.0513\dots 3.00 & < 1.35 & 0.00196\dots 1.92 & < 2.33 \cr 
& 1.0 & < 1.30 & None & < 2.30 & 0.0513\dots 3.00 & < 1.35 & 0.00196\dots 1.92 & < 1.61 \cr 
& 5.0 & < -2.70 & None & < 2.30 & 0.0513\dots 3.00 & < 1.35 & 0.00196\dots 1.92 & < 0.770 \cr 
9 & 3.7 & < 10.5 & 0.995 \dots 12.0 & < 10.5 & 1.82\dots 12.0 & < 9.49 & 0.775\dots 11.0 & 1.49\dots 11.6 \cr 
9 & 3.7 & < 12.0 & 0.415 \dots 13.4 & < 12.0 & 1.24\dots 13.4 & < 11.0 & 0.306\dots 12.3 & 0.994\dots 13.1 \cr 
9 & 3.7 & < 7.49 & 2.37 \dots 9.39 & < 7.51 & 3.25\dots 9.40 & < 6.51 & 2.22\dots 8.39 & 2.64\dots 9.09 \cr 
}
}
\caption{Some examples of 90\% confidence limits for various values}
\end{table}

It can be seen that where the signal is safely larger than the background, all
methods are in broad agreement.  Differences are a salutary reminder of the 
inevitable arbitrariness of any analysis.

  As signal sinks below background,
the conventional limits become manifestly erroneous.  The Bayesian limits
appear more sensible, though the differences between the two versions 
become larger.  The Feldman-Cousins value(s) flip nicely.

\subsubsection{\bf Some objections}

The new technique has been warmly welcomed by some, but has been 
criticised by others.

In the 1-sided regime the limit
decreases with the value of the expected background, as can be seen from the
table.  This  is
apparently counterintuitive: for 2 experiments with the same number of 
events seen the one with worse (i.e. higher) background has the 
better (i.e. lower) limit! 
The counterargument is that the inferior experiment has to be very lucky
to see so few events.  On average, an experiment with a higher background will
see more events and thus give a higher limit: some experiments will always
be more fortunate than others, and by increasing $B$ for fixed $N$ you are
increasing the experimental luck. 

The method does not cope cleanly with the case of zero observed events. This
is always going to be a special case because, uniquely, if you observe 
0 events you know the number of signal events (zero), and all the background
information is irrelevant.  Modifications of the method have been suggested
to get over this problem \cite{rjb:rw}.

The method can flip to the two sided form in cases the physicist knows are
inappropriate. When counting decays forbidden by the standard model, or
even by basic conservation laws, the algorithm will give a two sided limit
in cases (like 10 events on a background of 5)  which are obviously just
upward fluctuations of background. Here the experimenter is probably 
overstretching the applicability of the method. For more plausible decay
channels  the same experimenter would probably be happy to report a
limit from the same data (10 on 5). The method, by its nature, 
knows only about the numbers
and not about the physics. 

Broadly speaking (and with some exceptions), frequentists accept the 
Feldman Cousins method as an 
improvement on the original form, though it does still have some drawbacks.
Bayesians dislike it, perhaps because it expands the scope of frequentist
analysis in an area where previously Bayesian technique had a monopoly. 

\subsubsection{\bf Some questions}

Can this method be adapted to the setting of limits on the ratio
of two numbers, as was done by James and Roos \cite{rjb:jr} for the conventional
frequentist confidence level? 

When one is tuning cuts in an analysis, one aims to optimise
signal/background or signal/(signal+background), depending on 
whether you are questioning the presence of a signal, or
measuring the size of a signal you know to be there. Could a similar
unified approach be found here?

The unified approach prevents the experimenter flip-flopping between 
limit strategies, but still leaves the choice of confidence level
free.  An experimenter who normally used 90\% limits may choose to
present 68\% limits if they obtain a low result where they
believe there should be a signal, or 95\% or 99\% limits if they 
obtain a high result where they believe no signal should be.
Is there some natural way of incorporating this (undesirable?) freedom in
an algorithm?


\def\bvec#1{\ifmmode
\mathchoice{\mbox{\boldmath$\displaystyle\bf#1$}}
{\mbox{\boldmath$\textstyle\bf#1$}}
{\mbox{\boldmath$\scriptstyle\bf#1$}}
{\mbox{\boldmath$\scriptscriptstyle\bf#1$}}\else
{\mbox{\boldmath$\bf#1$}}\fi}

\subsection{\bf Statistical issues in heavy flavour physics}

{\it Glen Cowan, Royal Holloway, University of London}

\subsubsection{\bf Introduction}

In this paper two statistical issues are addressed that come up
in many physics analyses but which are particularly relevant
for heavy flavour physics.  The first, presented in Section~\ref{sec:brave},
is the problem of combining measurements of the branching ratios of 
rare decays where some of the measurements are based on only a few or perhaps 
no candidate events.  The solution recommended is a straightforward 
application of the method of maximum likelihood.  The next question,
discussed in Section~\ref{sec:theerr}, is much more open-ended, namely,
how to quantify theoretical uncertainties.  Both issues
are of course not unique to heavy flavour physics, and the approaches
recommended can be applied in many other contexts.

\subsubsection{\bf Combining branching ratios for rare decays}
\label{sec:brave}

Consider first a single experiment in which one searches for
rare decay and finds $n$ candidate events out of $N$ total decays.  
Suppose the efficiency for the signal is $\varepsilon$ and there are 
$b$ expected background events.  In general, $n$ should be treated 
as a binomially distributed variable where $N$ represents the total 
number of trials and the binomial probability ${\cal B}$ is the branching 
ratio.  If ${\cal B}$ is very small and $N$ very large, however, we 
can treat $n$ as following a Poisson distribution with expectation value 
$\nu$, i.e.,

\begin{equation}
\label{eq:pois}
P(n; \nu) = \frac{\nu^n}{n!} e^{-\nu} \;.
\end{equation}

\noindent  Here the number of events observed $n$ is the sum
of $n_s$ signal and $n_b$ background events and the expectation
value of $n$, here called $\nu$,  is also the sum of the signal
and background expectation values $s$ and $b$:

\begin{eqnarray}
\label{eq:nsb}
n & = & n_s + n_b \;, \\*[0.2 cm]
\nu & = & s + b \;.
\end{eqnarray}

\noindent The branching ratio ${\cal B}$, which is the parameter we
want to estimate, is related to the expected number of signal events $s$, 
the efficiency $\varepsilon$ and the total number of events $N$ by

\begin{equation}
\label{eq:brdef}
{\cal B} = \frac{ s }{\varepsilon N} \;.
\end{equation}

For a single experiment, the likelihood function $L({\cal B})$ is
simply the Poisson probability (\ref{eq:pois}) evaluated with
the observed number of events $n$ and expressed as a function of
the branching ratio ${\cal B}$.  Maximizing $L$ gives the
maximum likelihood (ML) estimator $\hat{{\cal B}}$,

\begin{equation}
\label{eq:mlest}
\hat{{\cal B}} = \frac{n - b}{\varepsilon N} \;,
\end{equation}

\noindent where here estimators will be denoted
by hats.  The statistical uncertainty in $\hat{{\cal B}}$ can be
quantified by giving an estimate of its standard deviation,

\begin{equation}
\label{eq:sigmaest}
\hat{\sigma}_{\hat{{\cal B}}} = \frac{\sqrt{n}}{\varepsilon N} \;,
\end{equation}

\noindent or by reporting a confidence interval
(see e.g.\ \cite{barlow}).

Now consider the case where we have $m$ independent measurements
of the branching ratio ${\cal B}$.  The $m$ experiments report
$n_1, \ldots, n_m$ candidate events out of $N_1, \ldots, N_m$
total decays; they have signal efficiencies $\varepsilon_1,
\ldots, \varepsilon_m$ and expected numbers of signal and background events
$s_1, \ldots, s_m$ and $b_1, \ldots, b_m$.  Assuming the measurements 
are statistically independent, the likelihood function is the 
product of the Poisson probabilities for each,

\begin{eqnarray}
\label{eq:totalL}
L({\cal B}) & = & \prod_{i=1}^m \frac{ \nu_i^{-n_i } }{n_i!} e^{-\nu_i} 
\nonumber \\*[0.2 cm]
& = & \prod_{i=1}^m \frac{ (\varepsilon_i N_i {\cal B} + b_i)^{n_i} }
{n_i!} e^{-(\varepsilon_i N_i {\cal B} + b_i)} \;.
\end{eqnarray}

\noindent The value of ${\cal B}$ that maximizes (\ref{eq:totalL})
is the ML estimator $\hat{{\cal B}}$, which in practice will need to be
determined numerically, e.g., by solving

\begin{equation}
\label{eq:deriv}
\frac{\partial \ln L}{\partial {\cal B}} = 
\sum_{i=1}^m \left[ \frac{ n_i \varepsilon_i N_i }
{\varepsilon_i N_i {\cal B} + b_i } - \varepsilon_i N_i \right] = 0 \;.
\end{equation}

As in the case of a single measurement, the
statistical uncertainty in $\hat{{\cal B}}$ can be quantified by
estimating its standard deviation using, for example, the curvature
of the log-likelihood function at its maximum,

\begin{equation}
\label{eq:sigmaest2}
\hat{\sigma}_{\hat{{\cal B}}} = \left( \left. -
\frac{\partial^2 \ln L}{\partial {\cal B}^2} \right|_{{\cal B} = 
\hat{{\cal B}}}
\right)^{-1} \;,
\end{equation}

\noindent or by constructing a confidence interval.

As the final result must be determined numerically it is difficult
to see what the properties of the ML estimator are.  If, however,
we consider the case where there is no background, i.e., $b_i = 0$
for all $i$, then the estimator $\hat{{\cal B}}$ takes on the simple
form

\begin{equation}
\label{eq:bhat_nob}
\hat{{\cal B}} = \frac{ \sum_{i=1}^m n_i }
{ \sum_{i=1}^m \varepsilon_i N_i } \;,
\end{equation}

\noindent for which the standard deviation can be estimated by

\begin{equation}
\label{eq:sigma_bhat_nob}
\hat{\sigma}_{\hat{{\cal B}}} =  \frac{ \sqrt{ \sum_{i=1}^m n_i } }
{ \sum_{i=1}^m \varepsilon_i N_i } \;.
\end{equation}

\noindent For this case, it is easy to see that there is no problem
if some of the $n_i$ are small or even zero.  Such problems would
arise, for example, in least squares averaging of individual estimates
of ${\cal B}$.  Furthermore, in the real case with nonzero backgrounds,
the ML estimate is not spoiled by including experiments that carry
less information than others because, for example, they have 
larger backgrounds or are based on smaller data samples.  

It should not be controversial that the ML estimator has many
advantages and should be used unless there is a good reason not to.
The important point to note here is that it is not sufficient for
experimenters to report only an estimate of the branching ratio
and a standard deviation.  In order to combine the measurements
using the method of maximum likelihood, one requires the number of 
observed events, the efficiency, the expected number of background 
events and the total number of events for all experiments.

\subsubsection{\bf Quantifying theoretical uncertainties}
\label{sec:theerr}

In many experimental analyses, the dominant uncertainty in the 
determination of a parameter does not stem from measurement errors
but from the approximations used in deriving the theoretical
prediction for the quantities observed.  An important example
from heavy flavour physics is the theoretical prediction for the
differential decay rates of B mesons into various final states.
Although the Standard Model makes a well-defined prediction for
the functional form of the probability distribution for measured
decay times, the exact form of this prediction cannot be derived
because of calculational difficulties, especially for multi-body
final states.  Instead, approximations
are used, based, for example, on a perturbation series calculated
only to some finite order.  As a result, the estimated parameter
values are in general uncertain even in the absence of
measurement errors, statistical or otherwise.  This is what
is meant here by the `theoretical uncertainty' in an estimated
parameter.

Here we will argue that the Bayesian approach to statistics provides
a potentially useful framework for quantifying theoretical uncertainties.
To keep the discussion as general as possible, suppose we measure
a vector of data $\bvec{x}$, whose probability distribution function
(pdf) is predicted as a function of a vector of parameters
$\bvec{\theta}$.  For example, the vector $\bvec{x}$ could represent
a set of measured decay times for B meson decays into 
$\mbox{J}/\psi\mbox{K}^0_{\rm S}$ and the vector $\bvec{\theta}$ would
then be related to CKM matrix elements as well as other Standard Model 
parameters that may enter.

Now suppose that we do not know the exact form of the pdf  
$f(\bvec{x}; \bvec{\theta})$ but the function that we have to work with
is only an approximation.  This approximation can be viewed as
belonging to a family of functions, one of which is the true
prediction of the theory, or at least close to it.  Suppose we can
parametrize this family by some vector of {\it nuisance
parameters} $\bvec{\mu}$.   The elements of $\bvec{\mu}$ could
represent, for example, higher order coefficients in a perturbation
series that are not included in the prediction actually available,
but which could be calculated in principle.  It is necessary in
the procedure presented here for one of the members of the family 
of functions to be sufficiently close (in principle, equal) to the true 
prediction.  The family can always be made arbitrarily large by including 
more nuisance parameters, but this will have the end result of inflating 
the uncertainties in the parameters of interest.

In Bayesian statistics, all knowledge about the parameters
of a theory is summarized by giving the {\it posterior} pdf,
$p(\bvec{\theta}, \bvec{\mu} | \bvec{x})$.  This is the conditional
pdf for the parameters given the data $\bvec{x}$.  
In our case, the parameters consist of those in which we
are interested, $\bvec{\theta}$, as well as the nuisance parameters
$\bvec{\mu}$.  According to Bayes' theorem, they are related to
the likelihood function $L(\bvec{x} | \bvec{\theta}, \bvec{\mu})$
and the {\it prior} pdf $\pi(\bvec{\theta}, \bvec{\mu})$ by

\begin{equation}
\label{eq:bayesthm}
p(\bvec{\theta}, \bvec{\mu} | \bvec{x}) \propto
L (\bvec{x} | \bvec{\theta}, \bvec{\mu}) 
\pi(\bvec{\theta}, \bvec{\mu}) \;.
\end{equation}

\noindent The likelihood function 
$L(\bvec{x} | \bvec{\theta}, \bvec{\mu})$
is the conditional probability for the data given the parameter
values, and the prior pdf $\pi(\bvec{\theta}, \bvec{\mu})$ summarizes
our knowledge of the parameters before carrying out the experiment.

It should be emphasized here that the probability associated with
the value of a parameter cannot be interpreted meaningfully as
a frequency of an outcome of a repeatable experiment.  Instead
it is understood to reflect the degree of belief that the parameters 
have particular values, an interpretation called {\it subjective}
probability.  In order to determine this degree of
belief given the data, we need to say what it was before carrying
out the experiment, i.e., we need the prior pdf 
$\pi(\bvec{\theta}, \bvec{\mu})$.  Bayesian statistics provides no 
fundamental prescription for determining the prior pdf.  Given
$\pi(\bvec{\theta}, \bvec{\mu})$, however, Bayes' theorem specifies how 
our knowledge about the parameters should change in the light of
the data.  

The usefulness of the Bayesian approach in the present
problem will depend on our being able to parametrize
the pdf $f(\bvec{x}; \bvec{\theta}, \bvec{\mu})$ in terms of
the nuisance parameters $\bvec{\mu}$, and to make a reasonable
statement about their probabilities.  Often this means interrogating
the theorist to the point where he or she tells you how large, say,
the next order coefficient in a perturbation series could be.
At first, the theorist may say something like `Until I've calculated
it, all values are equally likely'.  Under pressure, however, it will
be admitted that if it were to differ by many orders of magnitude
from the previous term, or from one's naive expectation of the value
based on past experience with perturbation theory or 
Pad\'e approximants or some other procedure, then this would be 
surprising.

The key now is to quantify this level of surprise, by {\it calibrating}
it against something for which probability can be defined in terms
of a frequency.  The theorist should state
the range for the nuisance parameter such that the degree of surprise
at finding out that it is outside this range is equal to,
say, the level of surprise that one would experience at drawing
a white ball out of an urn of balls, 32\% of which are white.
In most cases this level of surprise will be a smoothly varying
function of the parameter, which could be represented by some sort
of a bell-shaped curve like a Gaussian.  It would be rare, for example, 
that a uniform distribution between fixed limits would provide a reasonable
model for quantifying the uncertainty in such a parameter.  Assuming a 
Gaussian model, the central interval containing 68\% of the probability
corresponds to the one-sigma error.  In Bayesian parlance these are often
called 68\% {\it credible} intervals.  More information on subjective
probability and Bayesian statistics can be found in
\cite{ohagan}; its application to problems of data analysis in 
physics is treated at various levels of detail in \cite{eadie,
dagostini, sivia, cowan}.

It is clear that this calibration of a degree of belief 
will never be perfect, and we should
be happy if it is reasonable to within a factor of two or even
a factor of ten.  In principle the level of calibration could be
measured by forcing theorists to guess the 68\% credible intervals,
waiting long enough for the calculations to be made and then determining
the frequency with which they are right.  An alternative proposed at
this workshop would be to force them to bet (large) sums of money
on the values of nuisance parameters that at some point will be 
calculated.  The Bayesian framework will be useful if reasonable 
prior probabilities can be assigned; there are no doubt 
cases in which this is not possible.

The importance of making reasonable statements about prior probabilities
holds as well for the model parameters $\bvec{\theta}$, not only the 
nuisance parameters ${\bvec{\mu}}$.  Here the posterior will be insensitive
to the prior pdf for $\bvec{\theta}$ as long as it is a more slowly
varying function of $\bvec{\theta}$ than is the likelihood function.
If this is not the case, then the Bayesian approach is less likely
to yield useful results, since the posterior pdf will be determined
more by the prior beliefs than by the data.

Providing one can quantify the necessary prior probabilities, 
it is a straightforward exercise to apply Bayes' theorem to obtain
the posterior pdf $p(\bvec{\theta}, \bvec{\mu} | \bvec{x})$.  We are
then in a position to apply one of the most powerful tools of
Bayesian statistics:  we eliminate the nuisance parameters
$\bvec{\mu}$ from the problem simply by computing the {\it marginal}
pdf $p(\bvec{\theta} | \bvec{x})$, i.e.,

\begin{equation}
\label{eq:marginal}
p(\bvec{\theta} | \bvec{x}) = 
\int p(\bvec{\theta}, \bvec{\mu} | \bvec{x}) \,
d \bvec{\mu} \;.
\end{equation}

One can then summarize the posterior pdf by, for example, presenting contours
of constant probability in two-dimensional subspaces or by characterizing
it by various sorts of location parameters such as the mode.
Regions in the parameter space can be determined which contain specified
probabilities, e.g., 68\%, 90\%, etc.

The sizes of the contours of constant probability reflect both 
the statistical and systematic errors.  These two components of
the uncertainty can be separated by constructing estimators
$\hat{\bvec{\theta}}$, e.g., by taking the mode of the posterior pdf.
These will be specific functions of the data,
and as such they are random variables in the sense of frequentist
probability.  One can then ask, for example, how $\hat{\bvec{\theta}}$
would be distributed upon repetition of the experiment, and in this
way the statistical and systematic components can be separated, if 
desired.

Although up to now we have only taken $\bvec{\mu}$ to quantify errors
due to approximations made in deriving theoretical predictions, it
can just as easily represent other types of systematic uncertainties.
In either case, the assumptions made in quantifying the uncertainties
should clearly be reported along with the result.  It should be
emphasized that the posterior pdf by itself does not in general provide 
a useful summary of the result of the experiment,
as it combines both the data and the subjective beliefs needed
to construct the prior pdf.  If only the posterior pdf is reported,
it is impossible for a consumer of the result to separate the influence
of the experimenter's prior beliefs from the influence of the data.

The experimenter should therefore separate
as well as possible a summary of the observations from their interpretation.
In a paper, for example, this may be done in sections titled `results'
and `discussion' or `interpretation'.  In the section on results one
can report the values of observables that to
the greatest reasonable degree are free of systematic effects, theoretical
or otherwise.  For example, one should report the various numbers of events
observed that satisfy specific criteria, and not only the estimates
of parameters that are derived from these data.  In the interpretation
of the results, however, there is no reason for experimenters to shy 
away from making reasonable (stated) assumptions based on subjective 
probability and showing what they imply for the values of parameters 
of interest.

\subsubsection{\bf Conclusions}

A straightforward technique based on the method of maximum likelihood
has been presented for combining measurements of branching
ratios.  In order to employ this method, experimenters must report
the number of candidate events seen, the efficiency of the
selection procedure, the expected background
and total number of decays observed.  It cannot be used in general if
only an estimate of the branching ratio and its standard deviation
are reported.

The use of Bayesian statistics in quantifying systematic errors,
in particular those due to uncertainties in the functional
form of theoretical predictions, has been discussed.  In this framework
one must quantify the uncertainty of unknown constants using subjective
probability, i.e., probability based on a degree of belief.  The method
is useful to the extent that the uncertainty can be parametrized and
quantified, e.g., by interrogation of the theorist responsible
for the prediction.  Owing to difficulties in calibrating
one's degree of belief relative to frequentist probabilities, this
is necessarily an approximate procedure and all assumptions made in
quantifying the uncertainty should be reported along with the result.

%

\section{\bf Acknowledgements}

The authors thank the organizers for a wonderful conference and many
colleagues for discussions and help.

\end{document}